\documentclass[a4paper,fleqn,usenatbib,useAMS]{mnras}
\pdfoutput=1
\usepackage{newtxtext,newtxmath}
\usepackage[T1]{fontenc}
\usepackage{ae,aecompl}

\usepackage{graphicx}             %for PS/EPS graphics inclusion, new
\usepackage{amsmath,amssymb}
\usepackage{multirow}
\newcommand{\kms}{\,km\,s$^{-1}$}	% kilometres per second\newcommand\Ht{H$_2$}
\newcommand{\hcop}{HCO$^+$}
\newcommand{\lir}{$L_\text{IR}$}
\newcommand{\alphaco}{$\alpha_\text{CO}$}
\newcommand{\alphadense}{$\alpha_\text{dense}$}
\newcommand{\alphadenseG}{$\alpha_\text{dense}(G_0)$}
\newcommand{\sfedense}{SFE$_\text{dense}$}
\newcommand{\sfemol}{SFE$_\text{mol}$}
\newcommand{\fdense}{$f_\text{dense}$}
\newcommand{\mdense}{$M_\text{dense}$}
\newcommand{\mhcn}{$M_\text{HCN}$}
\newcommand{\mhcop}{$M_{\text{HCO}^+}$}
\newcommand{\lhcn}{$L'_\text{HCN}$}
\newcommand{\lhcop}{$L'_{\text{HCO}^+}$}
\newcommand{\lco}{$L'_\text{CO}$}
\newcommand{\ldense}{$L'_\text{dense}$}
\newcommand{\sigmasfr}{$\Sigma_\text{SFR}$}
\newcommand{\sigmastellar}{$\Sigma_\text{stellar}$}

\newcommand{\ncrit}{$n_\text{crit}$}
\newcommand{\rco}{$R_\text{31}$}
\newcommand{\rhcn}{$R_\text{HCN43}$}
\newcommand{\rhcop}{$R_\text{\hcop43}$}

\title[MALATANG: Dense Gas and Star-formation in NGC\,253]{The MALATANG Survey: Dense Gas and Star Formation from High Transition HCN and HCO$^+$ maps of NGC\,253}

\author[Xue-Jian Jiang et al.]{
Xue-Jian Jiang,$^{1}$\thanks{E-mail: 
\href{xjjiang@pmo.ac.cn}{xjjiang@pmo.ac.cn}; \href{x.jiang@eaobservatory.org}{x.jiang@eaobservatory.org}}
Thomas R. Greve,$^{2,3}$~
Yu Gao,$^{1,4}$~
Zhi-Yu Zhang,$^{5}$~
Qinghua Tan,$^{1}$
\newauthor
Richard de Grijs,$^{6,7,8}$~
Luis C. Ho,$^{9,10}$~
Micha{\l} J. Micha{\l}owski,$^{11}$~
Malcolm J. Currie,$^{12}$
\newauthor
Christine D. Wilson,$^{13}$~
Elias Brinks,$^{14}$~
Yiping Ao,$^{1}$~
Yinghe Zhao,$^{15,16}$~
Jinhua He,$^{15,17,18}$~ 
\newauthor
Nanase Harada,$^{19}$~
Chentao Yang,$^{20}$~
Qian Jiao,$^{1}$~
Aeree Chung,$^{21}$~
Bumhyun Lee,$^{9,21}$
\newauthor
Matthew W. L. Smith,$^{22}$~
Daizhong Liu,$^{23}$~
Satoki Matsushita,$^{19}$~
Yong Shi,$^{24}$
\newauthor
Masatoshi Imanishi,$^{25}$~
Mark G. Rawlings,$^{26}$~
Ming Zhu,$^{27}$~
David Eden,$^{28}$
\newauthor
Timothy A. Davis,$^{22}$~
and Xiaohu Li$^{29}$
\\
$^{1}$
Purple Mountain Observatory \& Key Laboratory for Radio Astronomy, Chinese Academy of Sciences, 10 Yuanhua Road, \\
~~~~Nanjing 210033, People's Republic of China\\
$^{2}$
Department of Physics and Astronomy, University College London, Gower Street, London WC1E 6BT, UK\\
$^{3}$
Cosmic Dawn Center (DAWN)\\ 
$^{4}$
Department of Astronomy, Xiamen University, Xiamen, Fujian 361005, China\\
$^{5}$
European Southern Observatory, Karl-Schwarzschild-Str 2, D-85748 Garching, Germany\\
$^{6}$
Department of Physics \& Astronomy, Macquarie University, Balaclava Road, Sydney NSW 2109, Australia\\
$^{7}$
Centre for Astronomy, Astrophysics and Astrophotonics, Macquarie University, Balaclava Road, Sydney NSW 2109, Australia\\
$^{8}$
International Space Science Institute--Beijing, Nanertiao, Zhongguancun, Hai Dian District, Beijing 100190, People's Republic of China\\
$^{9}$
Kavli Institute for Astronomy and Astrophysics, Peking University, Beijing 100871, People's Republic of China\\ 
$^{10}$
Department of Astronomy, School of Physics, Peking University, Beijing 100871, People's Republic of China\\
$^{11}$
Astronomical Observatory Institute, Faculty of Physics, Adam Mickiewicz University, ul.~S{\l}oneczna 36, 60-286 Pozna{\'n}, Poland\\
$^{12}$
RAL Space, Rutherford Appleton Laboratory, Harwell Campus, Didcot, Oxfordshire, OX11 0QX, UK\\
$^{13}$
Department of Physics and Astronomy, McMaster University, Hamilton, ON L8S 4M1, Canada\\
$^{14}$
Centre for Astrophysics Research, University of Hertfordshire, College Lane, Hatfield AL10 9AB, UK\\
$^{15}$
Yunnan Observatories \& Key Laboratory for the Structure and Evolution of Celestial Objects, Chinese Academy of Sciences,\\
~~~~Kunming 650011, People's Republic of China\\
$^{16}$
Center for Astronomical Mega-Science, Chinese Academy of Sciences, 20A Datun Road, Chaoyang District,\\
~~~~Beijing 100012, People's Republic of China\\
$^{17}$
Chinese Academy of Sciences South America Center for Astronomy, National Astronomical Observatories, CAS, Beijing 100101, China\\
$^{18}$
Departamento de Astronom\'{i}a, Universidad de Chile, Casilla 36-D, Santiago, Chile\\
$^{19}$
Institute of Astronomy and Astrophysics, Academia Sinica, 11F of Astronomy-Mathematics Building,\\
~~~~AS/NTU, No.1, Sec. 4, Roosevelt Rd, Taipei 10617, Taiwan, R.O.C.\\
$^{20}$
European Southern Observatory, Alonso de C\'{o}rdova 3107, Vitacura, Casilla 19001, Santiago de Chile, Chile\\
$^{21}$
Department of Astronomy, Yonsei University, 50 Yonsei-ro, Seodaemun-gu, Seoul 03722, Republic of Korea\\
$^{22}$
School of Physics and Astronomy, Cardiff University, The Parade, Cardiff CF24 3AA, UK\\
$^{23}$
Max-Planck-Institut f\"{u}r Astronomie, K\"{o}nigstuhl 17, D-69117 Heidelberg, Germany\\
$^{24}$
School of Astronomy and Space Science, Nanjing University, Nanjing 210093, People's Republic of China\\
$^{25}$
National Astronomical Observatory of Japan, 2-21-1 Osawa, Mitaka, Tokyo 181-8588, Japan\\
$^{26}$
East Asian Observatory, 660 N. A'oh\={o}k\={o} Place, Hilo, HI 96720-2700, USA\\
$^{27}$
National Astronomical Observatories \& Key Lab of Radio Astronomy, Chinese Academy of Sciences, Beijing 100012,\\
~~~~People's Republic of China\\
$^{28}$
Astrophysics Research Institute, Liverpool John Moores University, IC2, Liverpool Science Park, 146 Brownlow Hill,\\
~~~~Liverpool, L3 5RF, UK\\
$^{29}$
Xinjiang Astronomical Observatory, Chinese Academy of Sciences, Urumqi, Xinjiang 830011, P. R. China\\
}

\date{Accepted 2020 March 12. Received 2020 February 21; in original form 2019 November 22.}

\pubyear{2020}

\begin{document}
\label{firstpage}
\pagerange{\pageref{firstpage}--\pageref{lastpage}}
\maketitle

\clearpage

%----------------------------- Abstract  ----------------------------------
\begin{abstract}

To study the high-transition dense-gas tracers and their relationships to the
star formation of the inner $\sim$ 2\,kpc circumnuclear region of NGC\,253, we
present HCN $J=4-3$ and HCO$^+ J=4-3$ maps obtained with the James Clerk Maxwell
Telescope (JCMT). With the spatially resolved data, we compute the concentration
indices $r_{90}/r_{50}$ for the different tracers. HCN and \hcop\ 4-3 emission
features tend to be centrally concentrated, which is in contrast to the
shallower distribution of CO 1-0 and the stellar component. 
The dense-gas fraction ($f_\text{dense}$, traced by the velocity-integrated-intensity 
ratios of HCN/CO and HCO$^+$/CO) and the ratio \rco\ (CO 3-2/1-0)
decline towards larger galactocentric distances, but increase with higher SFR
surface density. The radial variation and the large scatter of \fdense\ and
\rco\ imply distinct physical conditions in different regions of the galactic
disc. 
The relationships of \fdense\ versus \sigmastellar, and \sfedense\ versus
\sigmastellar\ are explored. \sfedense\ increases with higher \sigmastellar\
in this galaxy, which is inconsistent with previous work that used HCN 1-0
data. This implies that existing stellar components might have different
effects on the high-$J$ HCN and \hcop\ than their low-$J$ emission. We also
find that \sfedense\ seems to be decreasing with higher \fdense\, which is
consistent with previous works, and it suggests that the ability of the dense
gas to form stars diminishes when the average density of the gas increases.
This is expected in a scenario where only the regions with high-density
contrast collapse and form stars.

\end{abstract}

%% Keywords should appear after the \end{abstract} command. 
%% See the online documentation for the full list of available subject
%% keywords and the rules for their use.

%-------------------- Intro  -----------------------
\section{Introduction} \label{sec:intro}

The important role of molecular gas in the star-formation process has received 
much attention from observational and theoretical studies
\citep{Kennicutt:2012,Krumholz:2014b}. In recent years, increasing samples
of molecular-gas measurements of galaxies in both the local and the early universe
\citep{Carilli:2013,Kruijssen:2014,Saintonge:2017,Tacconi:2018,Riechers:2019} have shed new
light on this
topic. The evolution of the molecular-gas fraction (defined as the gas mass per unit
stellar mass) is found to be surprisingly similar to that of the
evolution of the cosmic star-formation-rate density, in the sense that they both show a peak at redshift $z \sim$ 1 to 2 
and decline in the local universe
\citep{Walter:2014,Tacconi:2018,Riechers:2019}. This indicates that throughout most of
cosmic time, molecular gas is the crucial fuel for star formation and galaxy evolution. 

%--------------- Figure 1 observation positions--------------
\begin{figure}
  \centering
\includegraphics[width=1.02\linewidth, angle=0]{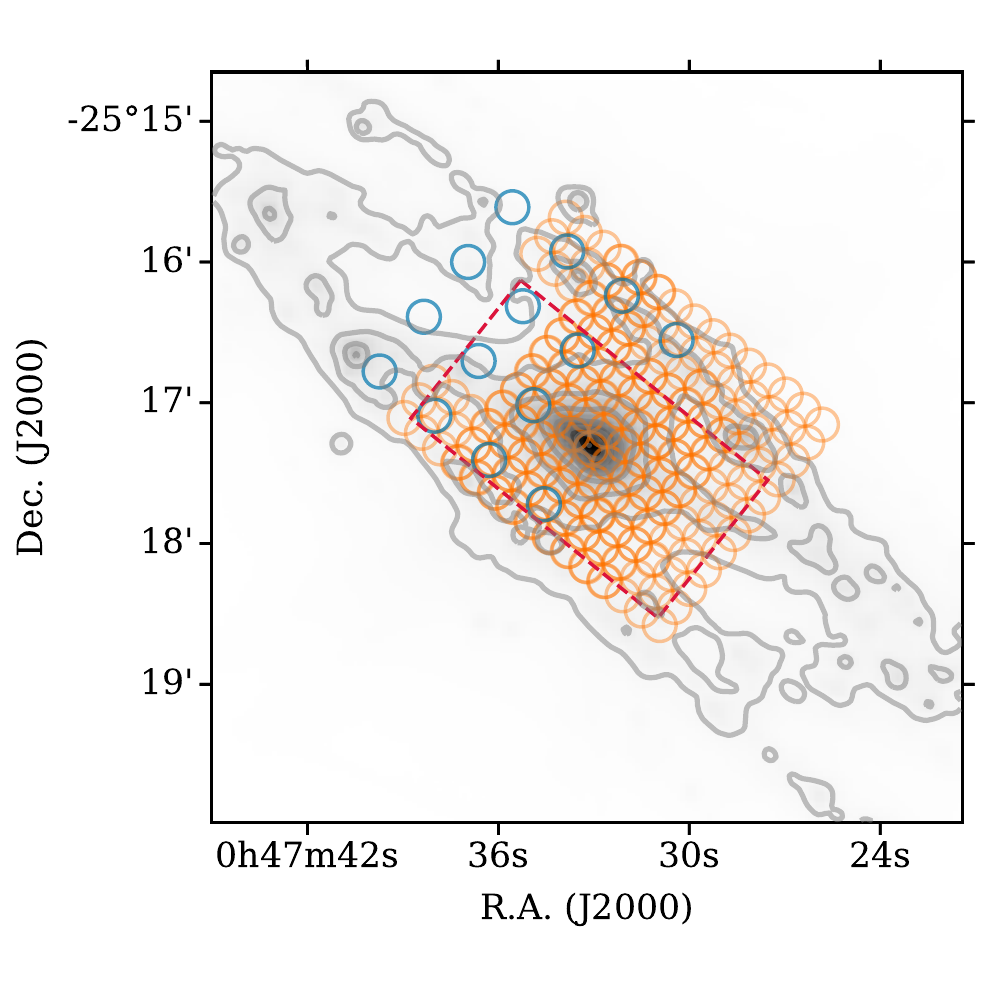}
\caption{JCMT observing positions for NGC\,253 overlaid on \textit{Herschel} PACS 70 \micron\
 emission on a logarithmic stretch. Orange circles denote positions
  observed in jiggle mode, and blue circles denote positions observed
  in stare mode. Diameters of the circles are 14\,arcsec representing the HPBW
  (half-power beam width)
  of JCMT at this frequency. The central 13$\times$7 positions (red
  box) outline those data points used for analysis in subsequent
  figures.
%The observation covers a 2\,arcmin$\times$2\,arcmin$ region ($\sim$1.6\,pc $\times$ 1.6\,pc).
\label{fig:obs_pos} 
}
\end{figure}
%\clearpage

However, it is still unclear how the physical and chemical parameters of 
the molecular gas affect the process of star formation. Specifically, how do we
interpret the variations observed in the Kennicutt--Schmidt (KS) star-formation (SF) 
relationship \citep{Kennicutt:1998}, and how do the
density and temperature of molecular gas relate to the star-formation rate
(SFR), or the star-formation efficiency (SFE, defined as the SFR per unit gas
mass)? Since the pioneering studies of 
\citet{Gao:2004a,Gao:2004b}, it has been shown that the amount of dense molecular gas
is tightly and
linearly correlated with the SFR of individual galaxies (in logarithmic scale), where 
dense gas is usually defined as gas with a volume density $n>10^4$ cm$^{-3}$,
and can be traced by molecular emission lines with high critical densities
(\ncrit) like HCN
\citep{Baan:2008,Gracia-Carpio:2008,Juneau:2009,lada:2010,Greve:2014,Liu:2015dz,
Liu:2015lj, Zhang:2014,Chen:2015,Braine:2017,Tan:2018}. Meanwhile,
observations on smaller
scales, such as of resolved galaxy structures or of molecular clouds in the
Milky Way, also showed that star-forming sites are mainly associated with 
dense structures \citep{Wu:2005,Andre:2014, Liu:2016, Stephens:2016,
Shimajiri:2017}. Together with the extragalactic results, these studies suggest
a simple threshold model where stars only form at gas densities above
$10^4$\,cm$^{-3}$, and this model might be universal across eight orders of
magnitude of the SFR \citep{Lada:2012,Evans:2014}. However, there is also a
compelling model that does not require threshold density, and the dense-gas 
SF correlation can be explained as the association between young stars and
nearby collapsing gas with a free-fall time comparable to the stars' ages
\citep{Elmegreen:2015,Elmegreen:2018}.

While the physical driver of the scatter along this dense-gas star-formation
relationship likely relates to the feedback from star-formation
or active galactic nuclei \citep{Papadopoulos:2014}, we still lack enough 
samples to quantify this effect in different types of galaxies.
More detailed
analyses based on high-spatial resolution observations
\citep{Usero:2015,Bigiel:2016, Gallagher:2018a,
Gallagher:2018b,Jimenez-Donaire:2019} have
revealed the variation of the SFR per unit dense gas mass (the dense-gas star-formation efficiency, \sfedense) in different regions of galaxy discs, and
also in luminous infrared galaxies \citep{Gracia-Carpio:2008}. 
Together, these results point
to an alternative model that is turbulence regulated
\citep{Krumholz:2005,Krumholz:2007}. 

Recent observations show that these molecules with high \ncrit\ 
can also be excited in extended translucent regions ($n \sim 10^3 \text{cm}^{-3}$)
that are not actively forming stars \citep{Pety:2017,Watanabe:2017,Kauffmann:2017,Nishimura:2017,Harada:2019}.
This is evidence for the scenario pointed out by \citet{Evans:1999}, that
the effective density to excite a certain line can be much lower 
than the critical density, 
and one should be cautious about 
interpreting single-line observations, namely they are not adequate to 
accurately estimate gas density.
So multiple transitions, especially high-$J$
lines, are necessary for a more-robust analysis of the relationship between dense 
gas tracers and SF.
Previous studies have mainly used the ground transition of dense-gas 
tracers and our work aims to explore the behaviour of their high-$J$ transitions,
allowing for the possibility to combine multiple transitions for the diagnosis of 
molecular gas in galaxies.

To bridge the gap between Milky Way clouds and galaxy-integrated
observations, and to explore systematically the relationship between dense gas and
star formation in nuclear versus disc regions, we carried out the MALATANG (Mapping
the dense molecular gas in the strongest star-forming galaxies) large program on
the James Clerk Maxwell Telescope (JCMT). MALATANG is the first systematic
survey of the spatially resolved HCN $J=4-3$ and \hcop\ $J=4-3$ emission in
a large sample of nearby galaxies. \citet{Tan:2018} presented the MALATANG data
of six galaxies.
They explored the relationship between the dense gas and the star
formation rate and show that the power-law slopes are close to unity
over a large dynamical range. They also imply that the variation of this
relationship could be dependent on the dense-gas fraction (\fdense) and dust
temperature. 

NGC\,253 is the nearest spiral galaxy hosting a nuclear starburst ($d$ =
3.5\,Mpc, based on the planetary nebula luminosity function,
\citealt{Rekola:2005}). This makes it an ideal laboratory for detailed studies
of the star-formation activity in an active environment, and it is accessible
to most major facilities, such as the JCMT and ALMA. 
In Table \ref{tab:n253} we list some adopted properties of NGC\,253.	
High-resolution
observations of molecular gas have found expanding molecular shells in the
starburst region \citep{Sakamoto:2006}, and star-formation activity may be
suppressed due to the expulsion of molecular gas \citep{Bolatto:2013}. The
molecular-gas depletion time
($\tau_\text{dep}$) of NGC\,253 is also found to be 5--25 times shorter than the
typical disc values of other galaxies. The molecular-line emission is
concentrated in the inner 1\,kpc, and the dense molecular-gas tracers show
clumpy and elongated morphologies, and they show excellent agreement with the
850-\micron\ continuum down to very small scales of few parsecs \citep{Leroy:2015,
Meier:2015, Walter:2017, Leroy:2018}.
Earlier studies have presented multiple HCN and \hcop\ observations in the nucleus
of NGC\,253 
\citep{Nguyen-Q-Rieu:1989,Nguyen:1992,Paglione:1995,Paglione:1997,Knudsen:2007},
and the average density of the gas is suggested to be
5$\times$10$^5$\,cm$^{-3}$ \citep{Jackson:1995}.
%obtained with APEX, 
An excitation
analysis revealed that the lines from both molecules are subthermally
excited. Submillimeter Array (SMA) observations of HCN 4-3 toward the central
kiloparsec of NGC\,253 have shown that the dense-gas fraction is higher in the
central 300\,pc region than in the surrounding area \citep{Sakamoto:2011}.

This paper is a follow-up to the MALATANG data of NGC\,253 (this galaxy was
included in the sample of \citealt{Tan:2018}). We focus on the
analysis of the variation in the parameters related to dense gas and
star formation. In Section~\ref{sec:data} the observations and data reduction
are introduced. In Section~\ref{sec:results} we present the spectra, images,
radial distributions, and line ratios of the molecular lines. In
Section~\ref{sec:discussion} we discuss the relationship between SFE, \fdense\ and
stellar components, and the dense-gas star-formation relationship.

%---------------  Table 1 NGC 253 properties--------------

\begin{table}
\caption{Adopted properties of NGC\,253. References:
%\multicolumn{3}{l}{
(1)  \citet{Best:1999},
(2)  \citet{Rekola:2005},
(3) \citet{Koribalski:2004},
%}\\
%\multicolumn{3}{l}{
(4) \citet{de-Vaucouleurs:1991},
(5) \citet{Jarrett:2003},
(6) \citet{Harrison:1999},
(7) \citet{Lucero:2015},
(8) \citet{Sanders:2003}.
%}
\label{tab:n253}
}
\centering
\begin{tabular}{lcl}
\hline
Parameters  &  Value  & Ref. \\
\hline
R.A. (J2000)& 00$^\text{h}$47$^\text{m}$33{\fs}1 & (1) \\ %1999MNRAS.310..223B
Dec. (J2000)& $-$25{\degr}17{\arcmin}19{\fs}7 & (1) \\ %1999MNRAS.310..223B
Distance    & 3.5 Mpc  	    & (2) \\ 
Velocity (Heliocentric) &  243 $\pm$ 2 \kms & (3) \\  %2004AJ....128...16K
Morphology  & SAB(s)c 		& (4)	\\   % 1991RC3.9.C...0000d
Diameter ($D_{25}$)  &  27.5 arcmin  	& (4) 	\\  %1989ESOLV.C...0000L
Inclination & 76$\degr$     & (5) 	\\
Position angle & 51$\degr$ & (5)	\\
nuclear CO column density & 3.5$\pm1.5 \times 10^{18}$\,cm$^{-2}$ &  (6) \\%
HI size (at 10$^{20}$ cm$^{-2}$ level) & 29 $\pm$ 2 kpc  & (7)\\ % doi.org/10.1093/mnras/stv856
SFR         &  	4.2 M$_{\sun}$yr$^{-1}$    &	(8)	\\
\hline
%\multicolumn{4}{l}{The last seven rows are positions that were only observed in JCMT-HARP's stare mode.}\\
\end{tabular}
\end{table}

%------------- Observations and Data reducation ----------------
\section{Observations and Data reduction} \label{sec:data}

\subsection{JCMT HCN 4-3 and HCO$^+$ 4-3 data}
A total of 390 hours was spent on MALATANG observations from 2015 December
and 2017 July (Program ID: M16AL007). 
The Heterodyne Array Receiver Program
(HARP, \citealt{Buckle:2009}) was used to observe HCN 4-3 and \hcop\ 4-3. Two adjacent receptors 
(H13 and H14)
on the edge of HARP were not functional, which caused a nonhomogeneous coverage 
for the jiggle-mode (see Fig. \ref{fig:obs_pos}).
In order to cover the major axes
of the galaxies, the orientation of HARP was adjusted for each galaxy according to
its position angle, so that four receptors lined up along the galaxy's major axis.
For NGC\,253, the inclination of its galactic disc is
76$\degr$ (north is the receding side), and the position angle of its major axis
(measured counter-clockwise from north) is about 51$\degr$ \citep{Jarrett:2003}.

The Auto-Correlation Spectral Imaging System (ACSIS) spectrometer was used as the
backend, with 1\,GHz bandwidth and a resolution of 0.488\,MHz, corresponding
to 840 and 0.41\,\kms at 354\,GHz, respectively. The half-power beam width
(HPBW) of each receptor at 350\,GHz is $\sim$ 14\,arcsec, corresponding to 240\,pc
linear resolution at the distance of NGC\,253. 
The telescope pointing was checked on R~Scl (R Sculptoris) before observing our target source and
subsequently every 60 to 90 minutes, using the CO 3-2 line at 345.8\,GHz. The uncertainty
in the absolute flux calibration was about 10 per cent for galaxies and was measured
using standard line calibrators. 
The details of the survey description, sample, and data are given in Zhang et al. 
(in prep., see also \citealt{Tan:2018}).

Two observing modes were used to observe NGC\,253. To fully map the central
2$\times$2\,arcmin region (centred at R.A.(J2000.0) = 00$^\text{h}$47$^\text{m}$33{\fs}1,
Dec(J2000.0) = $-$25{\degr}17{\arcmin}19{\fs}7), the 3$\times$3
jiggle observing mode was used, with a grid spacing of 10\,arcsec. This was
mostly done in 2015 December in excellent weather conditions, i.e., mean
$\tau_\text{225 GHz} \sim 0.024$ and 0.036 for HCN 4-3 and HCO$^+$ 4-3,
respectively. The integration times spent in jiggle mode for the HCN and
\hcop\ lines were 142 and 100 minutes, respectively.
To reach deeper integrations in the outer parts of NGC\,253, the stare
observing mode was used and the tracking centre was at
(J2000.0) = 00$^\text{h}$47$^\text{m}$34{\fs}8,
Dec(J2000.0) = $-$25{\degr}17{\arcmin}00{\fs}8, which was shifted by
30\,arcsec to the north-east along the major axis of NGC\,253. The grid
spacing was 30\,arcsec (see Fig.~\ref{fig:obs_pos}). The integration times
spent in stare mode for HCN 4-3 and \hcop\ 4-3 were 7.8 and 6.1 hours,
respectively.

\subsection{Ancillary data}
Ancillary data of CO 1-0 and CO 3-2 were obtained for a more-comprehensive
analysis of the interstellar medium in NGC\,253. The CO 1-0 data are from the
Nobeyama CO Atlas of Nearby Spiral Galaxies \citep{Sorai:2000, Kuno:2007}, and the CO 3-2
data from the JCMT archive (project ID: M08AU14). We also include infrared
archival imaging data from \textit{Spitzer} (IRAC 3.6\,\micron~and MIPS 24\,\micron),
and \textit{Herschel} (PACS 70, 100, and 160\,\micron). The Nobeyama and
\textit{Herschel} data were convolved to 14-arcsec resolution and regridded to the same pixel scale (see
\citealt{Tan:2018} for more details of the processing of the ancillary data).
This allowed us to calculate the total infrared luminosity \lir\ (8--1000\,\micron)
in each pixel using a combination of the 24-, 70-, 100-, and 160-\micron\
luminosities \citep{Galametz:2013,Tan:2018}.

%--------------- Figure intensity maps --------------
\begin{figure}
\includegraphics[width=.47\textwidth, angle=0]{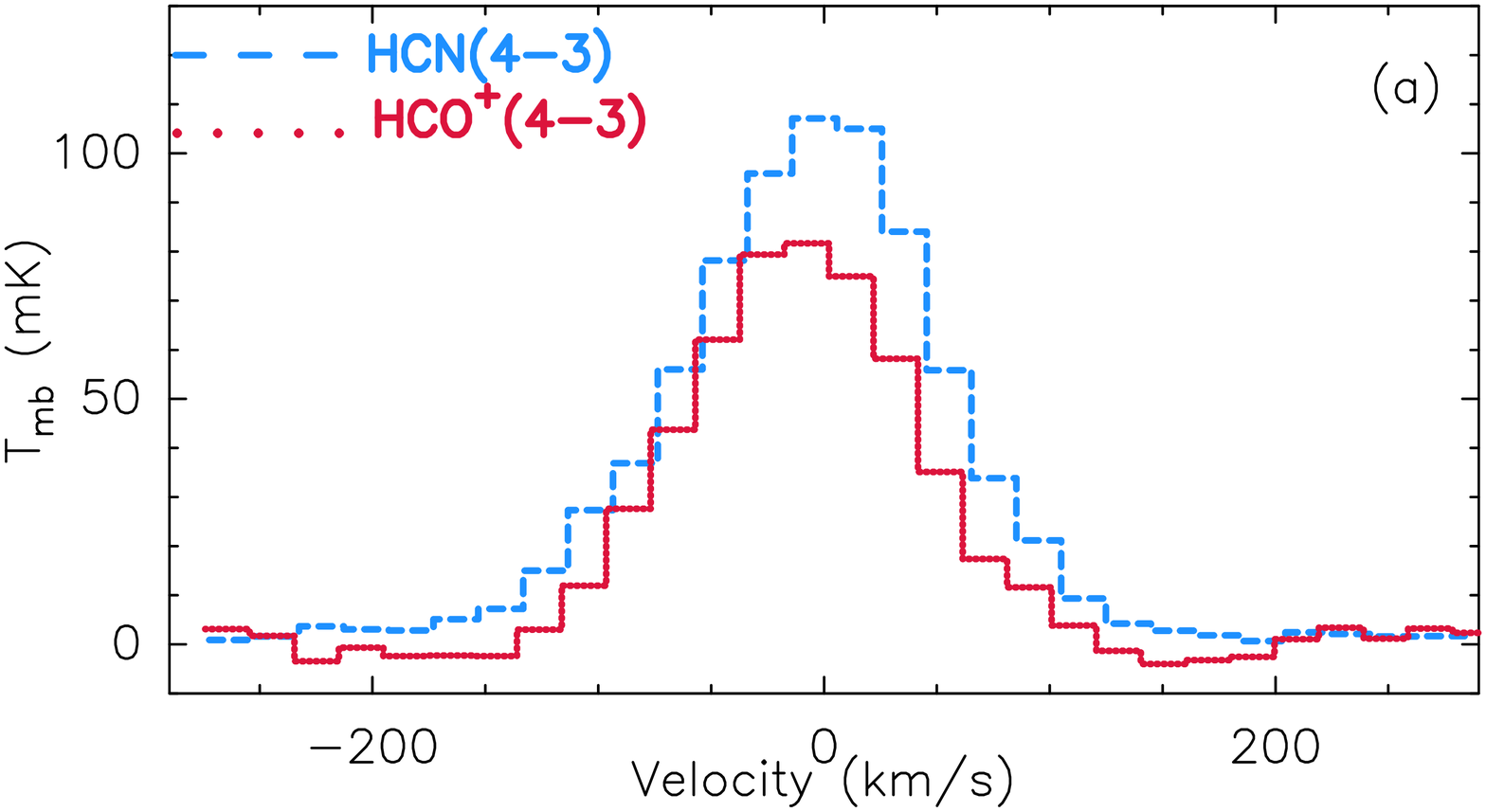}\\
\includegraphics[width=.47\textwidth, angle=0]{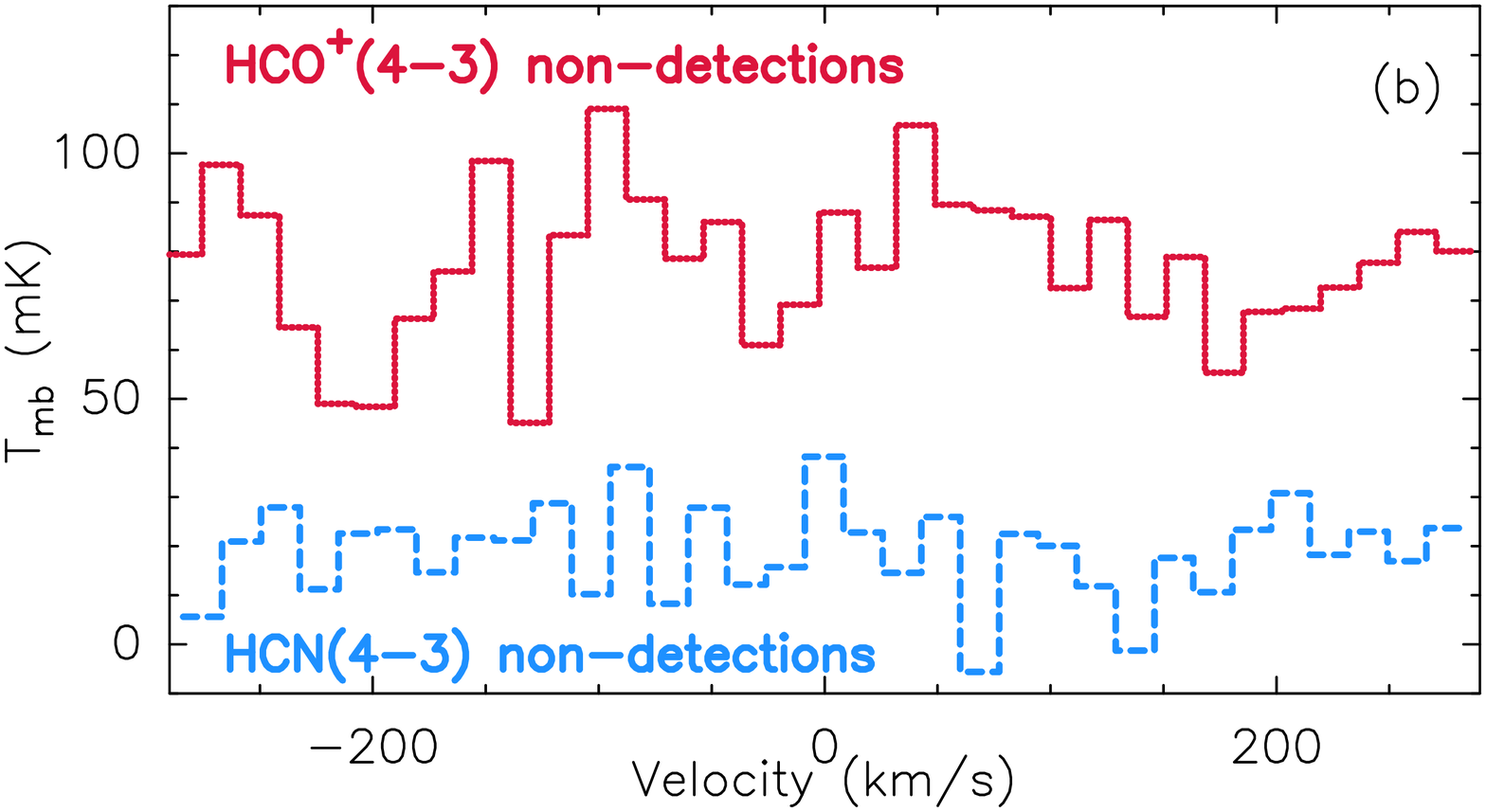}\\
\caption{\textit{top panel:} averaged spectrum of the central 5$\times$3 pixels 
($\sim$ 0.9\,kpc along the major axis).
\textit{bottom panel:} averaged spectrum stacked from all the non-detections, with their velocities corrected according to the CO 3-2 line centre. 
The spectrum of \hcop\ 4-3 is shifted by 4\,mK for clarity.
\label{fig:stack_spec} }
\end{figure}

%--------------- Figure intensity maps --------------

\begin{figure*}
\includegraphics[width=0.48\textwidth, angle=0]{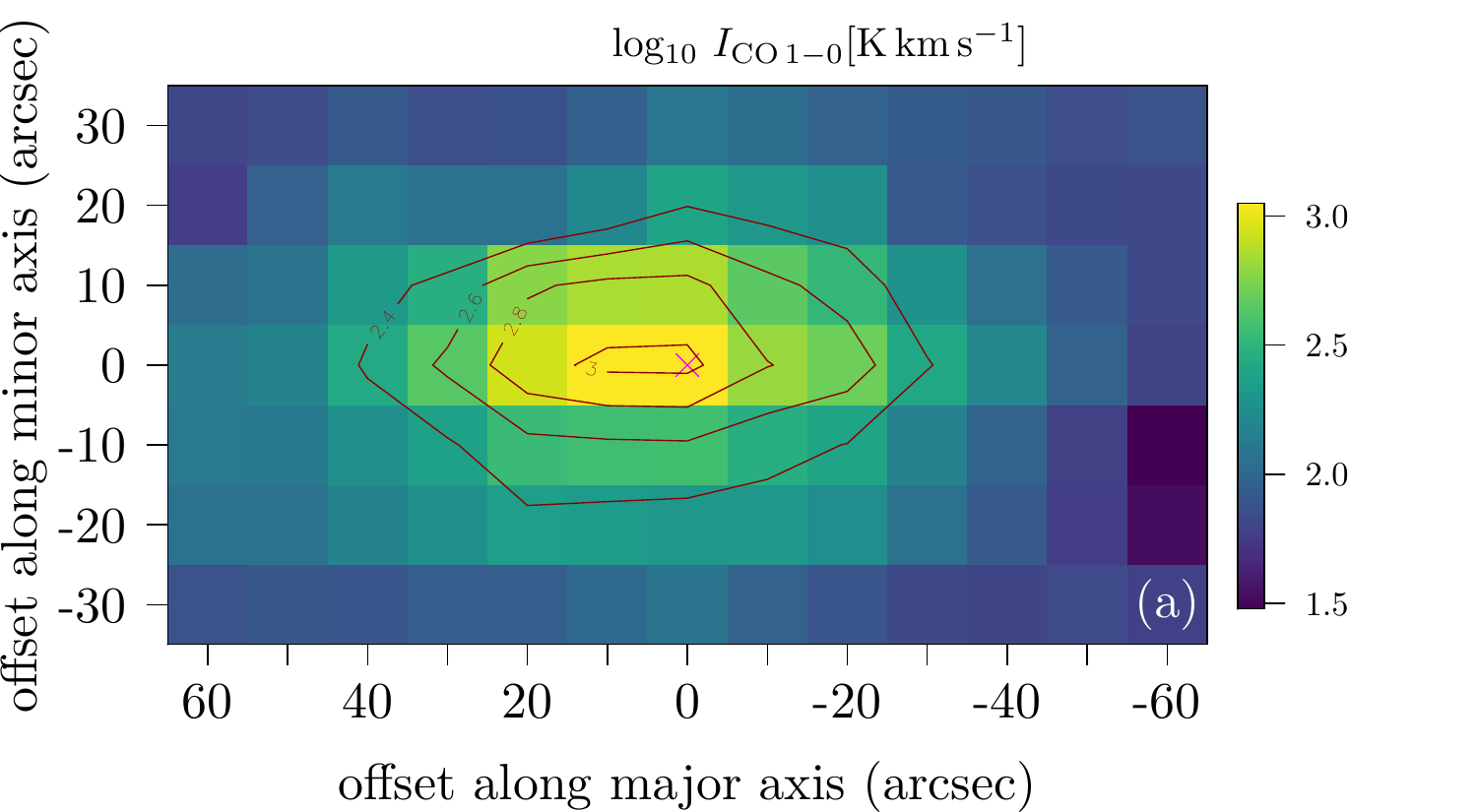}
\includegraphics[width=0.48\textwidth, angle=0]{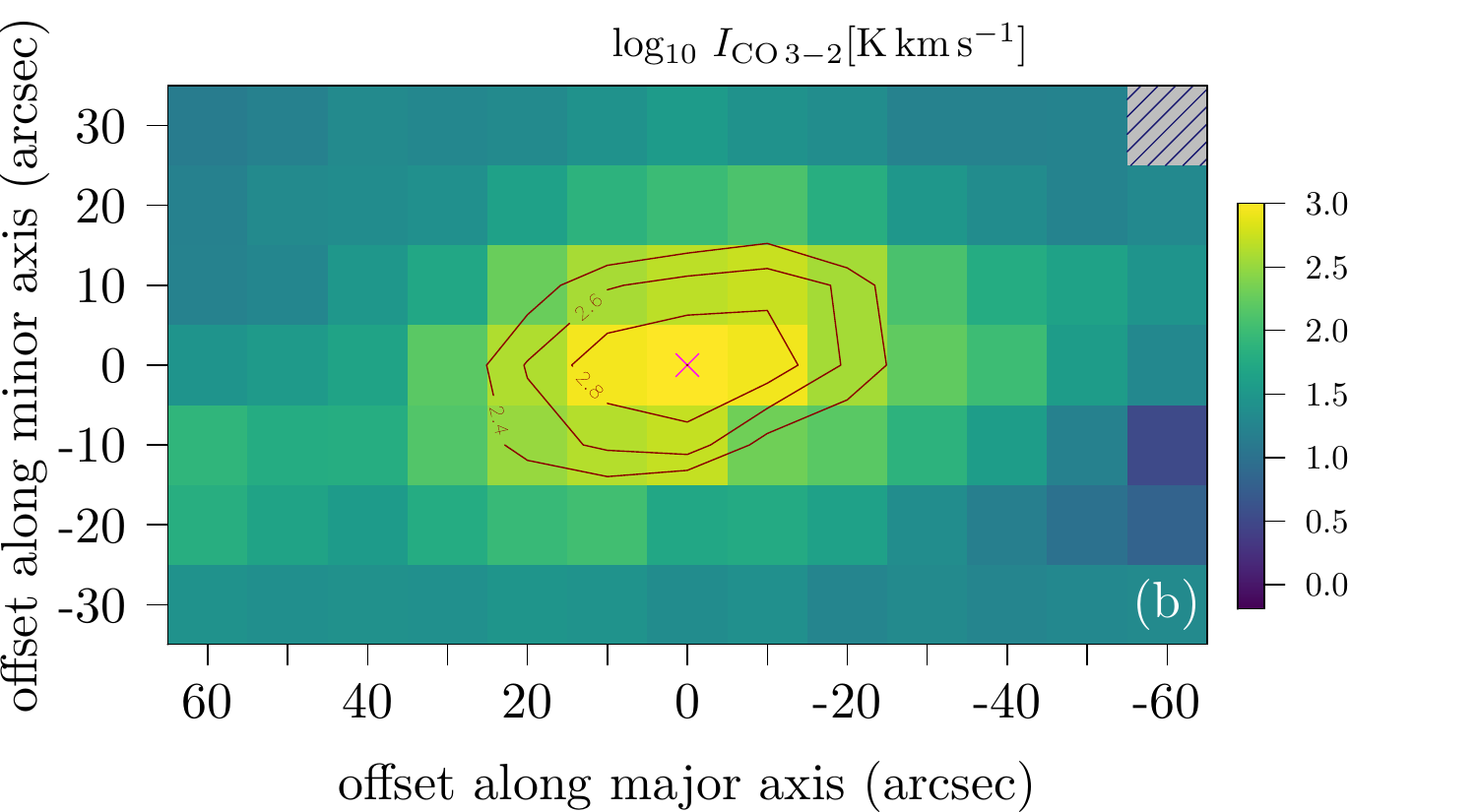}\\
\includegraphics[width=0.48\textwidth, angle=0]{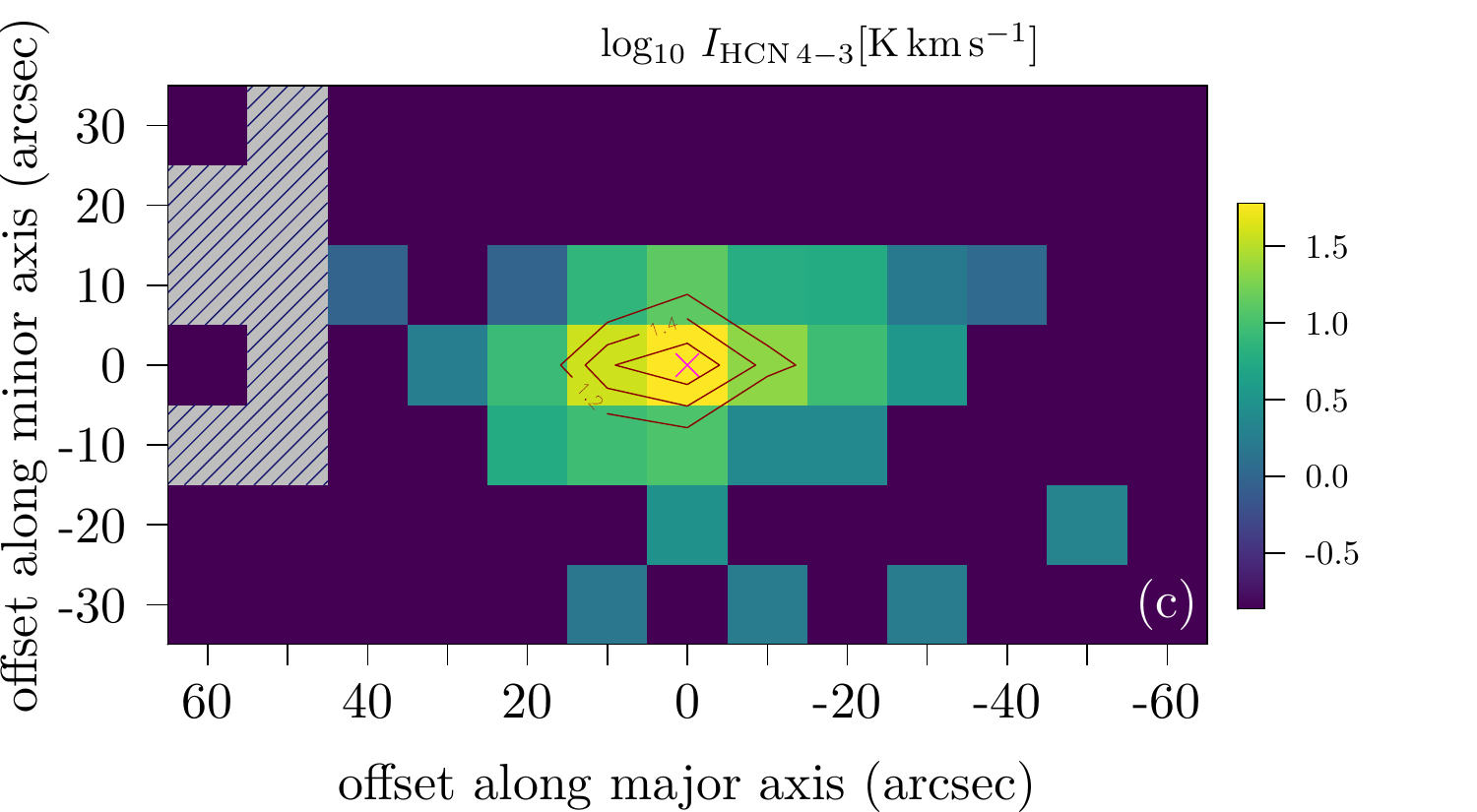}
\includegraphics[width=0.48\textwidth, angle=0]{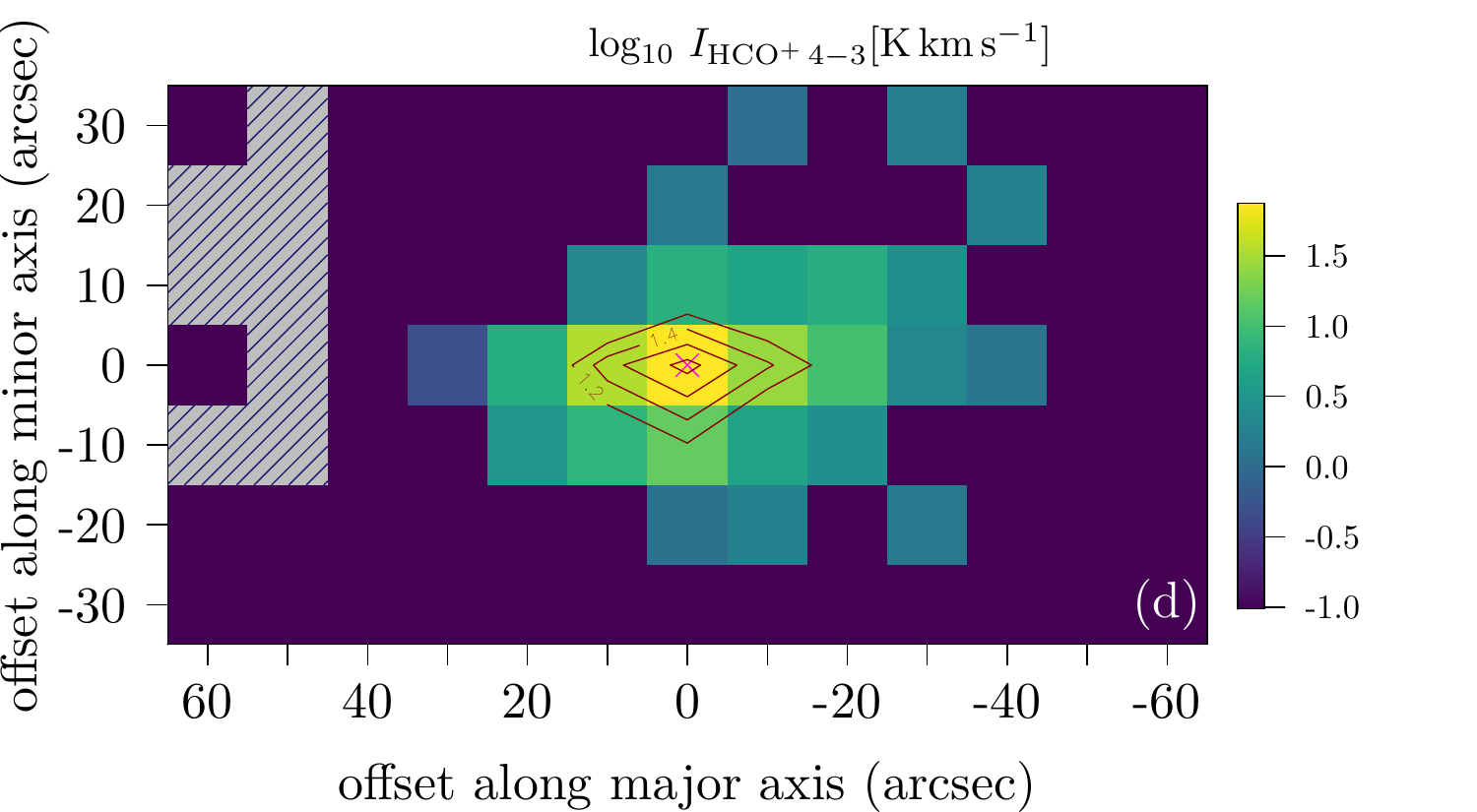}\\
\includegraphics[width=0.48\textwidth, angle=0]{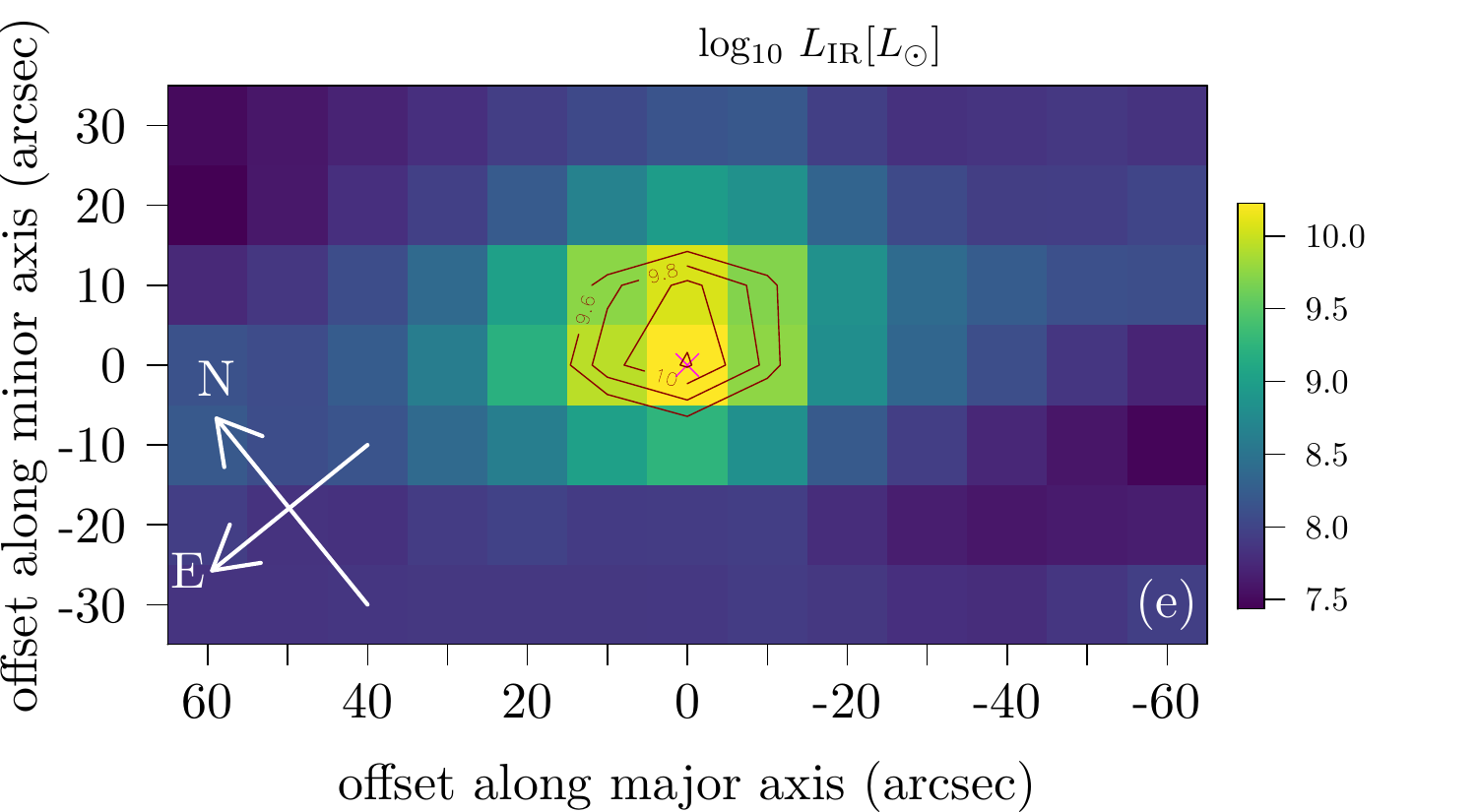}
\includegraphics[width=0.48\textwidth, angle=0]{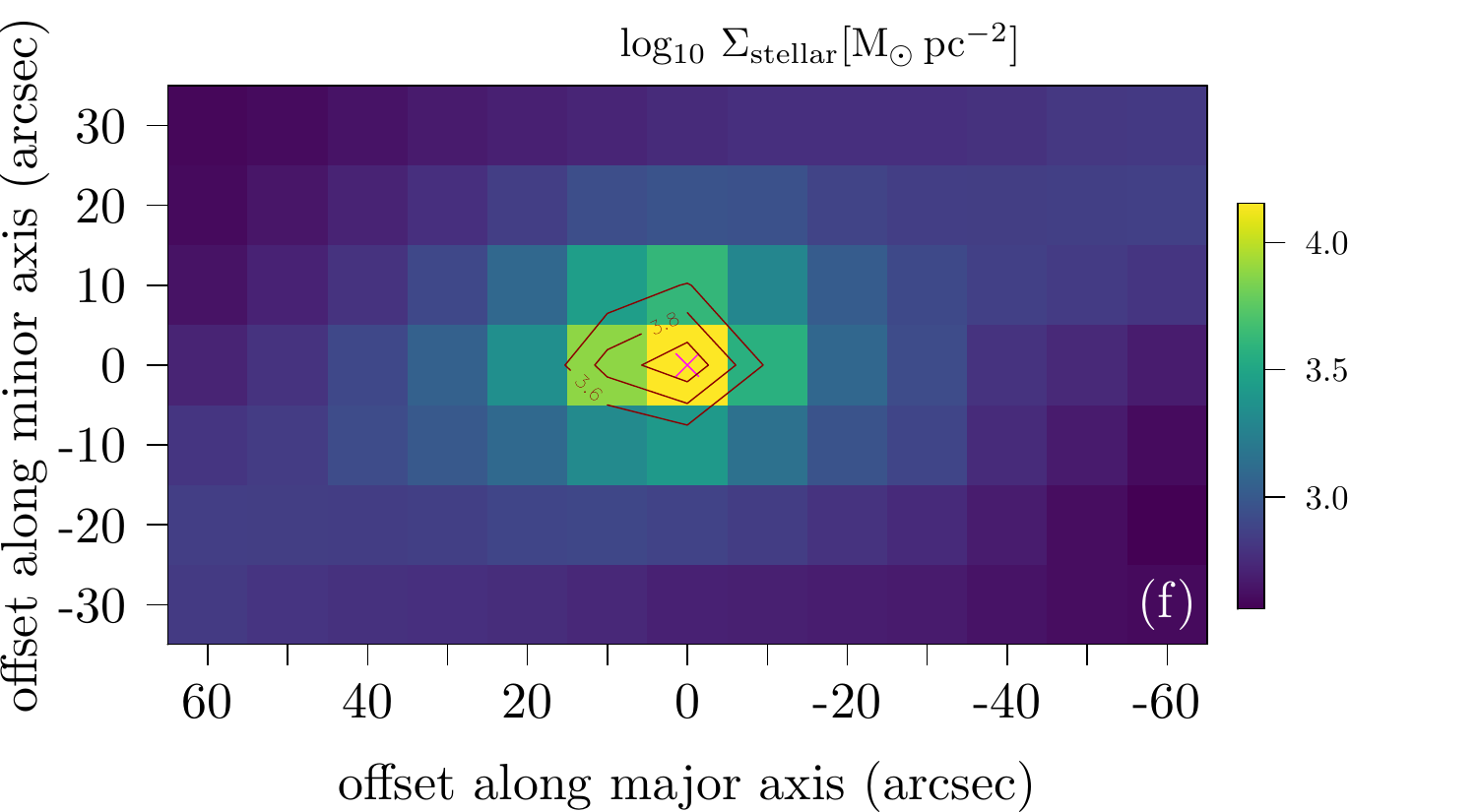}
\caption{Maps of the different molecular-line emissions, infrared luminosity
and stellar surface density. The units are given above each panel and
all maps are logarithmic. Infrared luminosity is from \citet{Tan:2018}.
Each pixel size is 10\,arcsec or 171\,pc. The hatched pixels are regions
without available data, and the typical RMS of each map is shown as the lowest
values. Contours
start from 30\% of the peaks, and the contour spacing is 0.2 dex.
The magenta cross marks the galactic centre.
Note that these images are rotated (see Fig.~\ref{fig:obs_pos}) to align with
the major axis of NGC\,253. \label{fig:int_map}} 
\end{figure*}

\subsection{Data reduction}
The \textsc{starlink} \citep{Currie:2014} software package was used to reduce the
JCMT data. Some of the receptors, mostly the outer ones, were not very stable
during some observing scans, and spikes and an unstable baseline can be seen in
some of the raw data. To enhance the signal-to-noise (SNR) ratio, we first
checked the raw data by eye with the \textsc{gaia} tool (part of \textsc{starlink}) and flagged the particularly bad 
sub-scans. Secondly, the parameters for the \textsc{orac-dr} pipeline
\citep{Jenness:2015} were adjusted to remove spikes and strong ripples further.
This step was especially necessary for the CO 3-2 data because this line is too
strong and wide, so there are not enough line-free channels for the default parameters to do proper baseline fitting,
and we need to adjust the parameters. Thirdly, the pipeline was run again to compare
with previous results. This way, we could better deal with the baseline
correction and reveal weak signals in some regions of the final products. The
final data are smoothed to a velocity resolution of $\sim 20$ \kms, and the
typical corresponding RMS noise is $\sim$ 7--10\,mK for spectra obtained in
jiggle-mode, and $\sim$ 2\,mK for spectra obtained in stare-mode (Fig.~\ref{fig:spec}).

The maps of NGC\,253 were regridded to a pixel size of 10\,arcsec. 
Fig.~\ref{fig:spec} shows the corresponding spectra of the central 13 $\times$ 7 pixels,
including CO 1-0, CO 3-2, HCN 4-3, and \hcop\ 4-3. Other pixels in the outer
regions of the galaxy disc are mostly non-detections with relatively high noise,
due to the relatively poor performance of the outer receptors of HARP.

The final maps were converted to the \textsc{CLASS} format and spectra were
measured with the \textsc{Gildas} package\footnote{http://www.iram.fr/IRAMFR/GILDAS/}.
The spectral intensity units were converted from antenna temperature $T_A^*$ to
main-beam temperature $T_\text{mb}$ using $T_\text{mb} = T_A^* / \eta_\text{mb}$,
where the main-beam efficiency $\eta_\text{mb}$ = 0.64
\footnote{https://www.eaobservatory.org/jcmt/help/workshops/}.

To identify detections of the emission lines used in this work, we adopted the
same criteria as \citet{Tan:2018}, i.e., 
we require that the integrated line intensity to be at least three times the "integrated noise"
(SNR > 3).
The uncertainty, $\sigma_I$, on the integrated line emission is:
\begin{equation}
\sigma_I = T_\text{RMS} \sqrt{\Delta v_\text{line} \delta v_\text{res}} \sqrt{1+ \Delta v_\text{line}/\Delta v_\text{base}},
\end{equation}
where $T_\text{RMS}$ is the RMS of the spectrum given a spectral velocity
resolution $\delta v_\text{res}$, $\Delta v_\text{line}$ is the velocity range
used to integrate the line emission, and $\Delta v_\text{base}$ is the velocity
width used to fit the baseline. Since CO lines are much stronger than HCN 4-3
and \hcop\ 4-3 lines, and the emitting region of CO is probably larger than the
regions dominated by these dense-gas tracers, but they still share the same
kinematics and are covered by similar telescope beam size in general, despite their
distinct spatial scales. So the velocity widths of the CO 1-0 and/or CO 3-2
lines are
used as a reference for the velocity ranges of the HCN 4-3 and \hcop\ 4-3 lines,
especially for positions with low SNR. This way we can use
the width to estimate upper limits (3$\sigma$) to
the integrated intensities. In Table~\ref{tab:1}, we list the integrated
intensities $I_\text{HCN 4-3}$, $I_{\text{HCO}^+ 4-3}$, $I_\text{CO 1-0}$, and 
$I_\text{CO 3-2}$, and their ratios, for every pixel in Fig.~\ref{fig:spec}. In
Table~\ref{tab:2} we list the line luminosities of HCN 4-3 and \hcop\ 4-3, the SFR,
and their dense-gas mass (\mdense). In accordance with \citep{Tan:2018}, we
calculate the line luminosities $L^{\prime}_{\text{gas}}$ (\lco, \lhcn, and
\lhcop in this paper) in units of K \kms pc$^2$ and SFR based on the following
equations:

\begin{equation}
\begin{aligned}
L^\prime_{\text{gas}} &= 3.25 \times 10^7 \left(\frac{S\Delta v}{{1\ \text{Jy km s}^{-1}}}\right)
\left(\frac{\nu_{\text{obs}}}{\text{1 GHz}}\right)^{-2} \\
& \times\left(\frac{D_{\text{L}}}{\text{1\ Mpc}}\right)^2 \left(1+z\right)^{-3}\ [\text{K km s}^{-1} \text{pc}^2]
\end{aligned}
\end{equation}

\begin{equation}
%\left(
\frac{\text{SFR}}{\text{1 M}_{\sun}\ \text{yr}^{-1}} = 1.50\times 10^{-10} \left(\frac{L_\text{IR}}{L_{\sun}}\right).
\end{equation}

The infrared luminosity is calculated from
\begin{equation}
L_\text{IR} = \Sigma\ c_i \nu L_\nu(i) L_{\sun}
\end{equation}
where $\nu L_\nu(i)$ is the resolved luminosity in a given band $i$ in units of $L_{\sun}$ and measured as $4\pi D_\text{L}^2(\nu f_\nu)_i$, and $c_i$ are the 
calibration coefficients for various combinations of $Spitzer$ and $Herschel$ \citet{Tan:2018}.
The stellar surface density \sigmastellar\ is calculated via:
\begin{equation}
\frac{\Sigma_{\text{stellar}}}{\text{M}_{\sun}~\text{pc}^{-2}} = 280 \frac{I_{3.6}}{\text{MJy sr}^{-1}} \text{cos}(i)
\end{equation}

%--------------- Figure ratio maps --------------

\begin{figure*}
\includegraphics[width=.48\textwidth, angle=0]{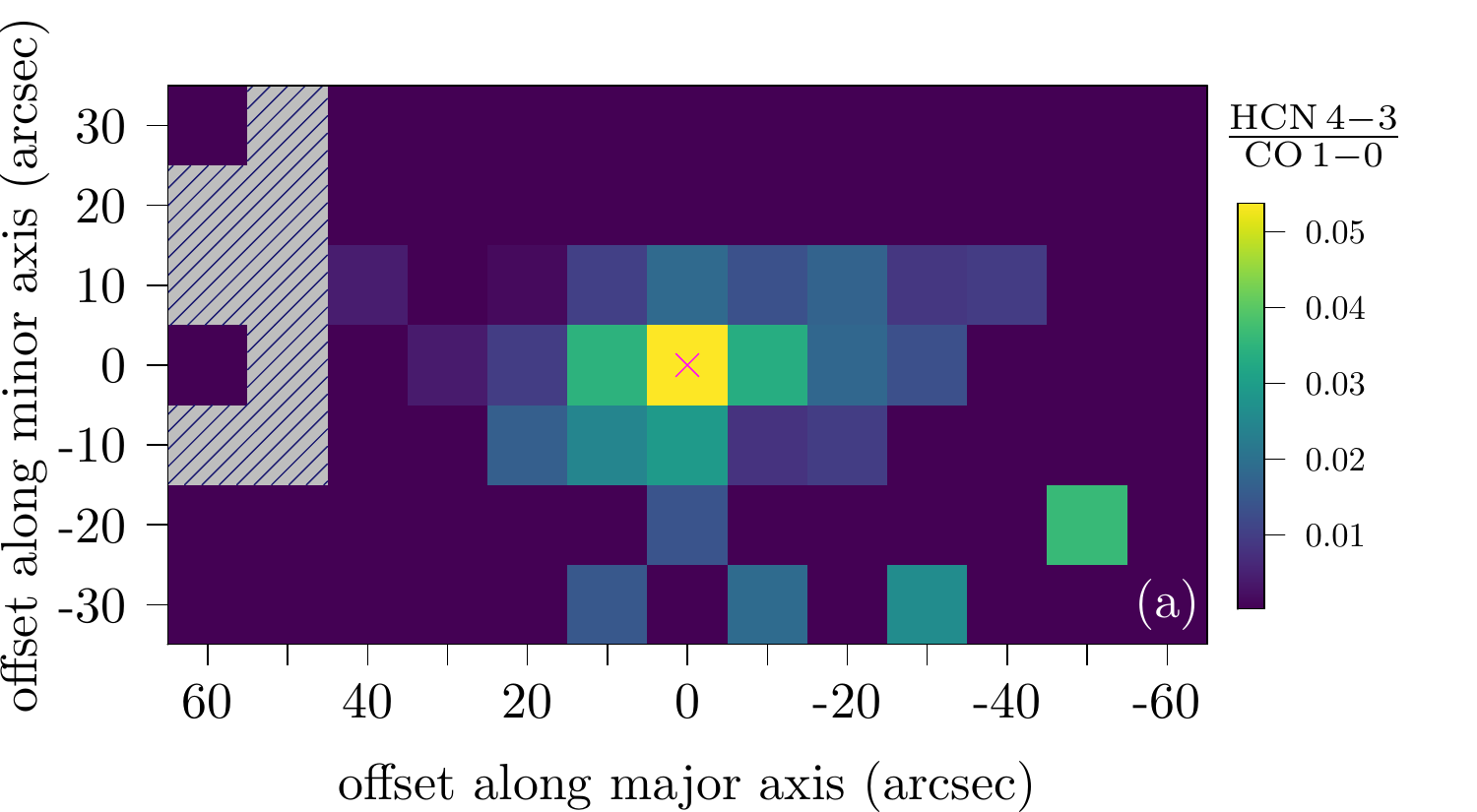}
\includegraphics[width=.48\textwidth, angle=0]{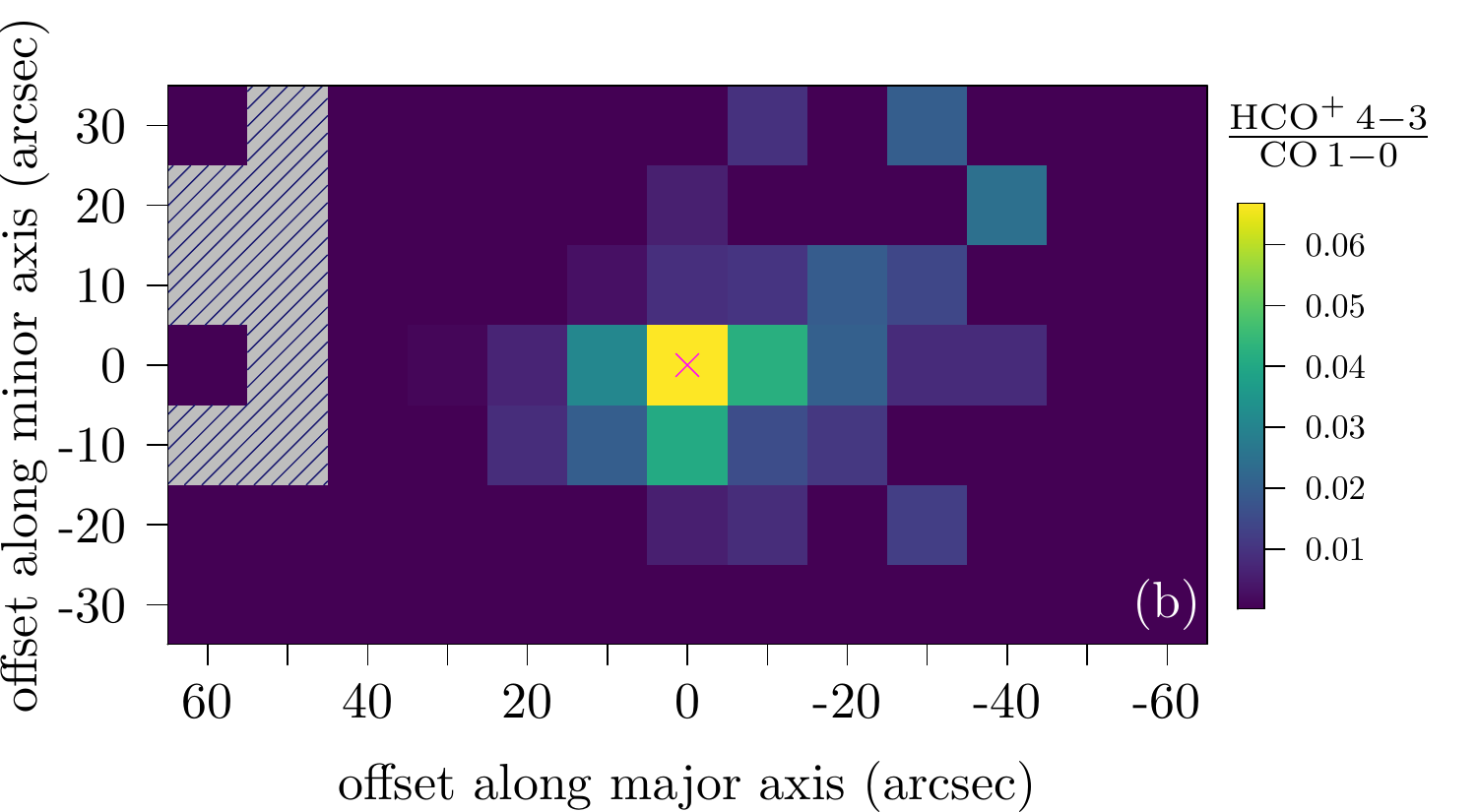}\\
\includegraphics[width=.48\textwidth, angle=0]{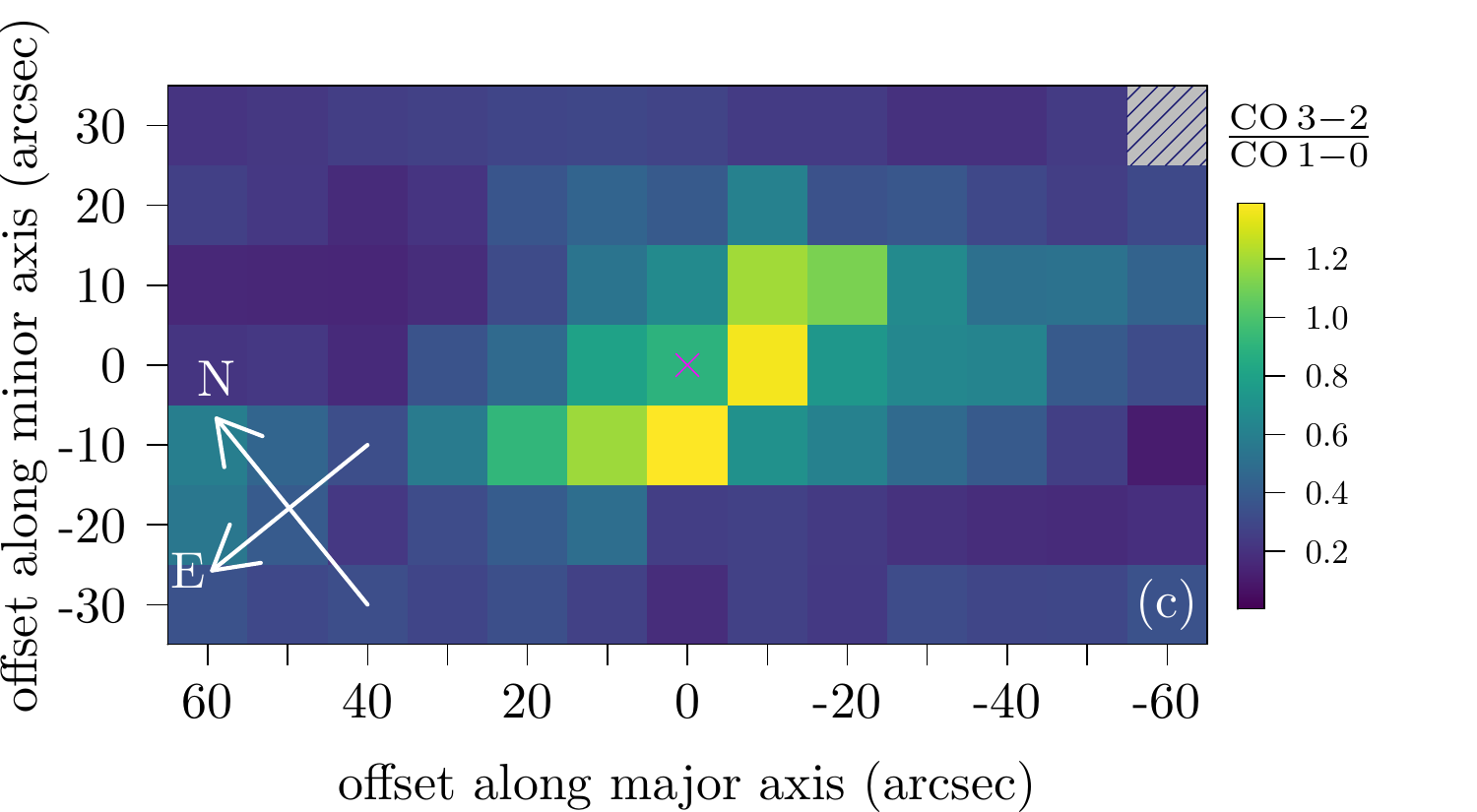}
\includegraphics[width=.48\textwidth, angle=0]{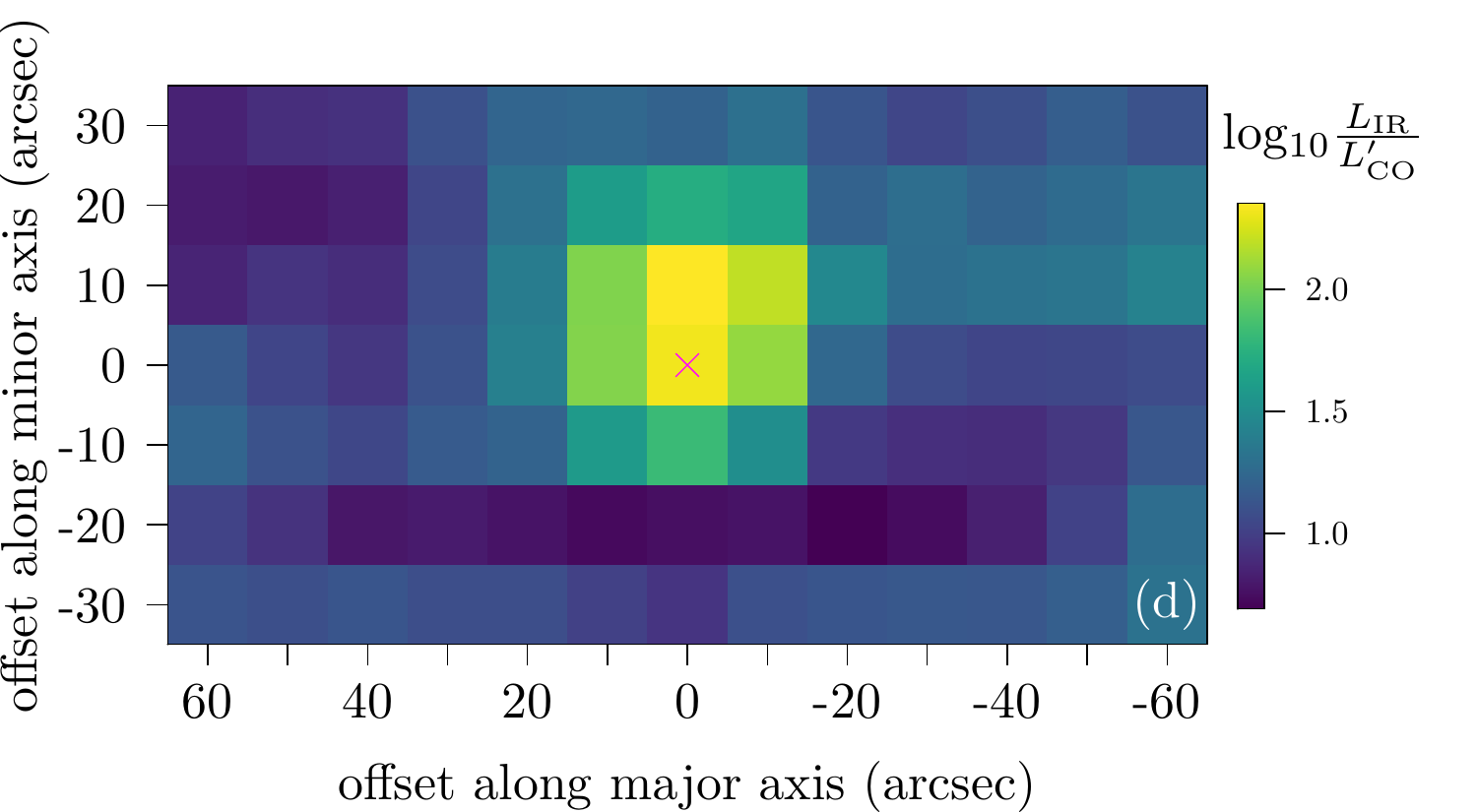}\\
\includegraphics[width=.48\textwidth, angle=0]{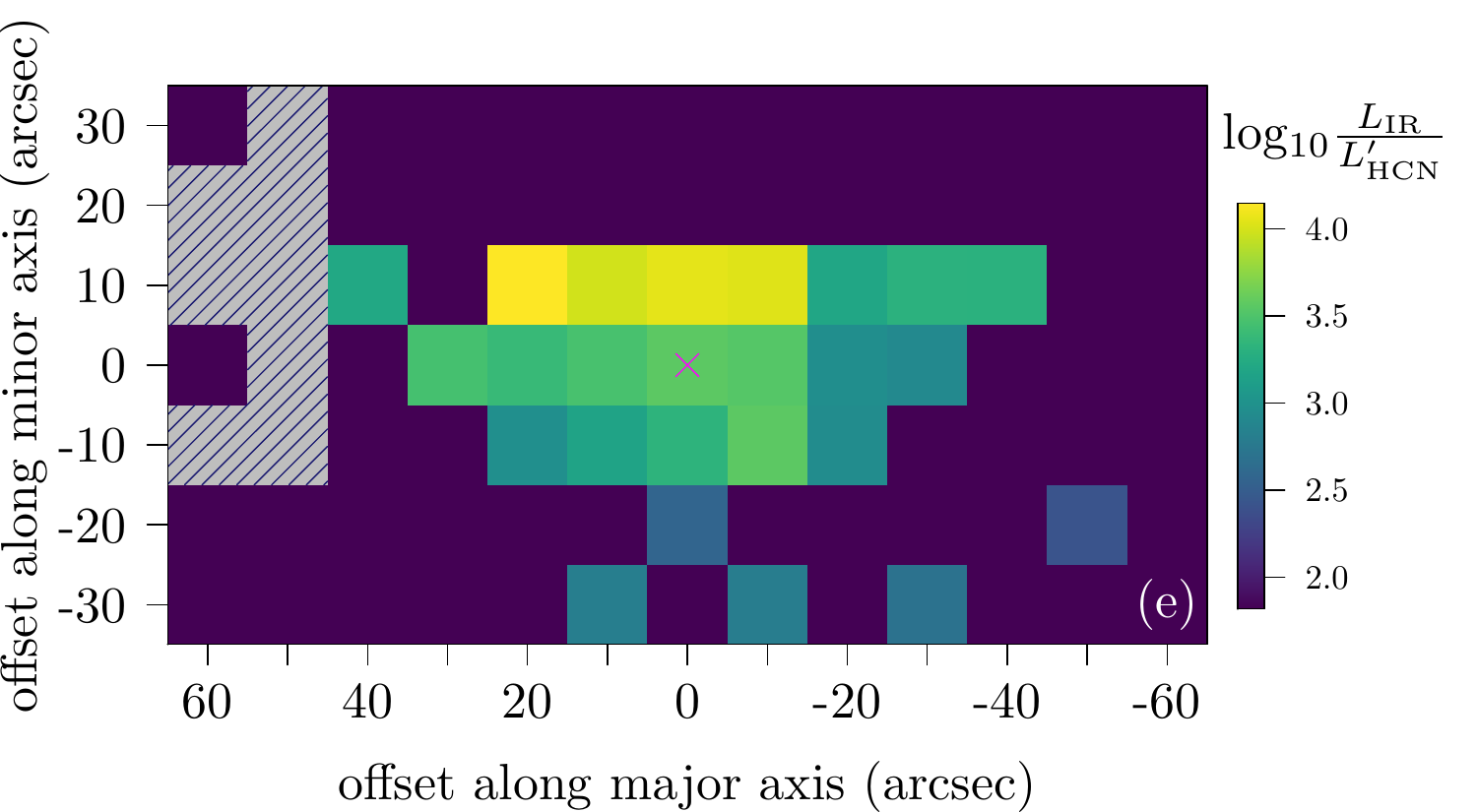}
\includegraphics[width=.48\textwidth, angle=0]{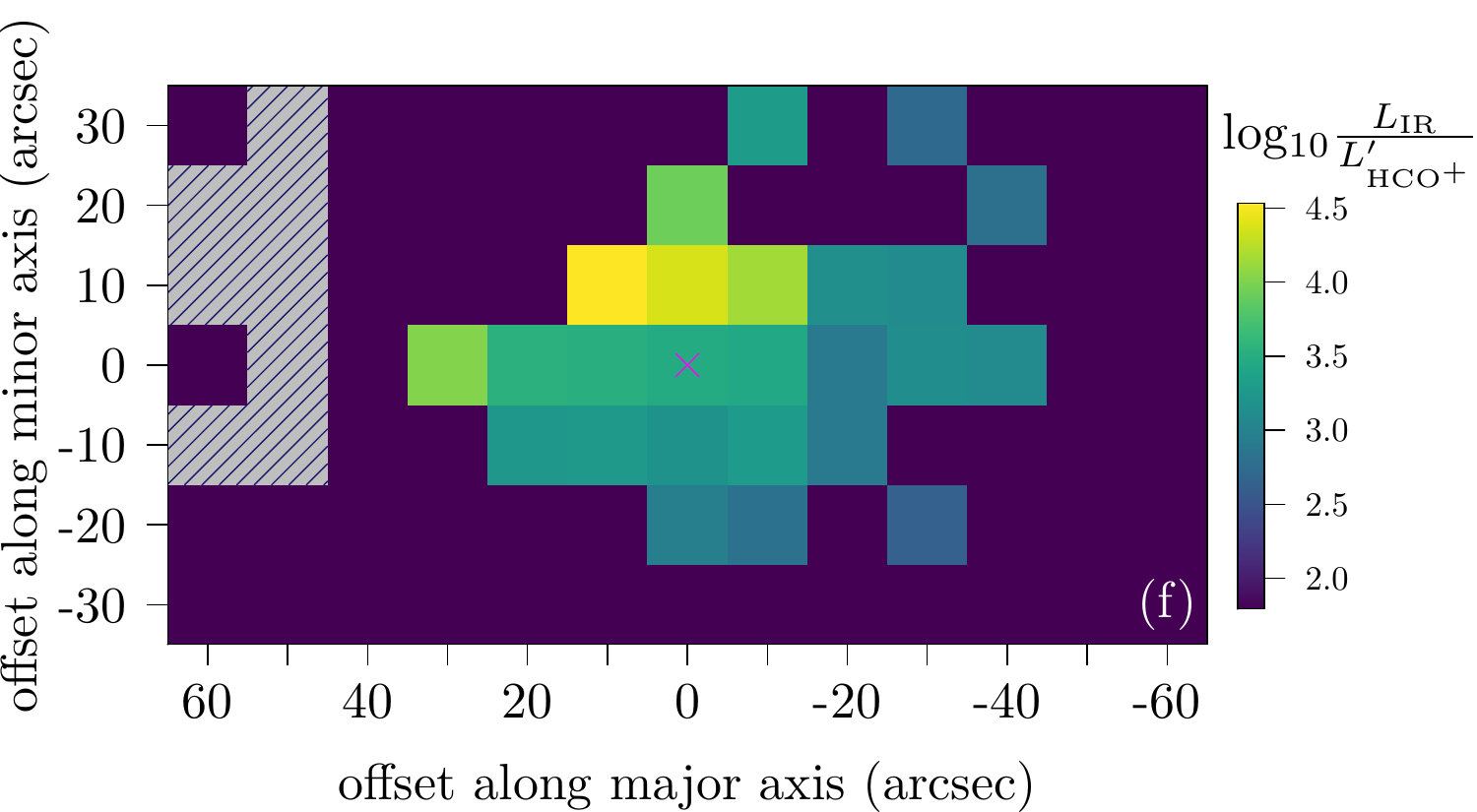}\\
\caption{Maps of the ratios of HCN 4-3/CO 1-0, HCO$^+$ 4-3/CO 1-0, CO 3-2/CO 1-0,
$L_\text{IR}/L'_\text{CO}$, $L_\text{IR}/L'_\text{HCN}$, and $L_\text{IR}/L'_{\text{HCO}^+}$.
The hatched regions indicate pixels without available data. Note that these images are rotated (see Fig.~\ref{fig:obs_pos}) to align with the major axis of NGC\,253.
\label{fig:ratio_map} }
\end{figure*}

% -------------- Results --------------------
\section{Results} \label{sec:results}

\subsection{Spectra}
\begin{figure}
\begin{center}
\includegraphics[width=\linewidth, angle=0]{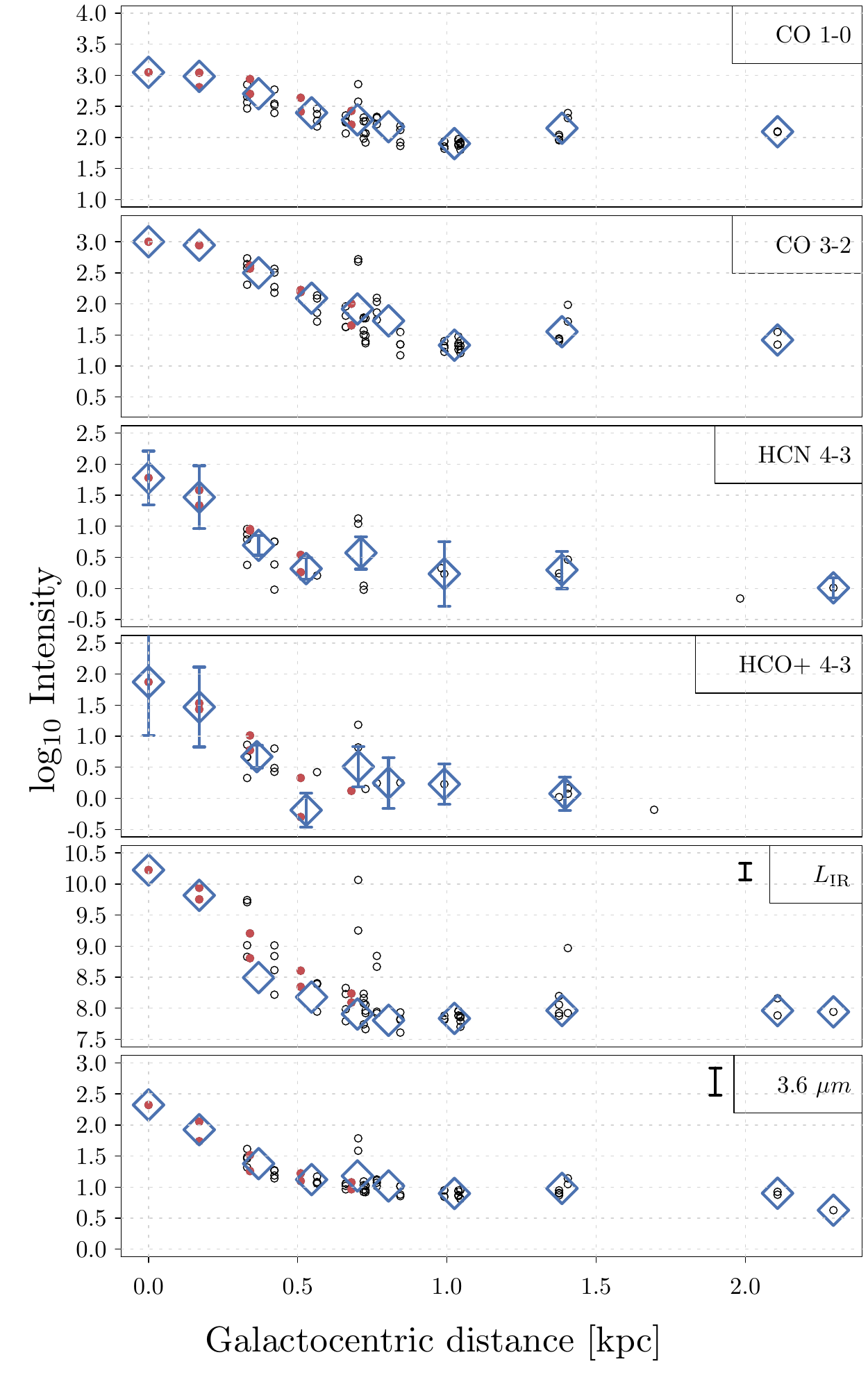}
\caption{Radial
distributions of different tracers used in this paper. 
The circles denote a data point extracted from each pixel in the final
13$\times$7 positions (Fig.~\ref{fig:spec}), and the red filled circles
indicate pixels situated on the major axis.
Blue diamonds denote the mean values within each 0.17-kpc radial bin. 
The units are $L_{\sun}$ for \lir, MJy\,sr$^{-1}$ for 3.6\,\micron, 
and K\,\kms\ for the molecular lines.
The measurement errors of most individual data points are typically $<$ 0.1\,dex.
For CO lines the uncertainty is dominated by the flux calibration (10 per cent).
For HCN 4-3 and \hcop\ 4-3 the uncertainty of the binning data points is
calculated based on the error from the individual pixels (see Section
\ref{subsec:profiles}). For \lir, the uncertainty is 30 per cent at most
\citep{Tan:2018}. For 3.6\,\micron, the uncertainty is considered to be
$\sim 50$ per cent. For the last two panels we denote the typical error bar in the
upper right corner.
\label{fig:int_profile}}
\end{center}
\end{figure}

Fig.~\ref{fig:spec} shows the HCN and HCO$^+$ 4-3 spectra at every
10-arcsec pixel position within the central 130$\times$70\,arcsec of
NGC\,253.  The figure also shows the CO 1-0 and 3-2 spectra at the same
positions. The CO spectra have been scaled down by a factor of 20 in order to
better compare the profiles with those of HCN and HCN$^+$. 
The 10-arcsec pixel size is the same as the beam spacing of the
observations, and results in a slightly undersampled map. In the central
pixel, the two dense-gas tracers are comparable to 1/20 of the peak intensity
of CO 1-0, but in pixels away from the centre, the relative strengths of HCN and
\hcop\ 4-3 quickly drop. The typical peak intensity ratio between HCN 4-3 (or
\hcop\ 4-3) and CO 1-0 that we can detect is about 1/100 in the outer parts.
In Section~\ref{subsec:profiles} and
Fig.~\ref{fig:int_profile} the radial profiles of these tracers are presented.

The different tracers have similar spatially integrated line centres and widths, indicating that
they originate from similar large-scale emitting regions. However, their line
profiles are different in many regions of the map. The line centres of CO 1-0
can be different from those of the other tracers by $\sim$100\,\kms\ in some
pixels, while the line centres of CO 3-2 are closer to those of HCN 4-3 and
\hcop\ 4-3.
While this difference is likely a result of the large optical depth of CO 1-0,
it could also be caused by the different excitation conditions of the emissions.
CO 3-2 traces warmer and denser gas than CO 1-0,
and its emission regions would be more similar to those of HCN 4-3 and \hcop\
4-3. High-spatial-resolution ALMA observations in the central 1.5\,kpc region
of NGC\,253 \citep{Meier:2015, Leroy:2015} have shown that the morphologies of
these molecules are quite different, in the sense that dense-gas tracers are
more compact and clumpy, while CO is much more diffuse and extended. The
measurements of the line intensities and their ratios are presented in Table~\ref{tab:1}.

Fig.~\ref{fig:stack_spec} shows the averaged spectrum (weighted by RMS) 
of the central 5$\times$3 pixels (top panel) and the stacked spectrum of those
non-detections (bottom panel). The central 5$\times$3 pixels contain $>$90 per
cent of the total flux of HCN 4-3 or \hcop\ 4-3. This total value will be used
as a global measurement later in Fig.~\ref{fig:sfr_M_all}. The stacking of
non-detections yields spectra with RMS $\sim$ 0.6 mK for HCN 4-3, 
and $1$\,mK for \hcop\ 4-3, respectively, and there is no sign of a robust
signal. Assuming a 200-\kms\ linewidth, the 3-$\sigma$ integrated intensities
correspond to \mhcn $<$ 10$^{6.0}$\,M$_{\sun}$ and \mhcop $<$
10$^{6.2}$\,M$_{\sun}$, respectively.

\subsection{Images of integrated intensities and their ratios}
\label{subsec:maps}

Fig.~\ref{fig:int_map} shows maps of the integrated intensities (moment
zero) of the different tracers used in this work. Contours starting from 30
per cent (0.2 dex) of the peak are overlaid. Although our observations are limited by
resolution and sensitivity, the differences in compactness between 
tracers are significant. \lir, HCN 4-3 and \hcop\ 4-3 show the most-compact
morphology, while CO 1-0 is the most extended. A quantitative comparison and
analysis are presented in Section~\ref{subsec:profiles}.

Fig.~\ref{fig:ratio_map} shows maps of the ratios of different tracers,
among which HCN 4-3/CO 1-0 (hereafter $R_\text{HCN43}$) and \hcop\ 4-3/CO 1-0
(hereafter $R_\text{\hcop43}$) can be treated as proxies
for the dense-gas fraction,
i.e., $R_\text{HCN43}$ and $R_\text{\hcop43}$ are proportional to
\fdense\ ($\equiv M_\text{dense}/M_\text{CO}$), but bear in mind that
the accuracy is largely limited by the conversion from line flux to mass
\citep{Usero:2015}, for both the dense gas (traced by HCN and \hcop), and CO.
We will discuss this in more detail in Section\ref{subsec:sfl}.
It is obvious that \fdense\ traced by either \rhcn\ or \rhcop\ is
much higher at the galaxy centre, and drops quickly toward the outskirts. The
integrated-intensity ratios in the centre are generally $<$ 0.06.
The integrated-intensity ratio of CO 3-2 and CO 1-0 (hereafter \rco) is
shown in Fig.~\ref{fig:ratio_map}c. This ratio shows an interesting
asymmetric morphology, which might be a result of the molecular outflow that 
was reported by \citet{Bolatto:2013}, though the outflow does not seem to
affect the dense gas, as judged from those images related to HCN or \hcop.
We will further discuss this in Section~\ref{sec:line_ratio}.

\lir/\lco\ can be treated as the molecular-gas star-formation efficiency, \sfemol 
$\equiv$ SFR/$M_\text{mol} = 1/\tau_\text{mol}$ ($\tau_\text{mol}$ is the depletion
time scale of the total molecular gas), while \lir/\lhcn\ and \lir/\lhcop\ can
be proxies of \sfedense. 
In Fig.~\ref{fig:ratio_map} they are shown in the last three panels (d-f).
We can see that, first, the peak of \sfemol\ is on the (0,10) pixel in panel d. Second,
\sfedense\ shows a more asymmetric morphology than \sfemol in panel e and f. Third,
regions to the
upper side (north-west) of the nuclei show about 0.5 dex higher \sfedense\ 
than the values in 
the centre. A similar behaviour was reported for M\,51, i.e.,
the \sfedense\ traced by \lir/$L_\text{HCN 1-0}$ shows a peak on its
northern spiral arm \citep{Chen:2015}. 
We can tell from Fig.~\ref{fig:int_map} that HCN 4-3 and \hcop\ 4-3
peak in the centre, so the asymmetric morphologies of \sfedense\ might be
mainly caused by the stronger \lir\ to the upper side (north-west).
This implies that the $\tau_\text{dep}$ of both the total molecular gas and
the dense gas in the north-west of the galaxy centre is shorter than that of
other regions of NGC\,253. The asymmetric distribution of \lir\ and SFE could
be attributed to the structure of the interstellar medium of the nuclei.
\citet{Bolatto:2013} reported an expanding molecular outflow in the centre of 
NGC\,253, and their data showed that the CO luminosity is approximately split
between the north (receding) and the south (approaching) sides of the outflow.
On the other hand, H$\alpha$ emission is predominantly seen on the south
side, and it is invisible on the north side probably due to obscuration.
Therefore, as shown in Fig.~\ref{fig:int_map}e, infrared
emission might dominate the north (receding) side, 
possibly because of more abundant dust than on the south 
(approaching) side. Our data are limited by modest resolution, and future
works combining H$\alpha$ and infrared data might help us more accurately
estimate \sfemol\ and \sfedense\ in the circumnuclear region of NGC\,253.

%--------------- Figure: line profile --------------

\begin{table}
\caption{$r_{90}$, $r_{50}$ and the concentration index $r_{90}$/$r_{50}$ derived from the radial profiles of the tracers
used in this paper. Statistical uncertainties (1$\sigma$) are given in parentheses.
}
\label{tab:cidx}
\centering
\begin{tabular}{cccc}
\hline
tracer & $r_{90}$ (kpc) & $r_{50}$ (kpc) &  $r_{90}$/$r_{50}$ \\
\hline
CO 1-0	   & 0.78 (0.15) & 0.38 (0.05) & 2.07 (0.48) \\ 
CO 3-2	   & 0.68 (0.07) & 0.30 (0.01) & 2.26 (0.27) \\
HCN 4-3	   & 0.58 (0.15) & 0.14 (0.01) & 4.00 (1.10) \\
\hcop\ 4-3 & 0.54 (0.18) & 0.10 (0.01) & 5.29 (1.91) \\ 
\lir       & 0.68 (0.12) & 0.33 (0.02) & 2.06 (0.39) \\ 
stellar    & 0.87 (0.13) & 0.37 (0.04) & 2.36 (0.44) \\ 
\hline
%\multicolumn{4}{l}{The last seven rows are positions that were only observed in JCMT-HARP's stare mode.}\\
\end{tabular}
\end{table}

% ---------- fit table -----------------
%\input{fit_results.tex}

% -------------------------------------------------------------------
% -------------------------------------------------------------------
% -------------------------------------------------------------------

%--------------- Figure line-ratio profiles --------------

\begin{figure}
\begin{center}
%In some cases \linewidth instead of \textwidth may be the better option. For
%example will it be the same as \textwidth in a single column document, but
%the same as \columnwidth in a two column document. \linewidth may also change
%in list environments, becoming smaller in nested lists.
\includegraphics[width=\linewidth, angle=0]{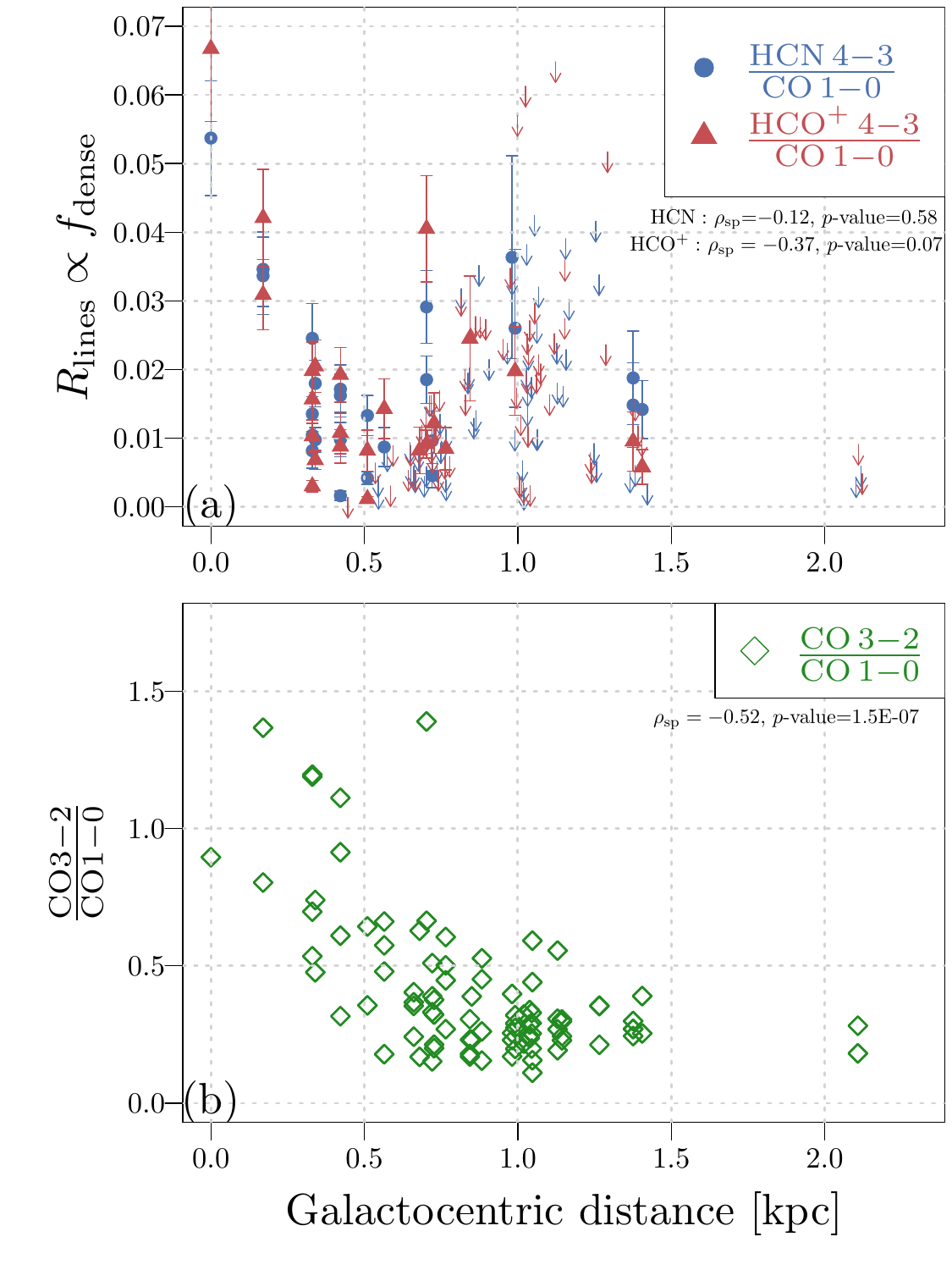}
\caption{
\textit{upper panel:} Radial distribution of dense-gas fraction
($f_\text{dense}$) traced by the integrated-intensity ratios $R_\text{HCN}$ =
HCN 4-3/CO 1-0 and $R_{\text{HCO}^+}$ = \hcop\ 4-3/CO 1-0. Arrows denote the
upper limits of \fdense. 
\textit{bottom panel:} Radial distribution of the ratio \rco\ between the two CO
transitions. 
\label{fig:fdense_profile} }
\end{center}
\end{figure}

% --------------   Figure dense-gas fraction vs. SFRsd  ---------------

\begin{figure}
\begin{center}
\includegraphics[width=\linewidth, angle=0]{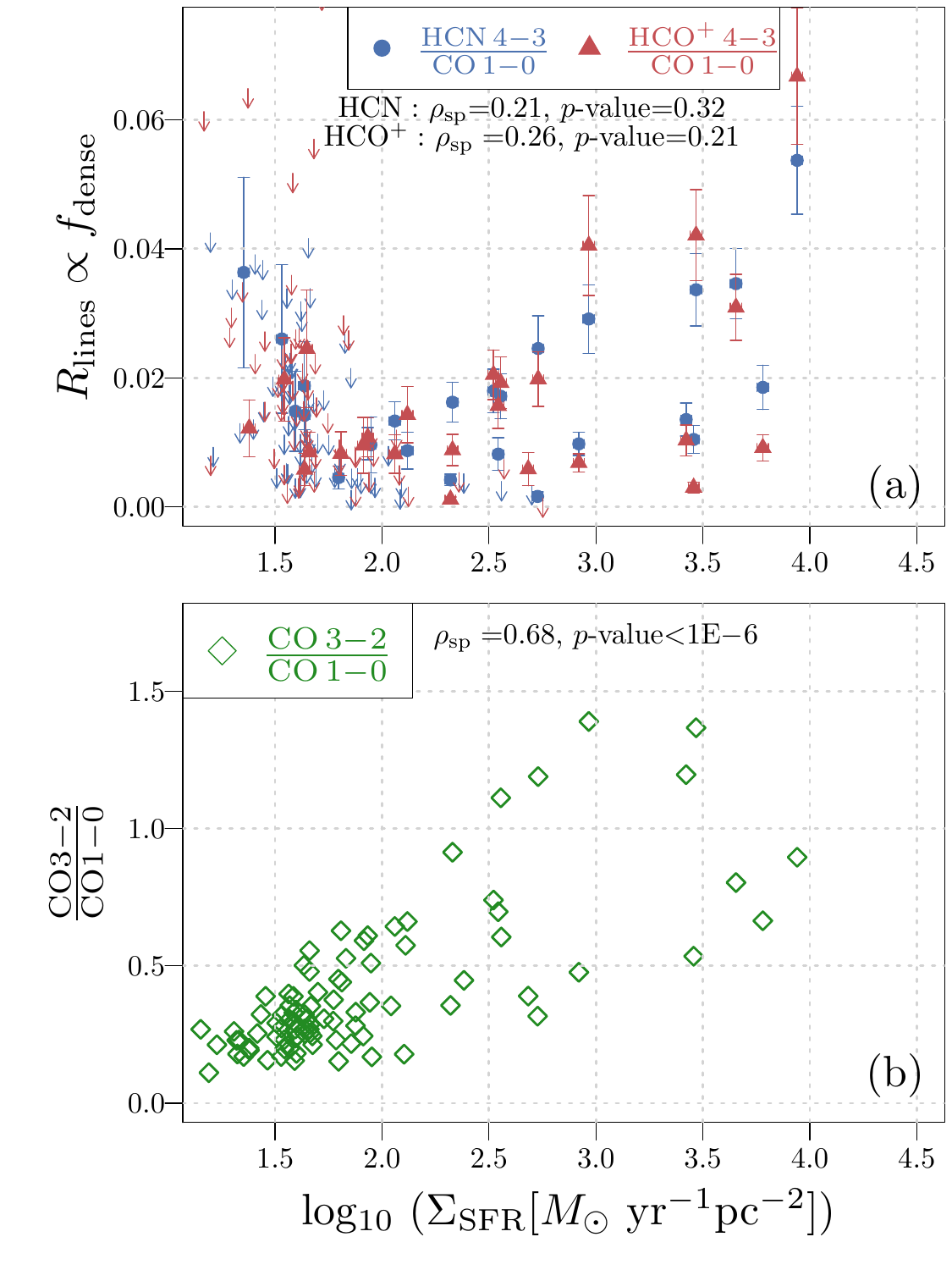}
\caption{
\textit{top panel:} $f_\text{dense}$ as a function of \sigmasfr (SFR per unit area).
 Arrows denote the upper limits of $R_\text{HCN}$ and $R_{\text{HCO}^+}$.
\textit{bottom panel:} \rco\ as a function of \sigmasfr. 
\label{fig:fdense_sfr}} 
\end{center}
\end{figure}

%--------------- Figure HCN/HCO+ profile --------------

% \begin{figure}[h]
% \begin{center}
% \includegraphics[width=\linewidth, angle=0]{./figures/hcn_hcop_ratio_profile.pdf}
% \caption{Radial profile of HCN/HCO$^+$ ratio along the major axis of NGC\,253.
% Arrows denote the upper and lower limits of the ratio.
% \label{fig:hcn_hcop} }
% \end{center}
% \end{figure}

\subsection{Radial profile and concentration index} 
\label{subsec:profiles}

Fig.~\ref{fig:int_profile} presents the radial profiles of all the tracers
used in this work. The radial distances are corrected for inclination, but note
that all pixels come from a highly elliptical beam 
and many are not independent measurements of emitting regions in the disc,	
so the profiles shown here are only indicative. 
Only pixels with SNR$>$3 are included.
We highlight those data points along the major axis with
red filled circles to guide the eye, and we find that they follow the main trend of most data points. 
%so at least in the inner $\sim 0.7$\,kpc region, the observed profiles do not seriously suffer from the inclination effects.

To obtain an averaged profile of each tracer, we calculate the mean intensity
within every 0.17-kpc distance bin (the beam spacing along the major axis), weighted by their measurement
error. The uncertainty of the mean value is calculated following equation 
(9) of \citet{Gallagher:2018a}, taking into account the error of each data
point, and the number of pixels within each bin:
\begin{equation}
\Sigma_{ii} = \frac{\sqrt{\sigma_1^2+\sigma_1^2+...+\sigma_N^2}}{N} \times \sqrt{O}
\end{equation}
where $\Sigma_{ii}$ is the uncertainty of the binned intensity, $N$ is the
number of pixels in the bin, and $O$ is the oversampling factor accounting for
the nonindependence of the pixels. $O$ = 1.4 in our work as the pixel size is 10$\arcsec$
and the resolution is 14$\arcsec$).
%The CO 1-0 distribution is the most extended along the galaxy disc. At $\sim$ 1 kpc from the centre, the integrated intensity of CO 1-0 is about 10 per cent of that in the galaxy centre.

It is interesting to see that, while all the tracers' intensity distributions
show significant decreasing trends towards larger radii, CO 3-2, \lir, HCN 4-3,
and \hcop\ 4-3 seem to have steeper gradients than CO 1-0 and the stellar
component traced by 3.6-\micron emission. For example, at $\sim$
0.5\,kpc from the galaxy centre, the integrated intensities of CO 3-2, \lir, HCN
4-3, and \hcop\ 4-3 are only approximately 10 per cent of those in the
centre. For comparison, the integrated intensities of CO 1-0 at 0.5\,kpc is
about 30 per cent of the central value (also see Fig.~\ref{fig:int_map}). The
difference indicates that, the spatial distribution of CO 1-0 is the most
extended among all the tracers.
While CO 1-0 emission comes from the cold and diffuse molecular gas, CO 3-2
and the two dense-gas tracers require denser and warmer gas, which tends to
reside in the galaxy centre. This is also the case for the infrared emission,
which traces cold dust emission that is closely associated with star-formation activity.

To establish a quantitative method to analyse the radial profiles of the
different tracers observed in NGC\,253, and to further apply this to the other
sources of the MALATANG survey in our future works, we fit the concentration
indices $C \equiv r_{90}/r_{50}$ to the radial distributions
\citep{de-Vaucouleurs:1977,Li:2011}. 
For HCN 4-3 and \hcop\ 4-3 only detections with radial distance $<1.4$\,kpc are 
included in the fits (see Fig.~\ref{fig:growth_curve}).
$r_{90}$ and $r_{50}$ are the radii that
encompass 90 and 50 per cent of the total flux of each tracer, respectively. The
total flux is derived from fitting the asymptotic intensity of a curve of
growth, following the method of
\citet{Munoz-Mateos:2009}. In Fig.~\ref{fig:growth_curve}
we show two examples of how the curve of growth and the asymptotic intensities are derived.

The fitted parameters are listed in Table~\ref{tab:cidx}. 
HCN 4-3 and \hcop 4-3 appear to have smaller $r_{90}$ than the other tracers,
and they also show significantly higher concentration indices.
Note that our resolution is 0.24\,kpc so the observed values are spatially 
smoothed, which might not reflect the intrinsic concentration parameters.
\citet{Leroy:2015} used high-resolution (FWHM $\sim 2\times$2\,arcsec) data of
NGC\,253 in the 3\,mm band, and they derived $r_{90}$ and $r_{50}$ for CO 1-0
(0.4 and 0.15\,kpc). These values are lower than ours ($\sim$0.78
and 0.38\,kpc), but the ratio $r_{90}$/$r_{50}$ is similar. 
For dense gas, they derive
$r_{90}$ and $r_{50}$ (0.3 and 0.1\,kpc, respectively) by averaging the 1-0
transitions of HCN, \hcop, and CS. Note that their dense-gas observations
lack short spacings data, so certain amount of emission on large spatial-scales
is missing. As a result their $r_{90}$ and $r_{50}$ might 
be overestimated. If we average the $r_{90}$ and $r_{50}$ 
for HCN 4-3 and \hcop\ 4-3, we get 0.54 and 0.12\,kpc, respectively.
We note that the telescope beam ($\sim 0.17$\,kpc) is larger than the
fitted $r_{50}$ of HCN 4-3 and \hcop 4-3, so they have large uncertainties 
and are likely overestimated by our data.
This implies that the dense-gas tracers might be too compact for the JCMT to
resolve their $r_{50}$.
Moreover, we speculate that, since the higher transition lines are
excited in more-compact clumps with higher gas density, their $r_{90}$ and
$r_{50}$ should be smaller while the ratio $r_{90}$/$r_{50}$ might be higher,
compared with their low-$J$ lines.

The $J=4-3$ emission lines require higher excitation temperatures and higher
critical densities, and in regions with different kinematic temperature
($T_\text{K}$), the effective critical densities ($n_\text{eff}$) for the
excitation of a certain line will change dramatically \citep{Evans:1999}. Thus
the large scatter between the centre and disc regions not only reflects the
change in dense-gas abundance, but it could also be a result of the distinct
excitation environments.
While few studies have explored the optical depths of dense-gas tracers in
galactic disc regions, they have been suggested to be optically thick in
galaxy centres \citep{Greve:2009,Jiang:2011,Jimenez-Donaire:2017} and nearby
ULIRGs \citep{Imanishi:2018}. At this
stage it is still unclear whether HCN and \hcop\ are optically thin in
galactic disc regions. So the profiles shown here do not necessarily reflect
their column densities.
%If we assume HCN and \hcop\ are optically thin in the disc regions, their intensity might be a good proxy of the mass or column density of dense gas. Thus the intensity ratio between the disc and the centre might suggest a upper limit of the column density ratio of dense gas. i.e., the column density of dense gas might be much less abundant in the disc. 
Accurate estimates of the gas temperature and optical depth of these lines are
needed to reveal the true population distribution of spectral energy levels
and their excitation conditions.

Note that the analysis using radial profiles is based on the simplified
assumption that regions with the same galactocentric distances have
similar properties. This is not always the case, especially substructures such
as rings, bars and spirals in the circumnuclear regions cannot be well recovered
by this approach. Radial profiles are useful diagnostics for samples with
intermediate-to-low spatial resolution, but we note that they can only serve
as a first-order approximation for the relationships between physical properties
and galactocentric distance. 
The result presented in this paper is uncertain and only indicative ,
and we look forward to investigating this method for large samples in future
studies  at higher resolution.

% Sect 3.4
\subsection{Line ratio and dense-gas fraction}\label{sec:line_ratio}

\citet{Tan:2018} reported the integrated-intensity ratios of $R_\text{HCN43}$
and $R_\text{\hcop43}$ in different regions of six galaxies. The ratios lie
between approximately 0.003 and 0.1, and they are higher in the centre and
lower in outer regions. Take $R_\text{HCN43}$ in galactic centres for example,
among their six galaxies, NGC\,1068 and NGC\,253 (a possibly weak AGN,
\citealt{Muller-Sanchez:2010}) have the highest ratios in the nuclei which are
slightly less than 0.1. In IC\,342, M\,83, and NGC\,6946, the ratios in the nuclei
are about 0.01 to 0.015. In the starburst galaxy M\,82, the nuclear ratio is
about 0.02 (see fig.~5 in \citealt{Tan:2018} for more details).
\citet{Gallagher:2018a} also resolved $R_\text{HCN10}$ and $R_\text{\hcop10}$
of four local galaxies, and their ratios are also in a similar range, while
$R_\text{HCN10}$ appears to be modestly higher than $R_\text{\hcop10}$ in
general (see their fig.~9). For more discussion of the line ratios and their
implications, refer to \citet{Izumi:2016} and references therein.

In Fig.~\ref{fig:fdense_profile} we show the radial profiles of the
integrated-intensity ratios of the molecular lines used in this work. The ratios of
$R_\text{HCN43}$ and $R_\text{\hcop43}$ are, to first order,
proportional to the dense-gas ratio \fdense, and we emphasize that \fdense\ 
here is different from that traced by HCN 1-0 and \hcop\ 1-0, since the
different transitions require different densities. Similar to
Fig.~\ref{fig:ratio_map}, \fdense\ peaks in the centre, but drops quickly to
only about one sixth of the central value at 0.5\,kpc.
This is consistent with the result from \citet{Jackson:1995} that the most 
highly-exited gas is confined to the inner 0.5\,kpc nuclear region. They
implied that the density distribution of the
molecular gas is distinct in the circumnuclear region, where the
high-density-gas fraction
is likely to be several times higher than in the outskirts. In other words,
compared with most of the galaxy disc, high-density gas (possibly in the form of
clumps) resides preferentially in the galaxy centre. This plot of the
decreasing \fdense\ with radius in the galactic disc also implies 
that the filling factor of dense gas is much smaller than that of the bulk of
molecular gas traced by CO \citep{Paglione:1997, Leroy:2015}. Thus we suggest that for
extragalactic observations telescopes with a smaller beam are more
suitable to detect dense-gas emission, as larger beams might suffer more from
beam dilution.

The CO 3-2/1-0 ratio (\rco) as a function of radius is also shown in
Fig.~\ref{fig:fdense_profile}. 
It is obvious that \rco\ drops quickly at larger radii, similar to
the decreasing trend of \fdense. 
\rco\ is close to one in the centre, which is several times higher than the \rco\
at 1\,kpc ($\sim 0.25$). We also note in Fig.~\ref{fig:ratio_map}c that on a few pixels
to the west side of the centre, \rco\ is higher than the value at the
(0,0) position. The pixel on the south-west (10 arcsec to the right side of
the disc) of the centre appear to be coincident with the position of the CO
outflow reported by \citet{Bolatto:2013}.
Compared with CO 1-0, CO 3-2 emission requires a
higher excitation temperature and a higher critical density, so \rco\ gives
important information on the excitation conditions of the molecular gas.

In previous studies, \rco\ and $R_\text{21}$ (CO 2-1/1-0 ratio) 
have been reported and are widely used to convert $I_\text{CO 3-2}$ and $I_\text{CO 2-1}$ to
$I_\text{CO 1-0}$ to estimate the total molecular-hydrogen mass
\citep{Leroy:2009, Mao:2010, Wilson:2012}.
\rco\ was reported to be in the range 0.2--1.9 (mean value = 0.81) 
in galaxy-integrated observations \citep{Mao:2010}. In resolved observations of nearby galaxies the mean
\rco\ is found to be 0.18 with a standard deviation of 0.06 \citep{Wilson:2012}.
Our values of \rco\ lie in the range from these works, and 
the variation of \rco\ in the central region of
NGC\,253 shows that this ratio is obviously dependent on galactic environments. 
In strongly star-forming
galactic centres where the conditions resemble those in luminous infrared
galaxies (LIRGs), CO 3-2 emission is easily enhanced and \rco\ is naturally
much higher than in the quiescent environment of discs \citep{Mao:2010}. We
suggest that the scatter of \rco\ among galaxies is a natural result of the
variation of \rco\ among different regions of any single galaxy. 
For single-dish observations that do not resolve the molecular gas of galaxies, one can
only obtain a spatially averaged \rco, and in galaxies with more denser and
warmer molecular gas, such as LIRGs, the averaged \rco\ tends to be higher.
Also, the filling factors of CO 3-2 and CO 1-0 for single-dish observations
are dependent on the physical scale covered by the beam, and this observed
effect can also contribute to part of the scatter of the observed \rco.
If \rco\ and $R_\text{21}$ are used to convert $I_\text{CO 3-2}$ and $I_\text{CO
2-1}$ to $I_\text{CO 1-0}$ to
estimate the total molecular-hydrogen mass,
we caution that one must take into account the variation of \rco\ or
$R_\text{21}$ as an important uncertainty in the conversion.

% \subsection{Line ratios and star-formation rate surface density} % (fold)
% \label{subsec:sfsd}

Fig.~\ref{fig:fdense_sfr} plots line ratios as a function of SFR surface
density, \sigmasfr, so as to explore the relationships of \fdense\
versus \sigmasfr, and \rco\ versus \sigmasfr. 
In Fig.~\ref{fig:fdense_sfr}a, \fdense\ appears to increase
for higher \sigmasfr, especially when we only look at data points along the 
major axis (circles). Limited by the SNR we could not obtain reliable 
\fdense\ in the lower \sigmasfr\ regime, where most of the data points come from the outer
disc region. On the other hand, we have relatively higher SNR for the two CO
lines, so in the bottom panel we are able to show \rco\ in all pixels used in 
this work, and to explore the lower-\sigmasfr\ regime.
We can see that \rco\ shows smaller scatter in the lower-\sigmasfr\ regime, 
and its scatter increases significantly with increasing \sigmasfr. 
For pixels with \sigmasfr\ $< 10^2 \text{M}_{\sun}\,\text{yr}^{-1}\,\text{pc}^{-2}$, 
\rco\ is mostly lower than 0.5, while near the galaxy centre \sigmasfr\ 
is about 100 times higher and \rco\ can be as high as 1.5. Again note that 
a few pixels have higher \rco\ than the central pixel, and they are
coincident with the position of
the CO outflow reported by \citet{Bolatto:2013}. In Fig.~\ref{fig:ratio_map}
we also note that the \rco\ map exhibits an asymmetric morphology.
Thus we speculate that the molecular outflow might
dominate the large scatter in the higher-\sigmasfr\ regime of the plot and the
more-active environment in the central $\sim 200$\,pc region.

The two plots in Fig.~\ref{fig:fdense_sfr} suggest that \fdense\ and \rco\ are more
likely to be higher in regions with higher \sigmasfr, i.e., the more-active
environment in the central $\sim 200$\,pc of NGC\,253. However, there are also
regions with high \sigmasfr\ accompanied with low \fdense\ and low \rco.
Considering that SFR is more directly related to the mass of dense/warm gas, 
we expect to see tighter correlations between HCN (or HCO$^+$) and \sigmasfr\
(see Section~\ref{subsec:sfl}),
as well as between CO 3-2 and \sigmasfr\ \citep{Wilson:2012}. 
So we speculate that the scatter in \fdense\
and \rco\ is likely caused by the variation of the CO intensity among
different regions.

% subsubsection _ (end)

% -------------------------------
% --------------- Discussions ----------------
% -------------------------------

\section{Discussion} \label{sec:discussion}

The spatially resolved dense-gas emission of NGC\,253 enables us to analyse how the
galactic environment affects the relationship between dense gas and star
formation \citep{Usero:2015}. In the following, we will first discuss the
dense-gas star-formation relationships, and how varying dense-gas conversion
factors affect the relationships. Then the role of stellar components in the
dense-gas fraction \fdense\ and the dense-gas star-formation efficiency,
\sfedense\ is discussed in Section~\ref{subsec:star}.

% --------------- Figure SF relationships -----------------
\begin{figure*}
\begin{center}
\includegraphics[width=\textwidth, angle=0]{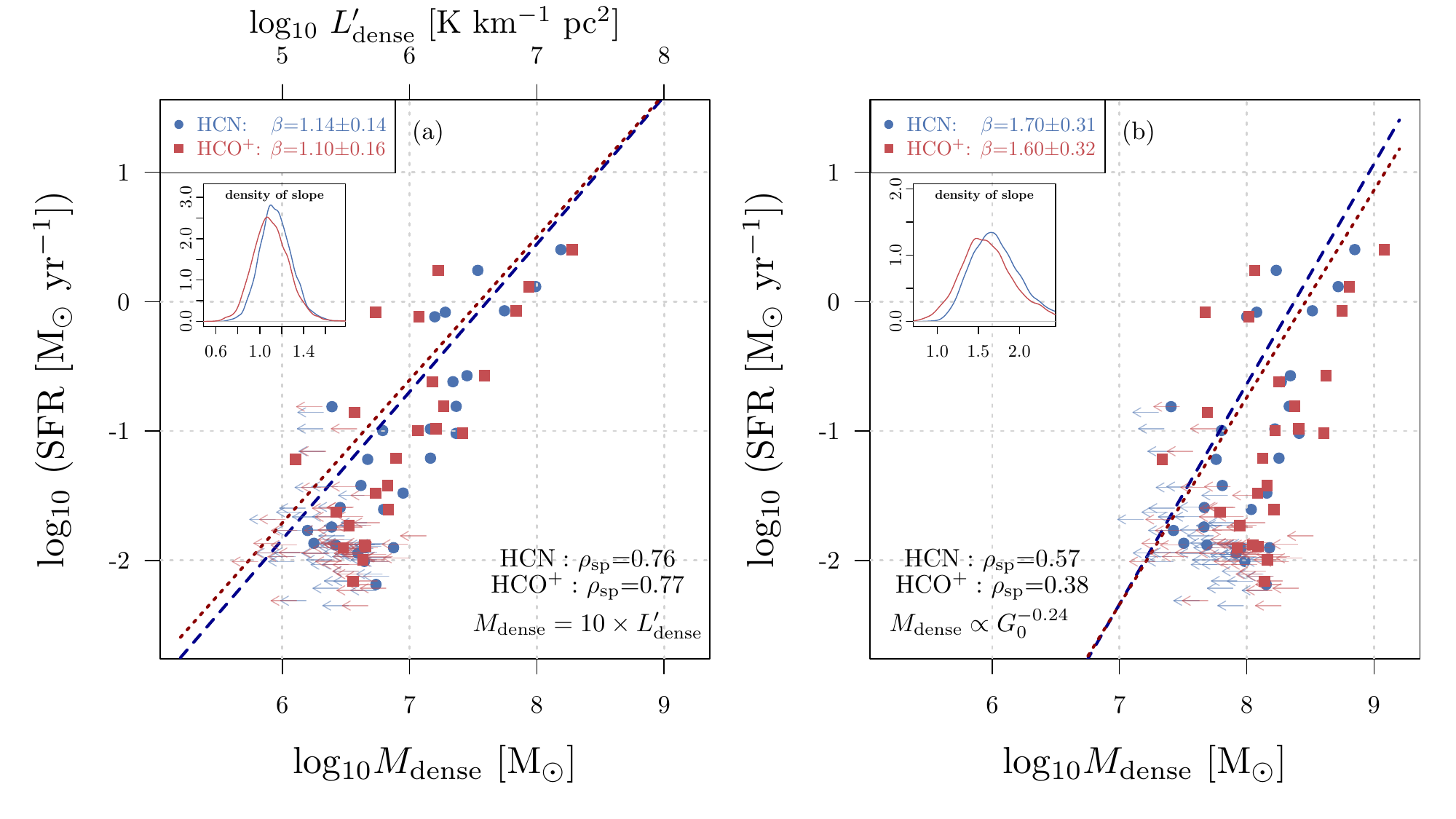}
\caption{Dense-gas star-formation relationships (SFR versus \mdense) using different
conversion factors (\alphadense). \mdense\ in the left panel is calculated using a
constant \alphadense, 
and the line luminosity $L'_\text{dense}$ is shown as the original observable on 
the top x-axis in panel (a).
 \mdense\ in the right panel is calculated
using a conversion factor \alphadenseG, which is a function of the radiation
field intensity $G_0$. 
Because in panel (b) the \alphadense\ depends on $G_0$
we do not show the axis for $L'_\text{dense}$.
The regression fits are plotted as blue and red dashed lines, matching 
the respective symbol colour of the HCN and \hcop\ data points.
The slopes $\beta$ are given by the Bayesian method for
linear regression \citep{Kelly:2007}, accounting for uncertainties and upper
limits. See the text for more details.
\label{fig:sfr_M} }
\end{center}
\end{figure*}

% --------------- Figure SF relationships for all sample-----------------
\begin{figure}
\begin{center}
\includegraphics[width=\linewidth, angle=0]{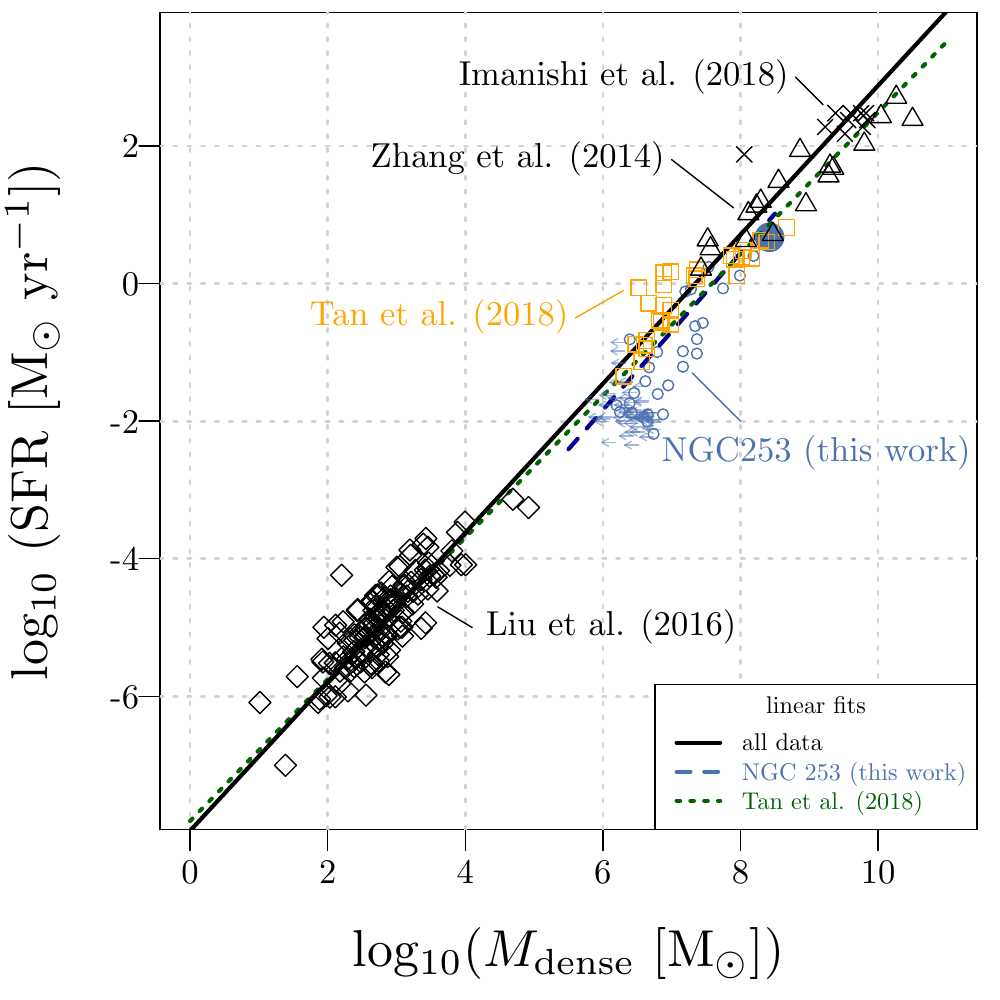}
\caption{Dense-gas star-formation relationships (SFR versus \mdense) for all HCN 4-3
samples compiled in this work. 
The filled blue circle is the sum value of the central $\sim 0.9$\,kpc of NGC\,253. 
The black solid line is a linear fit 
(in logarithmic scale) for
all data ($\beta=1.08\pm0.01 $). The blue dashed line is a fit for NGC\,253
data from Fig.~\ref{fig:sfr_M}a ($\beta=1.13\pm0.14$). The dotted line is
the fit ($\beta=1.03\pm0.01$) from \citet{Tan:2018}.
\label{fig:sfr_M_all} 
}
\end{center}
\end{figure}

%----------------------------------------------------------------
\subsection{Dense-gas star-formation relationship} \label{subsec:sfl}
%----------------------------------------------------------------

\citet{Tan:2018} discussed the dense-gas star-formation relationship, using 
the six galaxies of the MALATANG sample that were mapped by JCMT. The whole
sample shows a linear correlation between log$_{10}$\lir\ and 
log$_{10}$\ldense, and for HCN 4-3 the fitted slope $\beta$ is nearly unity.
However, the scatter about the relationship is not negligible, and 
the slopes are likely different among individual galaxies. 
Such a variation in the
ratio of the SFR and the mass of the dense molecular gas (\mdense) suggests
a varying star-formation efficiency, as
discussed in recent studies \citep{Usero:2015,Gallagher:2018a}. 

\citet{Shimajiri:2017} derived an empirical relationship between the dense-gas
conversion factors \alphadense\ $\equiv M_\text{dense}/L'_\text{dense}$
and the local far-UV radiation field, $G_0$, which can be
derived from \textit{Herschel} 70- and 100-\micron\ intensities. Here we adopt
this varying conversion factor \alphadenseG\ for our data and compare it with a constant
\alphadense\ to revisit the dense-gas SF relationship in NGC\,253,
using the full mapping data of HCN 4-3 and \hcop\ 4-3 with $\sim$ 0.24-kpc
resolution.
Hereafter we use \alphadenseG\ to distinguish it from the constant
\alphadense\ used by \citet{Gao:2004a,Gao:2004b} and most other studies. 
To explore how the dense-gas conversion factors
affect the relationship, two sets of \mdense\ are calculated based on
\alphadenseG\ and the constant \alphadense, respectively, then we compare
the relationships plotted for the different \mdense. 
Note that this approach relies on the assumption that NGC\,253 conforms to
Milky Way values in terms of gas and dust properties, but there is a large
uncertainty in this assumed similarity. Please see
\citet{Leroy:2018} and \citet{Knudsen:2007} for comparisons of star-forming
properties between NGC\,253 and the Milky Way.

We adopt the following
relationships from \citet{Shimajiri:2017}:
\begin{equation}
\alpha(\text{HCN}) = (496 \pm 94) \times G_0 ^{-0.24\pm0.07} 
[\text{M}_{\sun}\,\text{(K\,km}^{-1}\,\text{pc}^2\text{)}^{-1}]
\end{equation}

\begin{equation}
\alpha(\text{HCO}^+) = (689 \pm 151) \times G_0 ^{-0.24\pm0.08} 
[\text{M}_{\sun}\,\text{(K\,km}^{-1}\,\text{pc}^2\text{)}^{-1}].
\end{equation}

$G_0$ is calculated from \textit{Herschel}/PACS 70- and 100-\micron\ data
(here B is the bandwidth of the \textit{Herschel}/PACS
filters at 70 and 100\,\micron):
\begin{equation}
G_0 %= \frac{4\pi I_\text{FIR}}{1.6 \times 10^{-3} [\text{erg}\, \text{cm}^{-2}\, \text{s}^{-1}]} 
= \frac{4\pi I_\text{FIR}}{1.6 \times 10^{-26} [\text{Jy}\, \text{Hz}]},
\end{equation}
where:
\begin{equation}
\begin{aligned}
I_\text{FIR} 
%&= \left(\frac{F_{70\,\micron}}{[\text{erg\,cm}^{-2}\,\text{s}^{-1}\,\text{Hz}^{-1}\,\text{sr}^{-1}]} 
% \times \frac{B_{60\,\micron - 80\,\micron}}{[\text{Hz}]}\right) +\\
% &\quad \left(\frac{F_{100\,\micron}}{[\text{erg\,cm}^{-2}\,\text{s}^{-1}\,\text{Hz}^{-1}\,\text{sr}^{-1}]} 
% \times \frac{B_{80\,\micron - 125\,\micron}}{[\text{Hz}]}\right) \\
% &\quad \left[\text{erg\,cm}^{-2}\,\text{s}^{-1}\,\text{sr}^{-1}\right]\\
&= \left(\frac{F_{70\,\micron}}{[\text{Jy}\,\text{sr}^{-1}]} 
\times \frac{B_{60 - 80\,\micron}}{[\text{Hz}]}\right) +
\left(\frac{F_{100\,\micron}}{[\text{Jy}\,\text{sr}^{-1}]} 
\times \frac{B_{80 - 125\,\micron}}{[\text{Hz}]}\right) \\
&\quad \left[\text{Jy}\,\text{Hz}\,\text{sr}^{-1}\right]
.
\end{aligned}
\end{equation}

Our calculation shows that the median of $\alpha_\text{HCN}(G_0)$ in 
NGC\,253 is 
$\sim 25$\,M$_{\sun}\,(\text{K\,km\,s}^{-1}\,\text{pc}^2)^{-1}$, 
which is a factor of 2.5 higher than the constant conversion factor,
\alphadense\ = 10\,M$_{\sun}\,(\text{K\,km\,s}^{-1}\,\text{pc}^2)^{-1}$, used in
previous studies \citep{Gao:2004a,Wu:2005}.
We adopt a luminosity ratio of 0.3 to convert from $J=4-3$ to $J=1-0$ 
for HCN and \hcop\
following \citet{Tan:2018}. In Fig.~\ref{fig:sfr_M} we
plot SFR as a function of \mdense. In the left panel, \mdense\ is calculated
using the varying \alphadenseG, while the right panel uses
the constant \alphadense\ = 10. The
Bayesian method coded in the \textsc{idl} routine \textsc{linmix\_err} provided by
\citet{Kelly:2007} was used for linear regression of the data. The method accounts
for both uncertainties and upper limits. This is important especially for observations
of the weak molecular lines, of which a significant part of the measurements
have low SNR, as other methods not accounting for upper limits are biased to
high-SNR data. The density distributions of the fitted slopes are
shown as insets in the plot. Following \citet{Kelly:2007}, the posterior
median is adopted as an estimate for the parameter, 
and the median absolute deviation of the posterior distribution
is used as an error of the parameters.

Comparison between the two panels of Fig.~\ref{fig:sfr_M} shows that, 
when adopting \alphadenseG, slopes of the dense-gas SF relationship become much 
steeper than those based on the constant \alphadense. $\beta$(HCN) and 
$\beta$(HCO$^+$) are $1.14\pm0.14$ and $1.10\pm0.16$ for constant
\alphadense, 
while we obtain, $1.70\pm0.31$ and $1.60\pm0.32$, respectively, when
adopting \alphadenseG.
The difference between the two sets of data is mainly
caused by data points with lower SFR, for which \mdense\ values become
$\sim$0.5\,dex higher than \mdense\ estimated from fixed \alphadense.

Although the dynamic range of our data for NGC\,253 is limited to 
three orders of magnitude, Fig.~\ref{fig:sfr_M}a shows a
nearly linear correlation between SFR and \mdense for both HCN 4-3 and \hcop\ 4-3.
The difference between our fitted slopes and those demonstrated based on a much
larger dynamic range \citep[e.g.,][]{Tan:2018} is within the fitted errors.
However, while the relationship based on \alphadenseG\ shows a strong
correlation, it is obviously not a one-to-one relationship. In addition
to the uncertainty in the conversion factors themselves, 
the luminosity ratio between
4-3 and 1-0 transitions of HCN and \hcop ($L_{4-3}$/$L_{1-0}$) that we adopt
might also be a major source of uncertainty, since in the galaxy centre
$L_{4-3}$/$L_{1-0}$ is probably higher than in the disc \citep{Tan:2018}.
Also note that the excitation conditions and density distribution of the molecular
gas are not well constrained, so discussions based on high transition
emission line alone is far from accurate for revealing the physical properties of
star-forming gas. In Fig.~\ref{fig:sfr_M}a we therefore plot in addition the line
luminosity $L'$ on the top x-axis to show the actual measurements.

In Fig.~\ref{fig:sfr_M_all} we show a updated version of the SF relationship
from \citet{Tan:2018}, using NGC\,253 HCN 4-3 data from this paper, and new
HCN 4-3 data of LIRGs from \citet{Imanishi:2018}.
Instead of \lir\ and \ldense\ used in \citet{Tan:2018}, here we use SFR and \mdense.
In Fig.~\ref{fig:sfr_M_all}, the black solid line ($\beta$ = 1.08 $\pm$ 0.01) 
is given by the Bayesian method \textsc{linmix\_err}, and the blue dashed line is
the same fit for NGC\,253 data alone (from Fig.~\ref{fig:sfr_M}a, using the
constant \alphadense). We can see that the fit for NGC\,253 is slightly shifted
from the overall fit, suggesting that for certain \mdense\ they tend to have
lower SFR (by about 0.2\,dex), or the overall \sfedense\ in NGC\,253 is lower.
\citet{Jackson:1995} suggested that the overall 
gas density in NGC\,253 is at least 10 times higher than that in M\,82. 
So it is likely there is more dense gas in NGC\,253, and this could 
explain the higher \mdense\ at certain SFR comparing with M\,82 in this plot.
Finally, it is worth noting that with higher resolution we are looking at smaller 
regions in galaxies, and the scatter would become more prominent, 
due to the fact that star-formation events in these regions
take place during different epochs. Therefore, less dispersion is expected
in correlations using data from entire galaxies where we see averaging values of
parameters, although rare excursions can be also observed \citep{Papadopoulos:2014}.
% \subsection{Dense-gas conversion factor}\label{subsec:alpha-dense} 

The uncertainty associated with converting the CO emission to the mass or column density of the
total molecular gas (\alphaco\ or $X_\text{CO}$) has been extensively studied.
\alphaco\ is likely affected by a number of physical
conditions, such as the value of the gas surface density in giant molecular clouds,
$\Sigma_\text{GMC}$, the
brightness temperatures $T_\text{B}$ of the emitting gas, the distribution
of GMC sizes \citep{Bolatto:2013a}, the effect of cosmic rays and the metallicity. Although
\alphadense\ is poorly constrained, at least qualitatively, the environment
must also play a role in the uncertainty in the mass-to-light ratio of
dense-gas tracers. We note that adopting \alphadenseG\ in Fig.~\ref{fig:sfr_M}
relies on an assumption that the empirical
relationship between \alphadense\ and FIR intensity proposed by
\citet{Shimajiri:2017} holds for our observations on $\sim$ sub-kpc scales. 
However, it is unclear how to quantify the effect of the radiation field on
\alphadense\ across $>$ 200\,pc scales.
Calibrations of \alphadense\ are beyond the scope of this paper, but we note that high
spatial resolution and multiple-transition data of the dense-gas tracers
together might provide more information about the parameters necessary for
calibrating \alphadense, such as surface density, brightness temperature, and
the sizes of dense clumps. With regard to theoretical models that may explain 
variations in \alphadense\ and SFE, see \citet{Usero:2015} for a thorough
discussion.

% \begin{equation}
% \alpha_\text{CO} \propto \frac{\rho^{0.5}}{T_\text{B}}
% \end{equation} \label{eq:alphaco}

% --------------   Figure SFE vs. fdense  ---------------
%\renewcommand{\familydefault}{\sfdefault}
%\renewcommand{\familydefault}{\rmdefault}\normalfont
\begin{figure*}
%  \floatsetup{font={sf}}
% \input{./tex/SFE_fdense.tex}
\begin{center}
\includegraphics[width=\textwidth, angle=0]{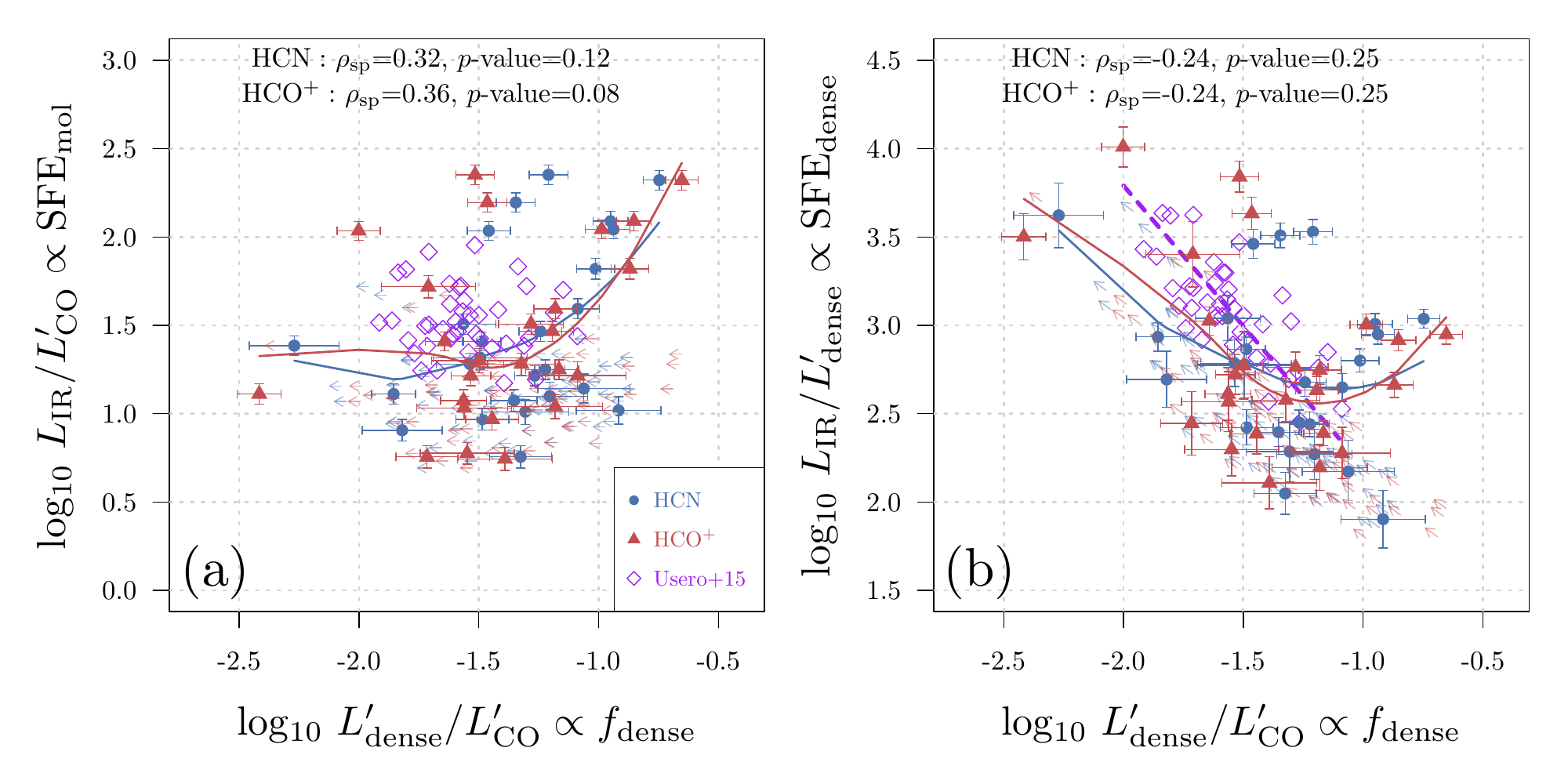}
\caption{\sfemol\ versus. $f_\text{dense}$ (left) and \sfedense\ versus. \fdense\ (right). The Spearman correlation coefficients $\rho_\text{sp}$ and $p$-values of the hypothesis test
are presented. Arrows show lower limits to the \sfedense\ and the corresponding upper limits to \fdense.
The lines are Local Polynomial Regression fits. This is essentially similar to binning points, just to guide the eye. 
\label{fig:sfe_fdense}}
\end{center}
\end{figure*}

%----------- Figure dense-gas ratio vs. SFR surface density ----------
\begin{figure*}
\begin{center}
\includegraphics[width=\textwidth, angle=0]{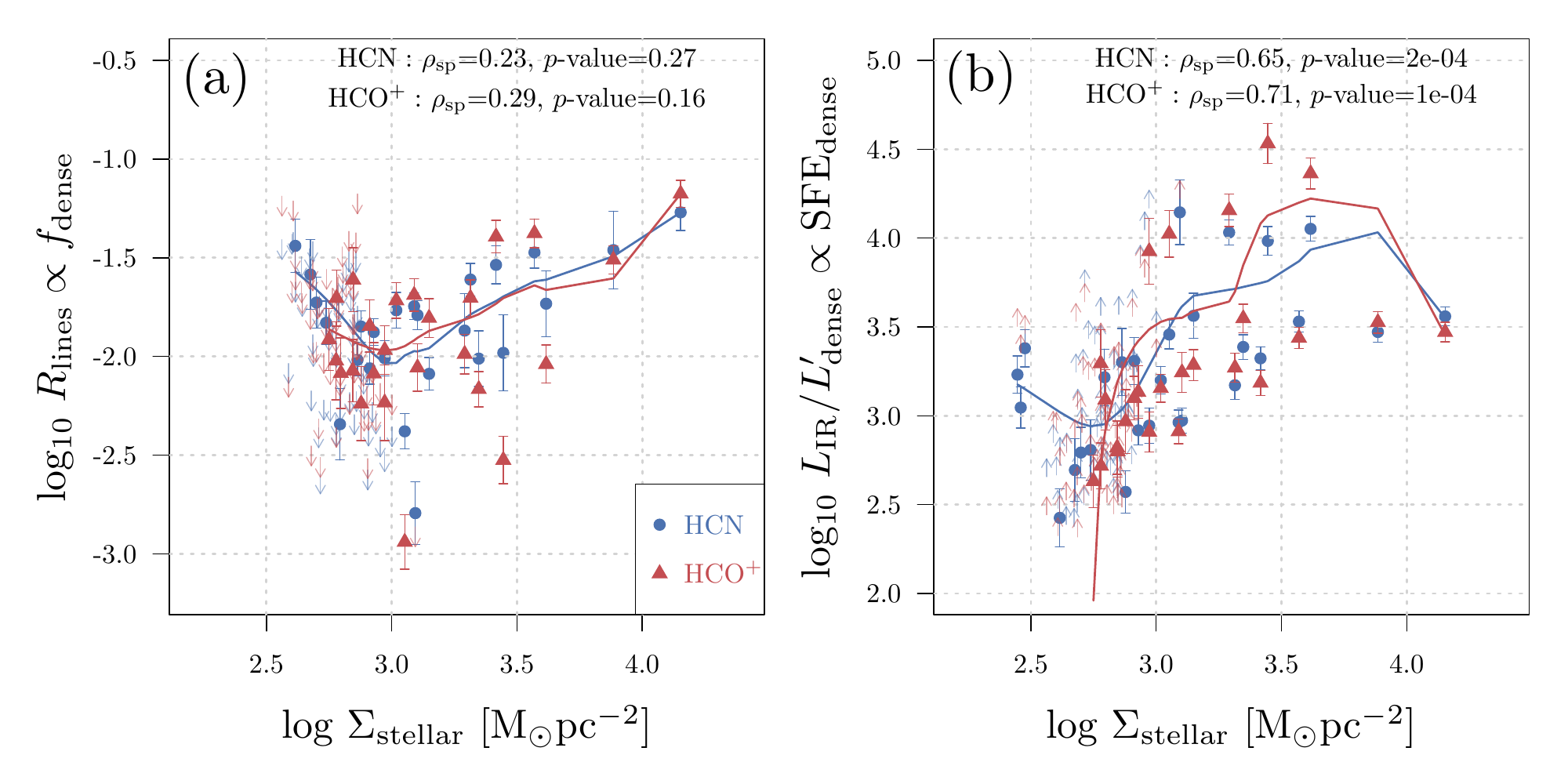}
\caption{\textit{left:} \fdense\ as a function of stellar surface density \sigmastellar. \textit{right:} dense-gas SFE (\sfedense $\equiv$ SFR/$M_\text{dense}$) as a function of \sigmastellar. 
$\rho_\text{sp}$ is the Spearman correlation coefficient and the $p$-value is for the hypothesis test as to whether they have zero correlation. 
The lines are Local Polynomial Regression fits.
\label{fig:sfe_star}} 
\end{center}
\end{figure*}

% 4.2
%----------------------------------------------------------------
\subsection{The relationships between star-formation efficiency, dense-gas fraction and stellar components}\label{subsec:star}
%----------------------------------------------------------------

Previous studies have used HCN 1-0 as a dense-gas tracer and showed that the
existing stellar component affects
the parameters related to dense-gas emission. \citep{Usero:2015, Bigiel:2016,
Gallagher:2018a,Jimenez-Donaire:2019}. They report decreasing trends
in both \sfedense\ versus \fdense, and \sfedense\ versus 
\sigmastellar, i.e., stellar surface density. With our JCMT 
HCN 4-3 and \hcop\ 4-3 data, and stellar surface densities derived from
\textit{Spitzer} 3.6-\micron\ data we can test such scaling relationships in this work.

In Fig.~\ref{fig:sfe_fdense} we plot \sfemol\ versus
\fdense\ and \sfedense\ versus \fdense\ to explore how \fdense\ affects the
SFE for the total molecular gas and dense molecular gas, respectively. 
The data from \citet{Usero:2015} is shown as purple diamonds in the plots. 
We convert the \ldense\ 
to the $J=1-0$ luminosity in these plots assuming $L_{4-3}$/$L_{1-0}$ = 0.3.
Considering the systematic differences that their sample used HCN 1-0 and CO 2-1,
the regression fits are only applied to our sample.
For the relationship of \sfemol\ versus \fdense\
(Fig.~\ref{fig:sfe_fdense}a), we do not see any statistically
significant correlation between the two
parameters. 
Note that in the subsequent analysis we only adopt data points with SNR$>$3.
This is because, unlike the star-formation relation (Fig. \ref{fig:sfr_M_all})
which is a rather simple log-linear fit (based on many previous studies), for the
other relations it is unclear what kind of relation holds between the two
parameters. So upper limits can not be include in the analysis, and we only
use spearman coefficients and hypothesis test to discuss whether they have potential
correlations. The lines in the plots are Local Polynomial Regression fits only to guide
the eye, which is similar to the binning method used in many other studies, and
they do not strongly suggest that our data points closely follow those fits.
This is consistent with \citet{Usero:2015} and
the larger galaxy sample compiled by \citet{Gallagher:2018a}, where 
they show a very weak correlation ($\rho_\text{sp}$ = 0.13) and
a large scatter about the relationships, and they suggest that the HCN/CO ratio 
is a relatively poor predictor of \sfemol.
Our plot shows a similar case using high-$J$ dense molecular lines, implying
that the SFE, if measured as SFR per unit total
molecular gas (\sfemol) is rather independent of the dense-gas fraction
(\fdense) that is measured with the $J=4-3$ transition. 
Furthermore, such a large scatter reported in our work
and other studies indicates that a higher \sfemol\ (or shorter
molecular-gas depletion time) does not necessarily correspond to a higher
\fdense, thus the real star-formation scenario might be more complicated than
the simple model proposed by \citet{lada:2010}.

In Fig.~\ref{fig:sfe_fdense}b we explore the relationship between the dense-gas
star-formation efficiency (\sfedense) and \fdense. We see plausible
decreasing trends, which are consistent with
\citet{Usero:2015}, and their fitted slope is shown as a purple line in
Fig.~\ref{fig:sfe_fdense}b for comparison. The anti-correlations are not
statistically significant, and we note that the trends might be partly
attributed to the correlation between \lir\ and \lco\ (similar to the
Kennicutt--Schmidt law relating the gas content and SFR),
since \fdense\ is partially cancelled out in this \sfedense--\fdense\
relationship. 
It is interesting to see the data points of \citet{Usero:2015} appear to be
consistent with our data. In Fig.~\ref{fig:sfe_fdense}a they together suggest
that there is no significant correlation between \sfemol\ and \fdense. 
In Fig.~\ref{fig:sfe_fdense}b their sample also lies in the same range, although
our HCN data do not follow their fitted slope (purple dashed line).
Note that they derived the CO 1-0 luminosity from 
converting CO 2-1 emission assuming $R_{21}$ = 0.7, and that our data also
rely on the assumption of $L_{4-3}$/$L_{1-0}$ = 0.3. It is therefore likely
that the two samples have systematic differences caused by the conversion
between different transitions.
Our data follow the same prediction given by the model used in
\citet{Usero:2015} for a 
fixed average gas volume density $\bar{n} = 100$ cm$^{-3}$. 
As pointed out in Section~\ref{sec:line_ratio}, dense-gas tracers are so 
compact compared with the single-dish beam that \fdense\ is dependent on the
observing resolution, and we need high-resolution data, e.g., from ALMA, 
to truly understand these scaling relationships related to \fdense.

In Fig.~\ref{fig:sfe_star} we explore how \fdense\ and \sfedense\ are affected by
\sigmastellar. In \citet{Bigiel:2016}, where HCN 1-0 was used, an increasing
trend between \fdense\ and \sigmastellar, and a decreasing trend between
\sfedense\ and \sigmastellar\ are reported.
They were interpreted as the effect of interstellar-gas pressure
(traced by the stellar surface density) on the gas-density structure, in the sense
that high pressure would increase the overall mean density of the interstellar
gas. Thus the HCN 1-0 intensity and \fdense\ are both elevated, but the density
contrast, which is defined as the difference between the high-density peaks and the
mean density, is reduced. \sfedense\ is controlled by the prevailing contrast, so it is
also reduced in environments with higher pressure. 

Fig.~\ref{fig:sfe_star}a shows that \fdense\ traced by both HCN 4-3 and
\hcop\ 4-3 seems to be weakly
correlated with $\Sigma_\text{stellar}$, with Spearman correlation
coefficients $\rho_\text{sp}$(HCN) = 0.23 and $\rho_\text{sp}$(\hcop) = 0.29,
respectively, but
hypothesis tests show that they are not statistically significant ($p >$
0.05). This is likely due to our limited data for only one galaxy, and the
variation of \fdense\ is large, especially in the log$_{10}$\sigmastellar\
range between 2.5 and 3. If we compare our data points with previous 
studies
\citep{Gallagher:2018a,Usero:2015}, they seem to lie in a similar range in
the plot, though we are using high-$J$ lines.

With regard to the relationship between \sfedense\ and \sigmastellar, 
previous studies using the $J=1-0$ transition of HCN and/or \hcop\ all showed
anti-correlations \citep{Usero:2015, Gallagher:2018a,Jimenez-Donaire:2019}, so it is somewhat surprising to see an increasing trend 
between \sfedense\ and $\Sigma_\text{stellar}$ in Fig.~\ref{fig:sfe_star}b. The Spearman correlation
coefficients are $\rho_\text{sp}$(HCN) = 0.65 and $\rho_\text{sp}$(\hcop) = 0.71, respectively, and they are
statistically significant ($p \ll$ 0.05). Running lines from
polynomial fits are shown just to guide the eye.
The increasing trend between \sfedense\ and \sigmastellar\ in 
Fig.~\ref{fig:sfe_star}b is consistent with the star-formation scenario that 
the stellar component plays an important role in regulating the local SFR, as 
the gravitational potential is dominated by the stellar mass and therefore is
deeper near the centre. Thus it increases the hydrostatic gas pressure in the disc
and reduces the free-fall time for gas to collapse. Therefore the SFE is 
likely enhanced in star-dominated regions \citep{Ostriker:2011,Shi:2011,Meidt:2016,Shi:2018}.

% Therefore, we have reproduced the relationships among \fdense, \sfemol,
% \sfedense\ and \sigmastellar\ using the $J=4-3$ transition of HCN and \hcop. 
% Comparing Fig.~\ref{fig:sfe_fdense} and \ref{fig:sfe_star} with previous
% studies using HCN 1-0 and \hcop\ 1-0, we see that only the relationship of
% \fdense--\sigmastellar\ is not consistent with that of
% \citet{Usero:2015}, \citet{Bigiel:2016} and \citet{Gallagher:2018a}. 
We speculate that the inconsistency between Fig.~\ref{fig:sfe_star}b and
similar results in previous studies could be explained by observational and/or
physical effects. Observationally, the argument by \citet{Bigiel:2016} relies
heavily on the large number of  low-SNR data points and upper/lower limits, 
at least in the low \sigmastellar\ regime ($\la$10$^{2.5}$ M$_{\sun}$ pc$^{-2}$).
In contrast, in Fig.~\ref{fig:sfe_star} our statistics only include
data points with SNR $>$ 3.
However, their latest result based on a larger sample 
shows more promising trends \citep{Jimenez-Donaire:2019}, so the difference
is more likely a physical effect.
The inconsistency implies the possibly different effects of
\sigmastellar\ on HCN 4-3 and \hcop\ 4-3. At this stage, it is difficult
to quantify the relationship between stellar surface density (gas pressure)
and average gas density or density structure, and the actual relationship between
\sfedense\ and local physical conditions, such as \sigmastellar\ might be 
more complicated. Qualitatively, high-$J$ lines
require higher critical densities \ncrit\ and higher excitation temperatures, 
which are only met in a small part of the gas structure, i.e., the highly
concentrated molecular clumps. Thus they might be less sensitive to the change
in the overall density structure compared with low-$J$ lines. As proposed by
\citet{Bigiel:2016} high \sigmastellar\ means high gas pressure, which would
raise the overall average gas density and decrease the density contrast traced
by HCN 1-0. As a consequence, low-$J$ dense molecular lines no longer only
trace the gas undergoing star formation, and this is their explanation
for the decreasing \sfedense (HCN 1-0) with increasing \sigmastellar. 
On the other hand, the high-$J$ lines might still trace the "density contrast"
properly thus they are good tracers of the molecular gas undergoing star formation.

\citet{Leroy:2017} show that the emissivity of the high-density tracers
($J=1-0$) depends strongly on the density distribution of the gas, and
the corresponding line ratios can reflect the change in the gas-density structure.
Therefore the difference between Fig.~\ref{fig:sfe_star}(b) and the graphs by
\citet{Bigiel:2016} and \citet{Gallagher:2018a} might reflect the change of
emissivity associated with the $J=1-0$ and $J=4-3$ transitions of HCN and \hcop. 
However, \citet{Leroy:2017} do not include the $J=4-3$ models and we cannot fully
explain the discrepancy based on our current information. 
Also, we note that our \sigmastellar\ values are at the
higher end of those derived by \citet{Usero:2015}, and the different \sigmastellar\
range might also partly contribute to the discrepancy. We hope that more 
high-$J$ data with a larger dynamical range of \sigmastellar\ will help us to
clarify and better understand the relationship between the stellar 
components and the dense-gas phase of the molecular gas.

%----------------------------------- 
%--------------- Summary -------------------- 
%----------------------------------- 

\section{Summary}\label{sec:summary}
In this paper we present JCMT HCN 4-3 and \hcop\ 4-3 maps of
the inner $\sim$ 2\,kpc of NGC\,253, the nearest nuclear starburst galaxy,
obtained with HARP on the JCMT as part of the MALATANG survey results. Archival
CO 1-0, CO 3-2 and infrared data
are incorporated for a multi-line analysis. 
At $\sim$0.24-kpc spatial resolution, 
we derive radial profiles of the different gas tracers, and
analyse the variation in gas parameters in the disc, including the dense-gas
fraction and the dense-gas star-formation efficiency. 
Their relationships with stellar surface density are also discussed. Here are our 
main findings.

\begin{enumerate}

\item Both HCN 4-3 and \hcop\ 4-3 show more concentrated emission
morphologies than CO, but are similar to that of the infrared distribution. This
is consistent with HCN and \hcop being faithful tracers of the dense gas
responsible for the on-going star formation.
%Curve of growth is adopted to fit the concentration index 
%$r_{90}$/$r_{50}$ for the tracers. HCN 4-3 and \hcop\ 4-3 shows the highest
%concentration indices. More data of higher resolution will be needed for
%more-accurate fits and for comparisons among galaxies.

\item Using HCN-to-CO and \hcop-to-CO ratios we derive dense-gas fractions,
\fdense, and using ratios of CO 3-2 and CO 1-0 we derive the CO-line ratio,
\rco, an indication of the excitation condition pertaining to the total gas.
We show that \fdense\ and \rco\ both decline towards larger radii. At 0.5\,kpc
from the centre, \fdense\ and \rco\ are several times lower than their
values in the galaxy centre. The radial variation, and the large scatter of
these parameters, imply distinct physical conditions in different regions
of the galaxy disc. We suggest that, when estimating the total gas mass using CO
3-2 alone, one should take into account the uncertainty induced by the
inherent variation and scatter of \rco\ within a galaxy.

\item We discuss the star-formation relationship (SFR versus \mdense) and use two
kinds of dense-gas conversion factor \alphadense\ to estimate \mdense\ for
comparison. When adopting the variant \alphadenseG\ that is dependent on the
radiation-field intensity, the power-law slopes of SFR versus \mdense($G_0$)
are super-linear, with slopes $\beta$(HCN) = 1.70 and $\beta$(HCO$^+$) = 1.60.
When the
fixed \alphadense\ is adopted to calculate \mdense, we obtain $\beta$(HCN) =
1.14 and $\beta$(HCO$^+$) = 1.10, which are more consistent with the linear
correlation derived in other works. 
% More data are needed for better calibration of the dense-gas conversion factor.

\item We explore the relationships between total molecular-gas star-formation
efficiency \sfemol\ and \fdense, and the relationships between \sfedense\ and
\fdense. We do not see any significant correlation for \sfemol\ and \fdense,
although a weak anti-correlation is obtained for \sfedense versus \fdense.
These results are consistent with \citet{Usero:2015}, and they follow the
same prediction as for a fixed average gas volume density $\bar{n} = 100$
cm$^{-3}$.

\item We explore the relationship between \fdense\ and the stellar surface density
\sigmastellar, and the relationship between \sfedense\ and \sigmastellar. They both show weak increasing
trends, but only the \sfedense versus \sigmastellar\ relationship is statistically
significant. While the \fdense\ versus \sigmastellar\ relationship is consistent
with that presented in previous works using HCN 1-0 emission, it is
intriguing to see an increasing trend in the \sfedense versus \sigmastellar\
relationship, which is inconsistent with other works. It remains unclear how to
interpret this trend, but we speculate that this might
be a result of the different transitions used from other works, 
since the existing stellar components may have a different effect on the gas traced by HCN 1-0 than by HCN 4-3, and in regions with higher \sigmastellar\ 
the high-$J$ dense lines of HCN and \hcop\ might be less sensitive to the change of the overall density and they could still trace the densest gas undergoing star formation.

\end{enumerate}

Our results show that JCMT observations can resolve the central
$\sim$ kpc scale of nearby galaxies, allowing analysis of the variations of
dense-gas parameters among
different regions of galactic discs. The variation of gas properties, such as \fdense\ and \sfedense\ in
different environments of individual galaxies is important for the
understanding of star-formation activity that regulates galaxy evolution.
Other galaxies in the MALATANG sample
will be studied in future papers, and deeper integration will be needed to detect 
the weak lines of dense gas in most disc regions. While other works have
demonstrated the power of high-resolution
observations using facilities like ALMA, more galaxies have to be observed in
a similar manner, to reveal the true structures and properties of dense gas
in the sub-structure of galaxies.

% --------------  Acknowledgement  -----------------
\section*{Acknowledgements}
%
%nls2 , linmix\_err, survival, 
We thank the anoynomous refereree for the very helpful comments.
We thank Antonio Usero for kindly providing the data used for 
Fig.~\ref{fig:sfe_fdense}. 
We also thank Padelis P. Papadopoulos for his contribution to the observing
effort for the project and for providing helpful discussions and feedback on the draft.
This research is supported by the National Key R\&D Program of China with no. 2017YFA0402704, and no. 2016YFA0400702. It is also supported by NSFC grants nos.
11861131007, 11420101002, 11603075, 11721303, U1731237, 11933011 and 11673057, 
and Chinese Academy of Sciences Key Research Program of Frontier Sciences grant
no. QYZDJ-SSW-SLH008. 
M.J.M.~acknowledges the support of the National Science Centre, Poland through the grant 2018/30/E/ST9/00208.
The research of CDW is supported by grants from the Natural Sciences and
Engineering Research Council of Canada and the Canada Research Chairs program.
JHH is supported by NSFC grants nos. 11873086 and U1631237, and by Yunnan Province of China (No.2017HC018).
SM is supported by the Ministry of Science and Technology (MOST) of Taiwan, MOST 107-2119-M-001-020.
This work is sponsored (in part) by the Chinese Academy of Sciences (CAS), through a grant to the CAS South America Center for Astronomy (CASSACA) in Santiago, Chile.
The James Clerk Maxwell Telescope is operated by the East Asian Observatory on behalf of The National Astronomical Observatory of Japan; Academia Sinica Institute of Astronomy and Astrophysics; the Korea Astronomy and Space Science Institute; Center for Astronomical Mega-Science (as well as the National Key R\&D Program of China with No. 2017YFA0402700). Additional funding support is provided by the Science and Technology Facilities Council of the United Kingdom and participating universities in the United Kingdom and Canada.

This work made use of \textsc{r} \citep{R:2008}, and \textsc{astropy}\footnote{http://www.astropy.org}, a community-developed core Python package for Astronomy \citep{astropy:2013, astropy:2018}.
% Grants

%\clearpage
\bibliographystyle{mnras}
\bibliography{bibtex}

%%%%%%%%%%%%%%%%%%%%%%%%%%%%%%%%%%%%%%%%%%%%%%%%%%

%%%%%%%%%%%%%%%%% APPENDICES %%%%%%%%%%%%%%%%%%%%%

\appendix

\section{Spectra and data tables}

Fig.~\ref{fig:spec} shows the CO 1-0, CO 3-2, HCN 4-3, and \hcop\ 4-3 spectra of
the central 13$\times$7 pixels (based on Fig.~\ref{fig:obs_pos}) from HARP
observation towards NGC\,253. The grid size is 10\,arcsec for all tracers.
CO 1-0 is obtained from the archive of the Nobeyama 45-m
telescope, CO 3-2 is obtained from the JCMT archive, and HCN 4-3
and HCO$^+$ are MALATANG data (this work). 
In Fig.~\ref{fig:stare-spec} we show three examples of spectra obtained in stare mode that are not shown in Fig.~\ref{fig:spec}. Note that in Fig.~\ref{fig:stare-spec} the intensity of CO 1-0 spectra is divided by 100.
%----------------- Figure  spectra ---------------------
%\begin{landscape}
%\begin{figure}
%%\begin{center}
%\includegraphics[angle=-90,scale=0.88]{figures/spectra_l.pdf}
%\caption{Spectra of CO 1-0 (black), 
%CO 3-2 (green), HCN 4-3
%(blue) and HCO$^+$ 4-3 (red) emission in the central $\sim$ 1\.kpc region of NGC\,253.
%The $T_\text{MB}$ (in unit of mK) range on the $y$-axis of the central three
%rows is set to be [-30, 450],  while this range for the other rows is
%[-30,100]. On the central three rows, CO 1-0 and CO 3-2 lines are scaled by a
%factor of 0.05 for clearer comparison with HCN 4-3 and \hcop\ 4-3. on the other rows we offset the CO spectra by 30\,mK for clarity.
%CO 1-0 is obtained from the archive of the Nobeyama 45-m
%telescope, CO 3-2 is obtained from the JCMT archive, and HCN 4-3
%and HCO$^+$ are MALATANG data (this work). The velocity resolution is 10\,\kms\
%for the two CO lines, and 20\,\kms\ for HCN 4-3 and \hcop\ 4-3. All data are
%resampled with a pixel size of 10\,arcsec, corresponding to $\sim$ 170\,pc.
%}
%\label{fig:spec}
%%\end{center} 
%\end{figure}
%\end{landscape}

\begin{figure*}
\begin{center}$
\begin{array}{cccc}
\includegraphics[scale=0.168, angle=0]{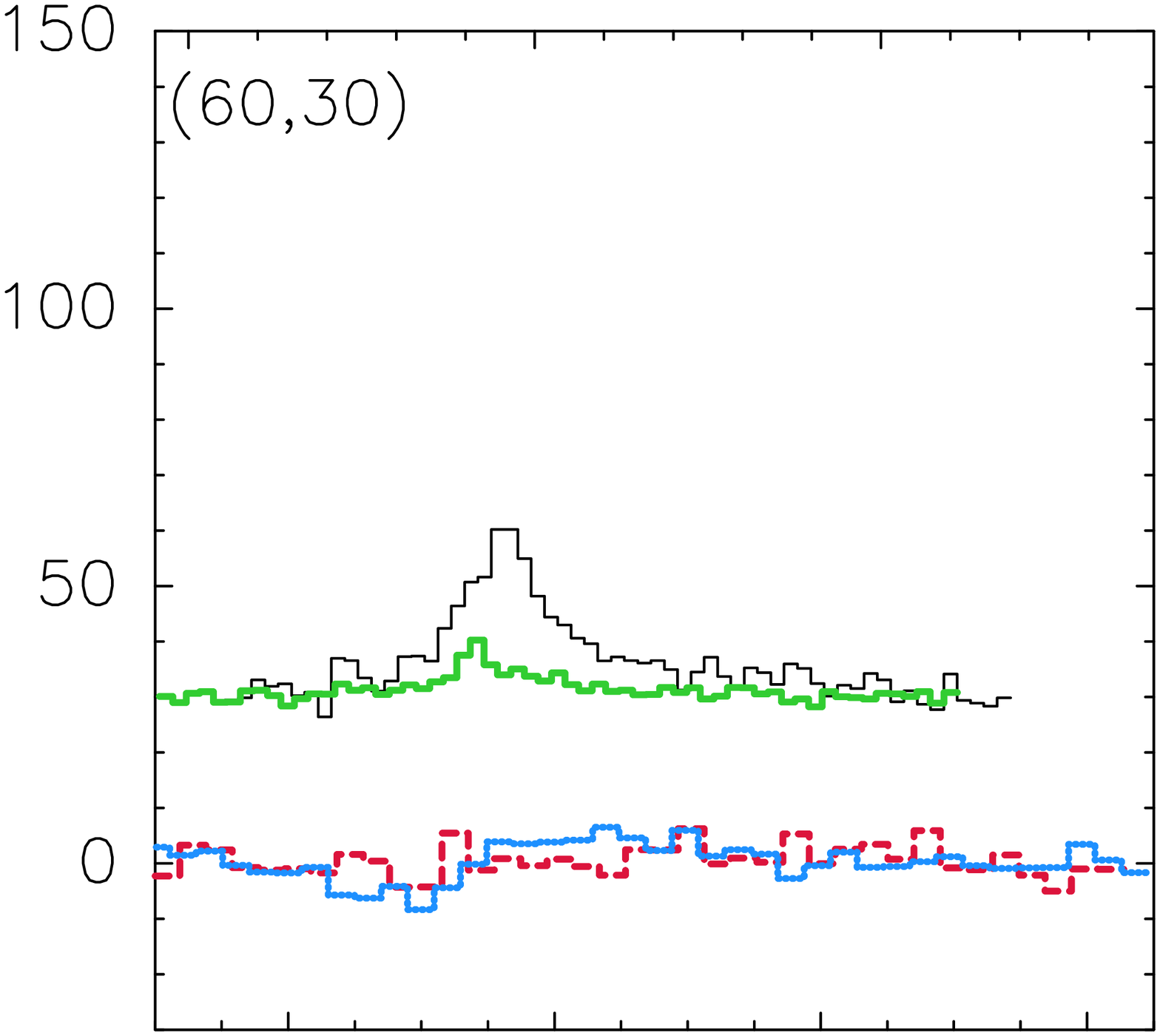}&
\includegraphics[scale=0.168, angle=0]{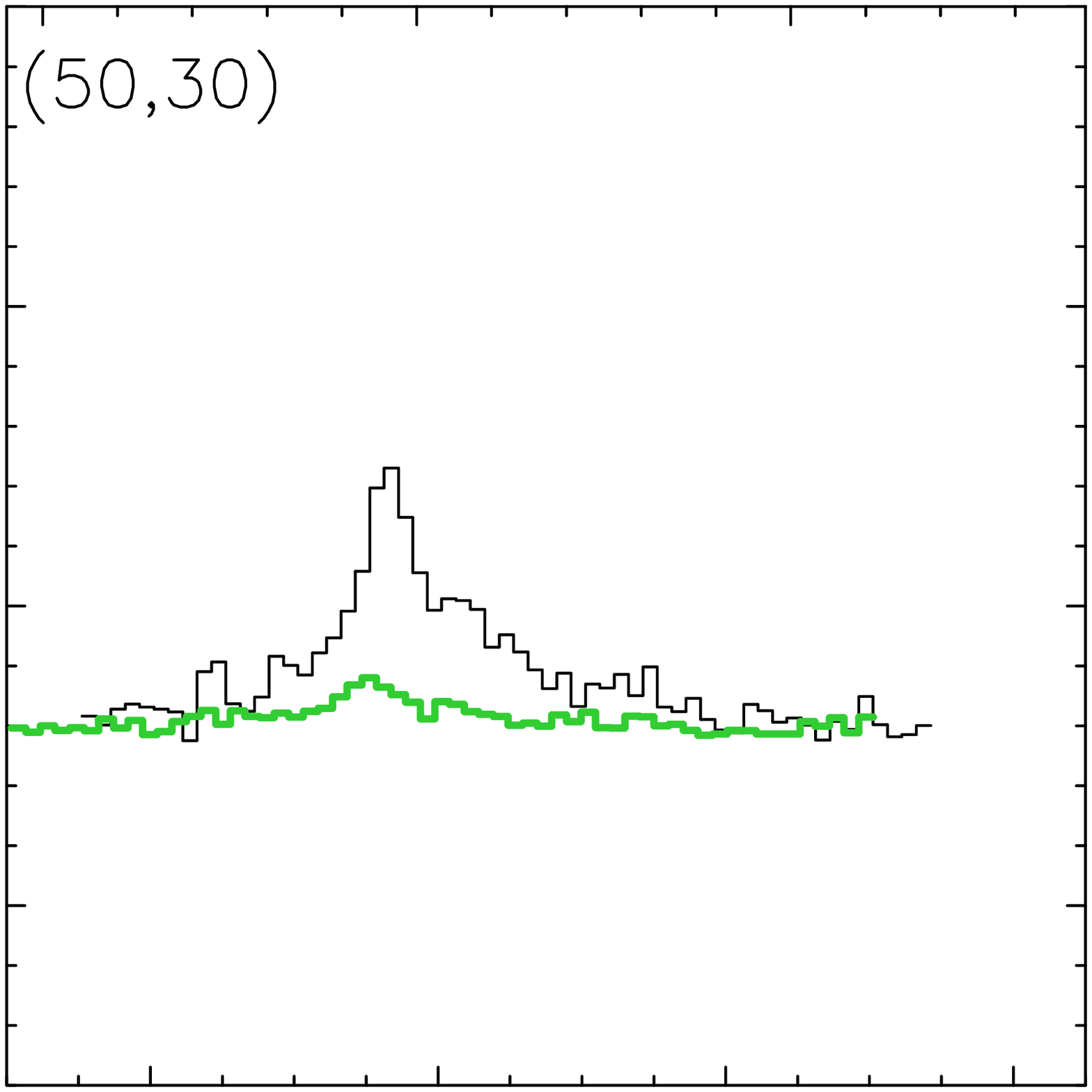}&
\includegraphics[scale=0.168, angle=0]{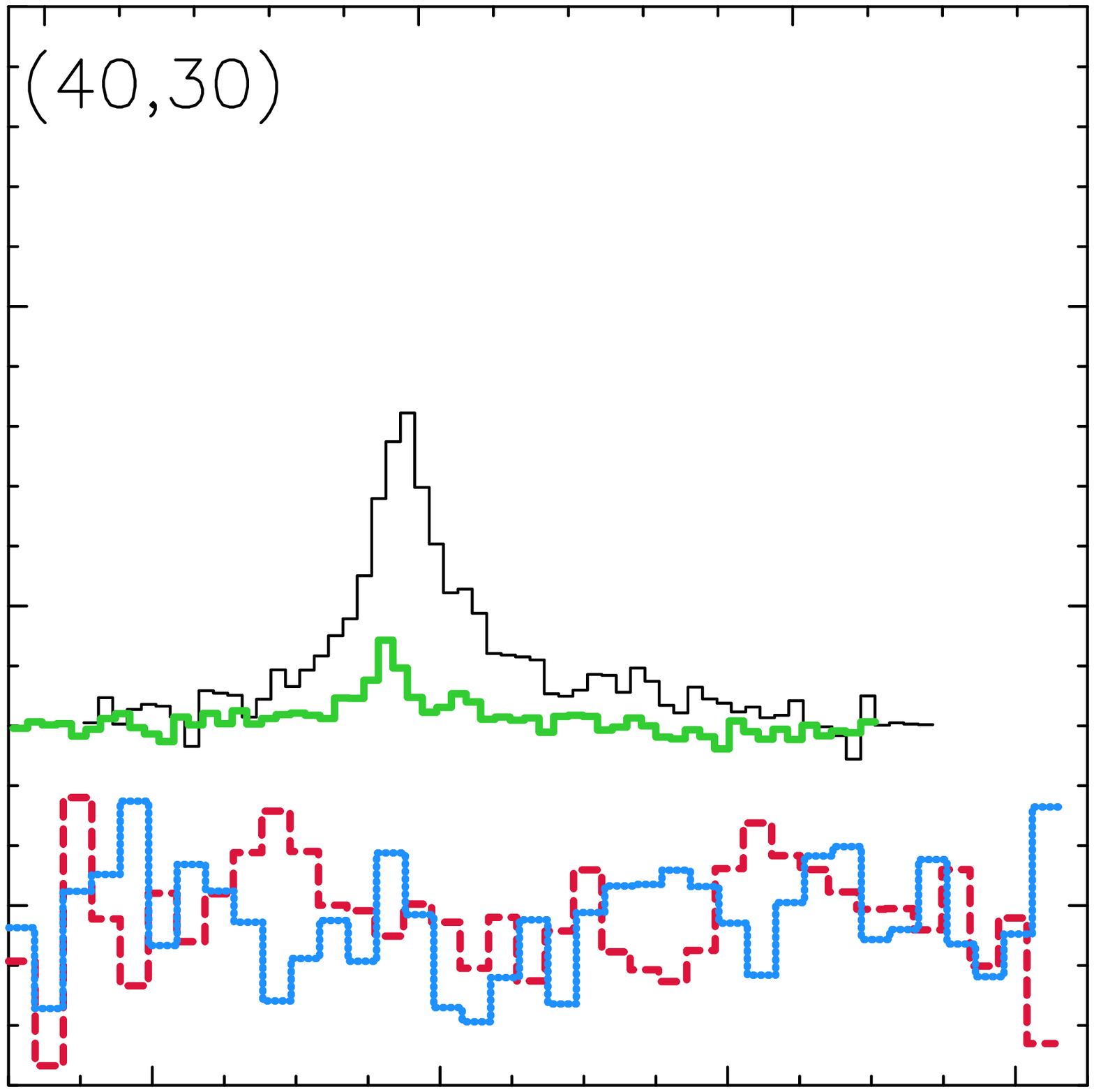}&
\includegraphics[scale=0.168, angle=0]{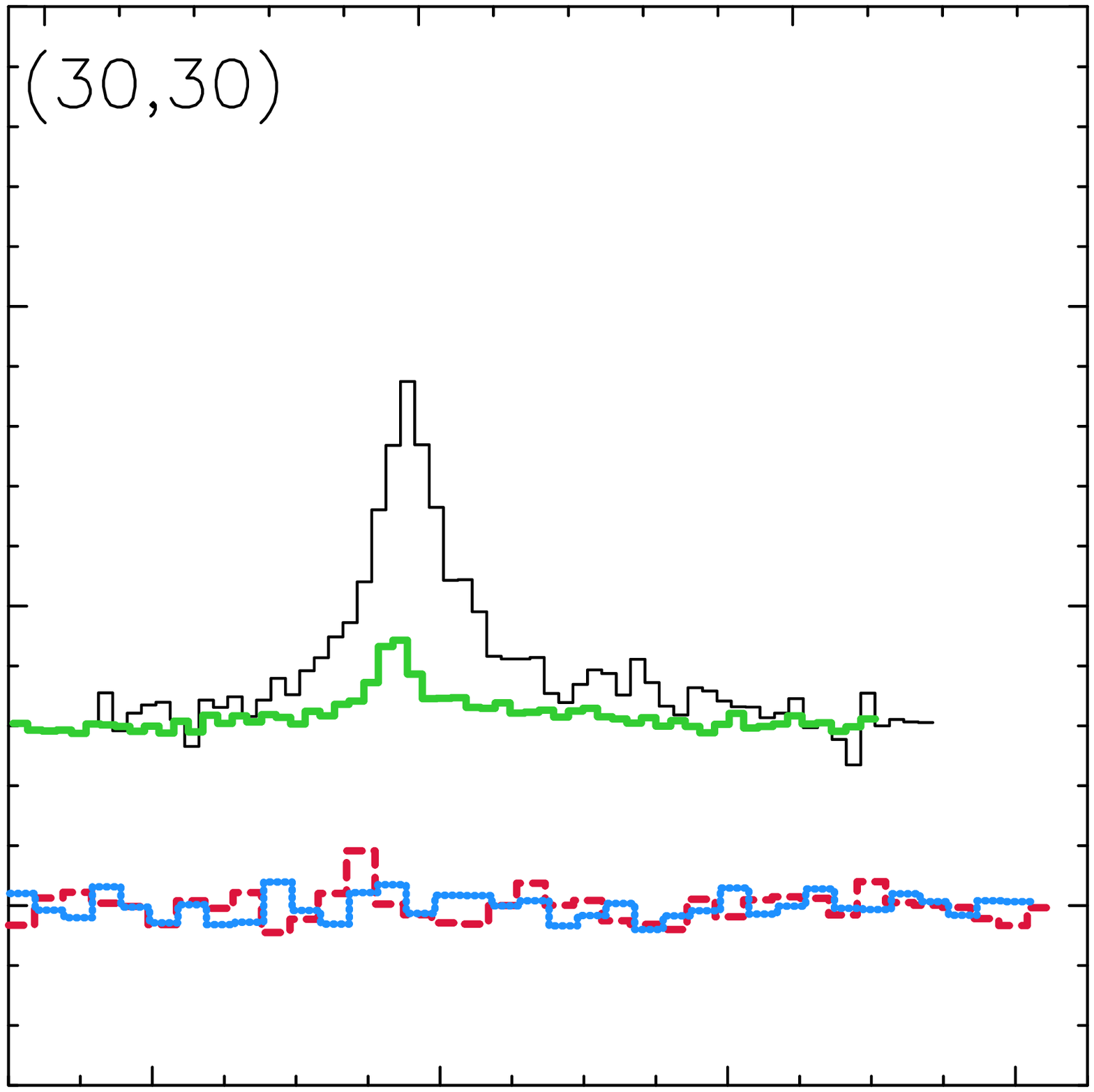}\\
\includegraphics[scale=0.168, angle=0]{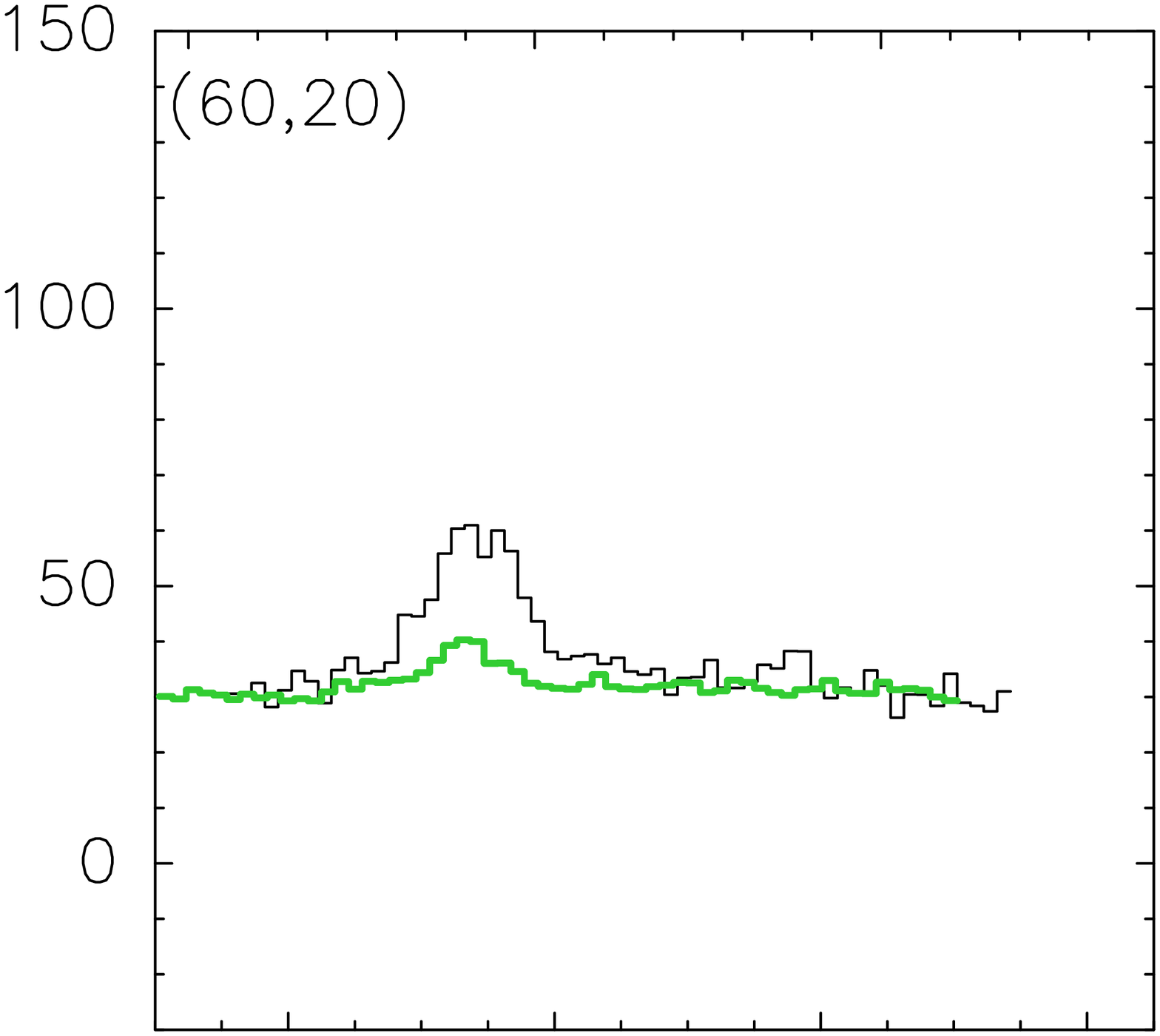}&
\includegraphics[scale=0.168, angle=0]{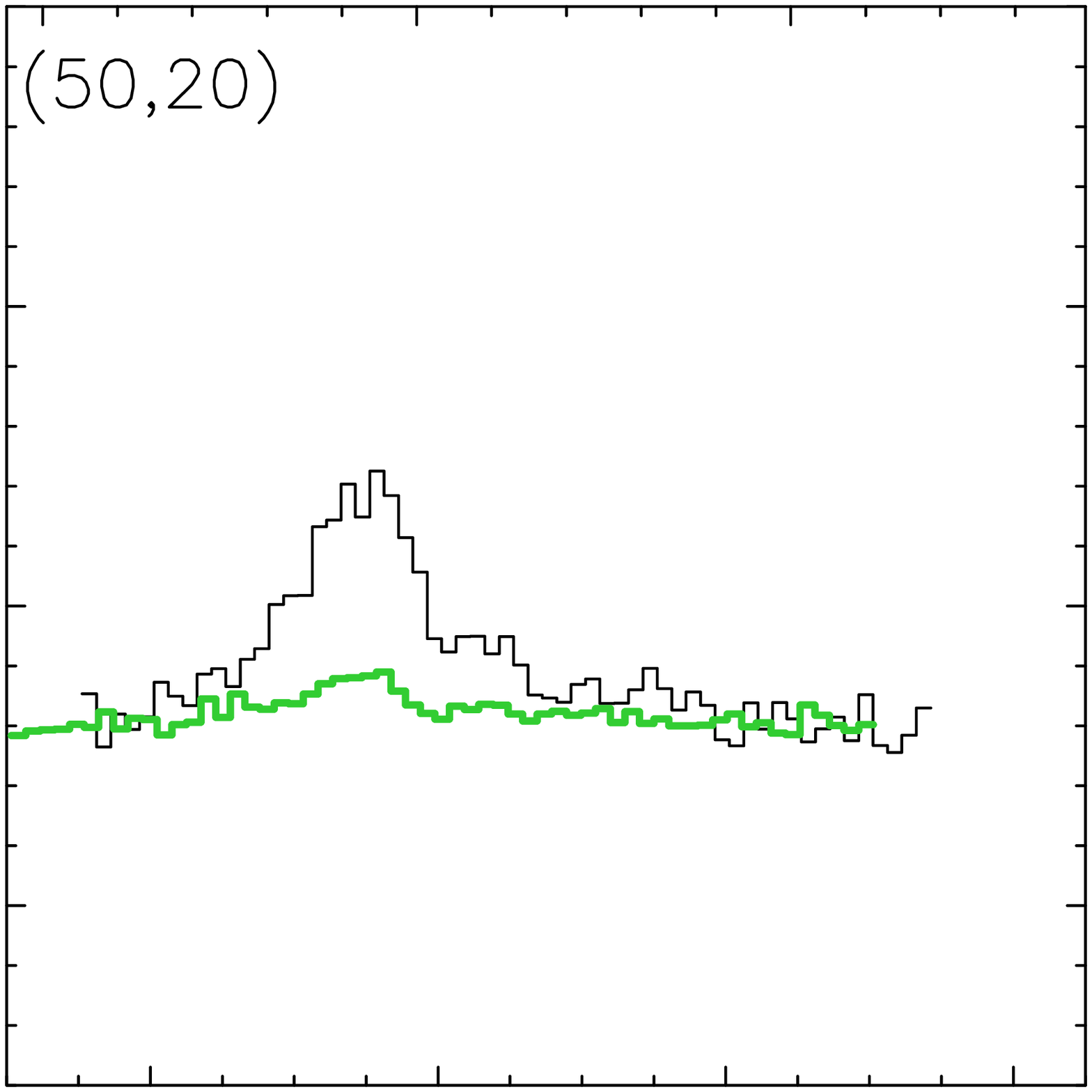}&
\includegraphics[scale=0.168, angle=0]{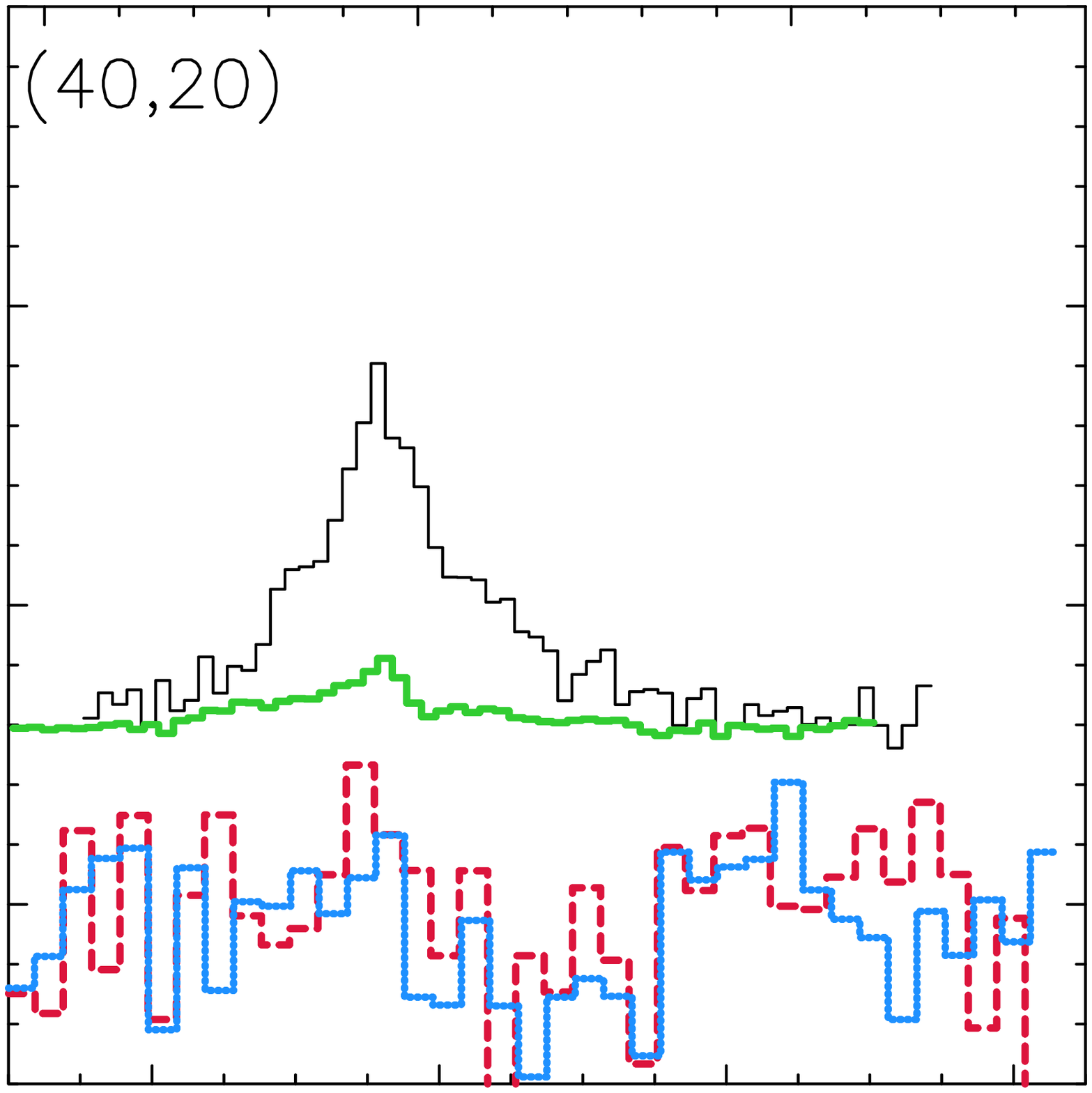}&
\includegraphics[scale=0.168, angle=0]{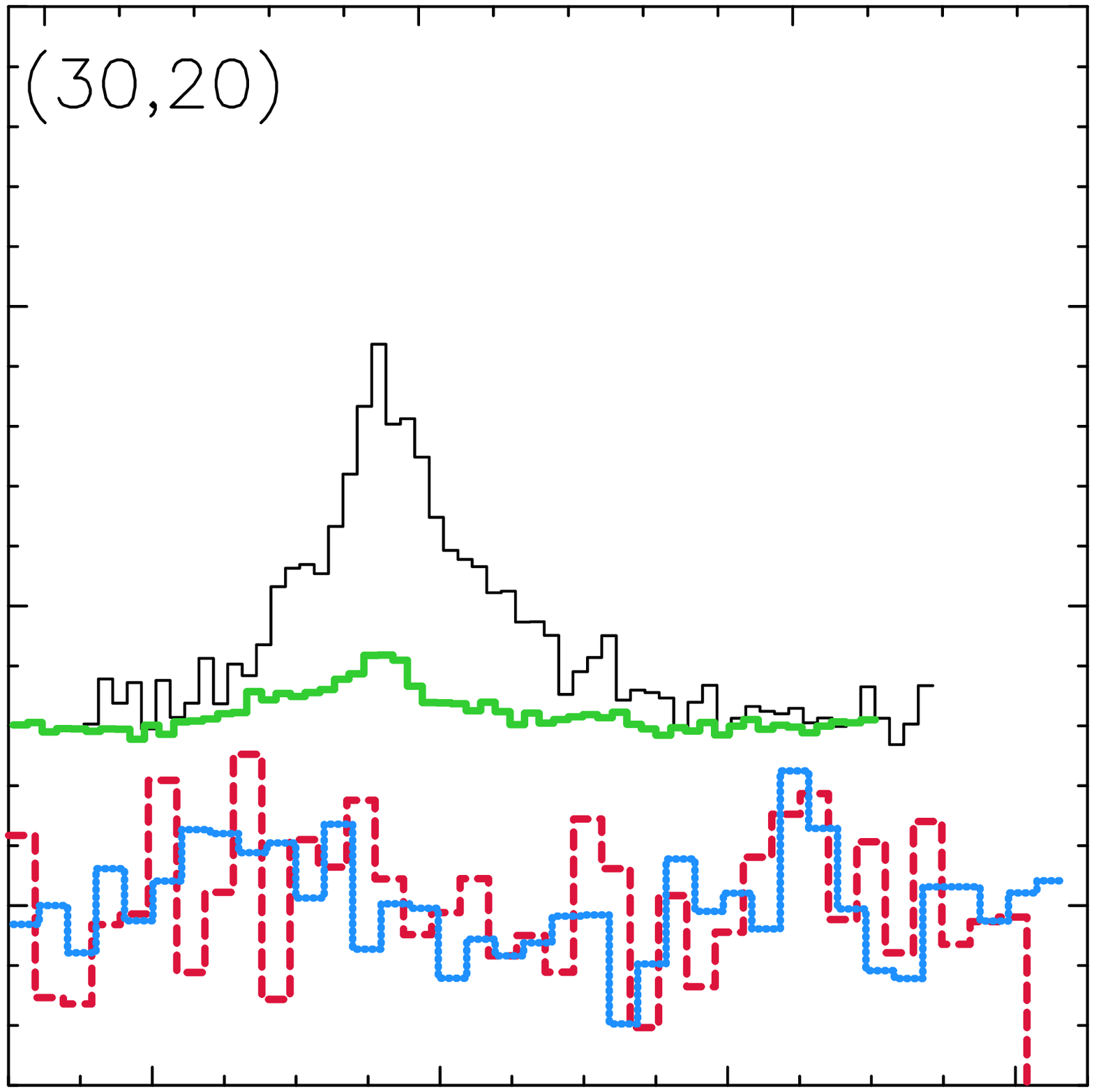}\\
\includegraphics[scale=0.168, angle=0]{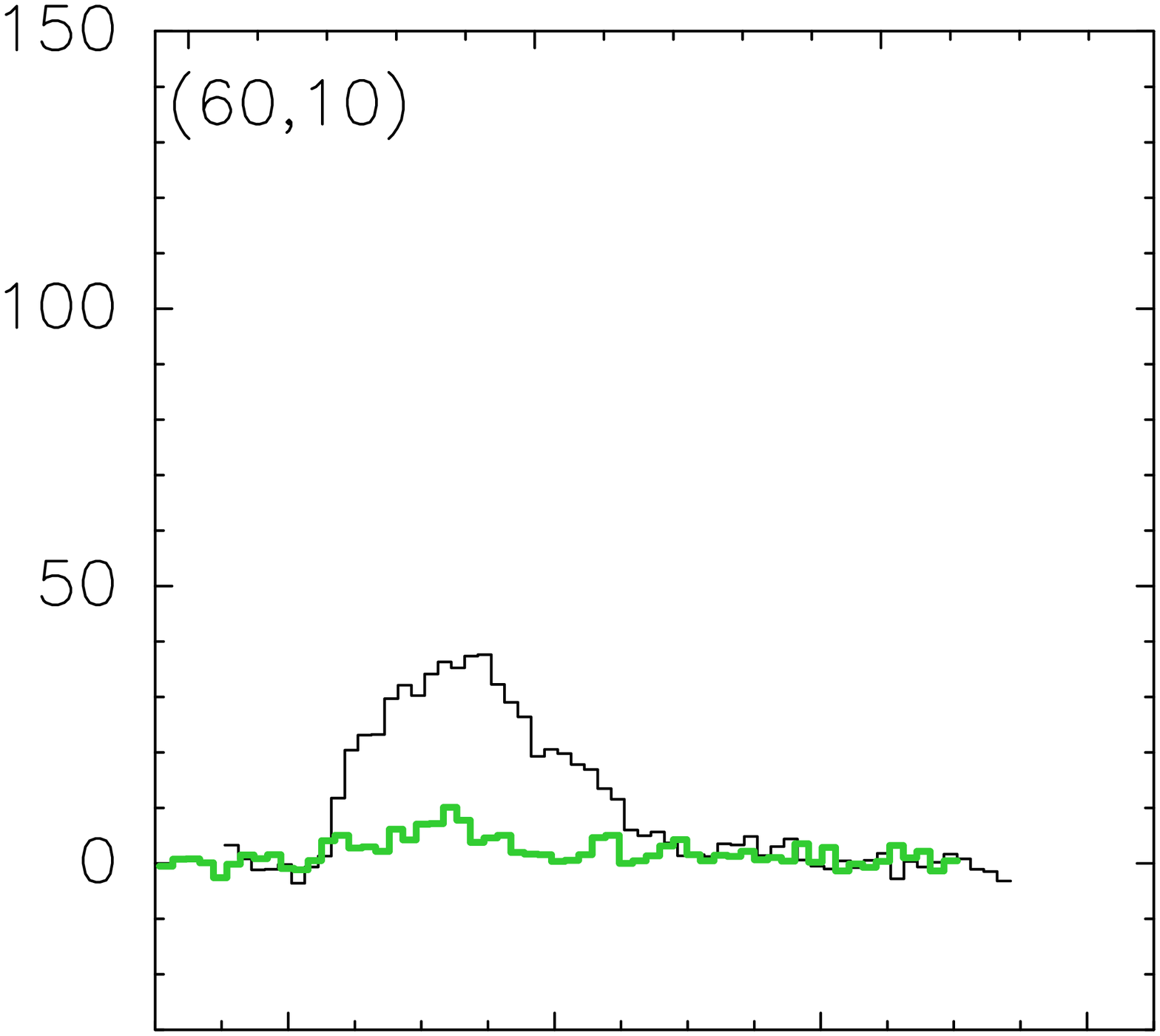}&
\includegraphics[scale=0.168, angle=0]{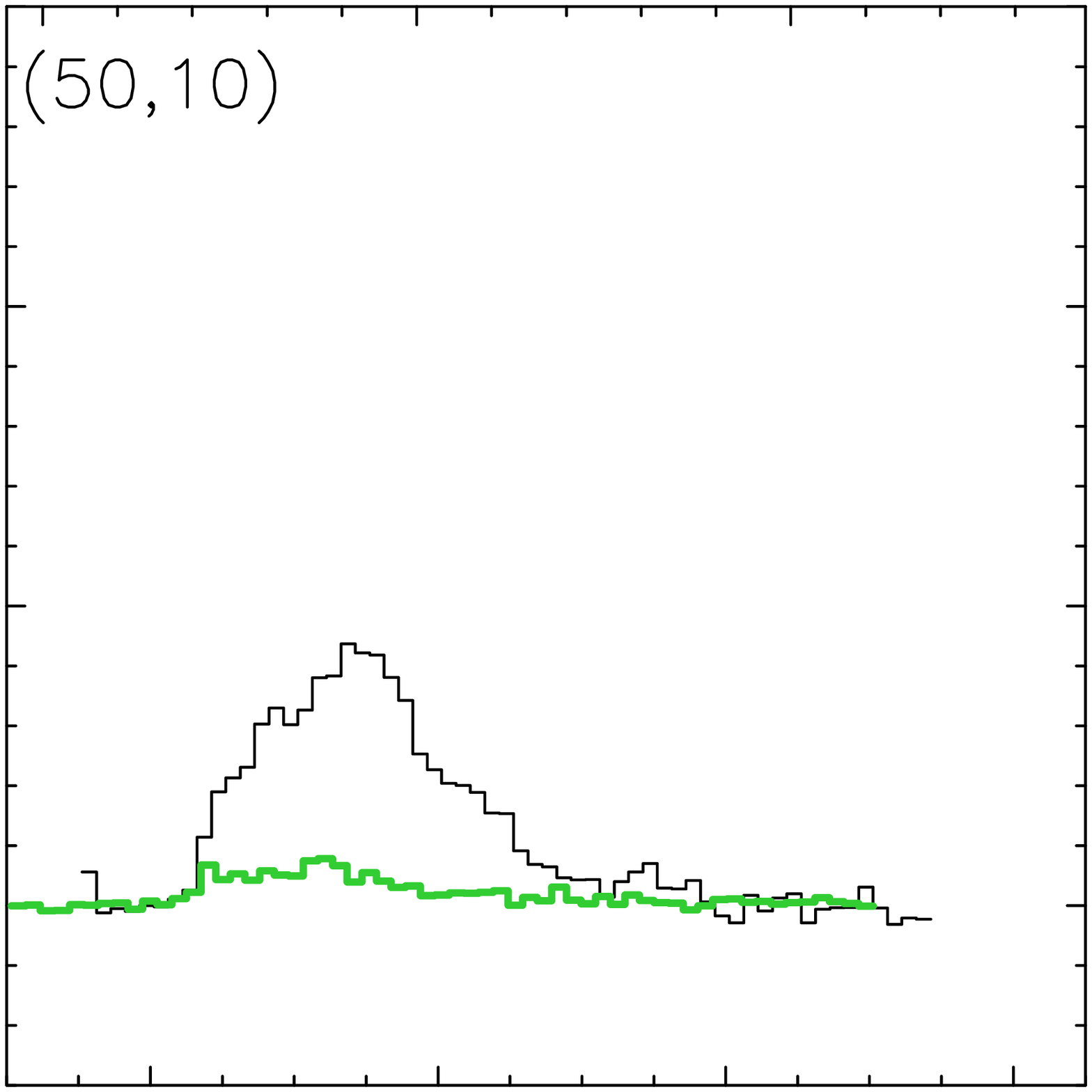}&
\includegraphics[scale=0.168, angle=0]{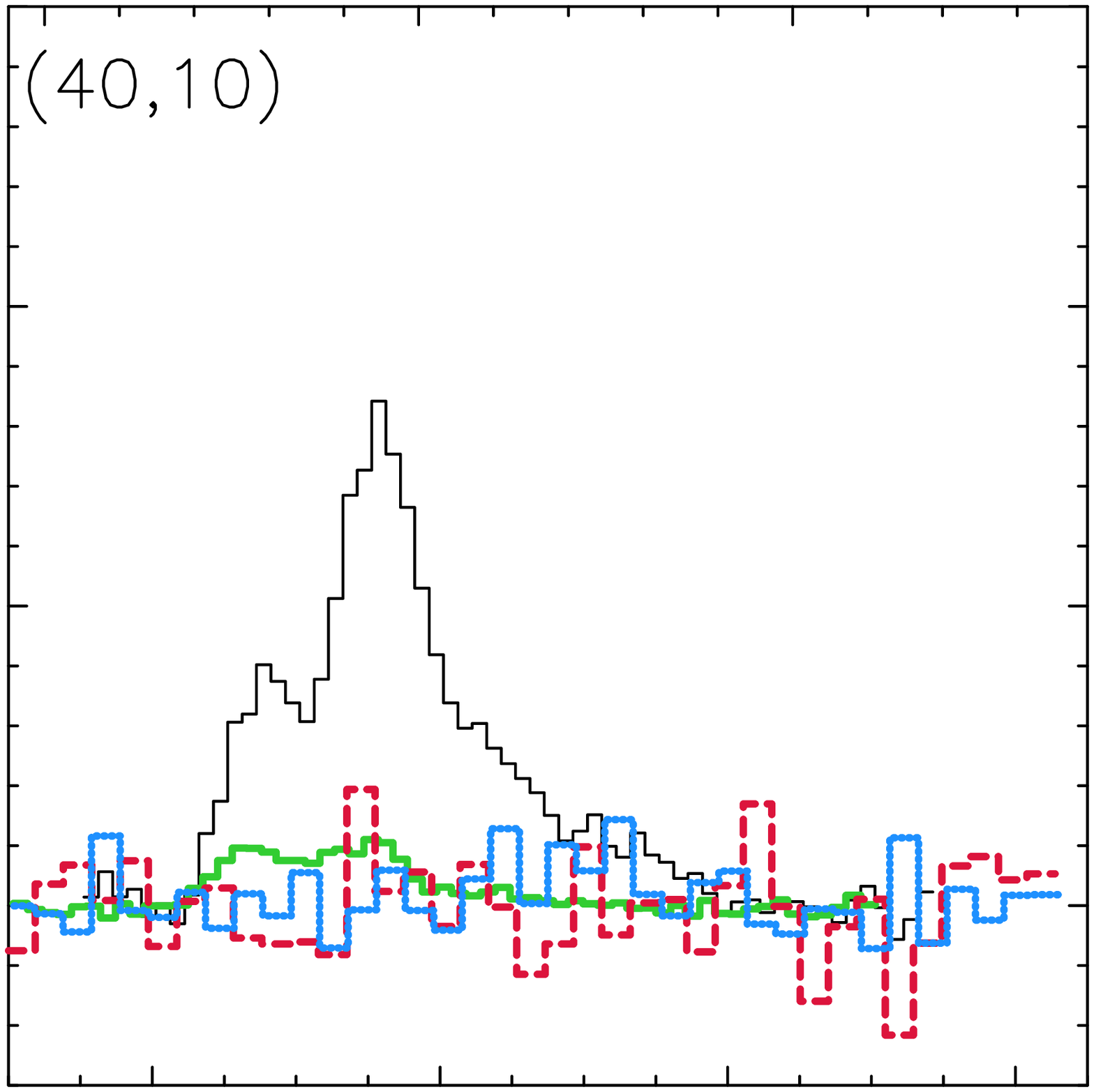}&
\includegraphics[scale=0.168, angle=0]{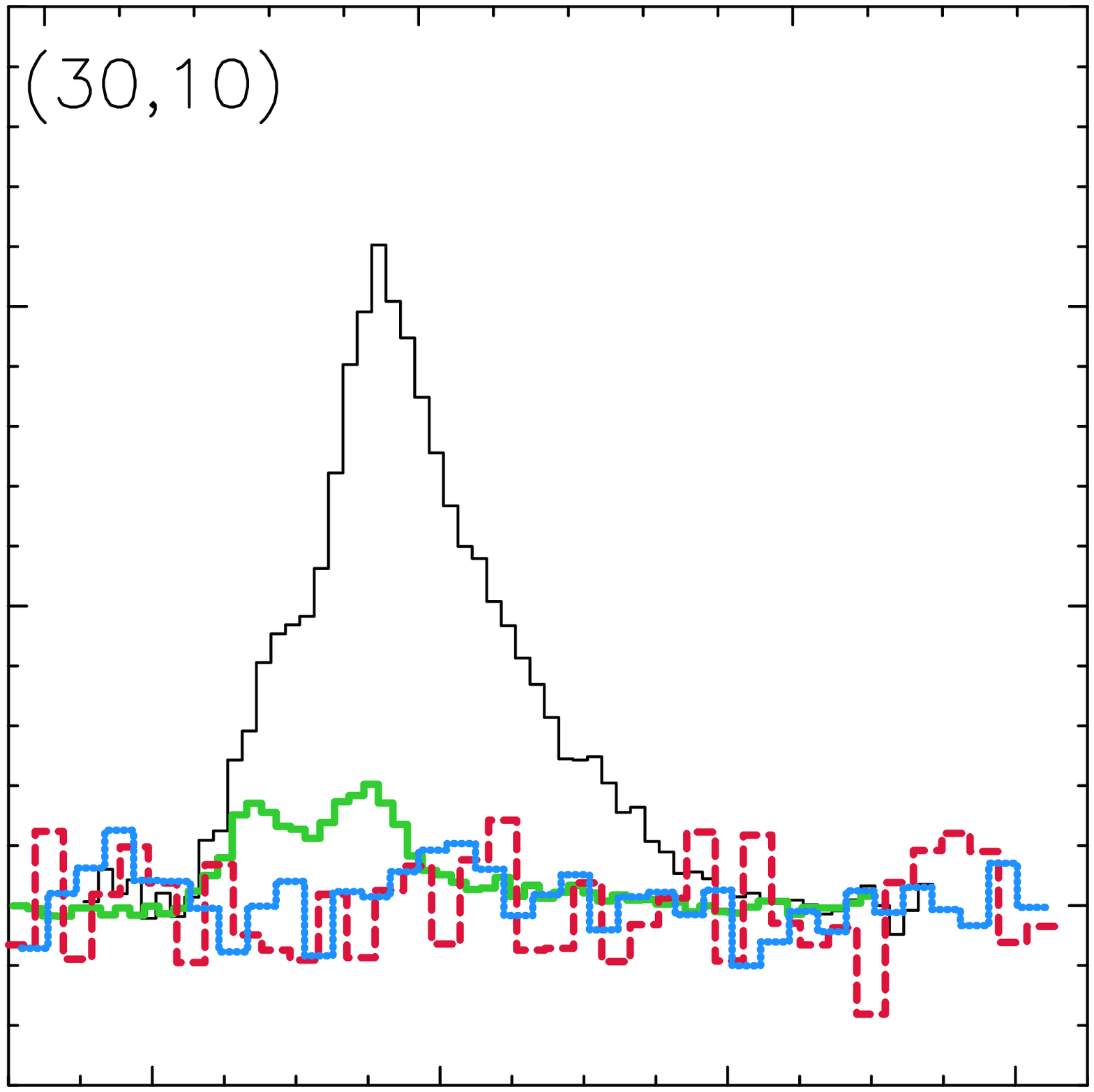}\\
\includegraphics[scale=0.168, angle=0]{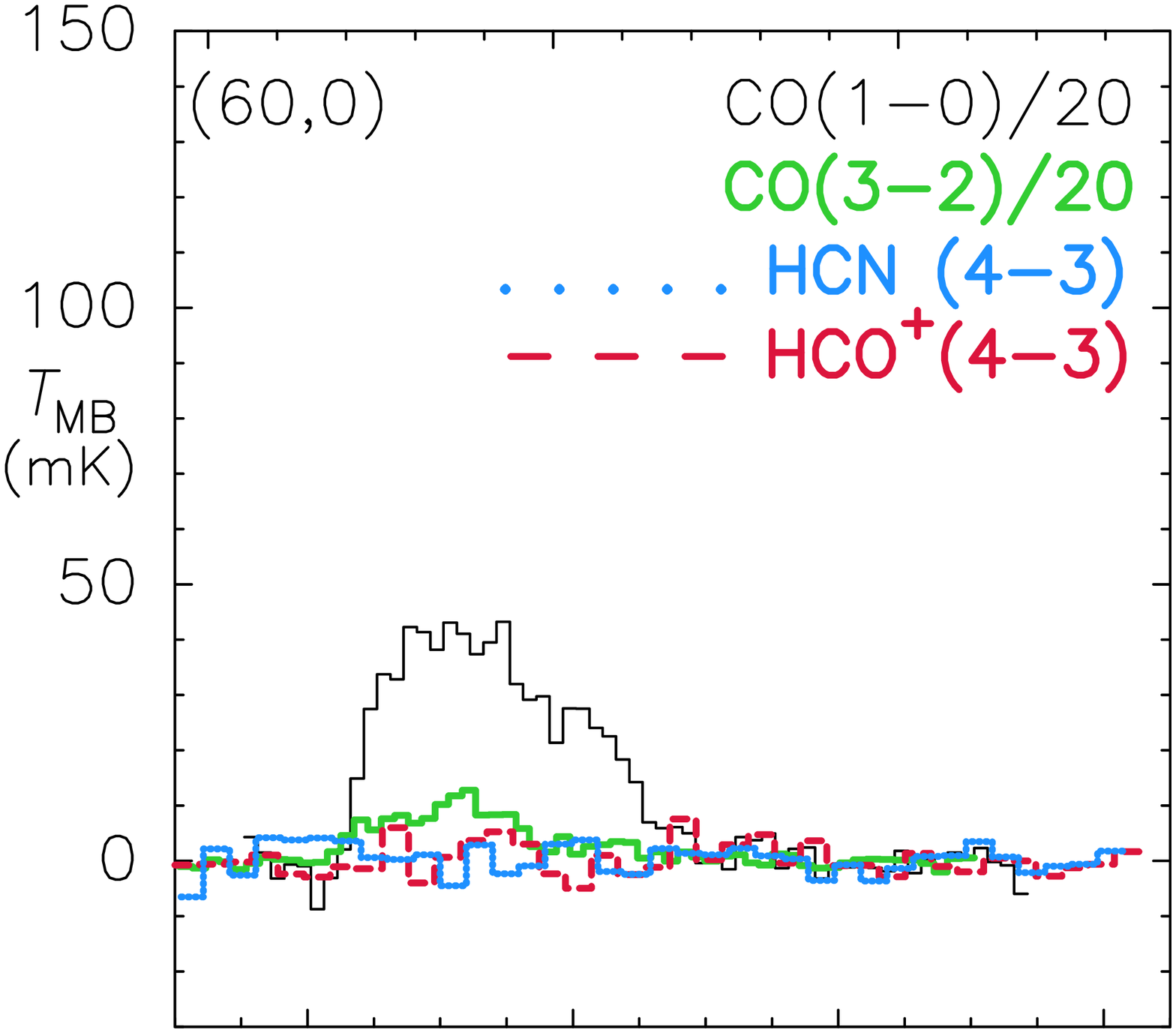}&
\includegraphics[scale=0.168, angle=0]{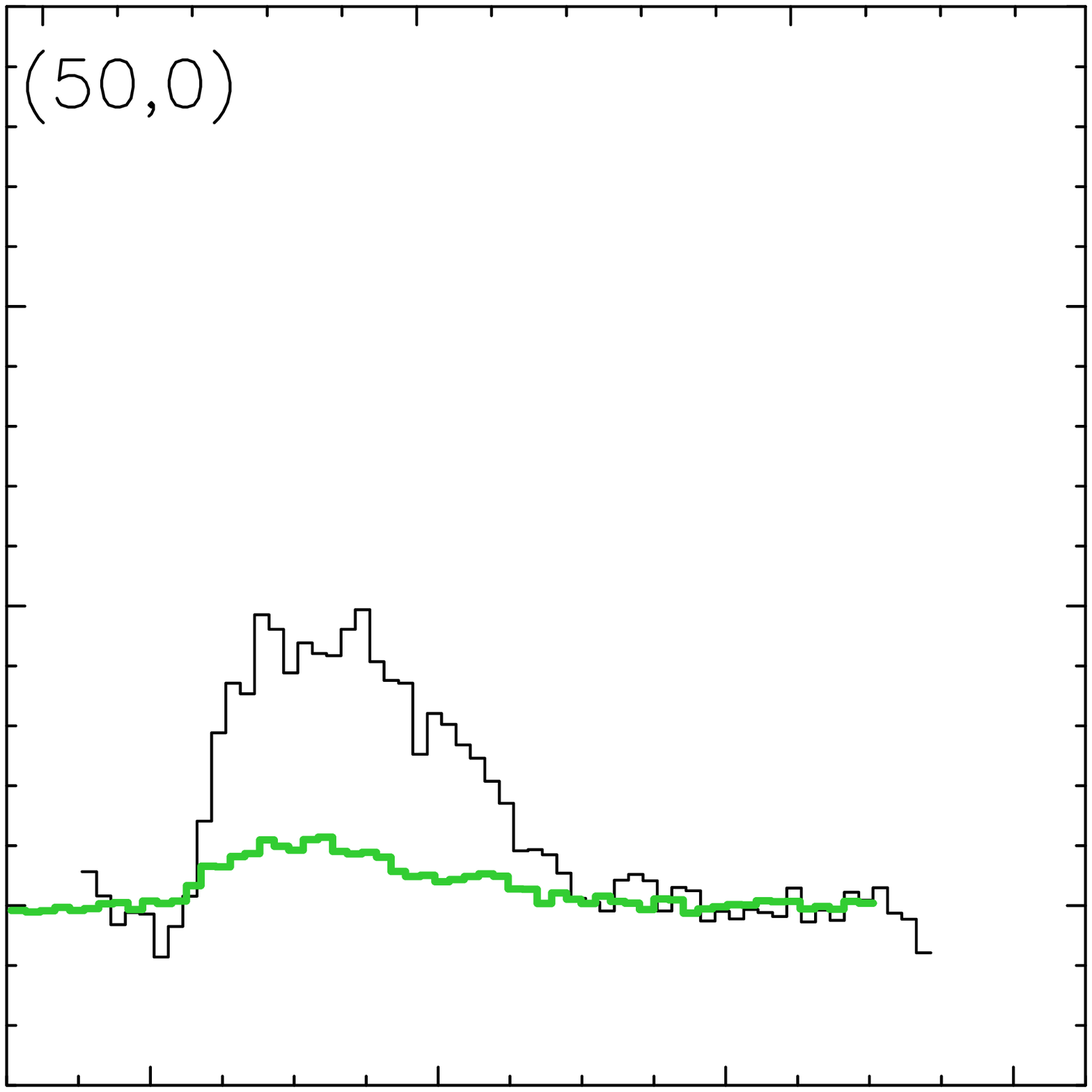}&
\includegraphics[scale=0.168, angle=0]{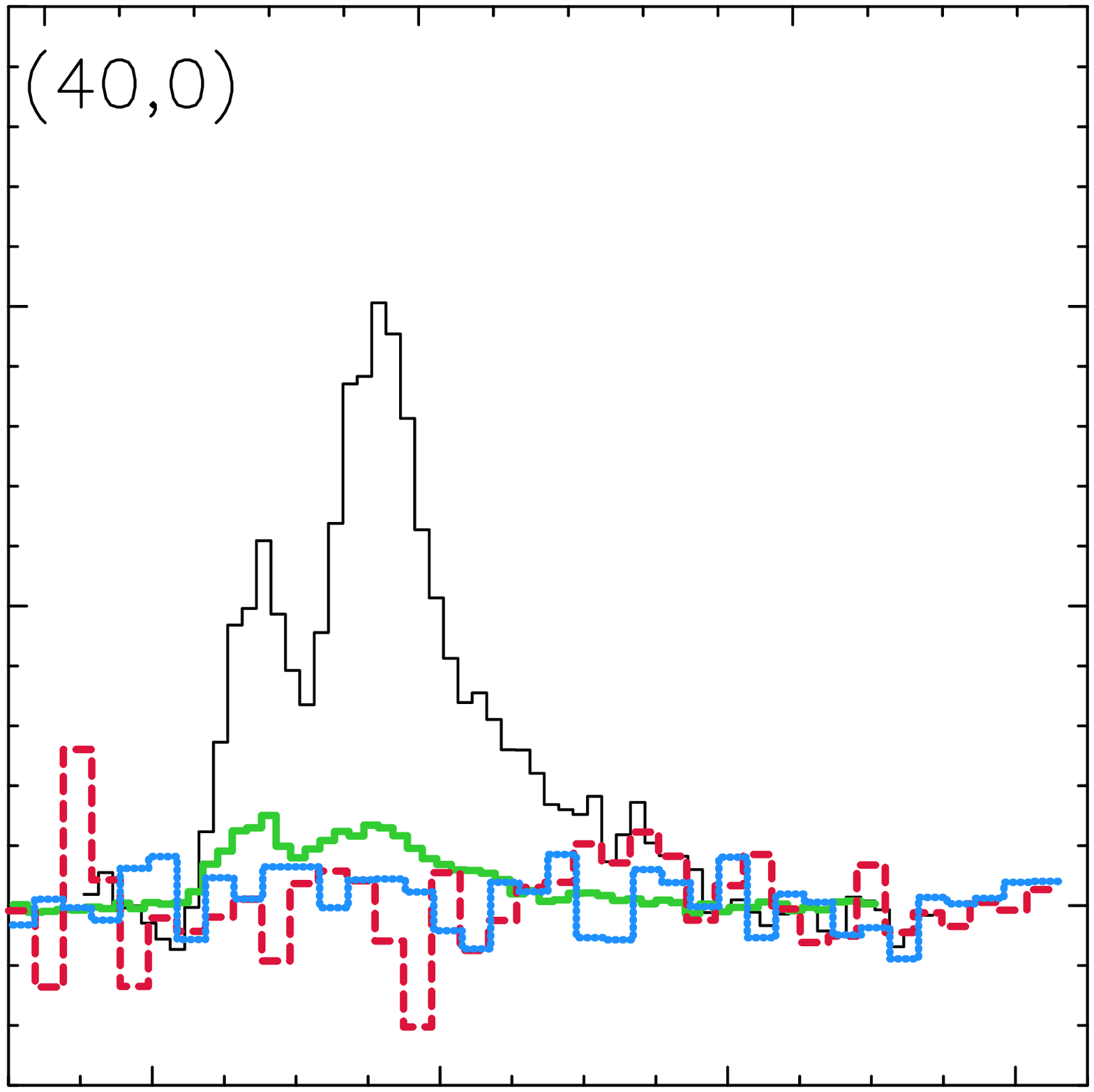}&
\includegraphics[scale=0.168, angle=0]{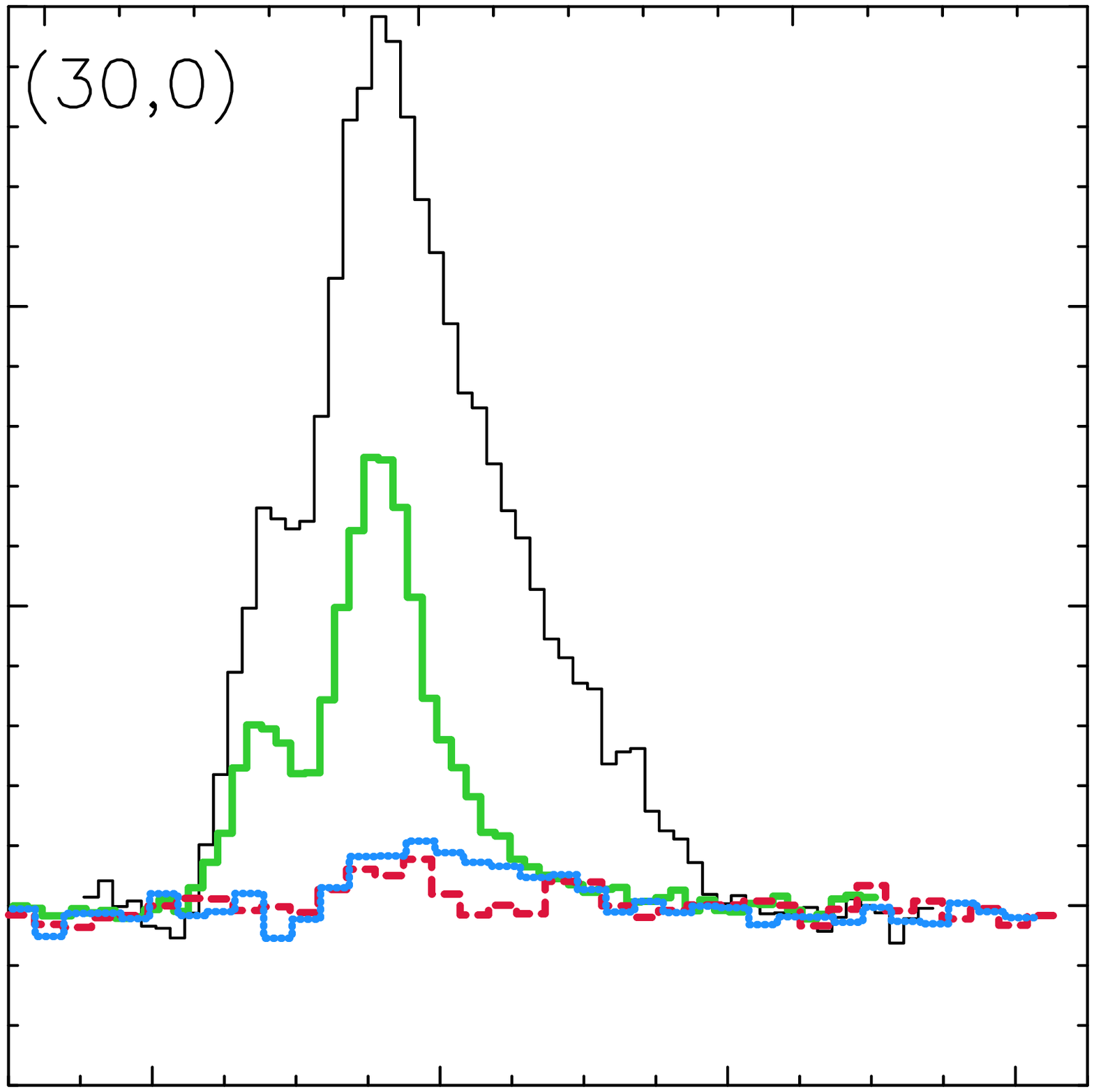}\\
\includegraphics[scale=0.168, angle=0]{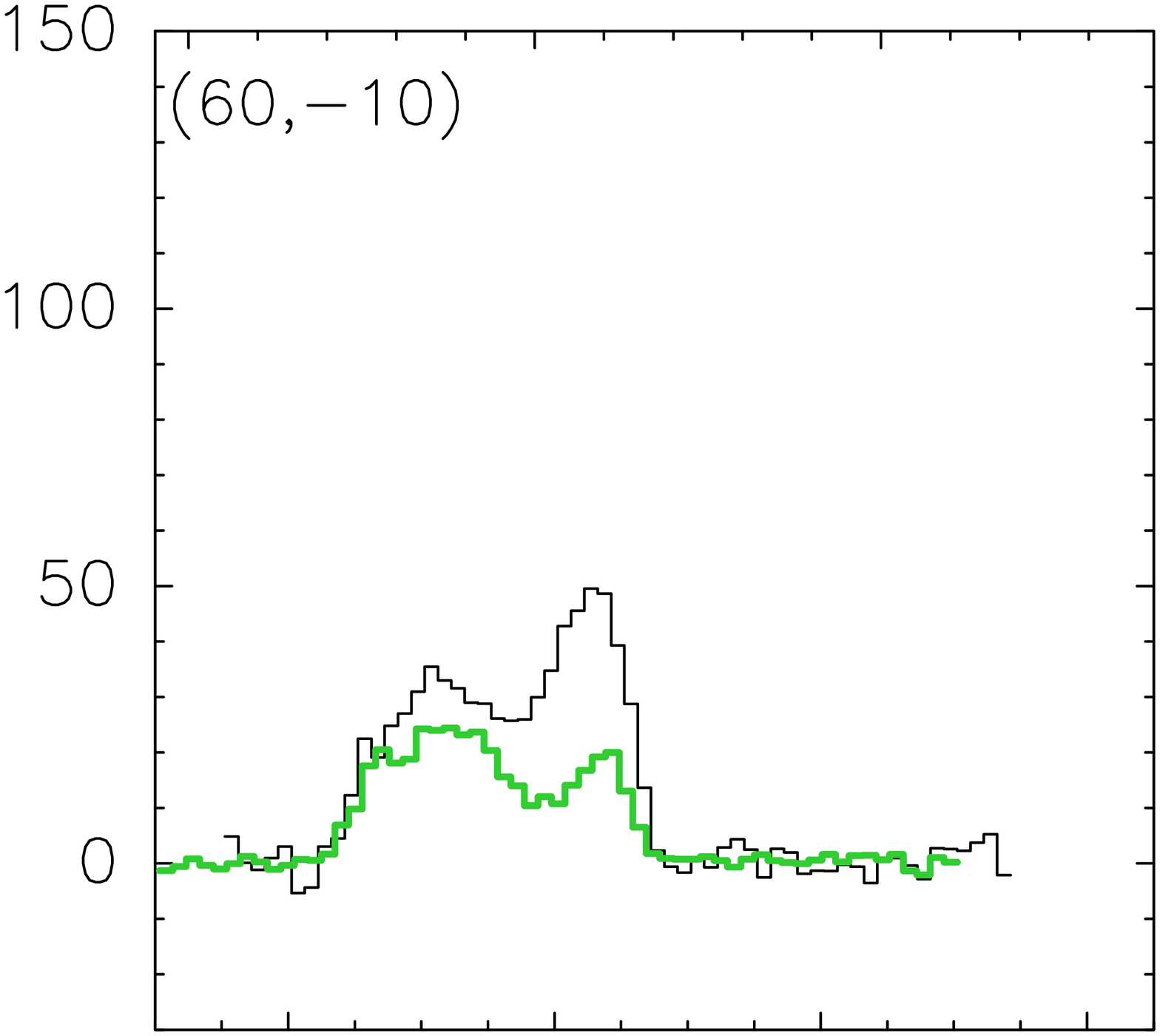}&
\includegraphics[scale=0.168, angle=0]{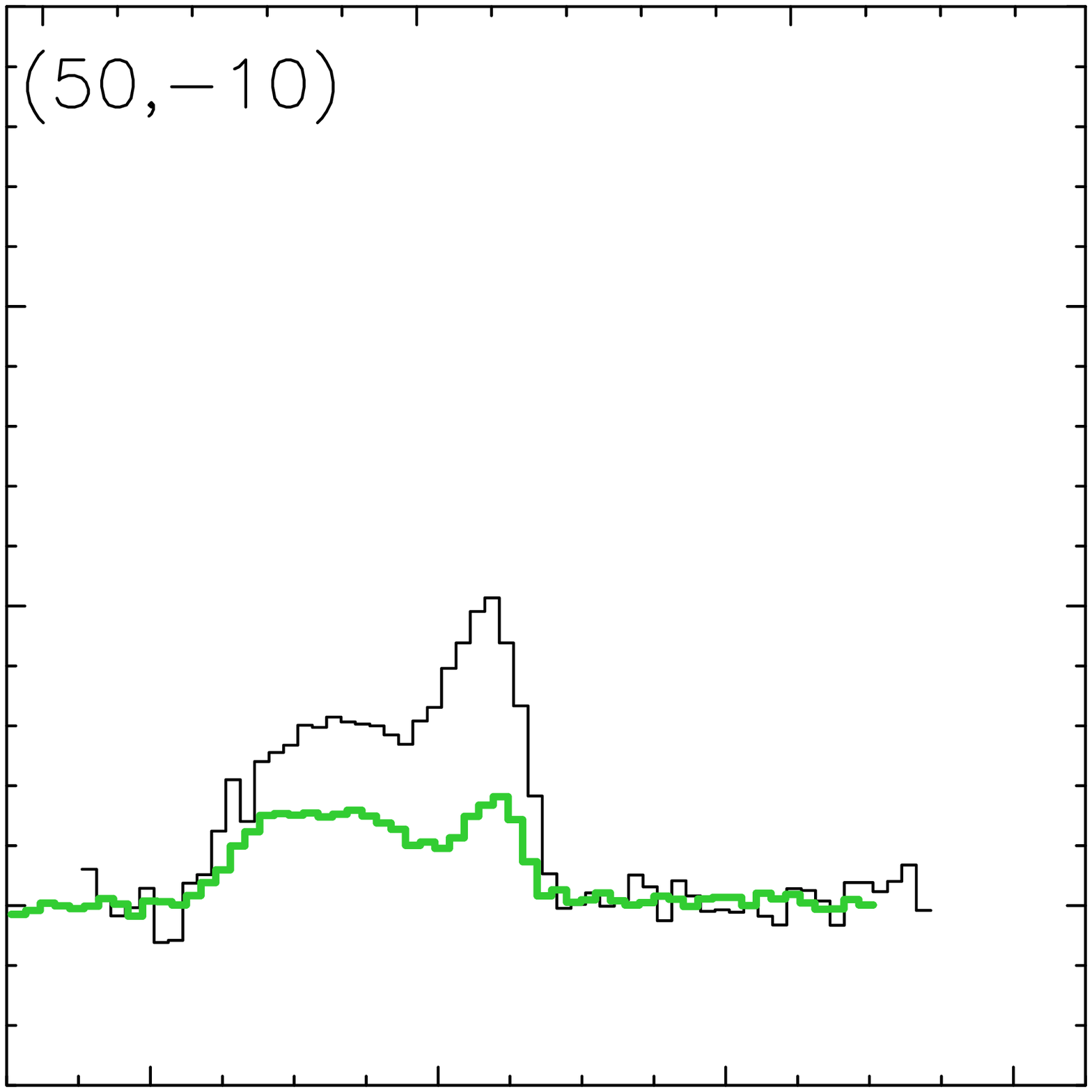}&
\includegraphics[scale=0.168, angle=0]{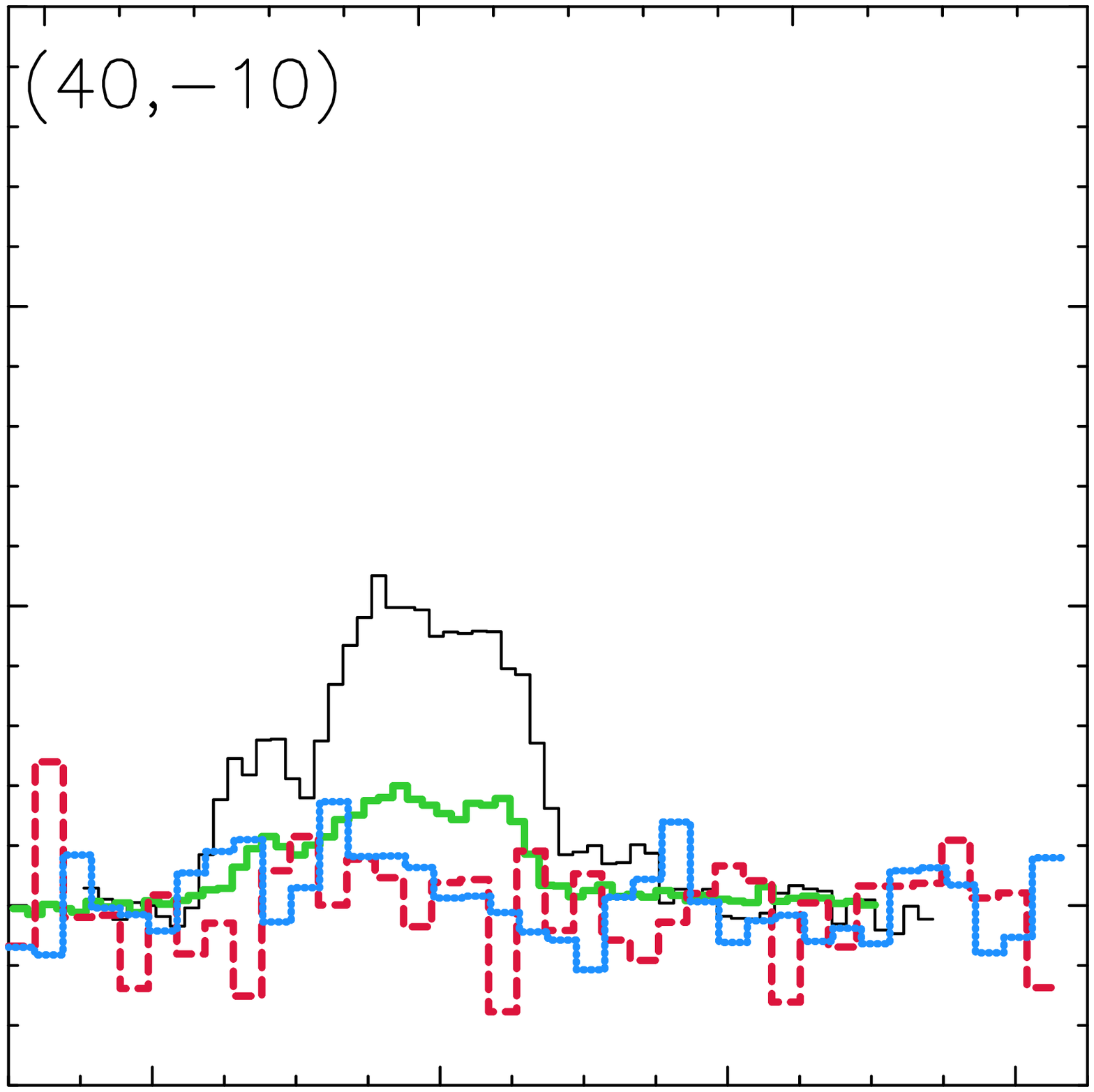}&
\includegraphics[scale=0.168, angle=0]{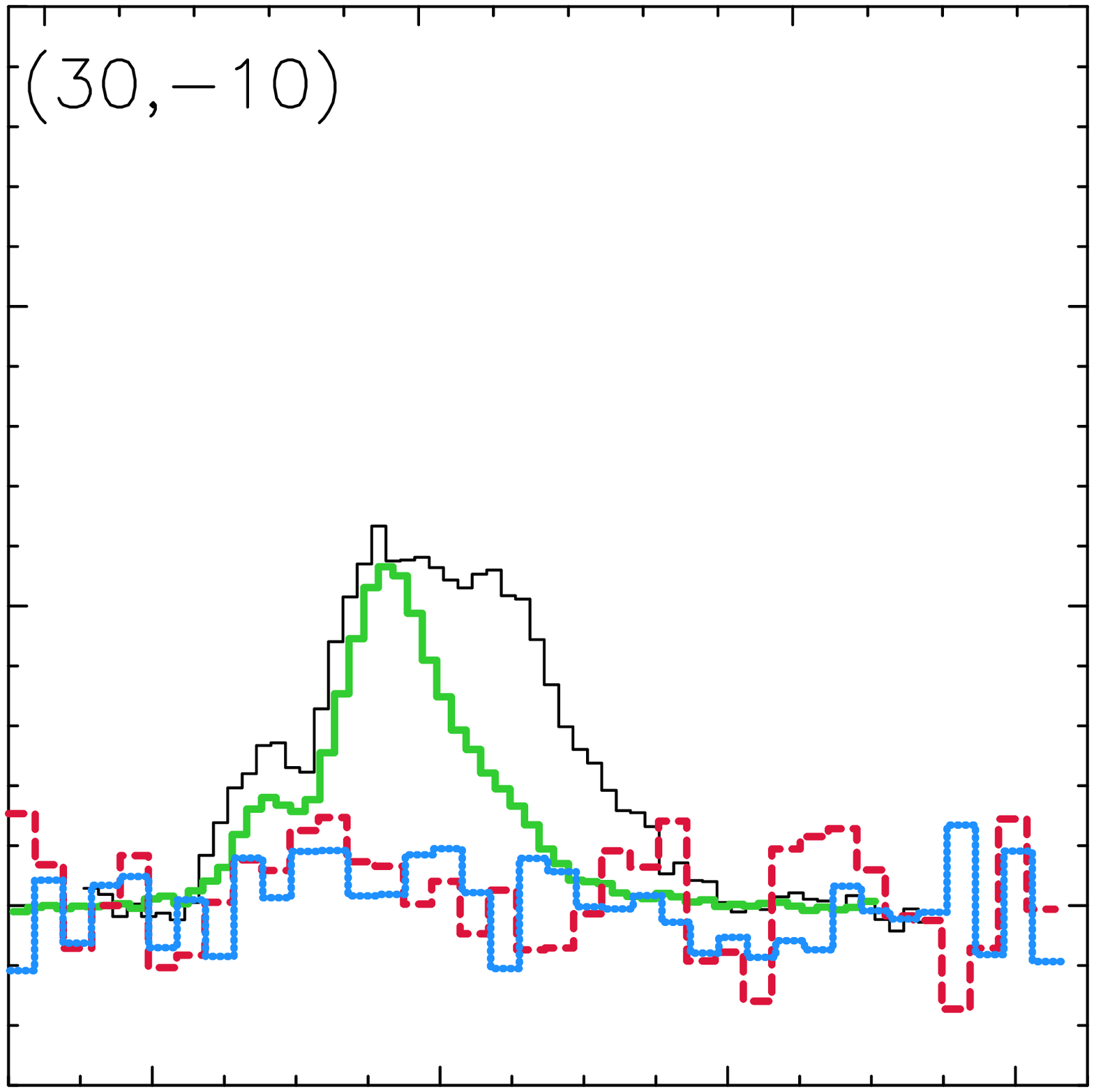} \\
\includegraphics[scale=0.168, angle=0]{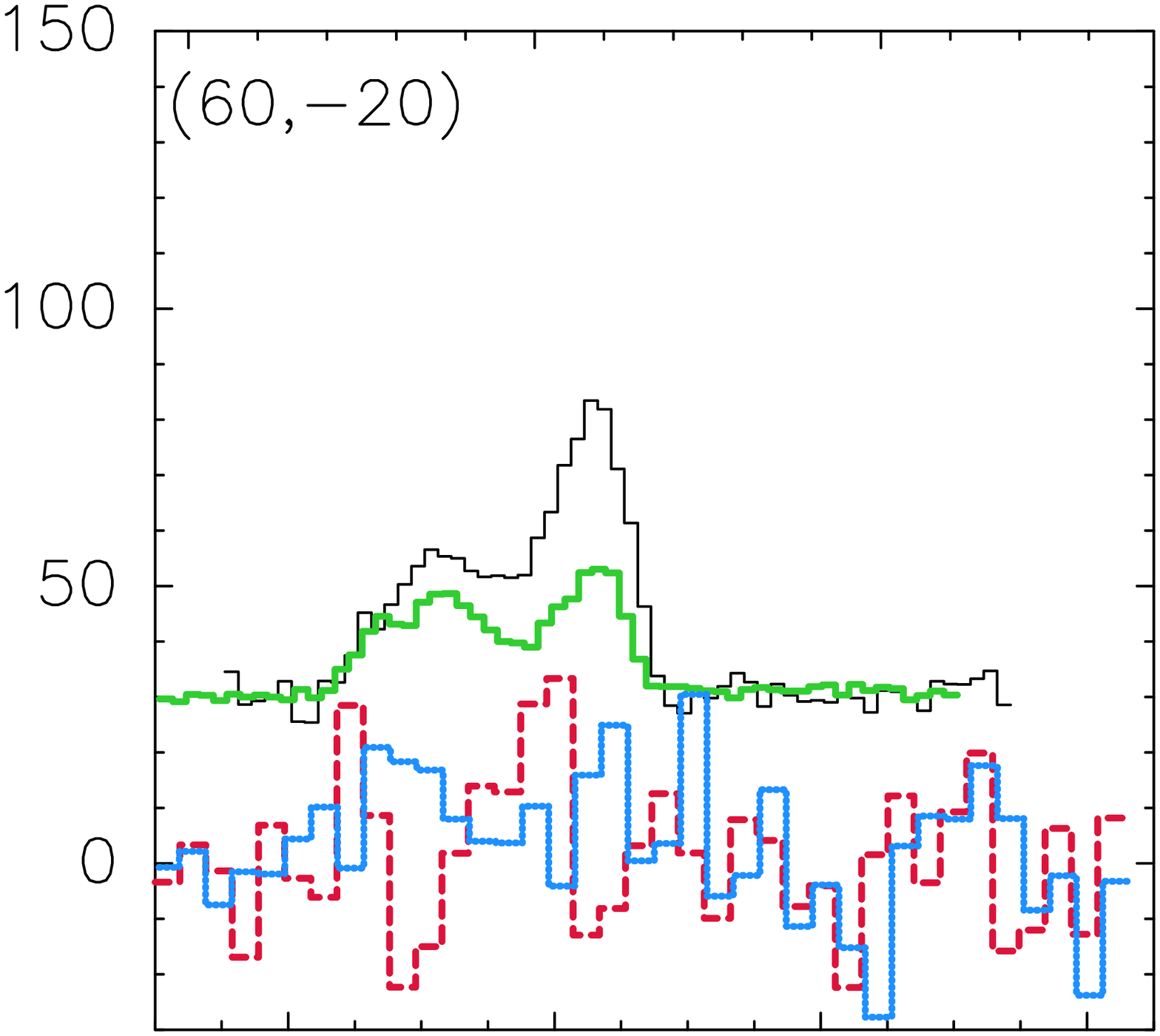}&
\includegraphics[scale=0.168, angle=0]{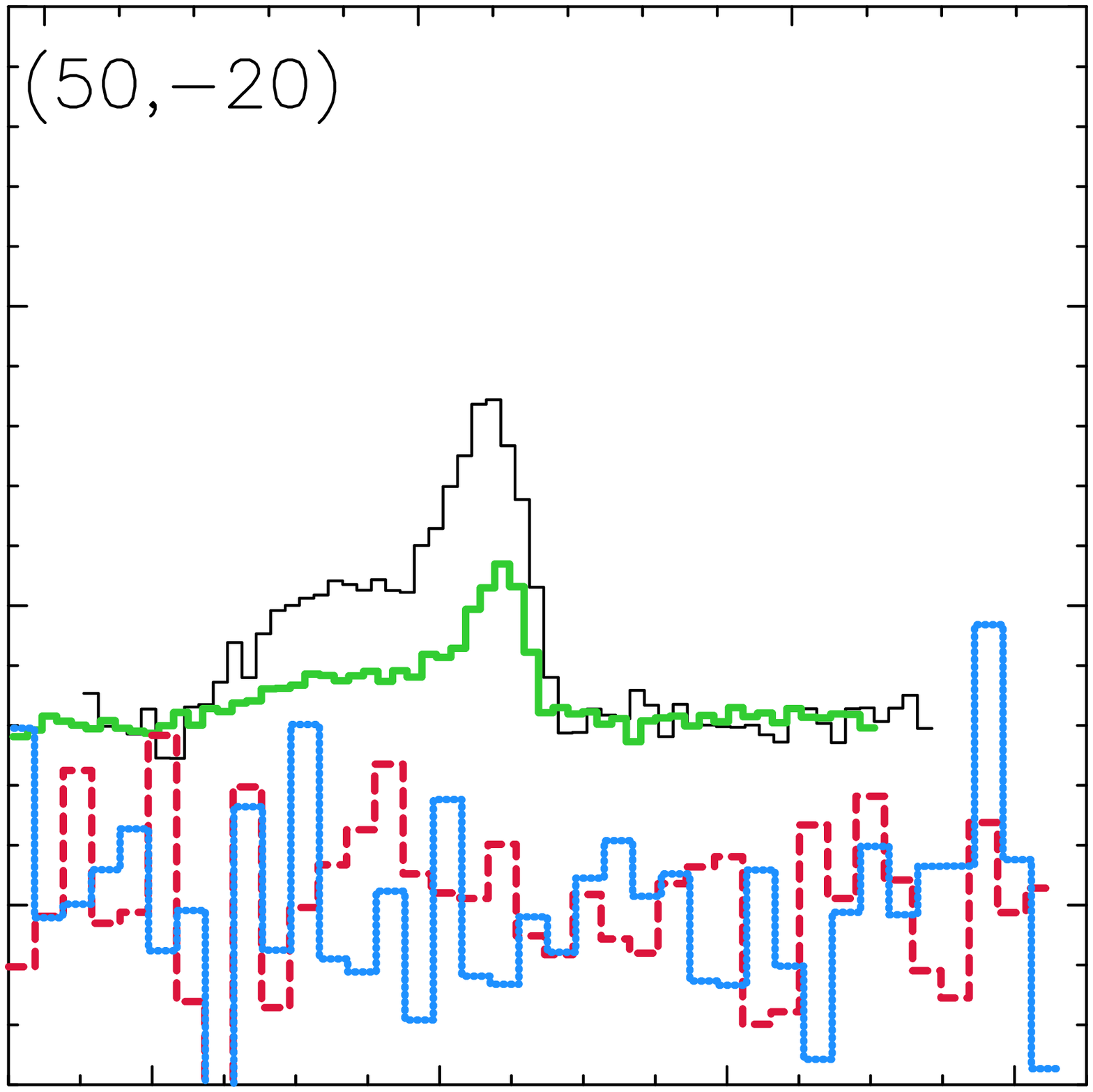}&
\includegraphics[scale=0.168, angle=0]{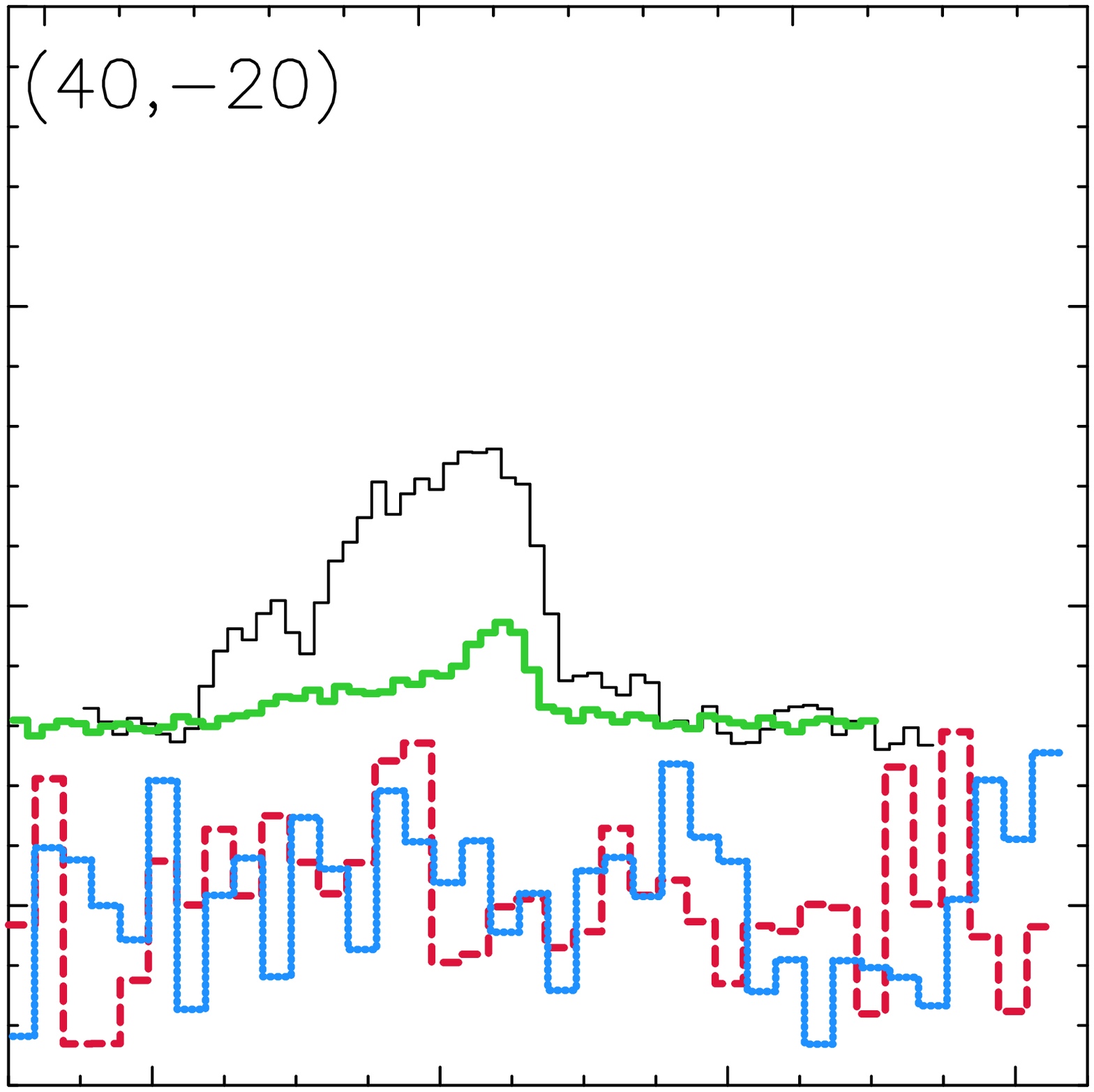}&
\includegraphics[scale=0.168, angle=0]{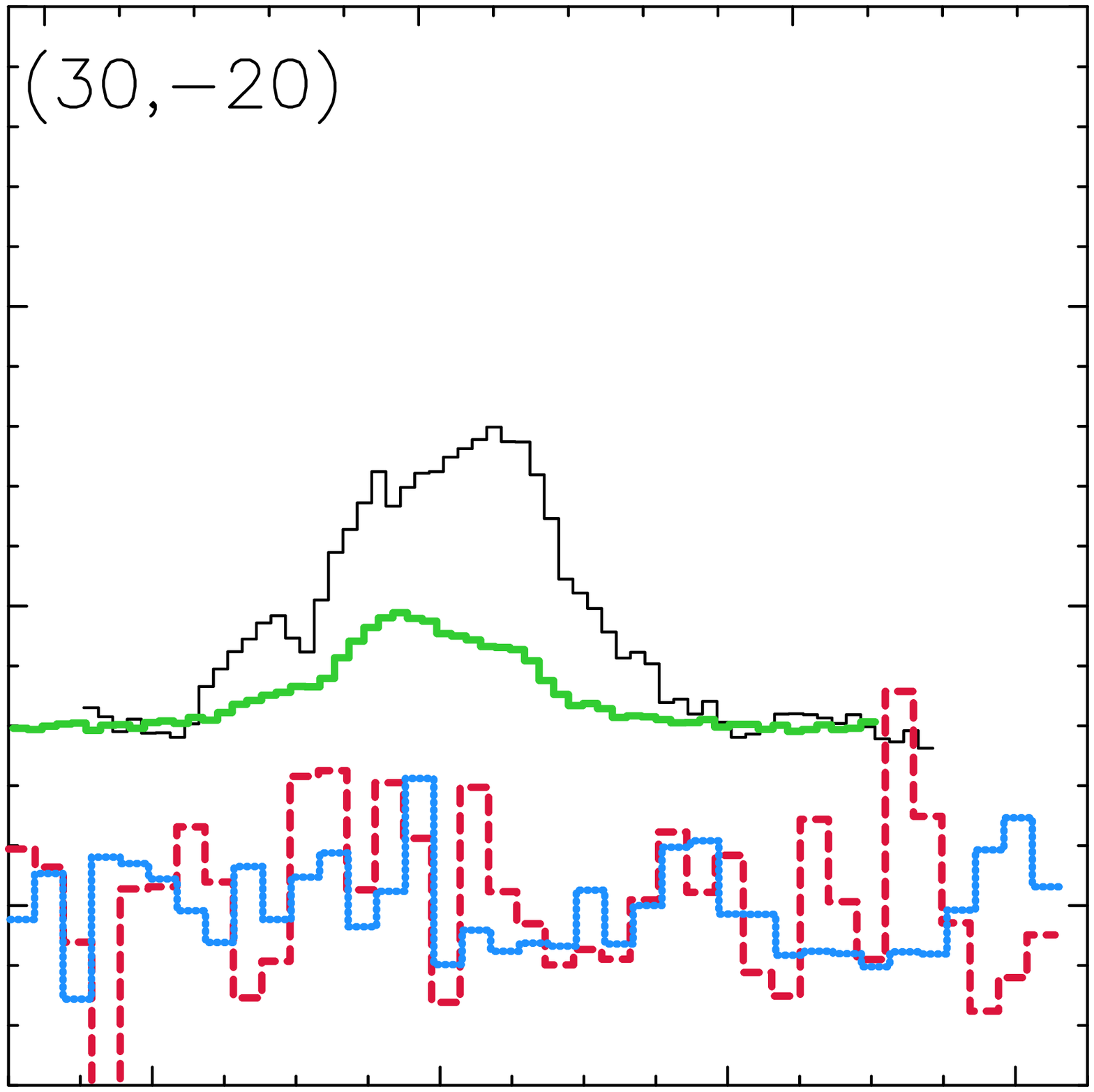}\\
\includegraphics[scale=0.168, angle=0]{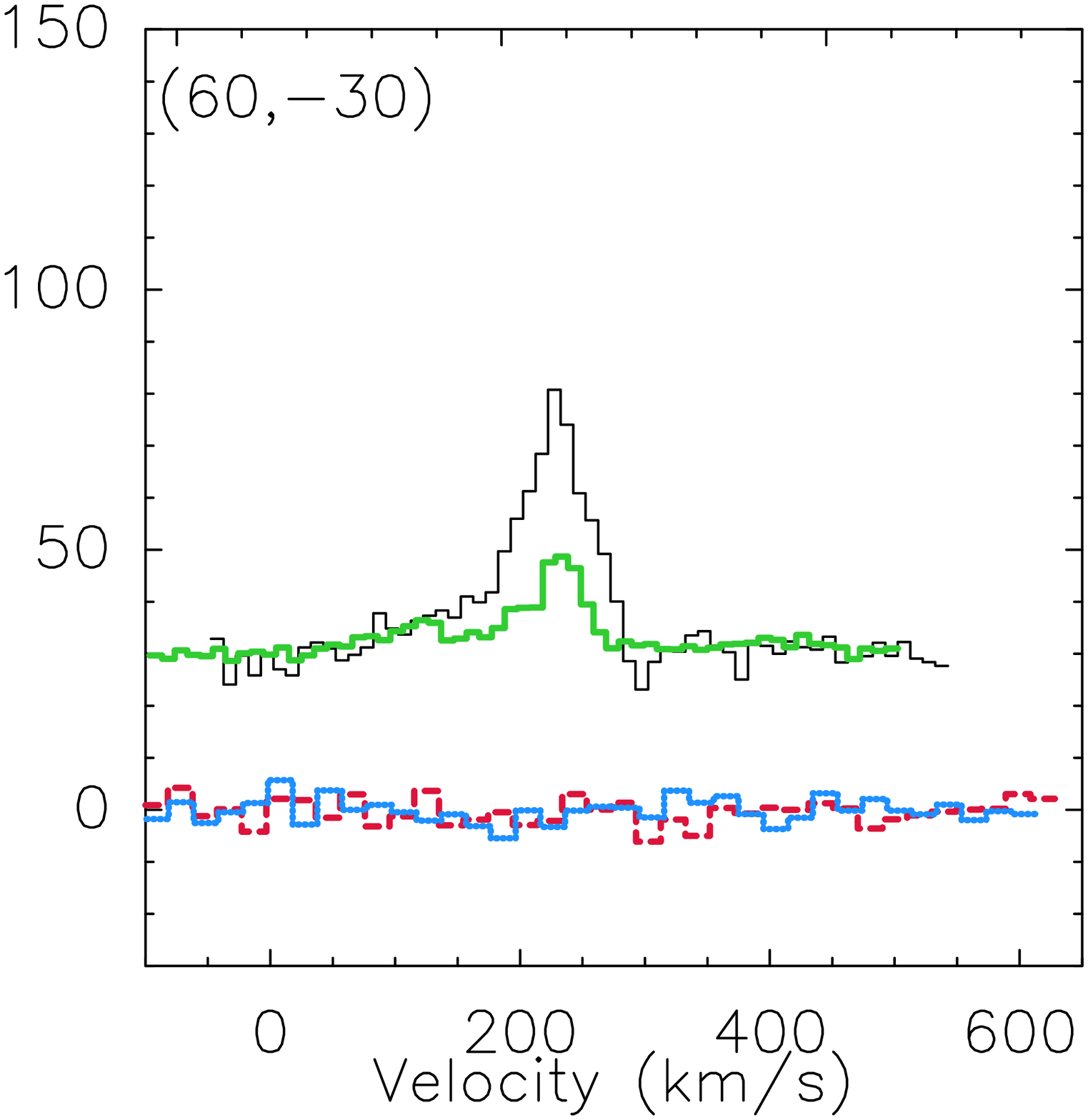}&
\includegraphics[scale=0.168, angle=0]{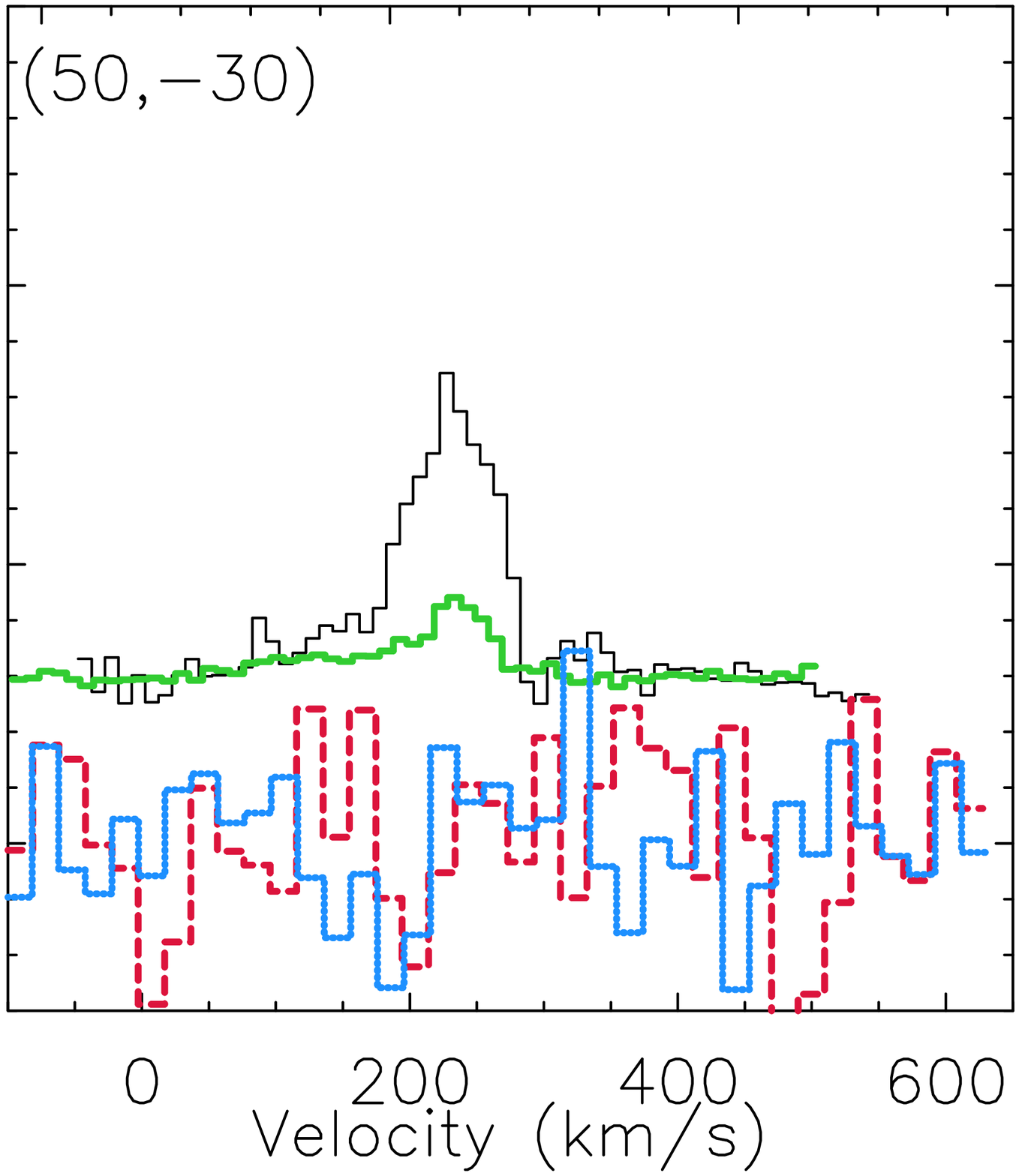}&
\includegraphics[scale=0.168, angle=0]{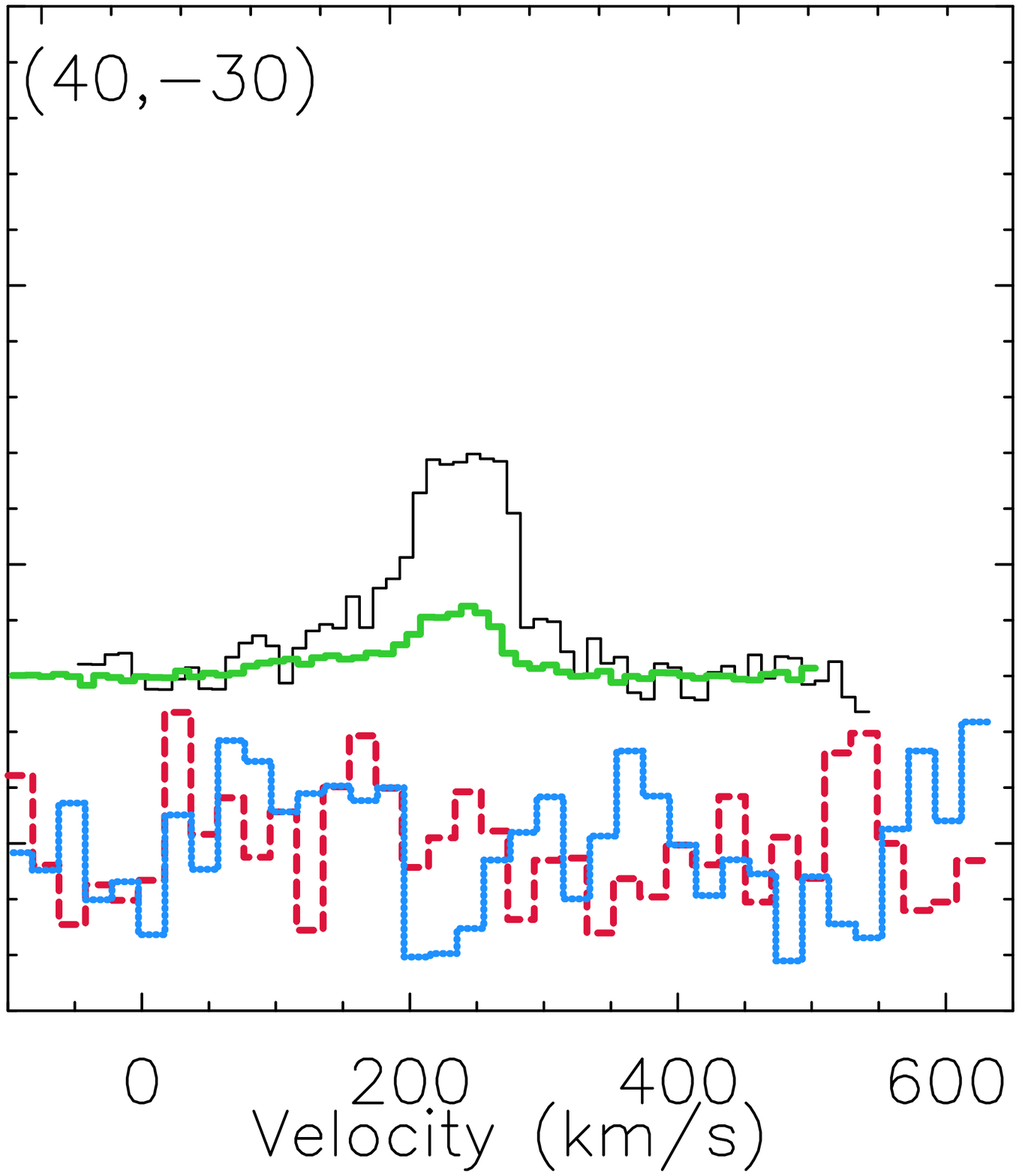}&
\includegraphics[scale=0.168, angle=0]{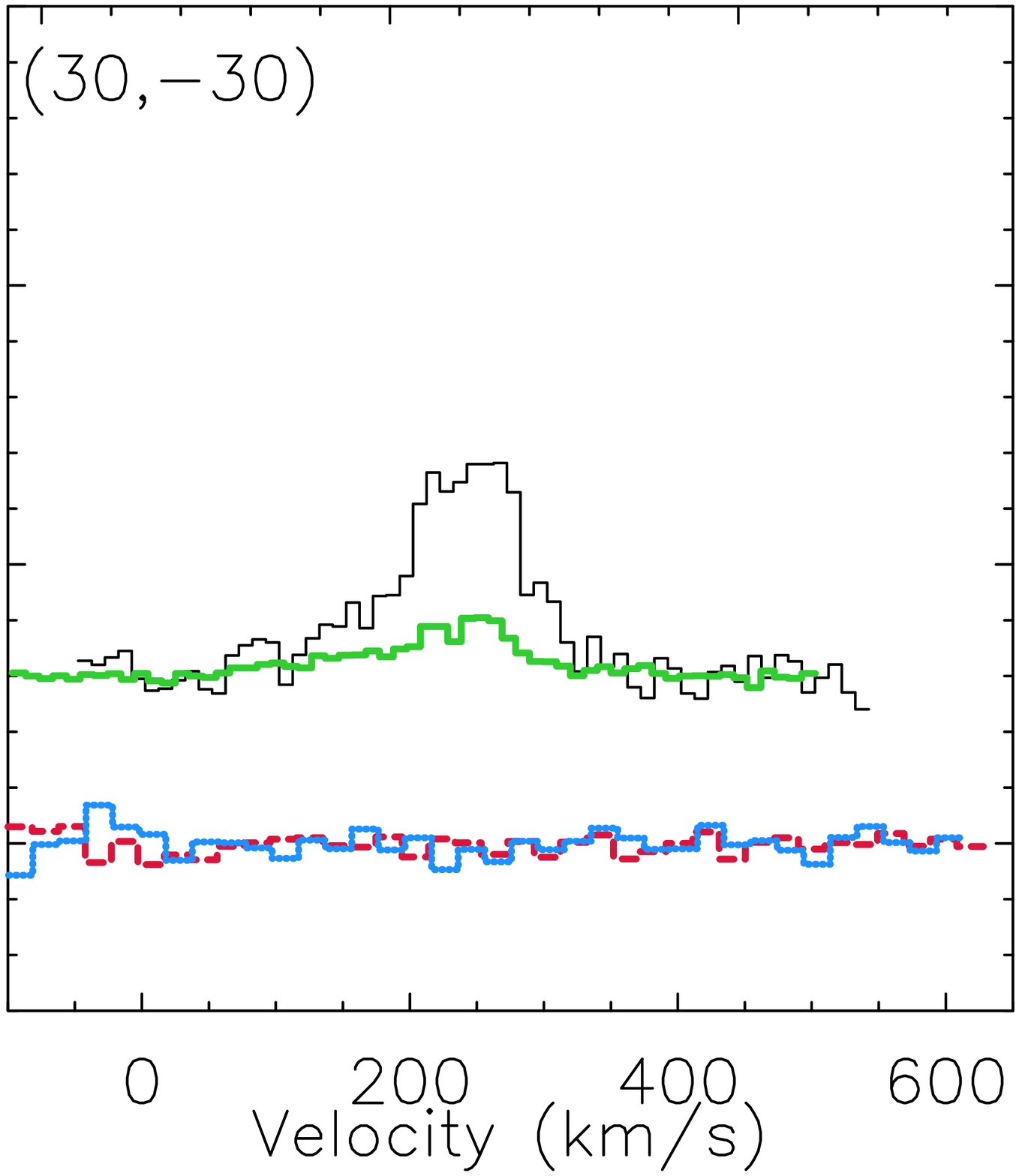}\\
\end{array}$
\end{center}
\begin{center}
\caption{Spectra of CO 1-0 (black), CO 3-2 (green), HCN 4-3
(blue) and HCO$^+$ 4-3 (red) emission in the central $\sim$ 1\,kpc region of NGC\,253.
The $T_\text{MB}$ (in unit of mK) range on the $y$-axis is [-30,150]. CO 1-0 and CO 3-2 lines are scaled by a factor of 0.05 for clearer comparison with HCN 4-3 and \hcop\ 4-3. On the top two rows and the bottom two rows we offset the CO spectra by 30\,mK for clarity.
The velocity resolution is 10\,\kms\ for the two CO lines, and 20\,\kms\ for HCN 4-3 and \hcop\ 4-3. All data are
resampled with a pixel size of 10\,arcsec, corresponding to $\sim$ 170\,pc.
}
\label{fig:spec} 
\end{center} 
\end{figure*}

\begin{figure*}
\begin{center}$
\begin{array}{ccccc}
\includegraphics[scale=0.168, angle=0]{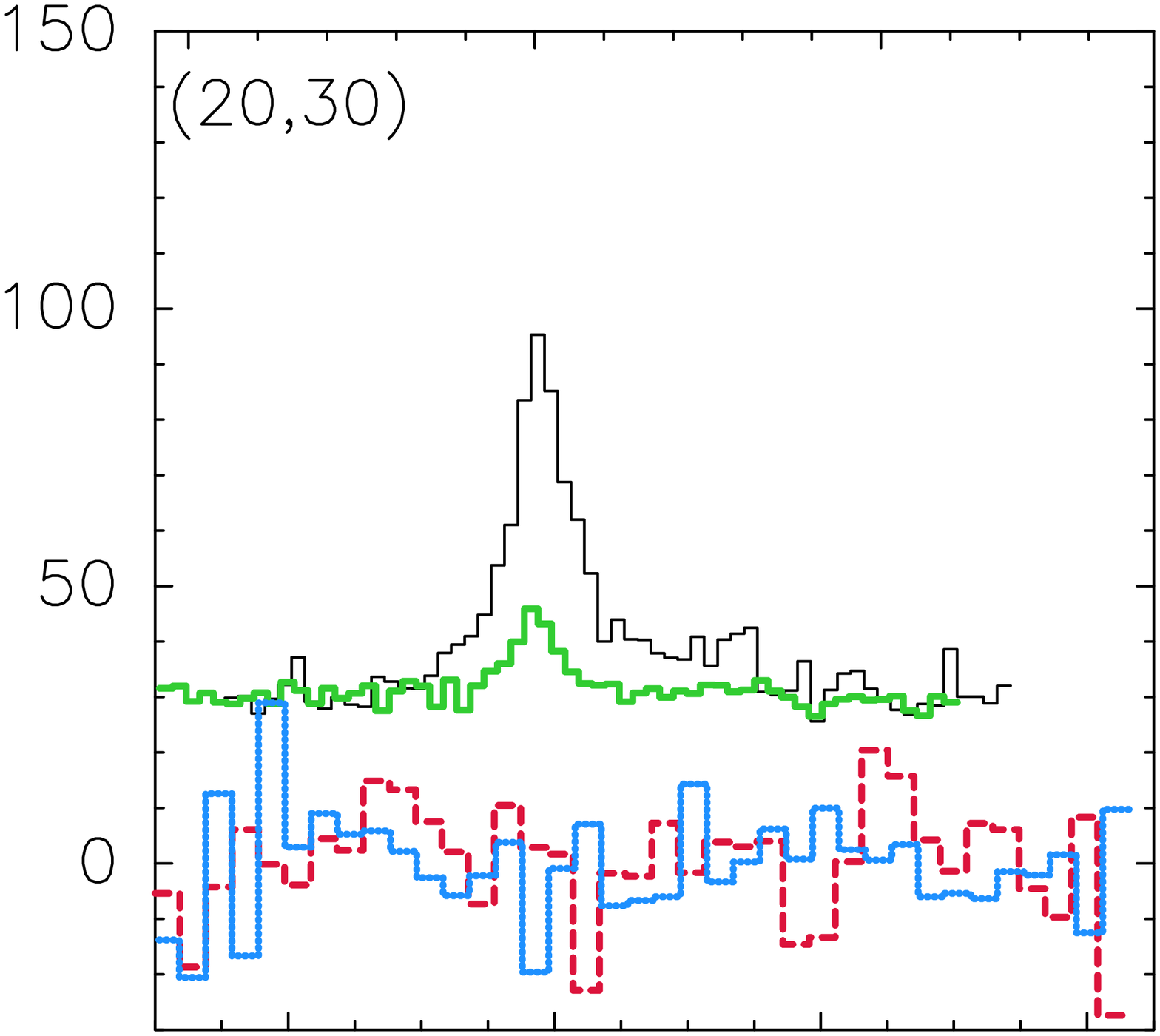}&
\includegraphics[scale=0.168, angle=0]{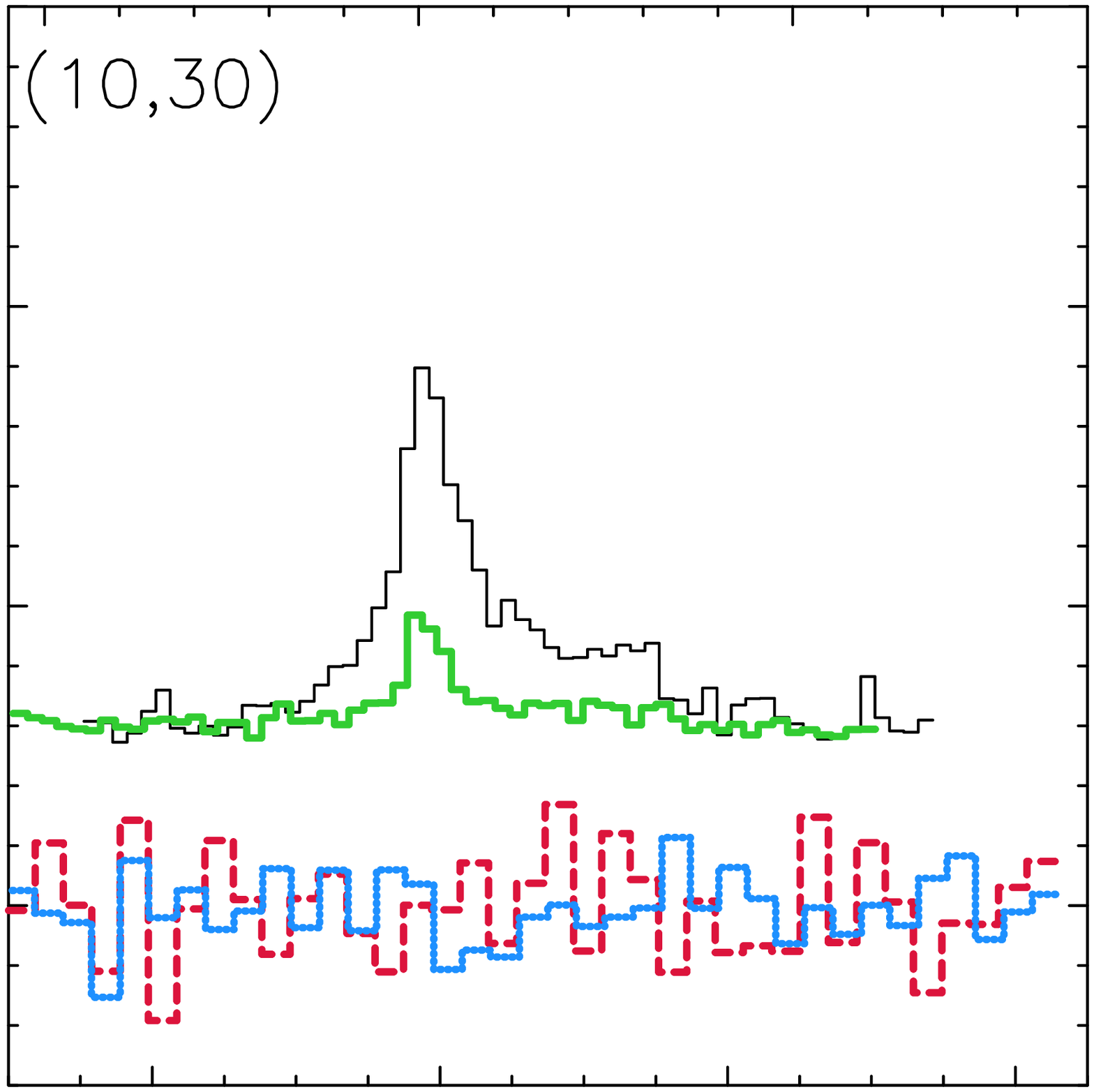}&
\includegraphics[scale=0.168, angle=0]{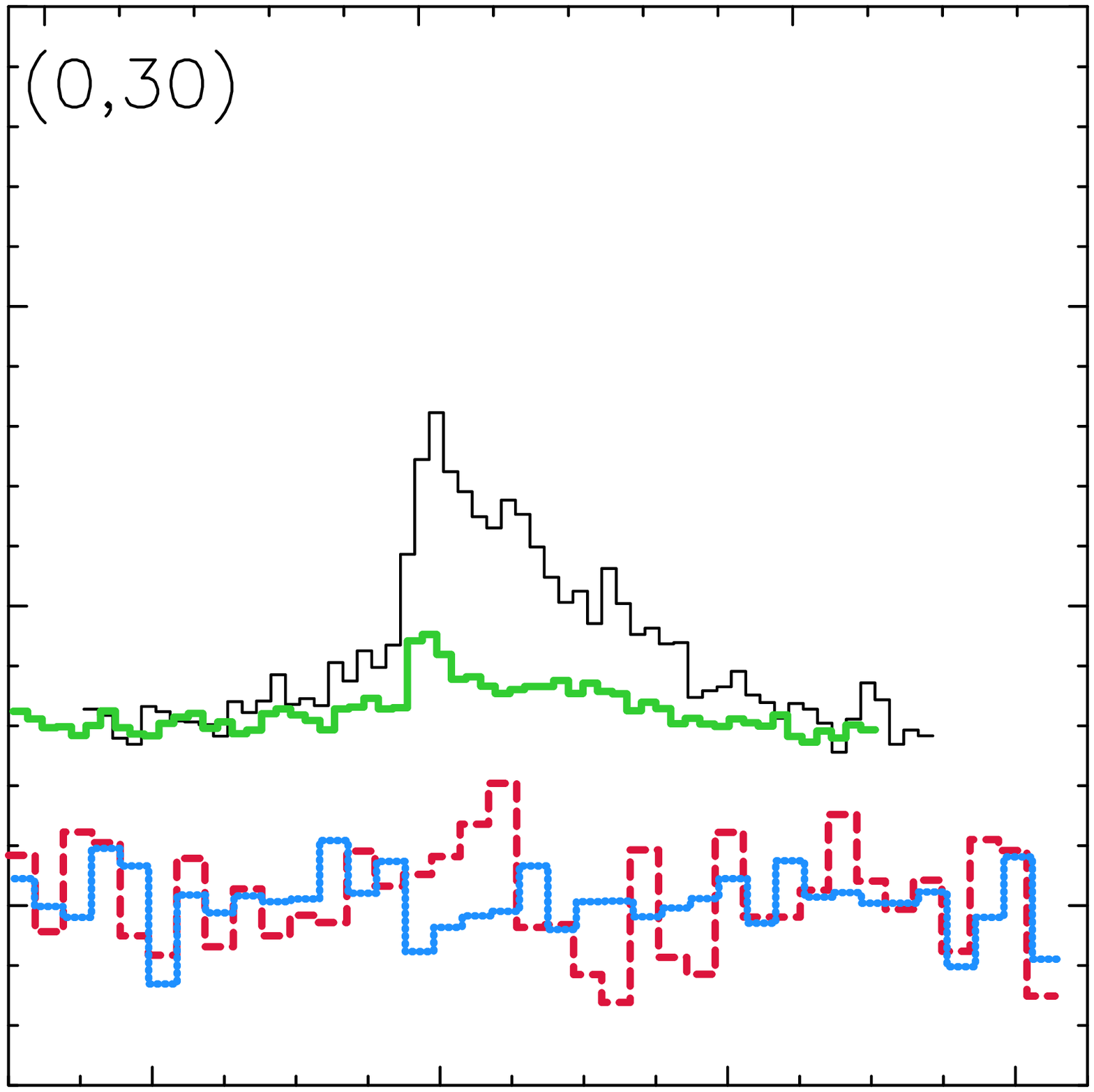}&
\includegraphics[scale=0.168, angle=0]{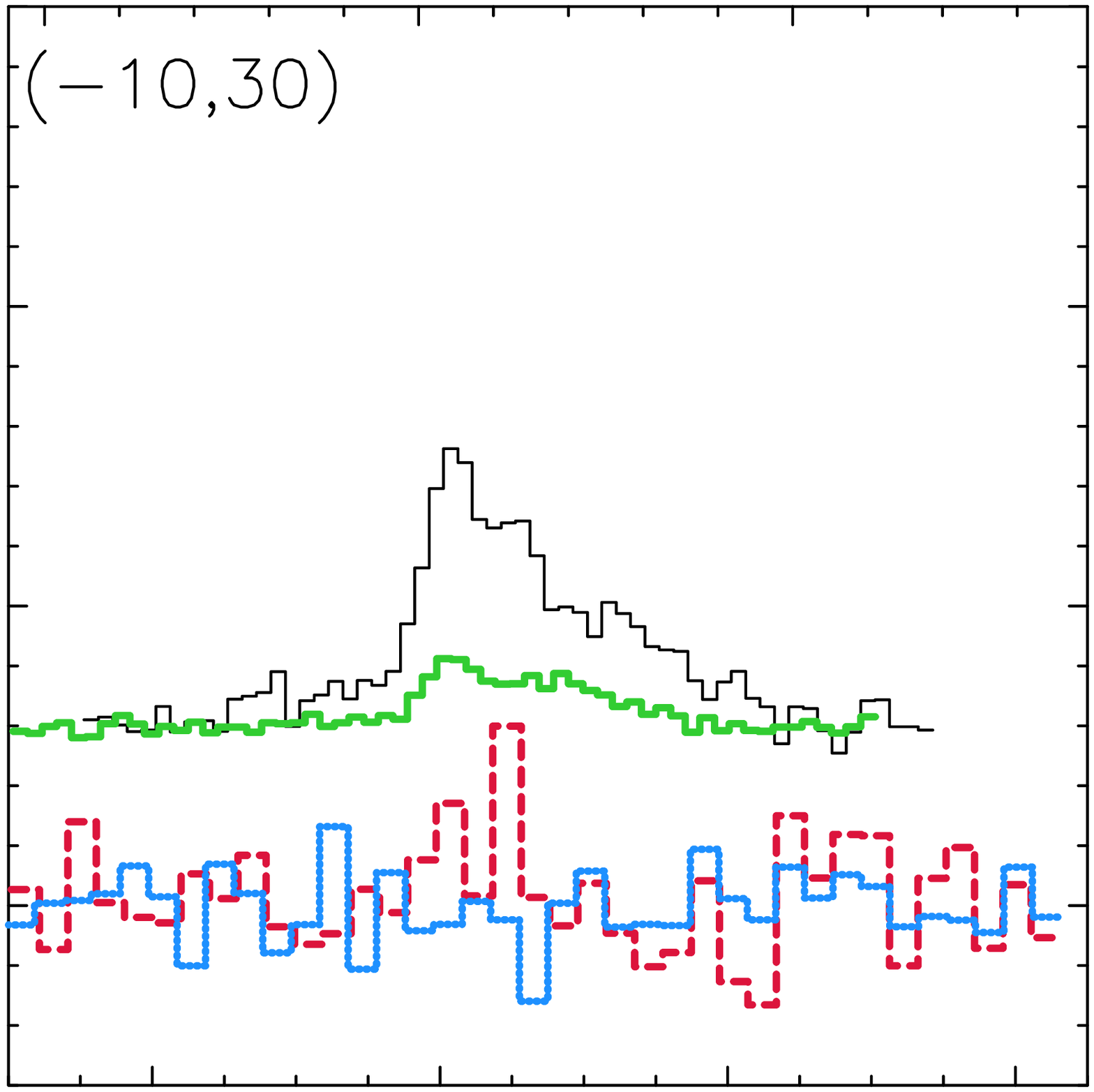}&
\includegraphics[scale=0.168, angle=0]{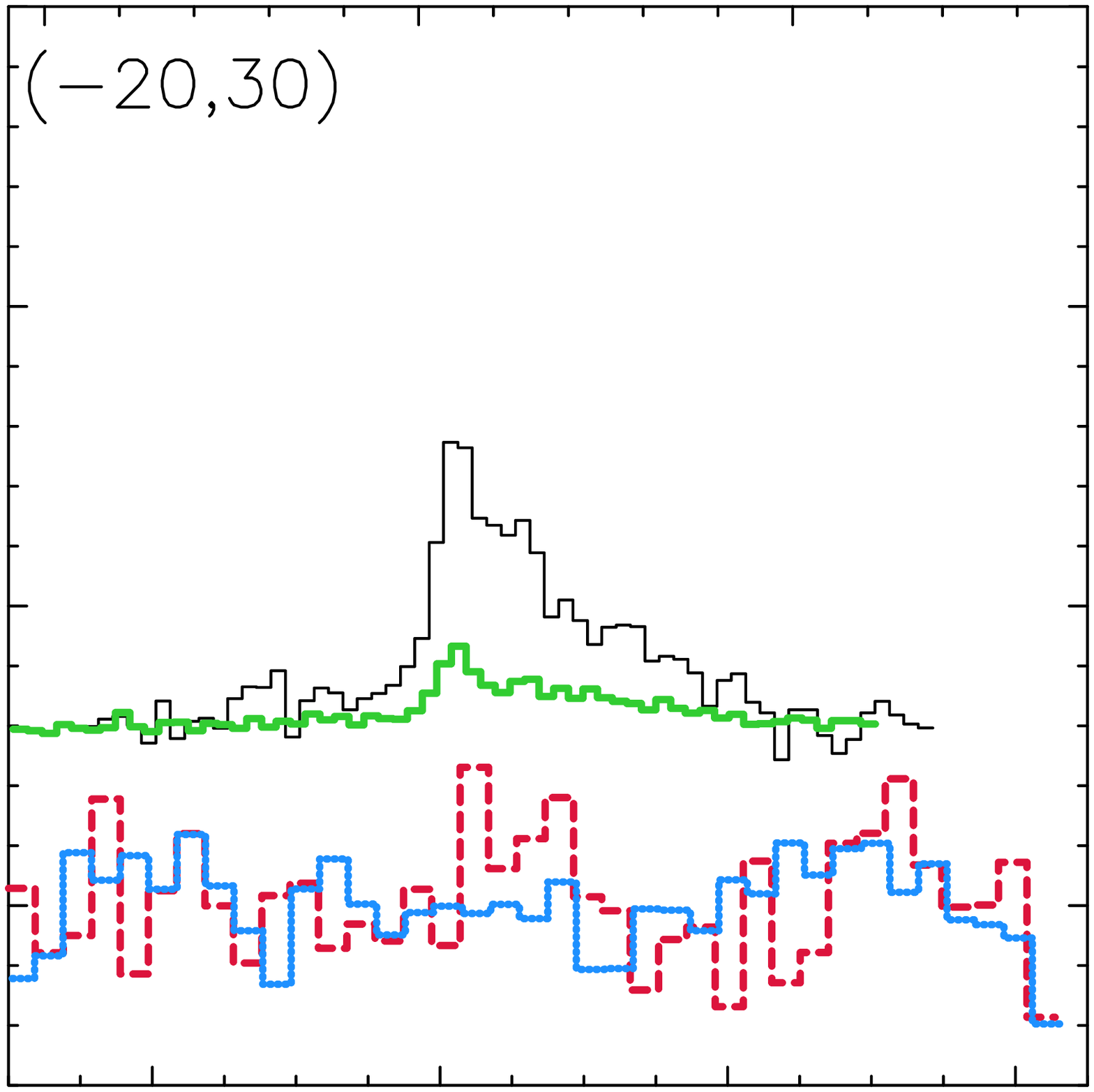}\\
\includegraphics[scale=0.168, angle=0]{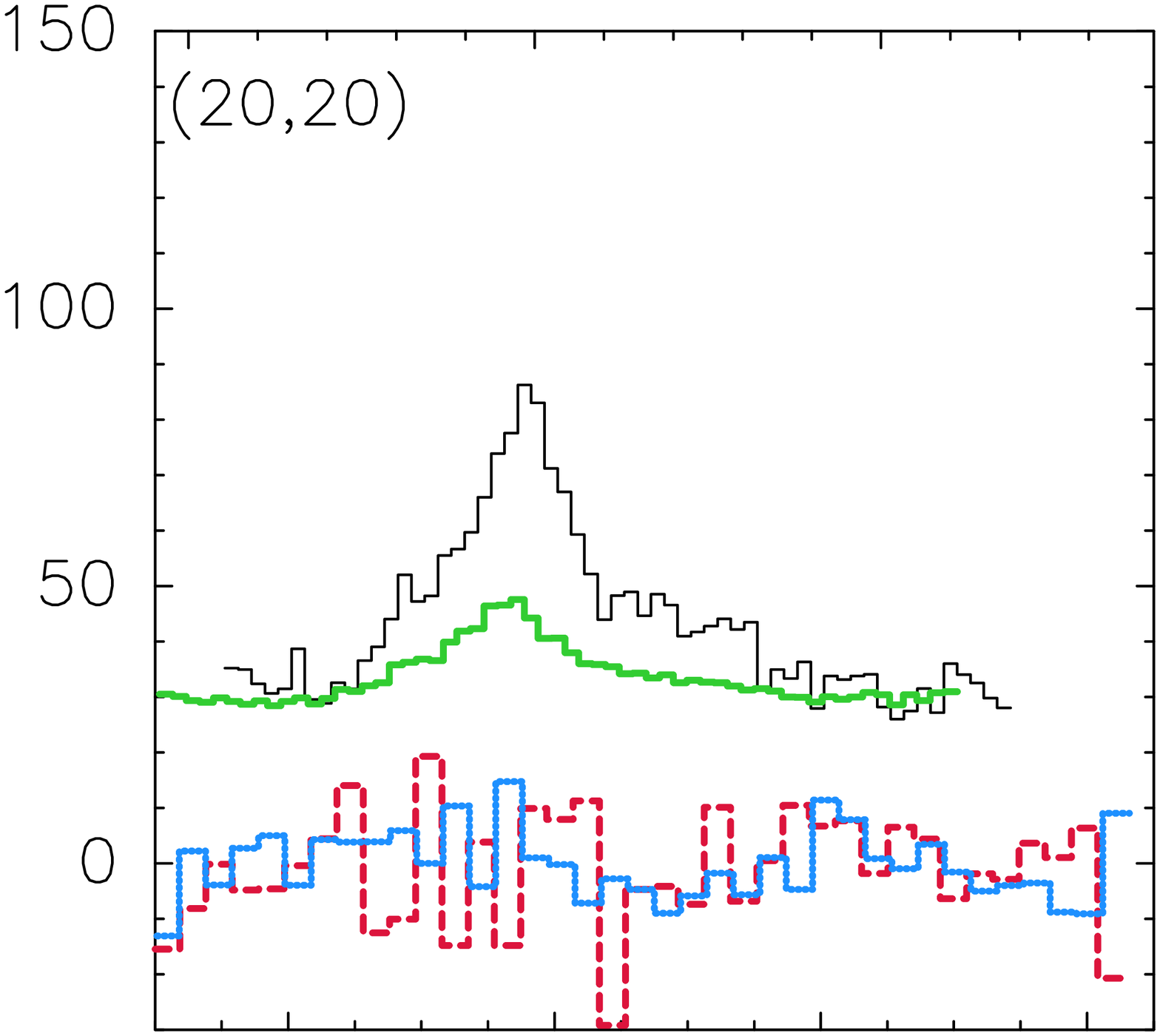}&
\includegraphics[scale=0.168, angle=0]{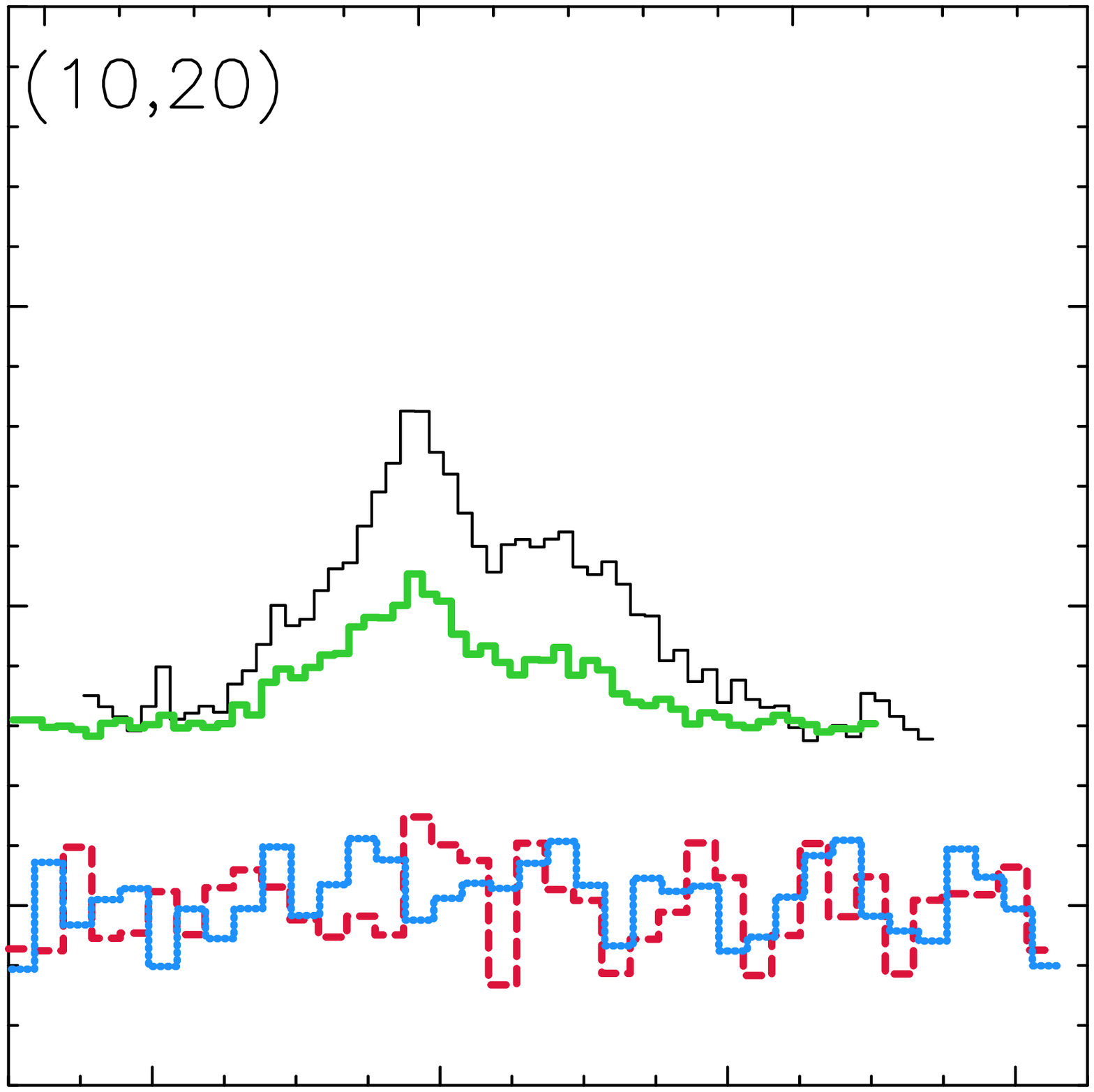}&
\includegraphics[scale=0.168, angle=0]{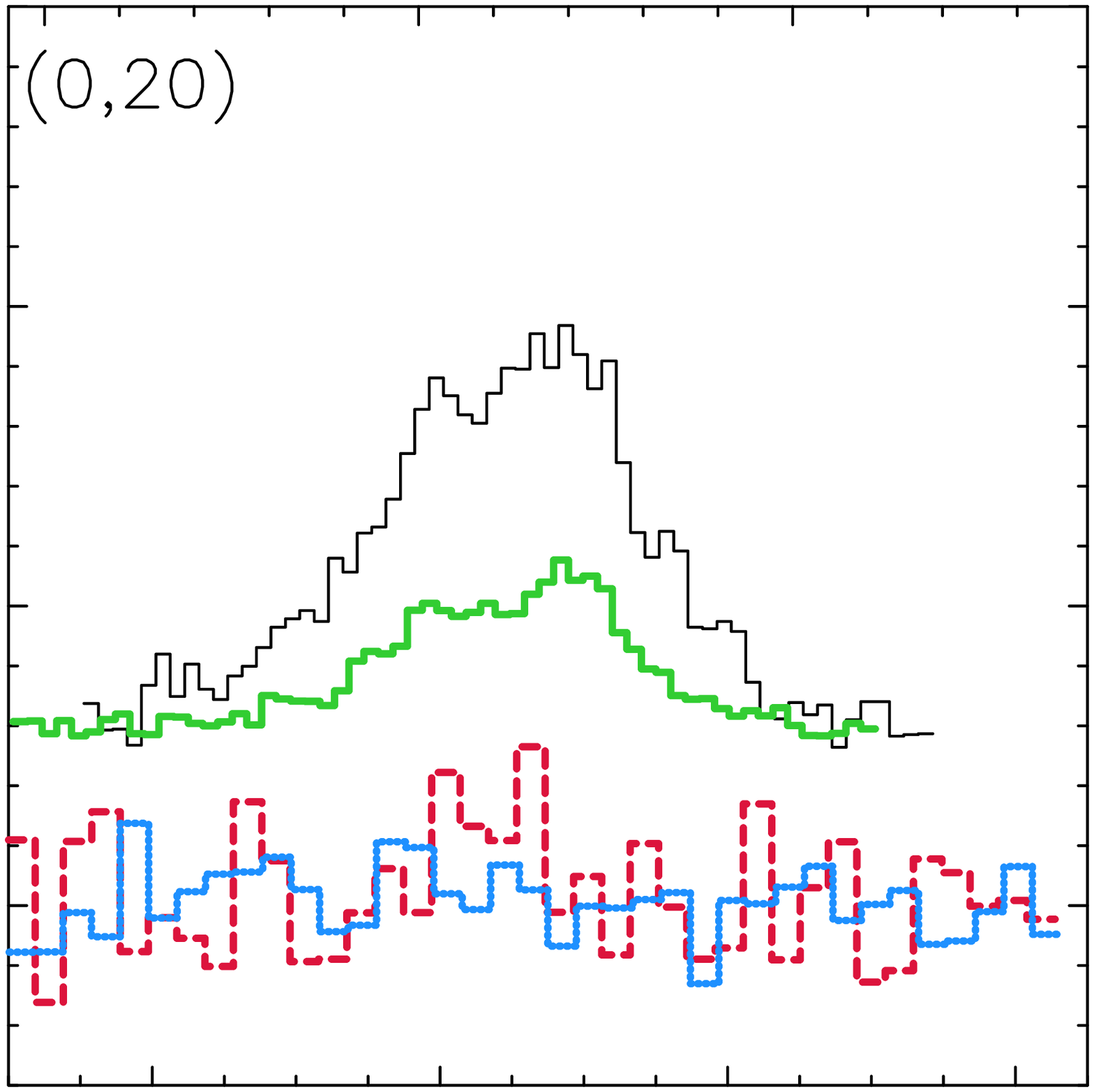}&
\includegraphics[scale=0.168, angle=0]{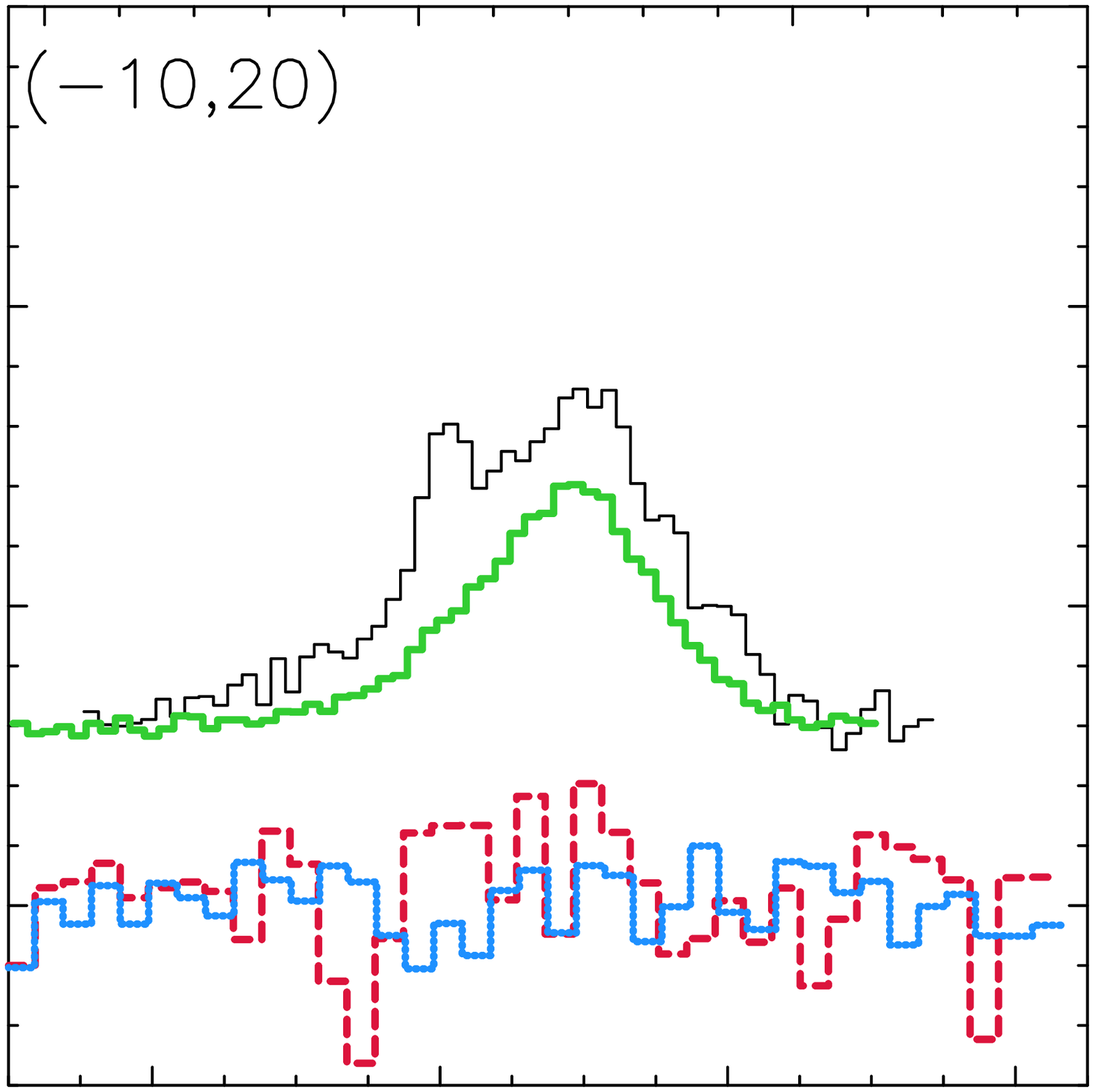}&
\includegraphics[scale=0.168, angle=0]{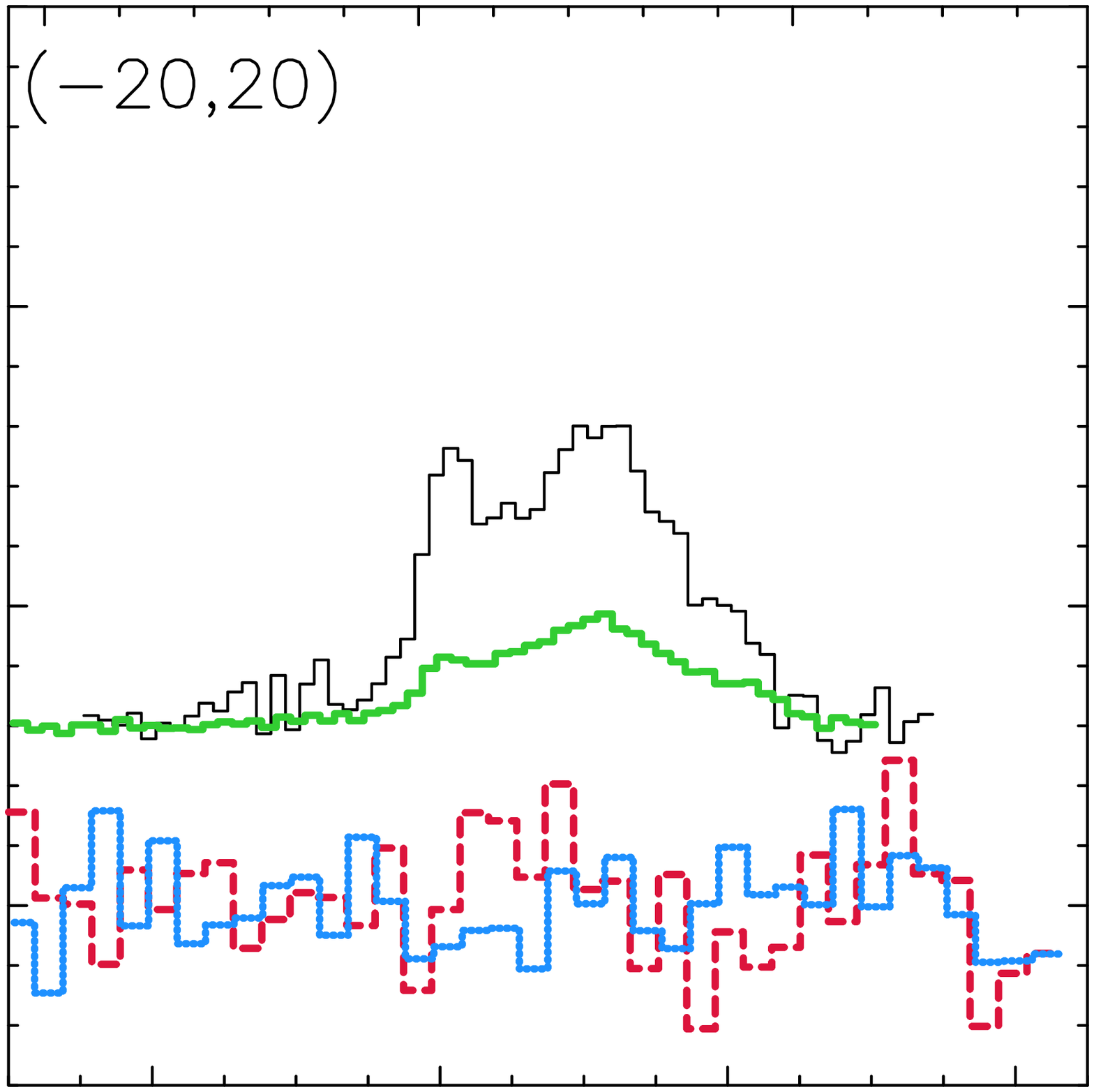}\\
\includegraphics[scale=0.168, angle=0]{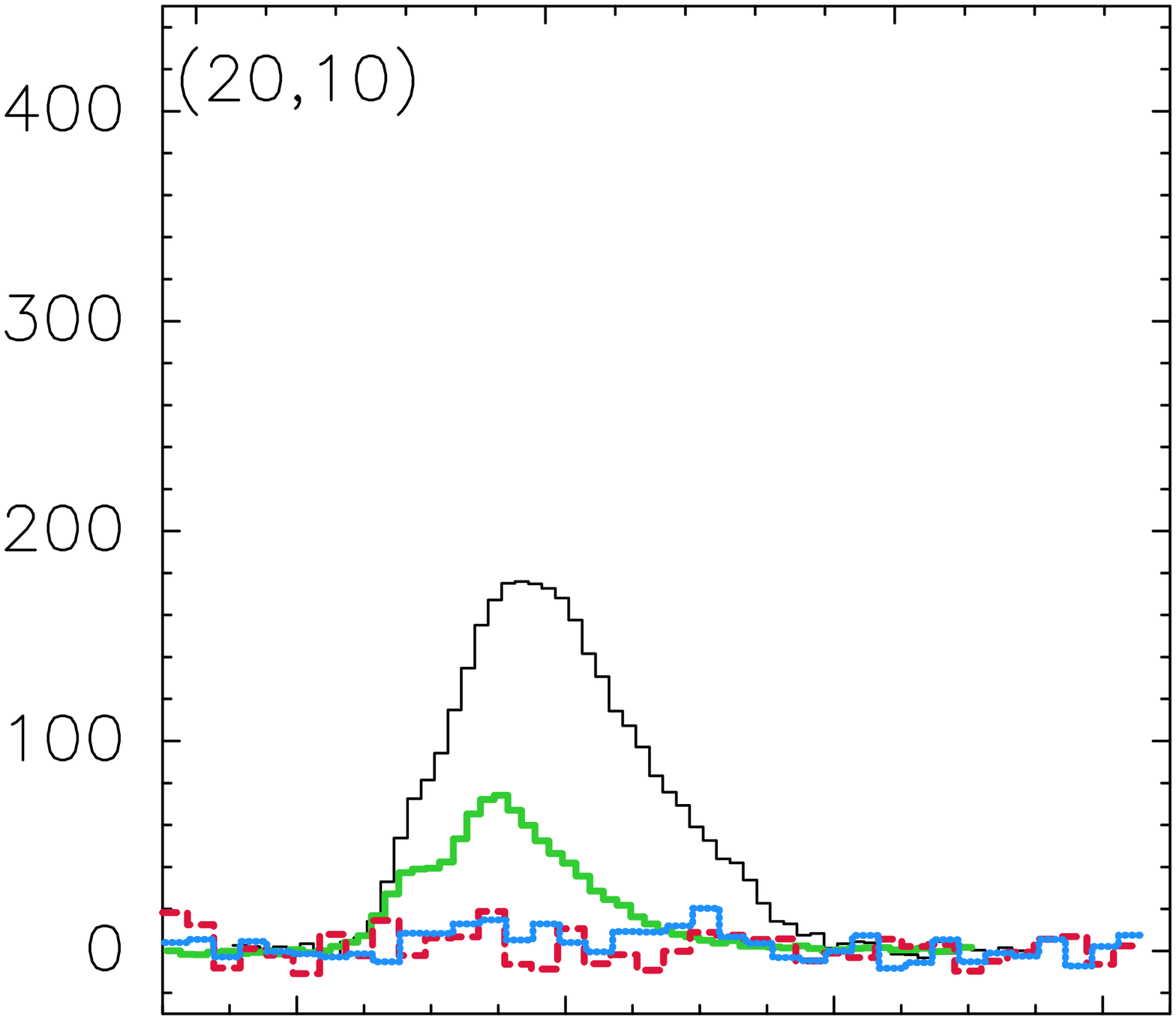}&
\includegraphics[scale=0.168, angle=0]{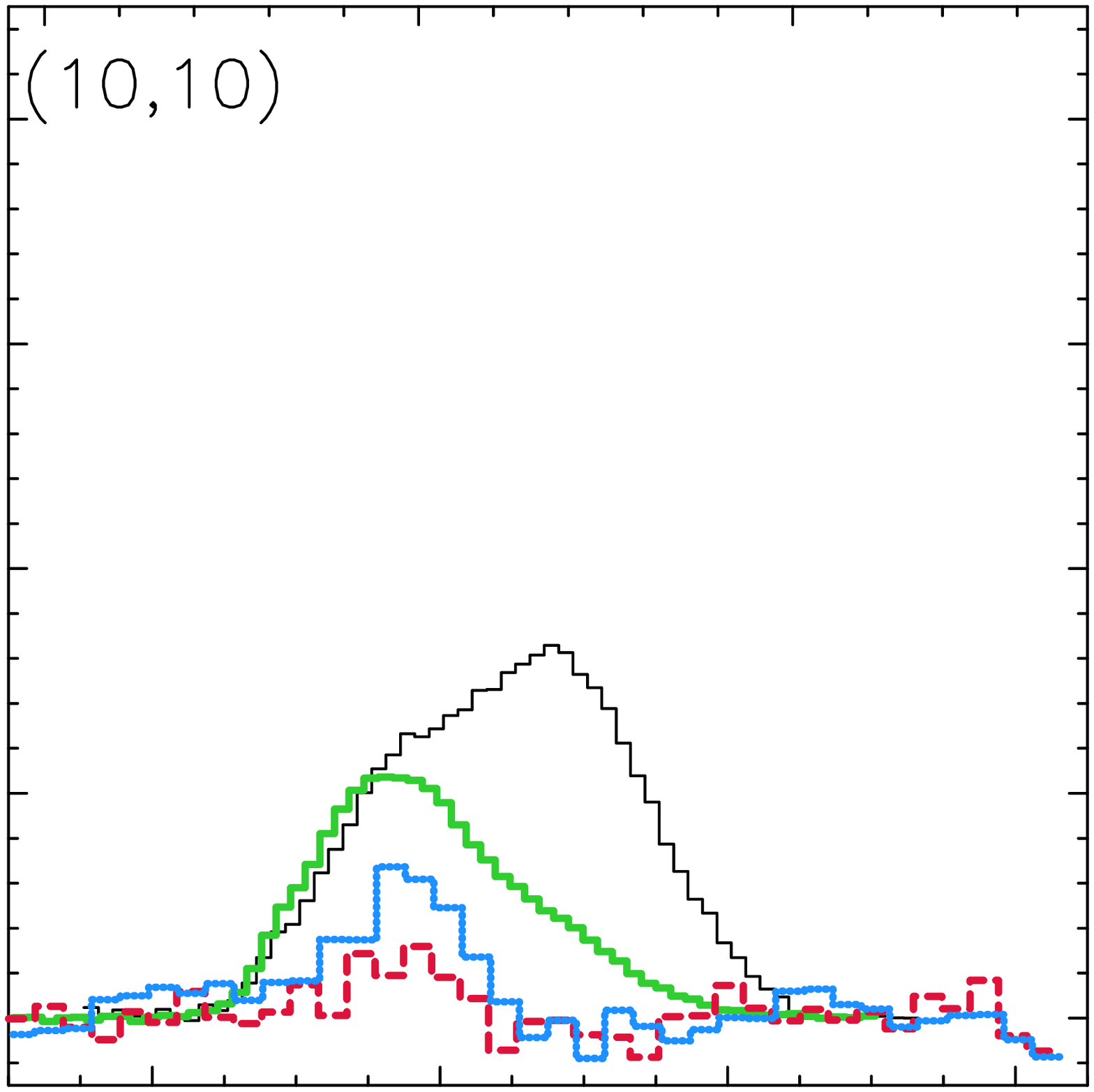}&
\includegraphics[scale=0.168, angle=0]{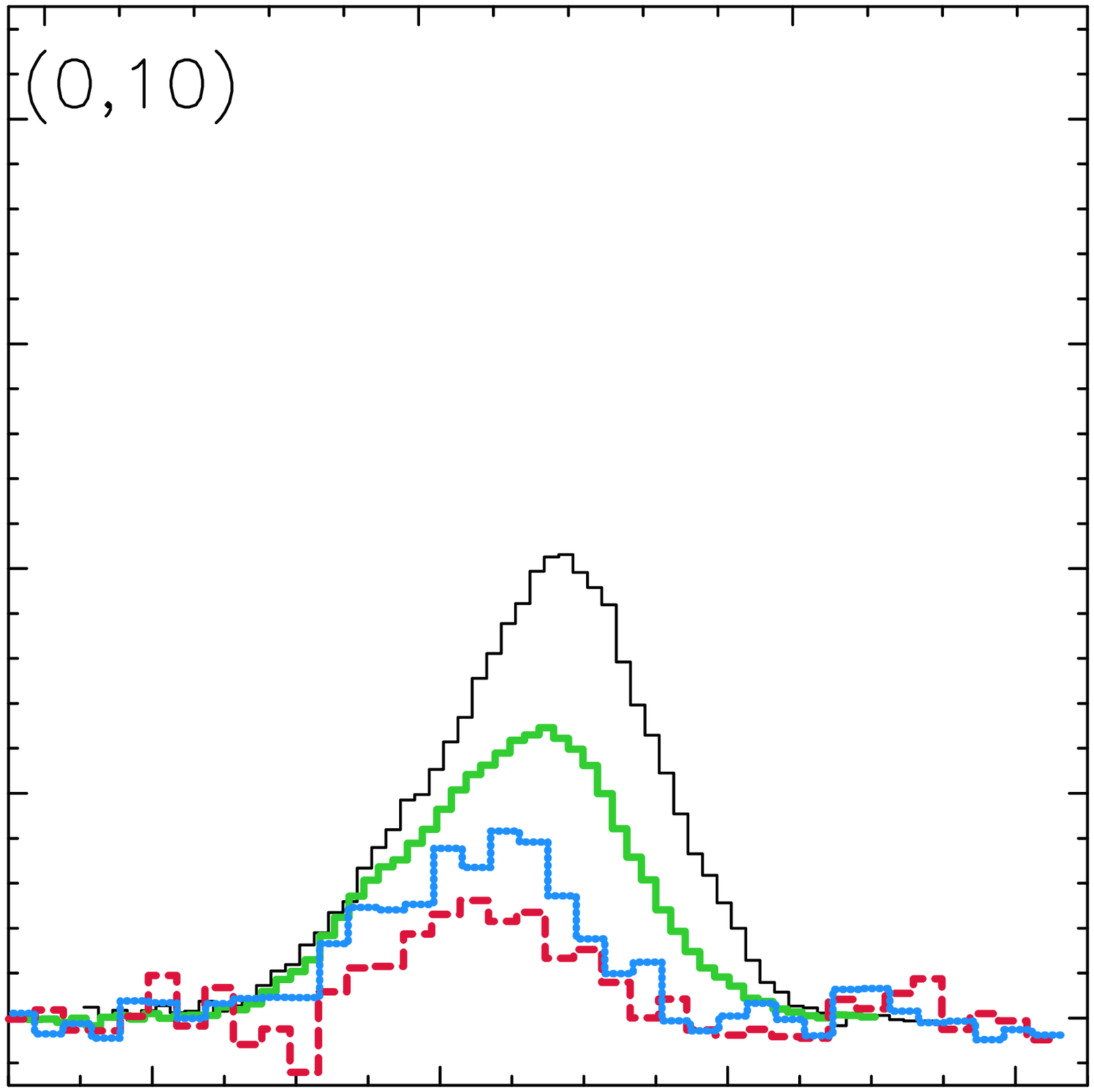}&
\includegraphics[scale=0.168, angle=0]{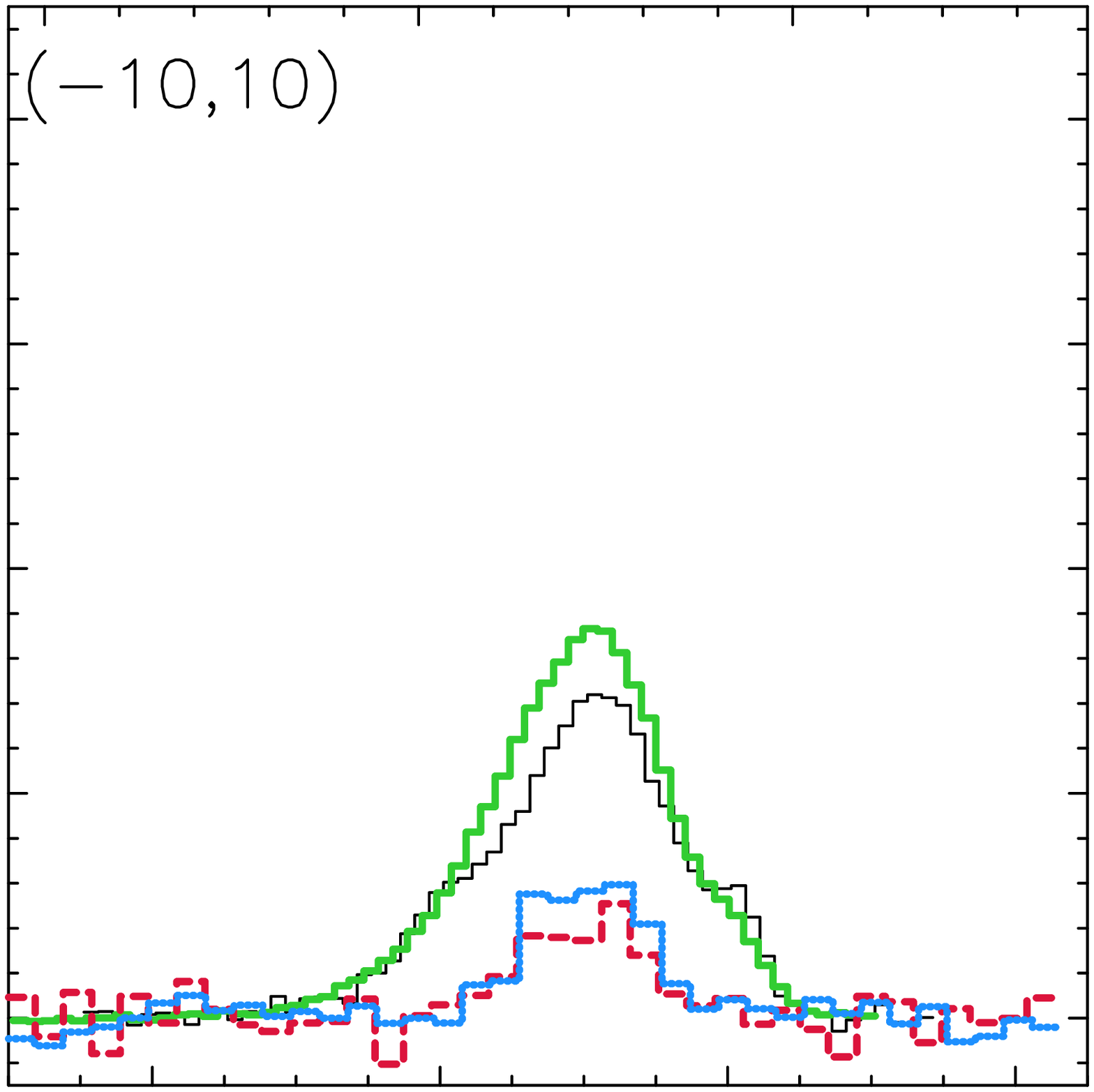}&
\includegraphics[scale=0.168, angle=0]{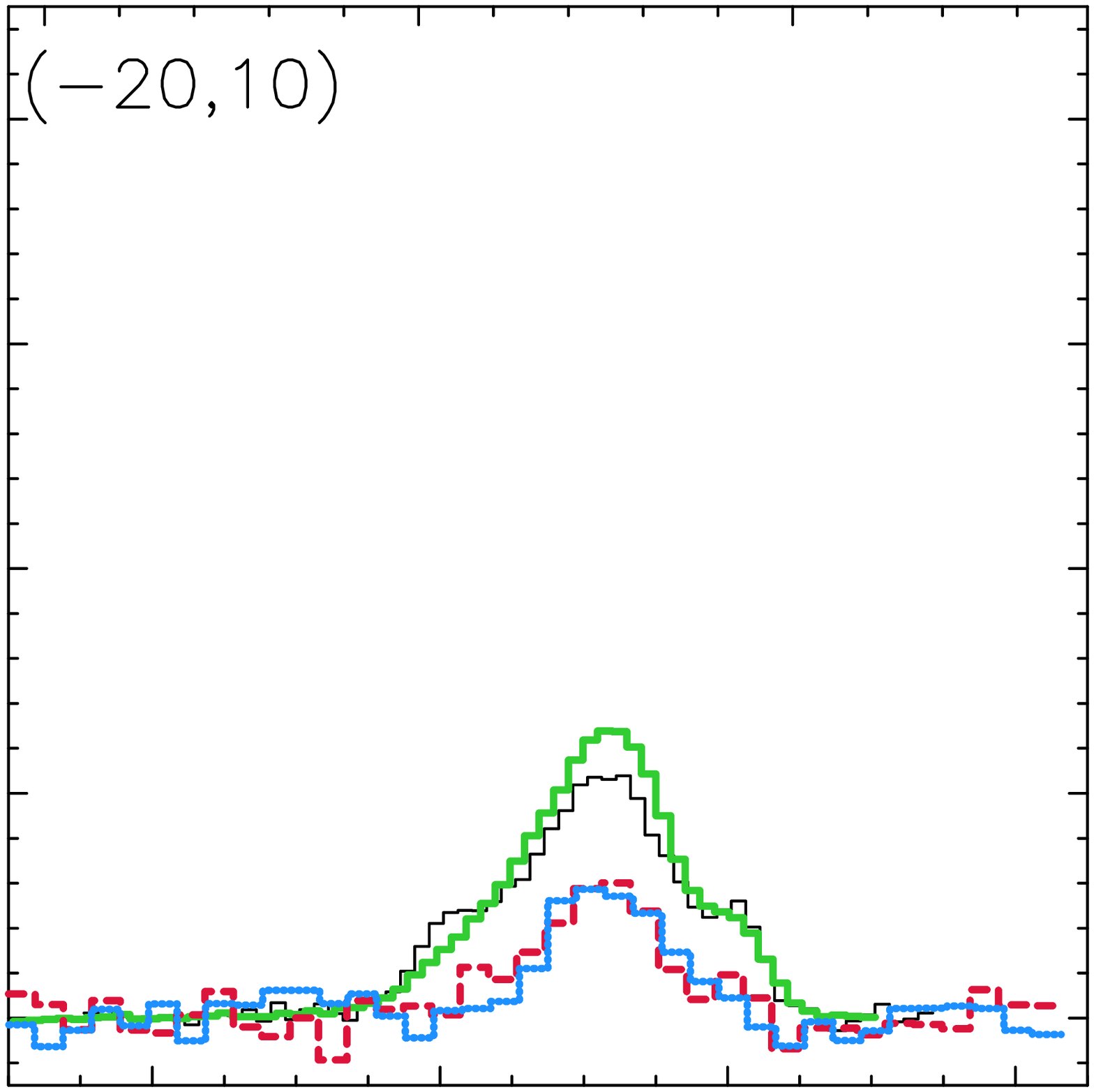}\\
\includegraphics[scale=0.168, angle=0]{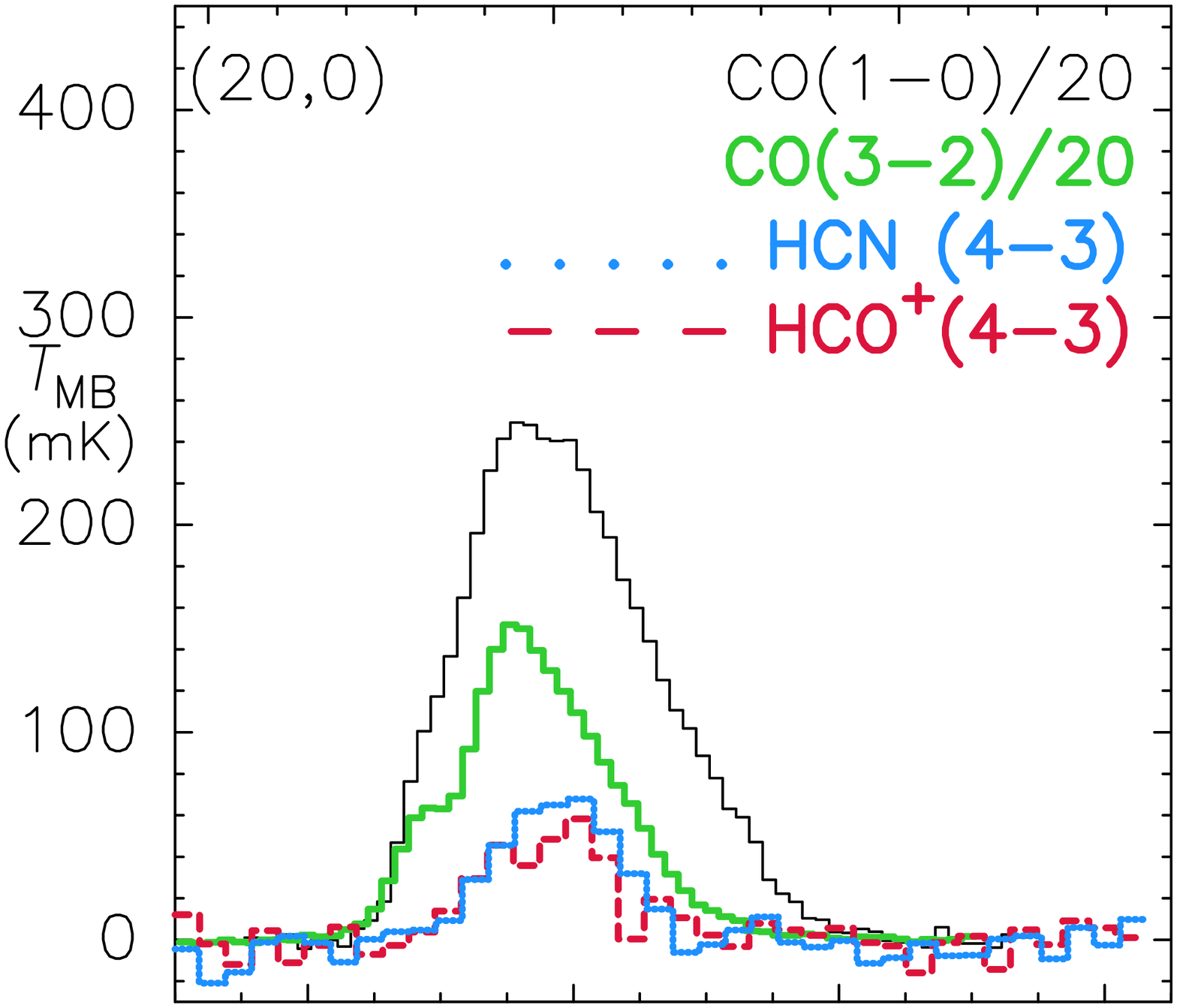}&
\includegraphics[scale=0.168, angle=0]{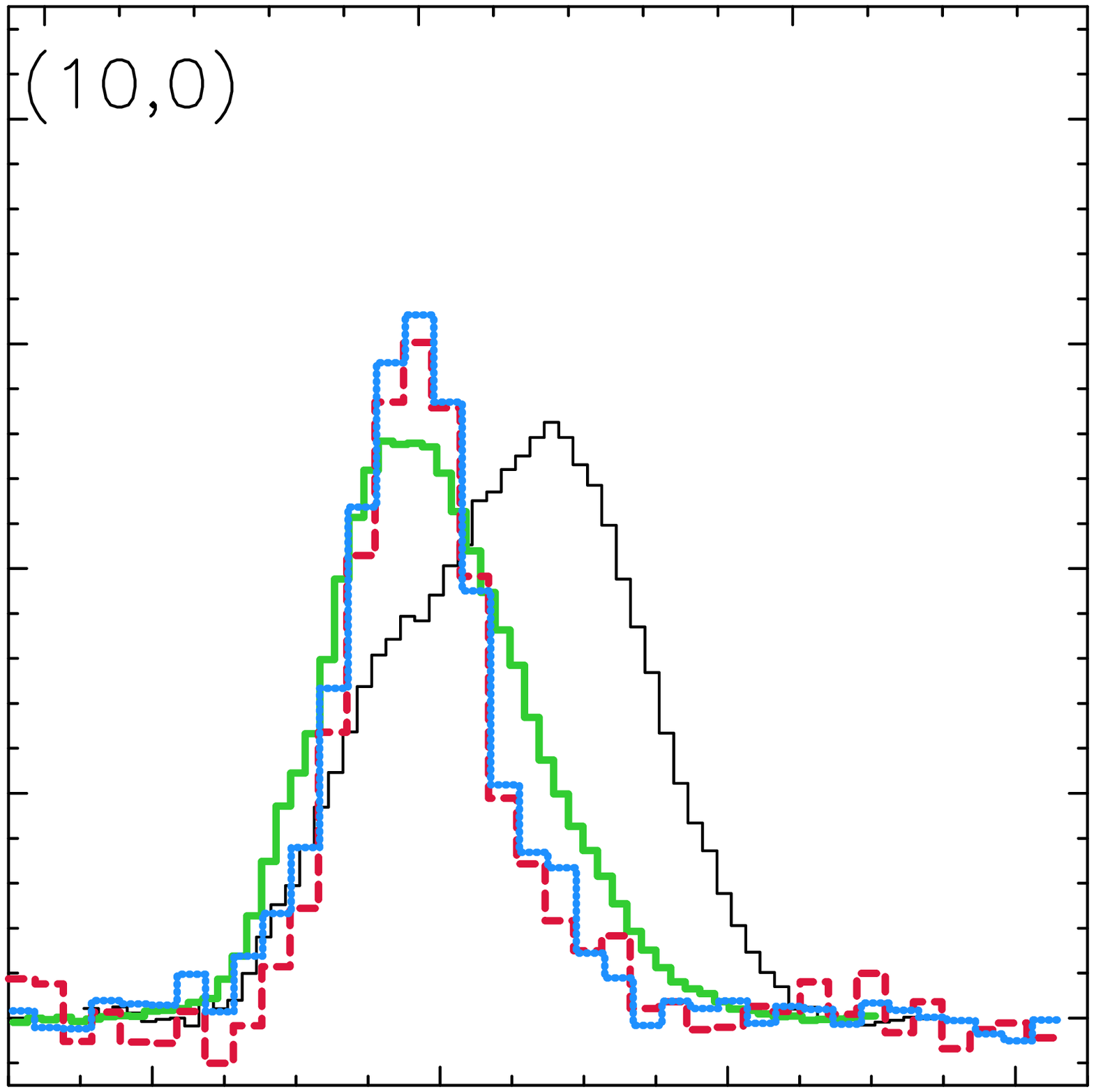}&
\includegraphics[scale=0.168, angle=0]{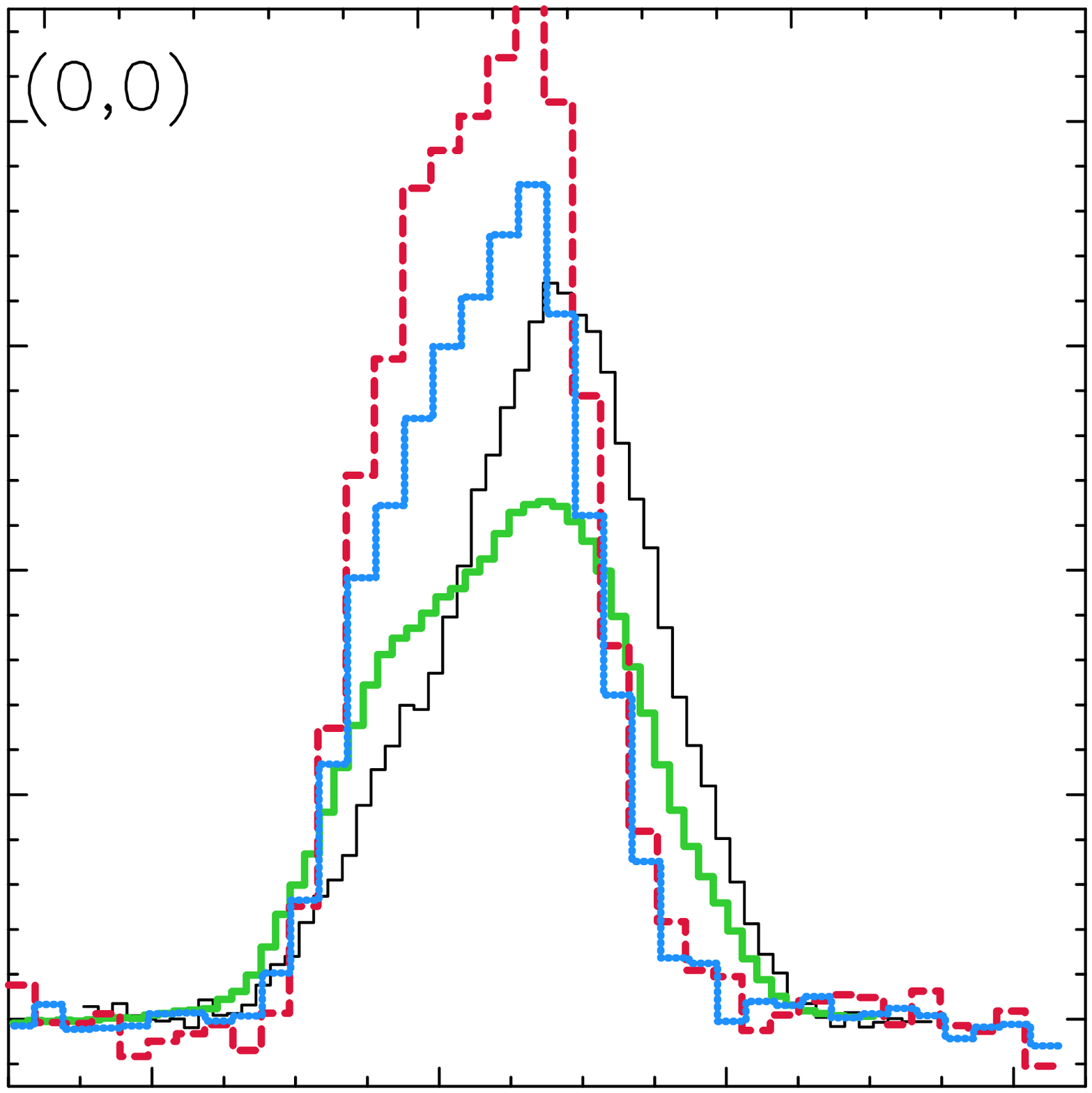}&
\includegraphics[scale=0.168, angle=0]{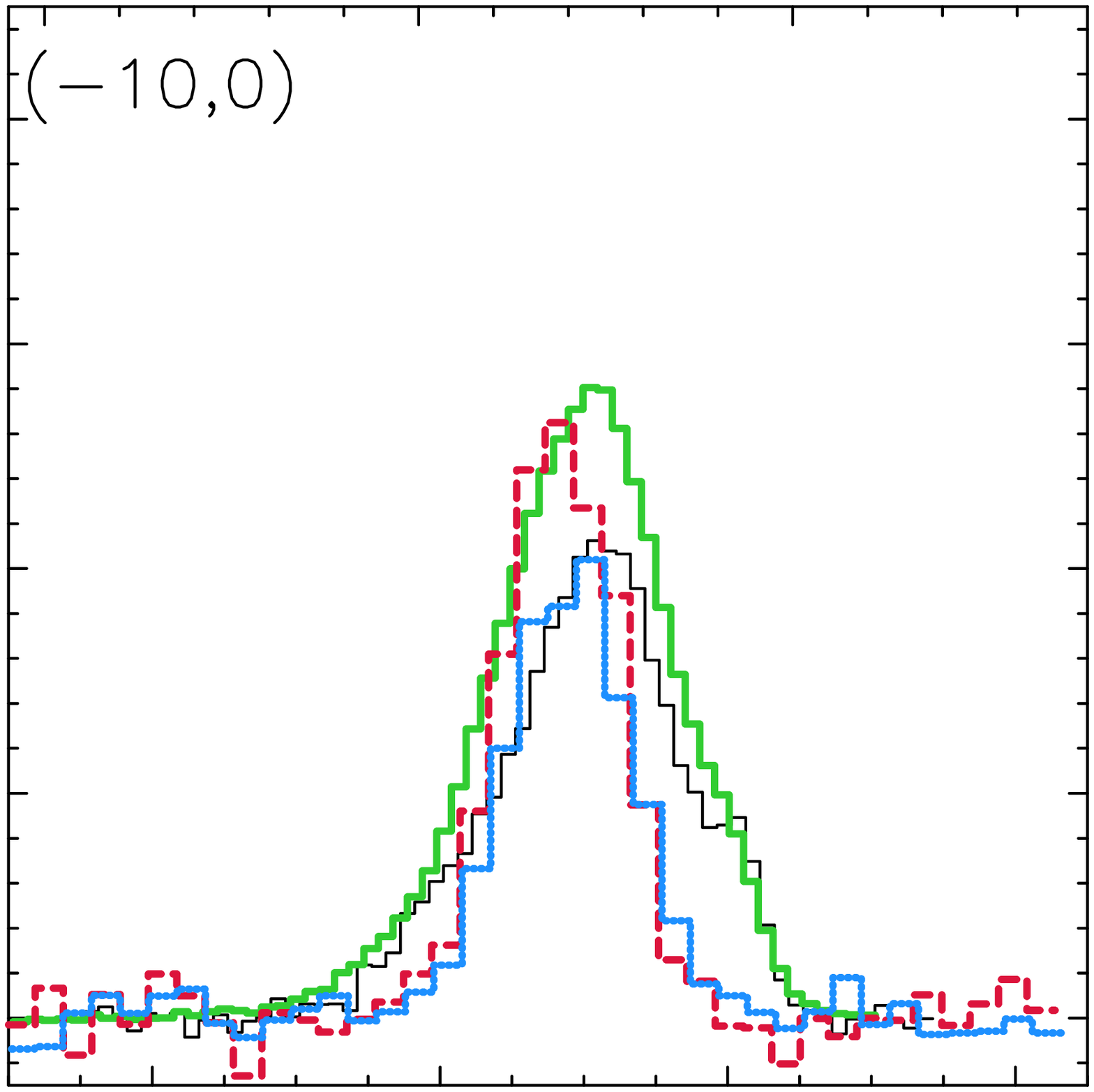}&
\includegraphics[scale=0.168, angle=0]{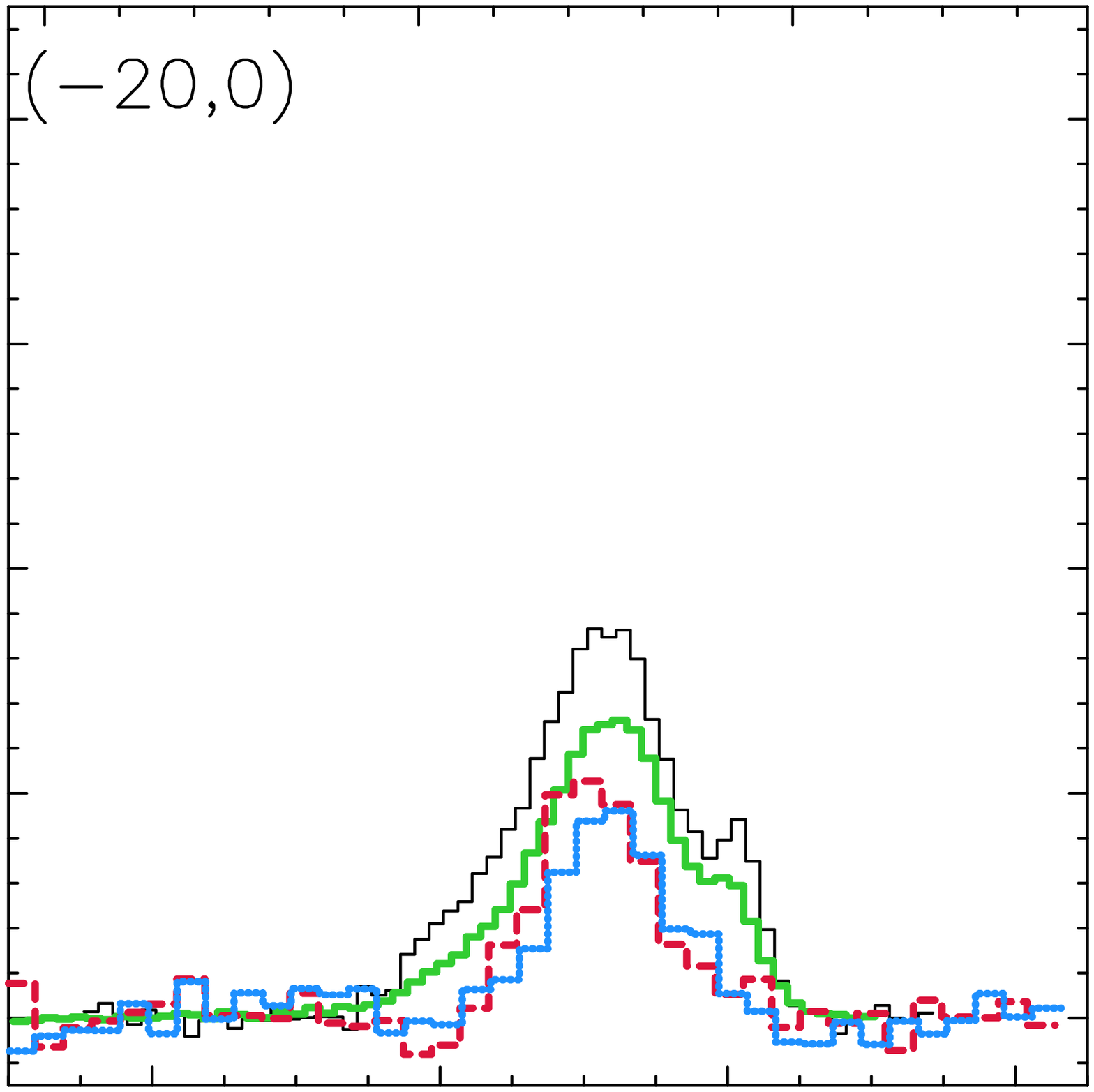}\\
\includegraphics[scale=0.168, angle=0]{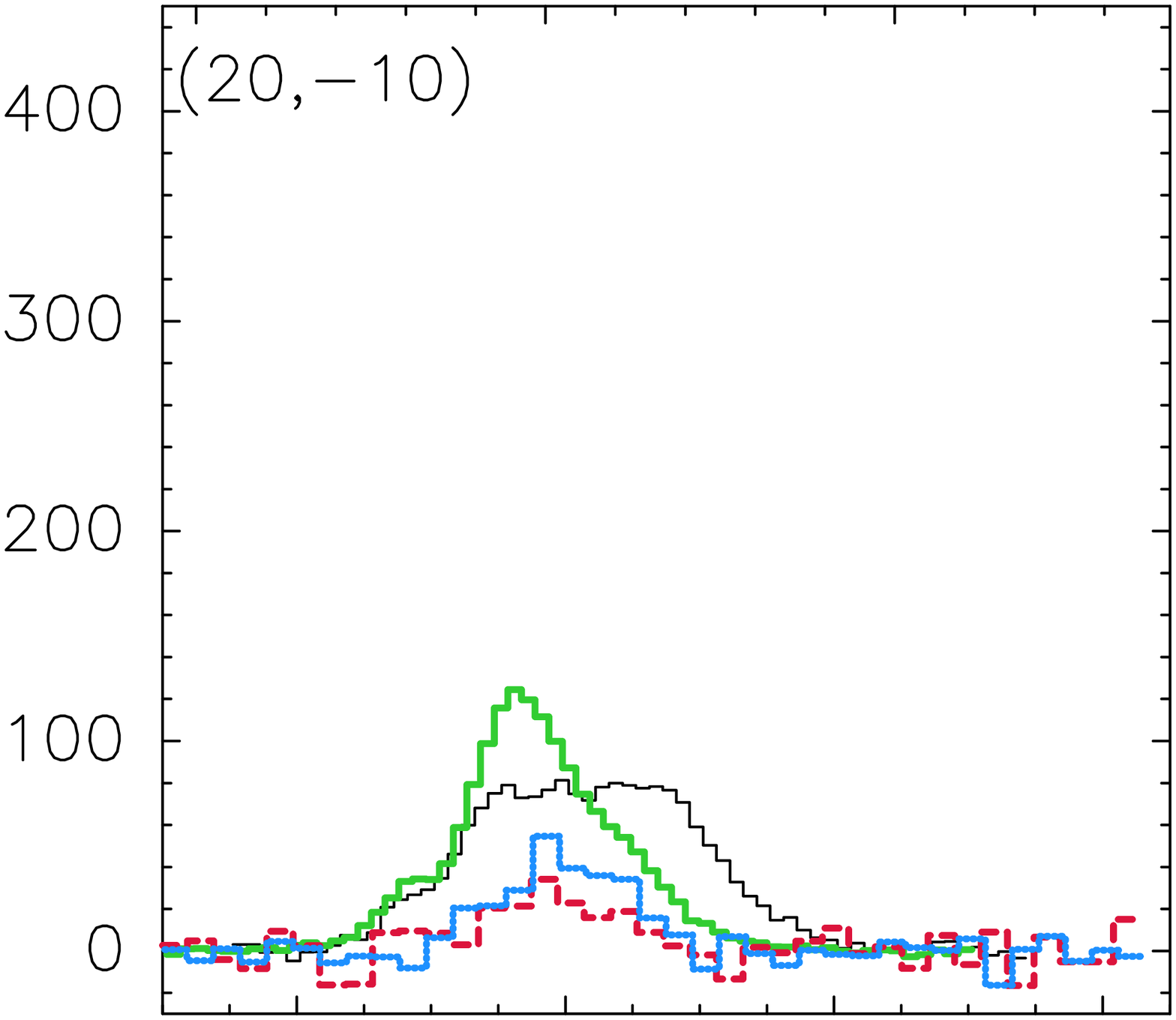}&
\includegraphics[scale=0.168, angle=0]{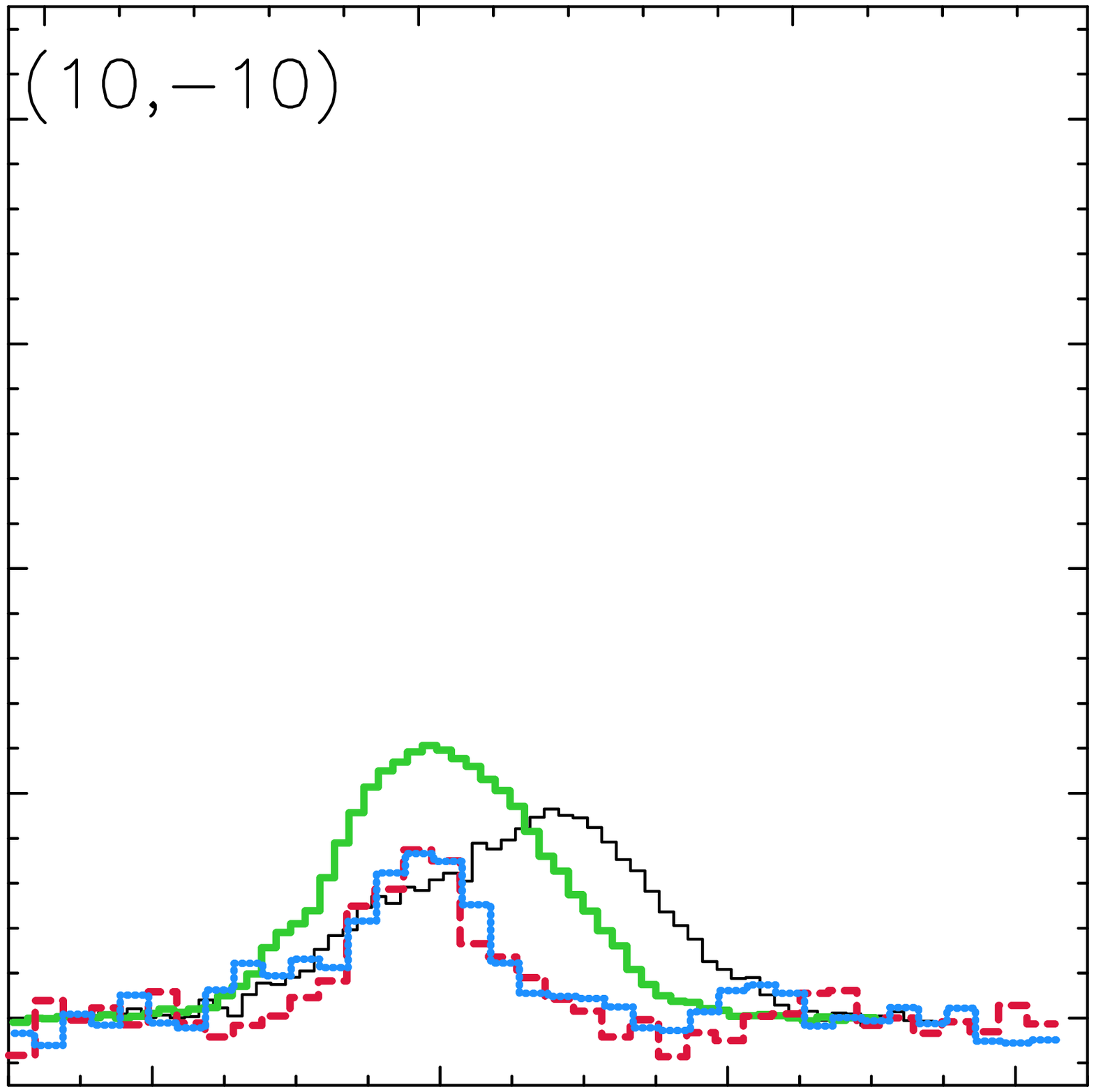}&
\includegraphics[scale=0.168, angle=0]{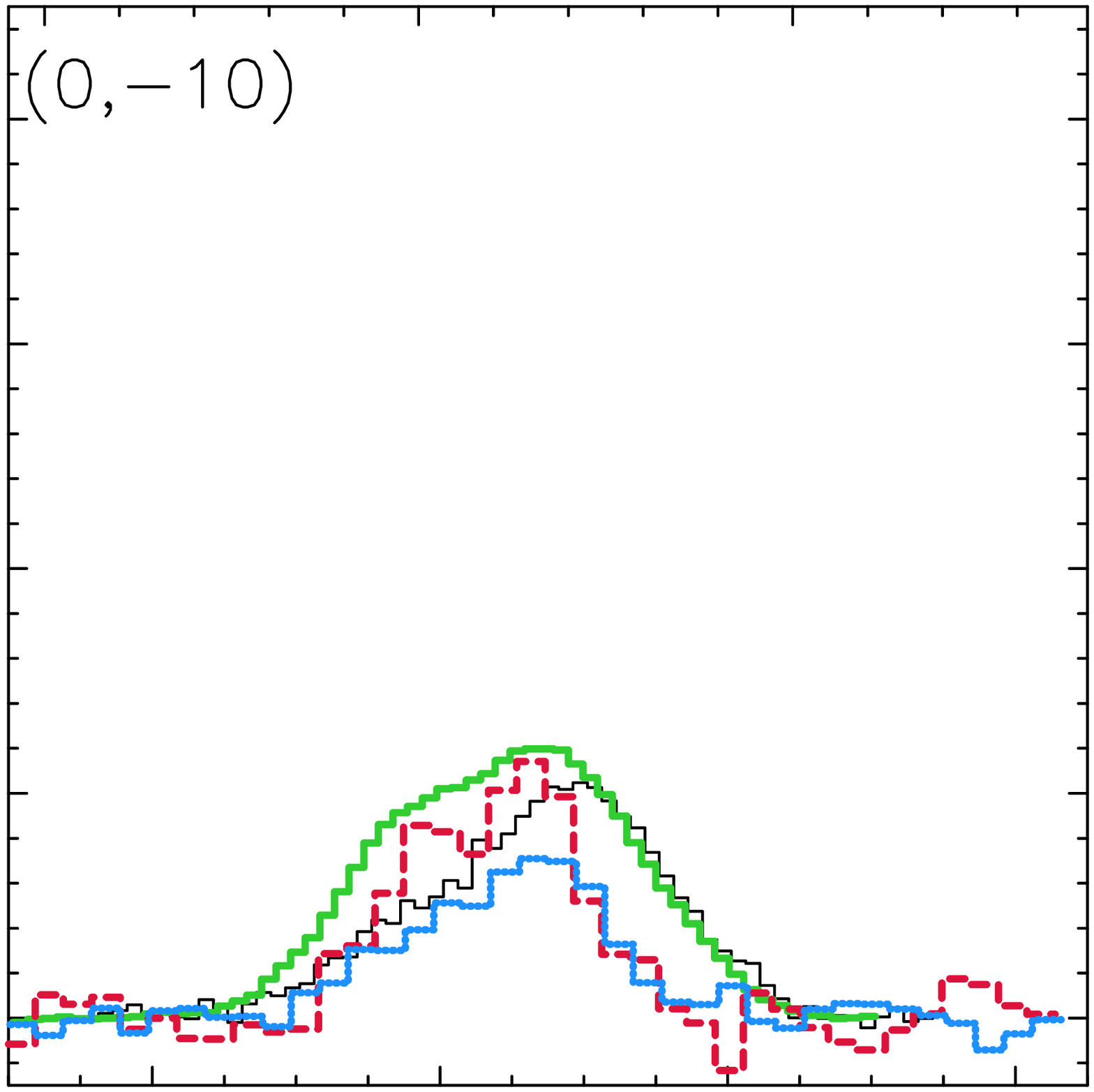}&
\includegraphics[scale=0.168, angle=0]{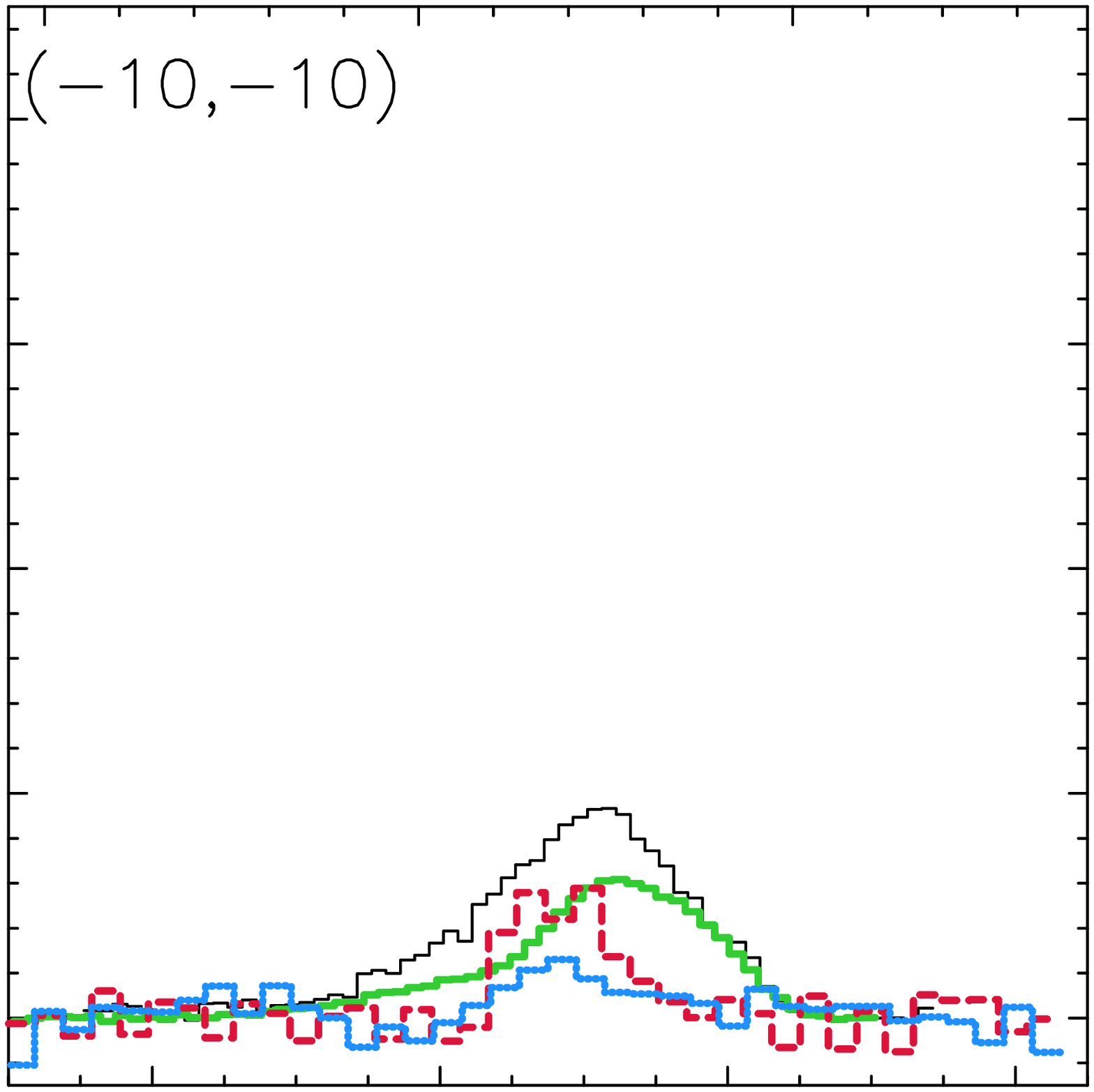}&
\includegraphics[scale=0.168, angle=0]{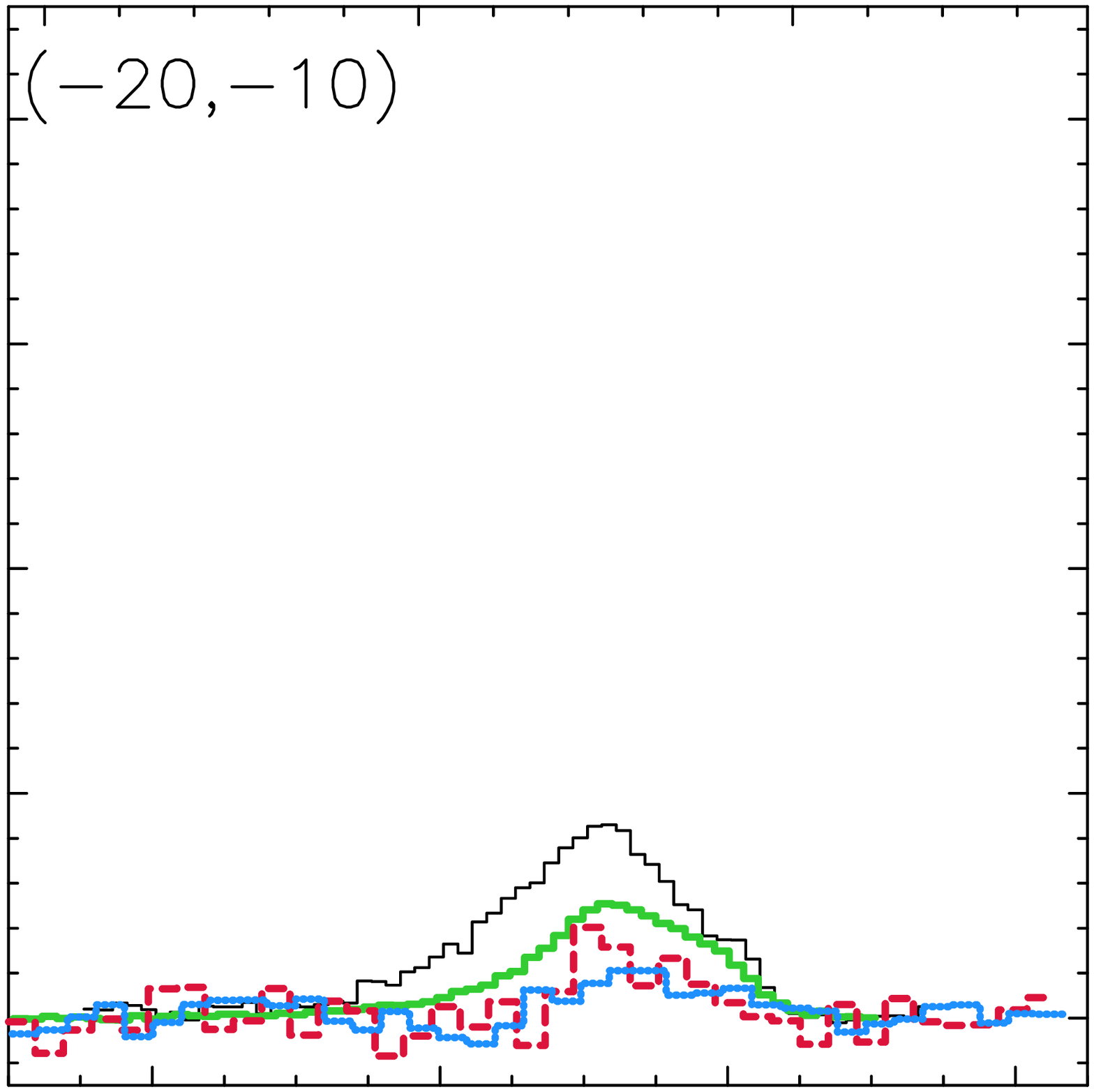}\\
\includegraphics[scale=0.168, angle=0]{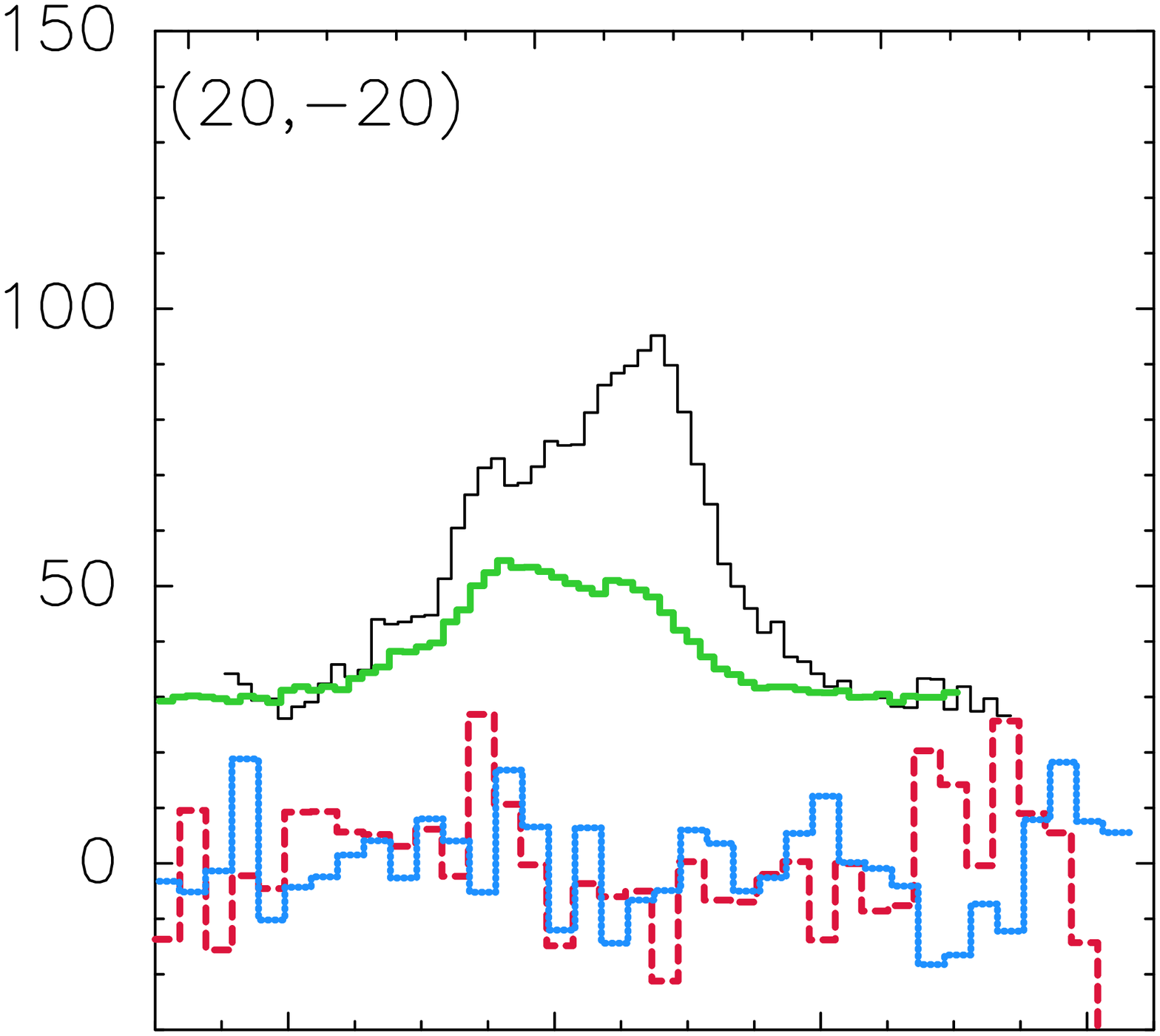}&
\includegraphics[scale=0.168, angle=0]{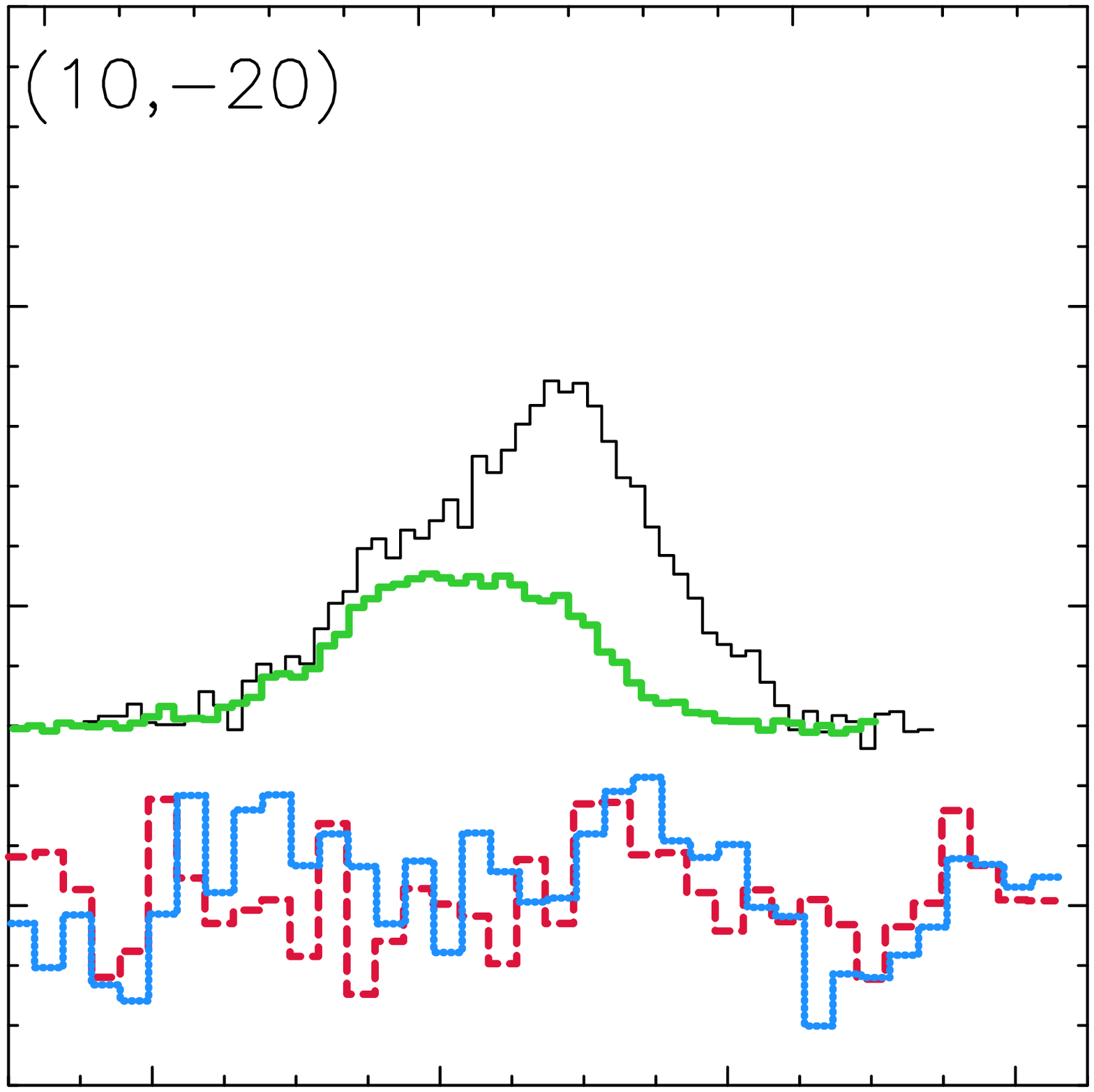}&
\includegraphics[scale=0.168, angle=0]{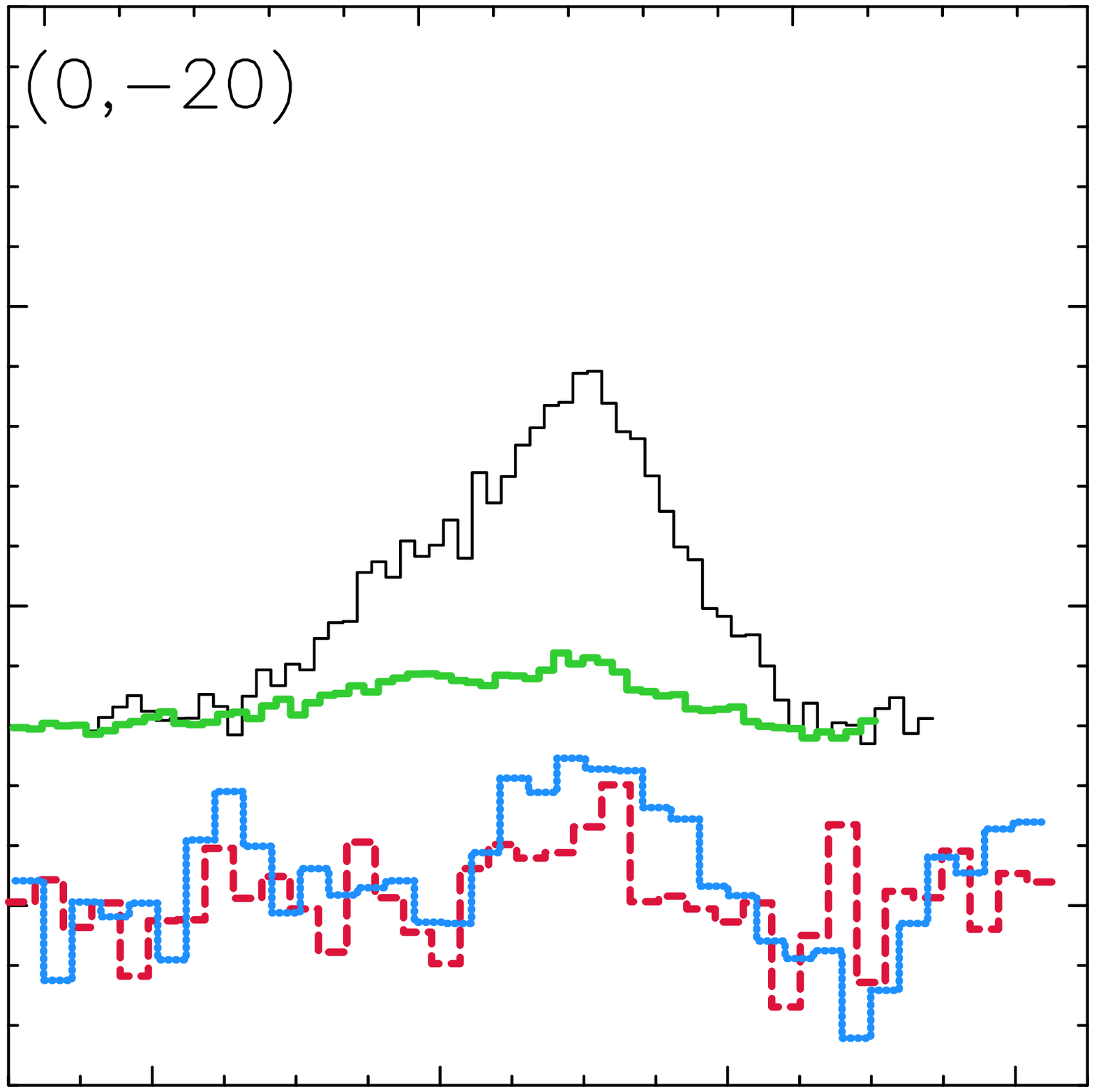}&
\includegraphics[scale=0.168, angle=0]{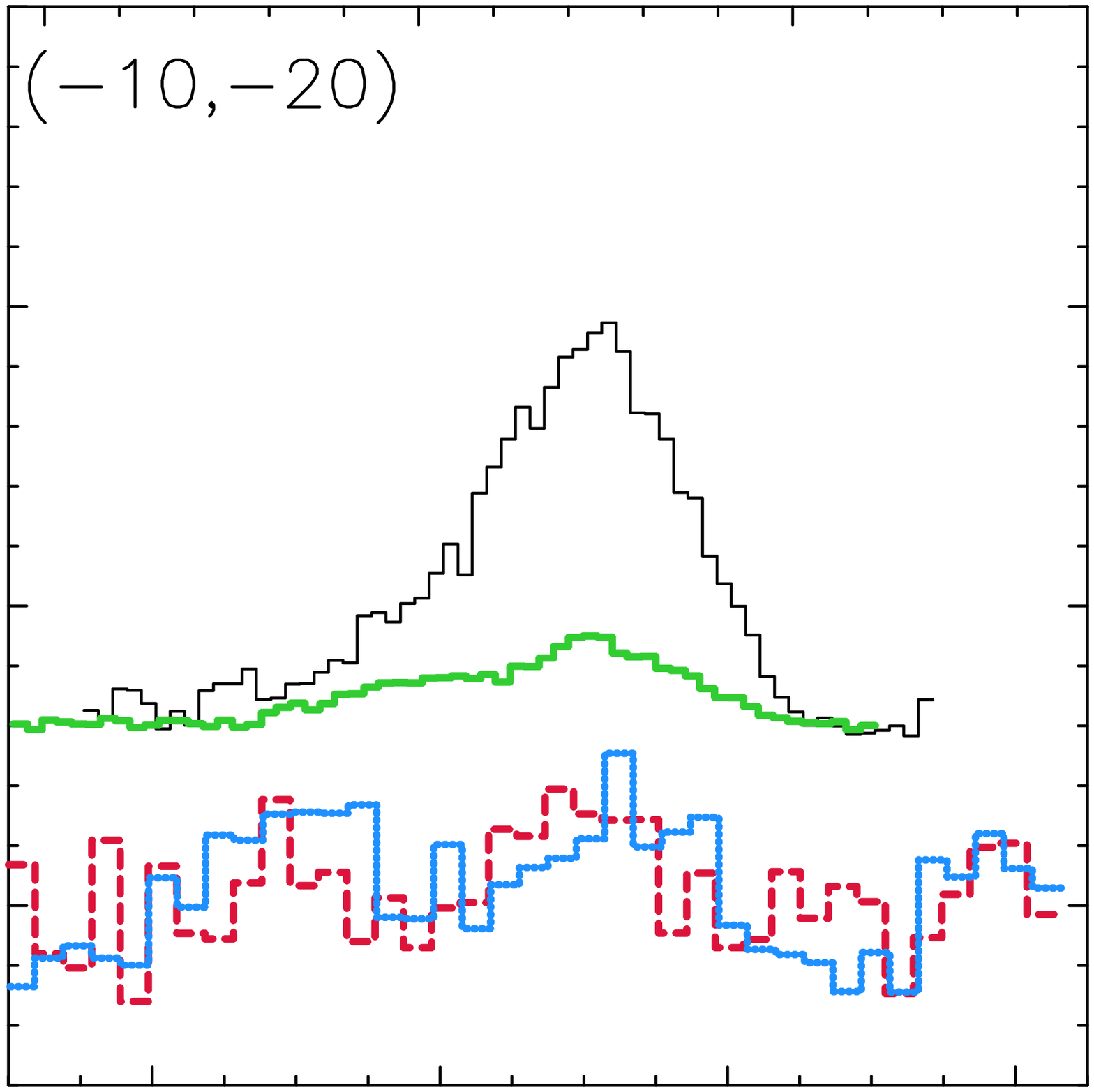}&
\includegraphics[scale=0.168, angle=0]{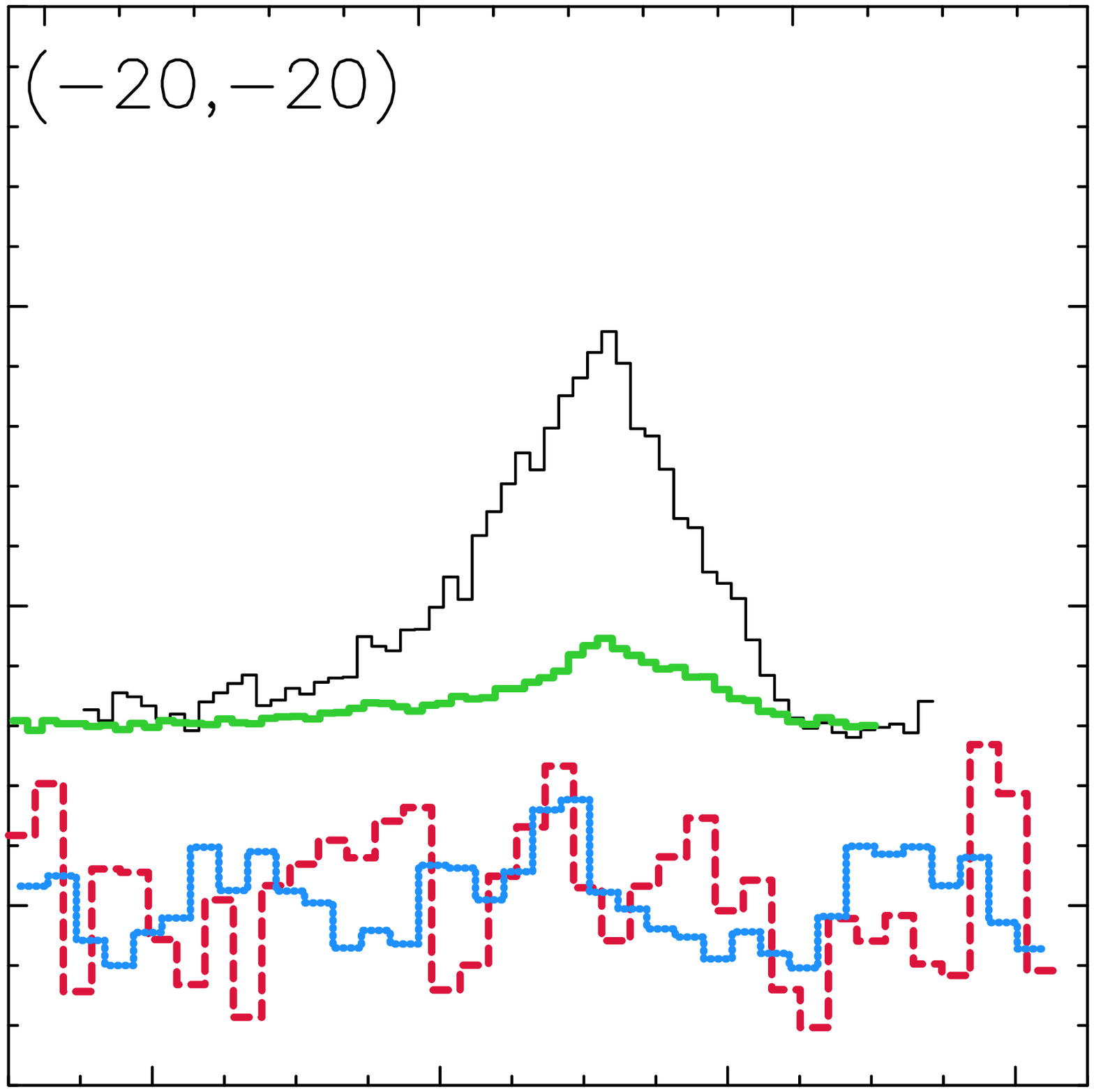}\\
\includegraphics[scale=0.168, angle=0]{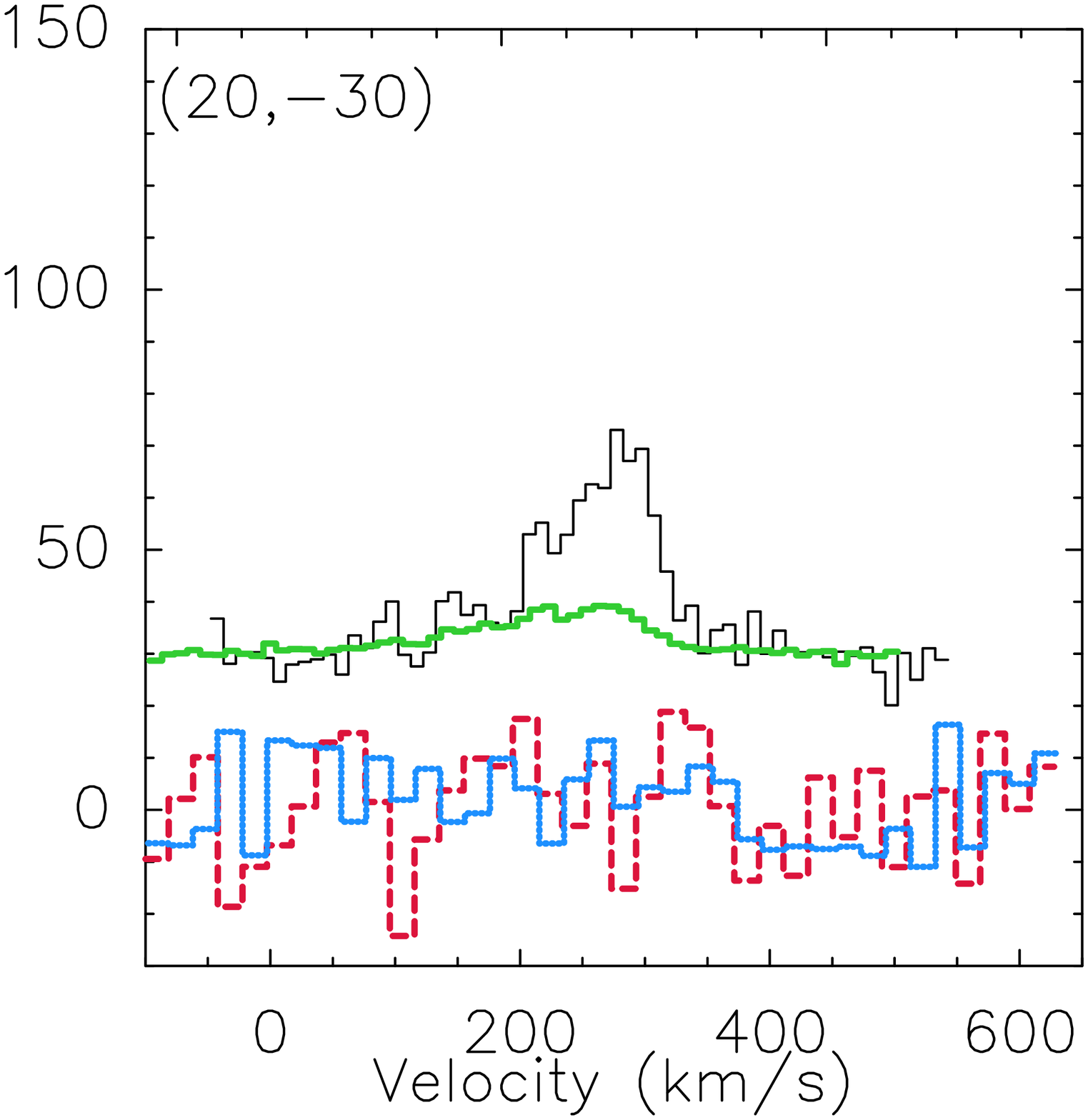}&
\includegraphics[scale=0.168, angle=0]{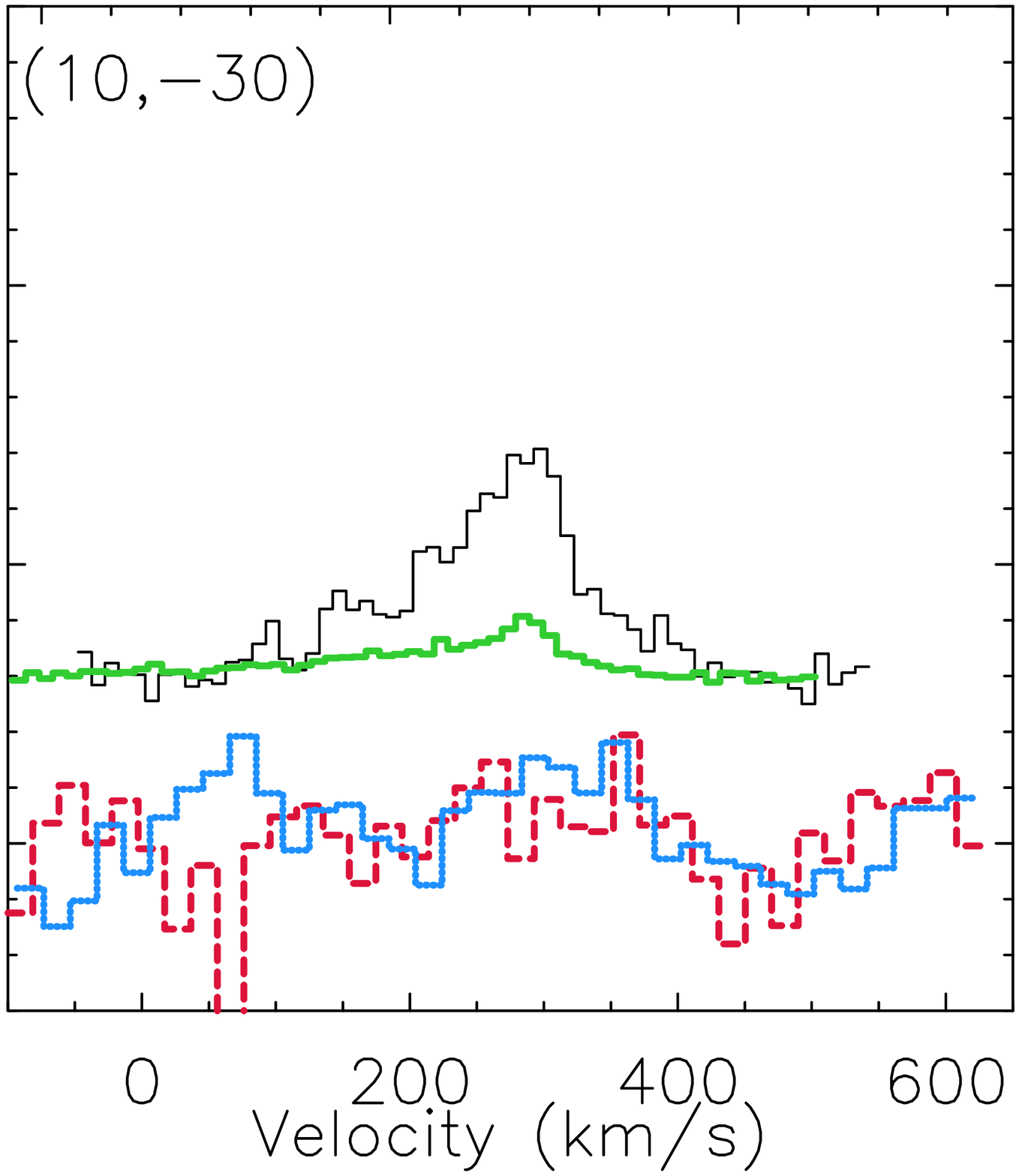}&
\includegraphics[scale=0.168, angle=0]{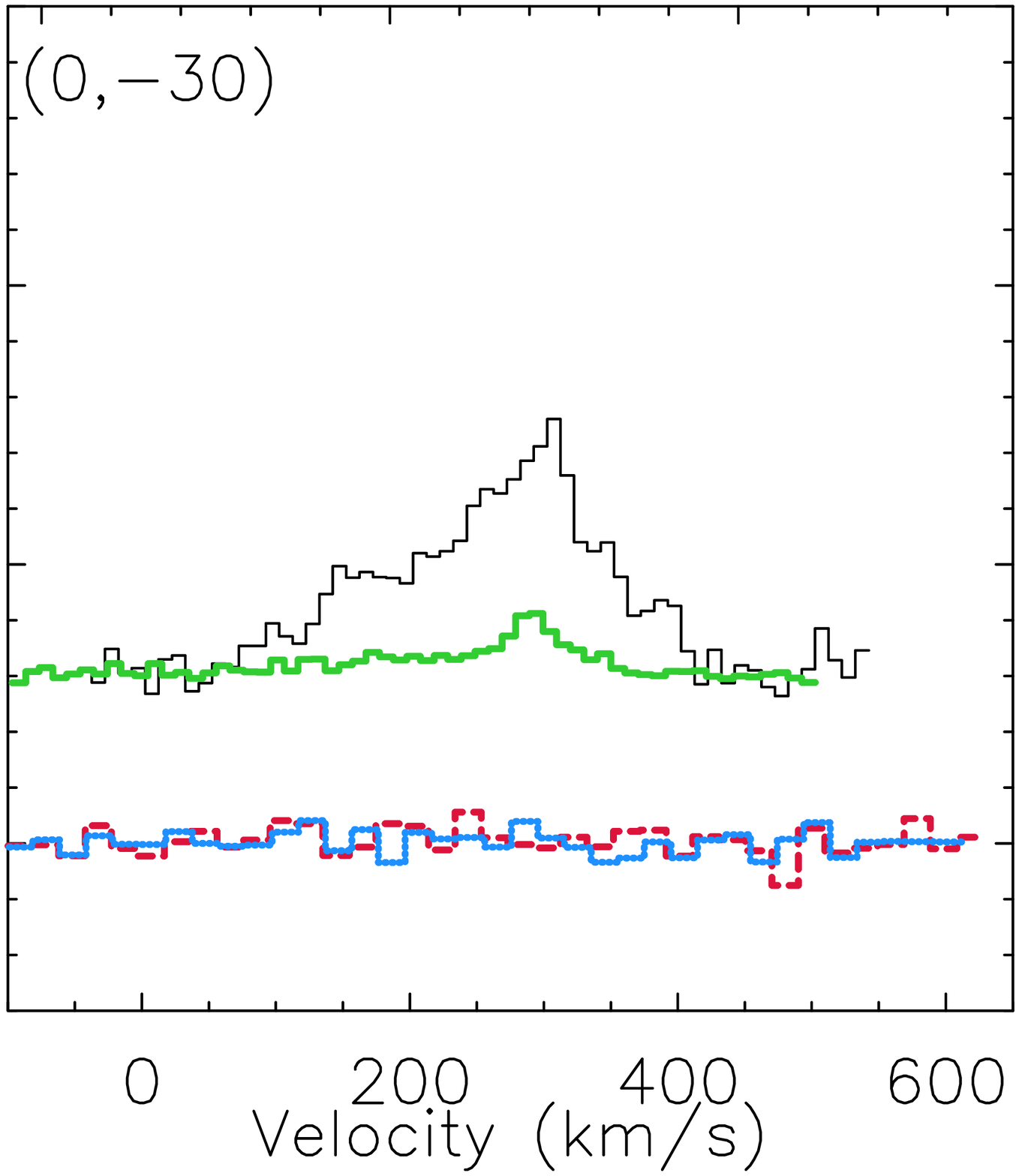}&
\includegraphics[scale=0.168, angle=0]{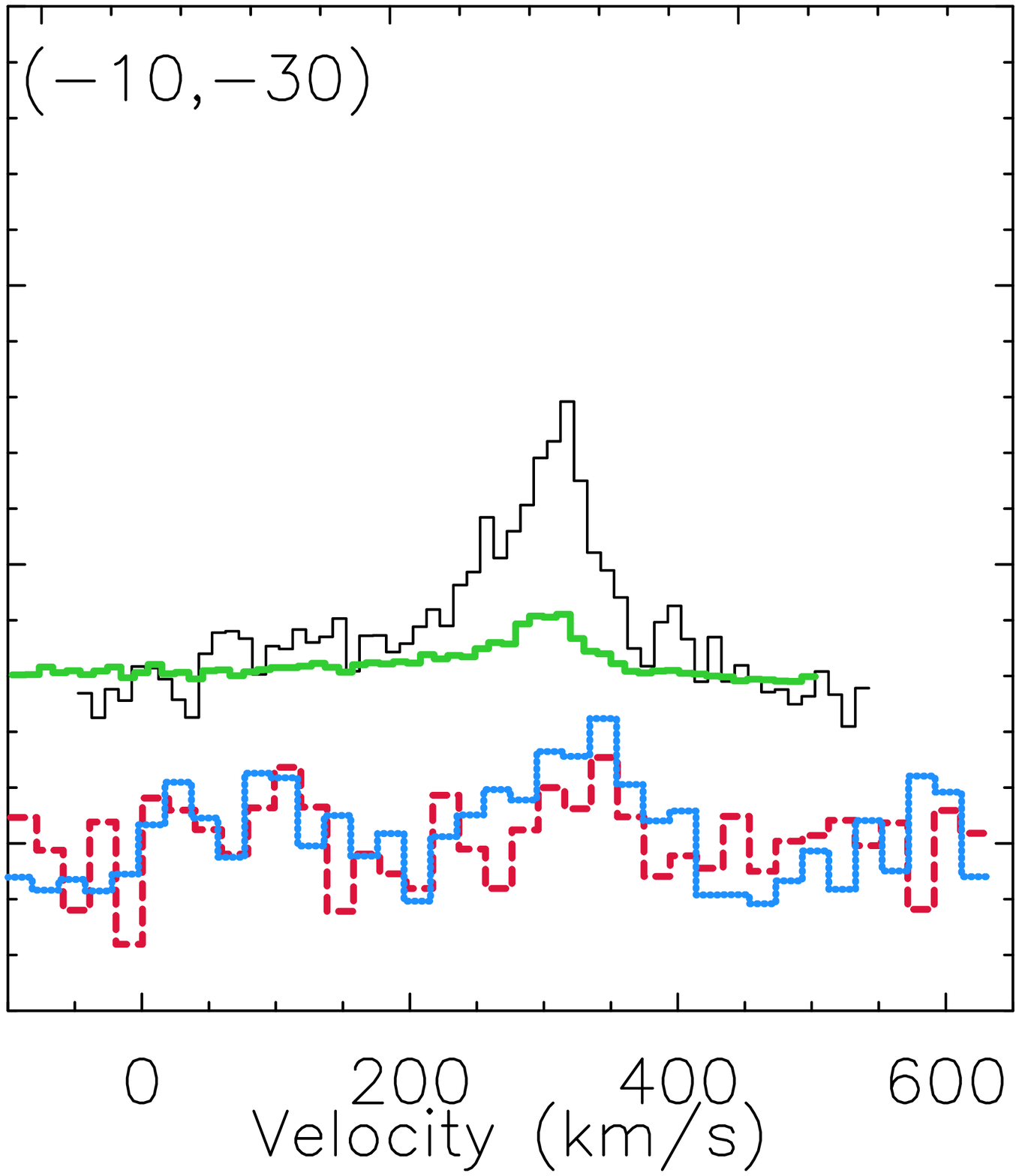}&
\includegraphics[scale=0.168, angle=0]{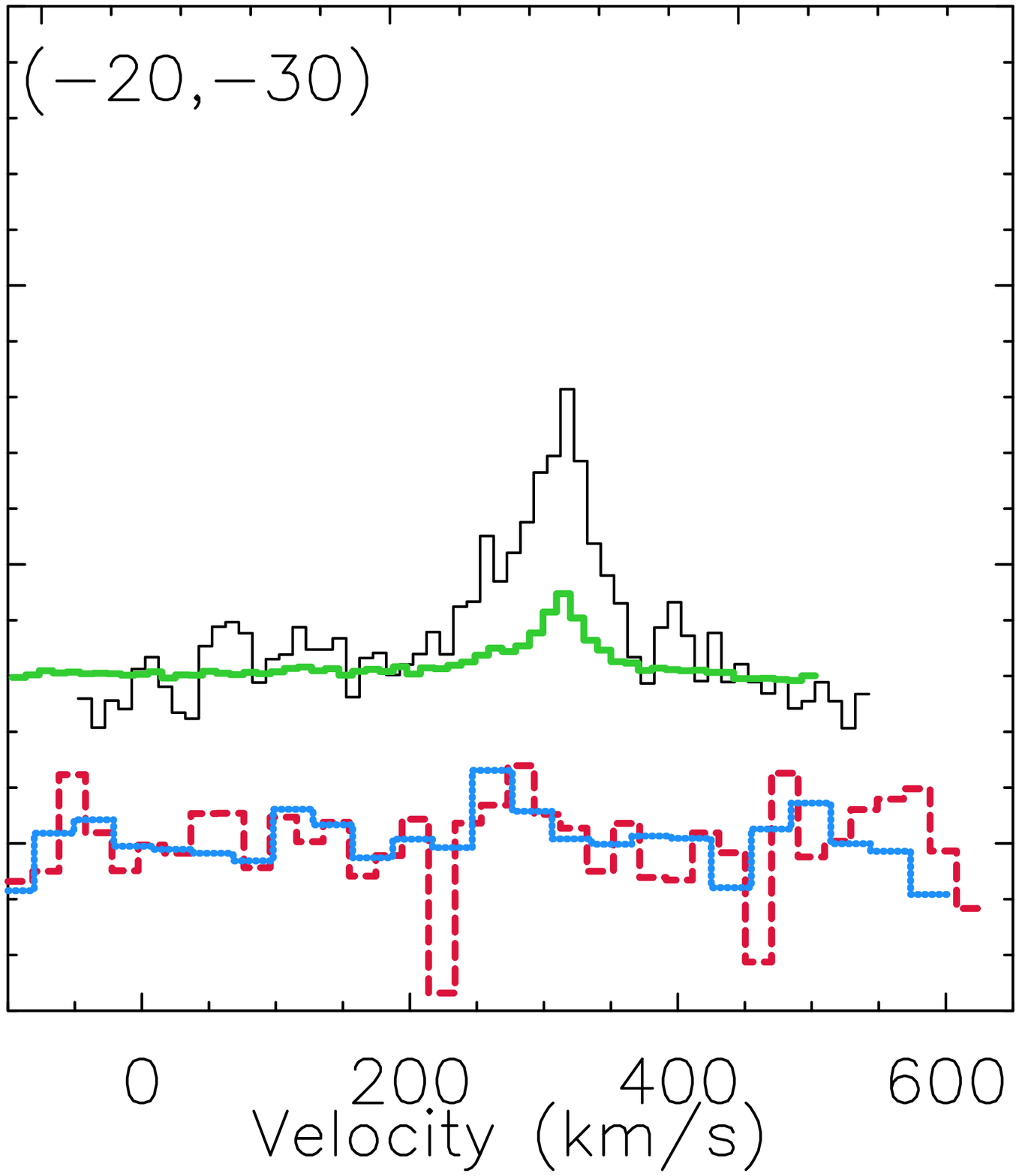}\\
\end{array}$
\end{center}
\begin{center}
\contcaption{Spectra of CO 1-0 (black), 
CO 3-2 (green), HCN 4-3
(blue) and HCO$^+$ 4-3 (red) emission in the central $\sim$ 1\,kpc region of NGC\,253. The $T_\text{MB}$ (in unit of mK) range on the $y$-axis of the central three
rows is set to be [-30, 450].
}
\label{fig:spec2} 
\end{center} 
\end{figure*}

\begin{figure*}
\begin{center}$
\begin{array}{cccc}
\includegraphics[scale=0.168, angle=0]{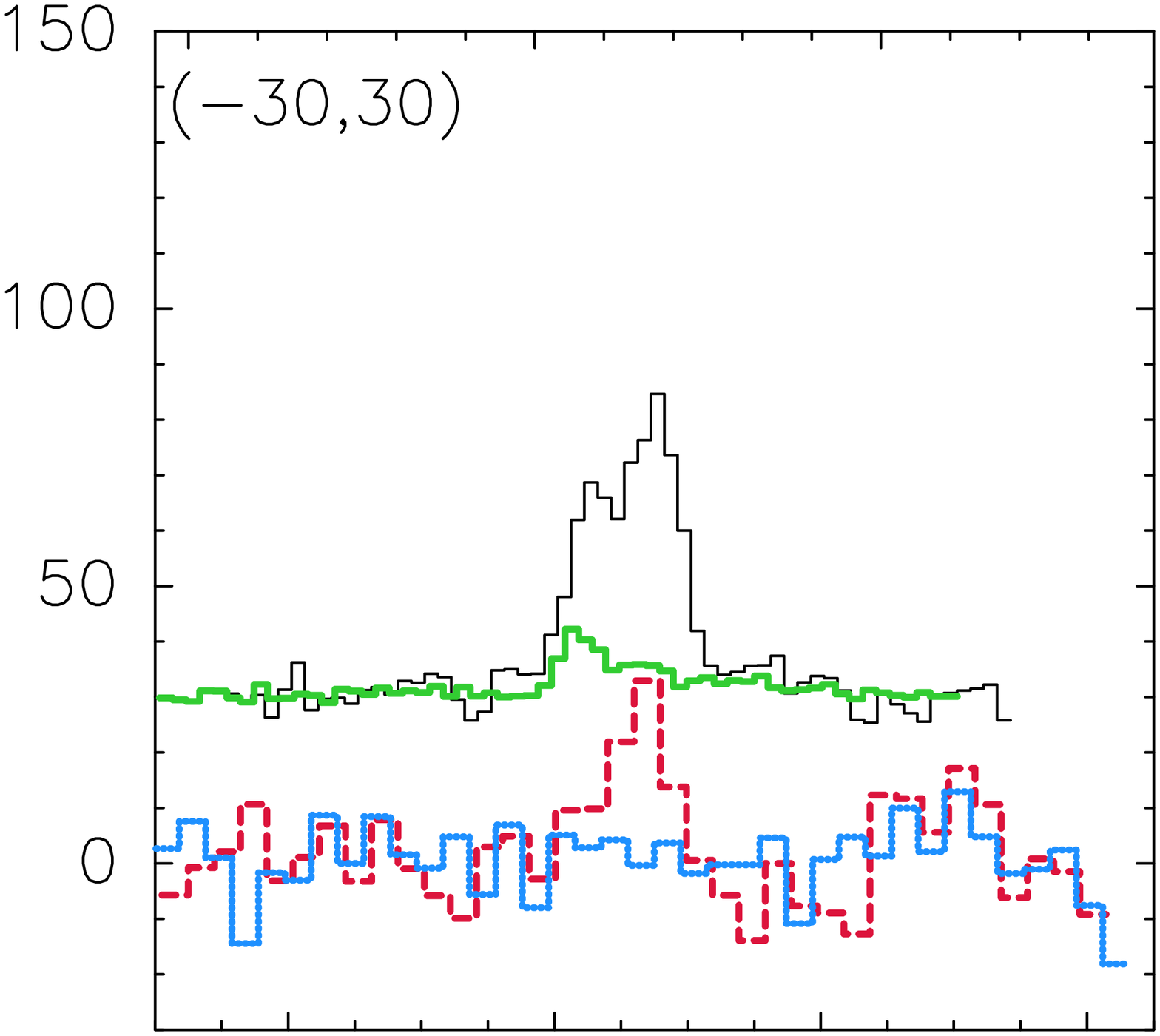}&
\includegraphics[scale=0.168, angle=0]{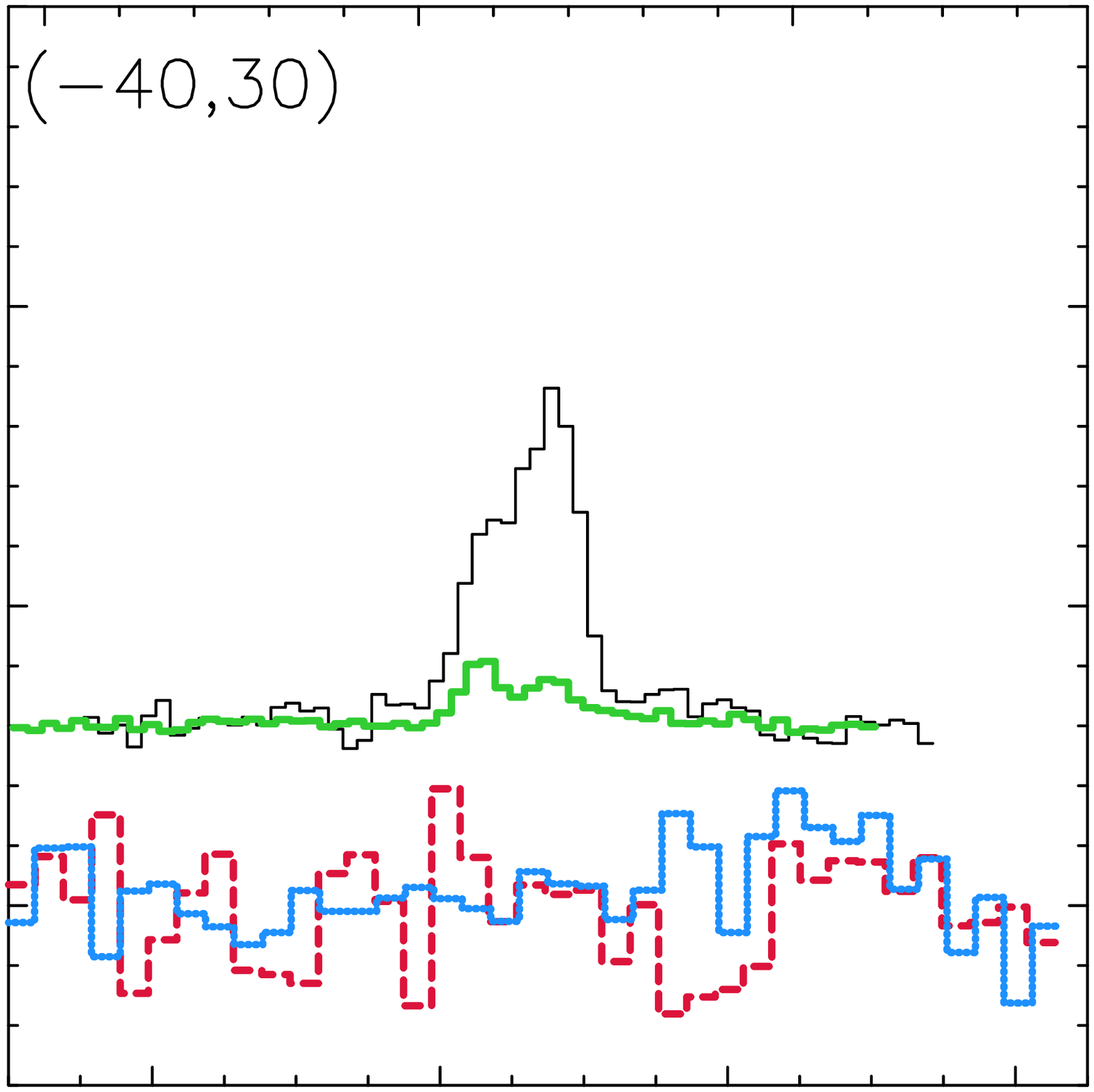}&
\includegraphics[scale=0.168, angle=0]{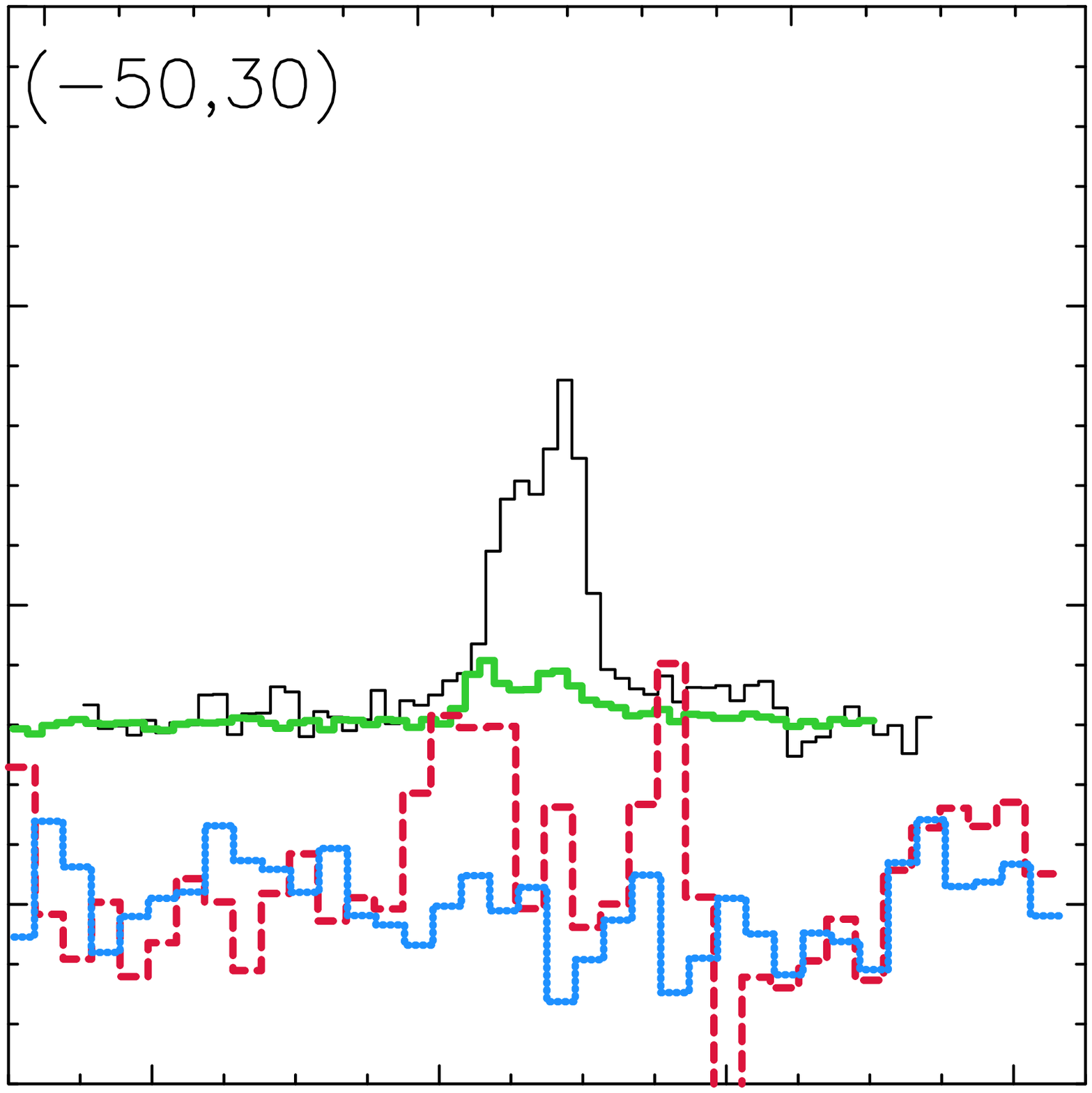}&
\includegraphics[scale=0.168, angle=0]{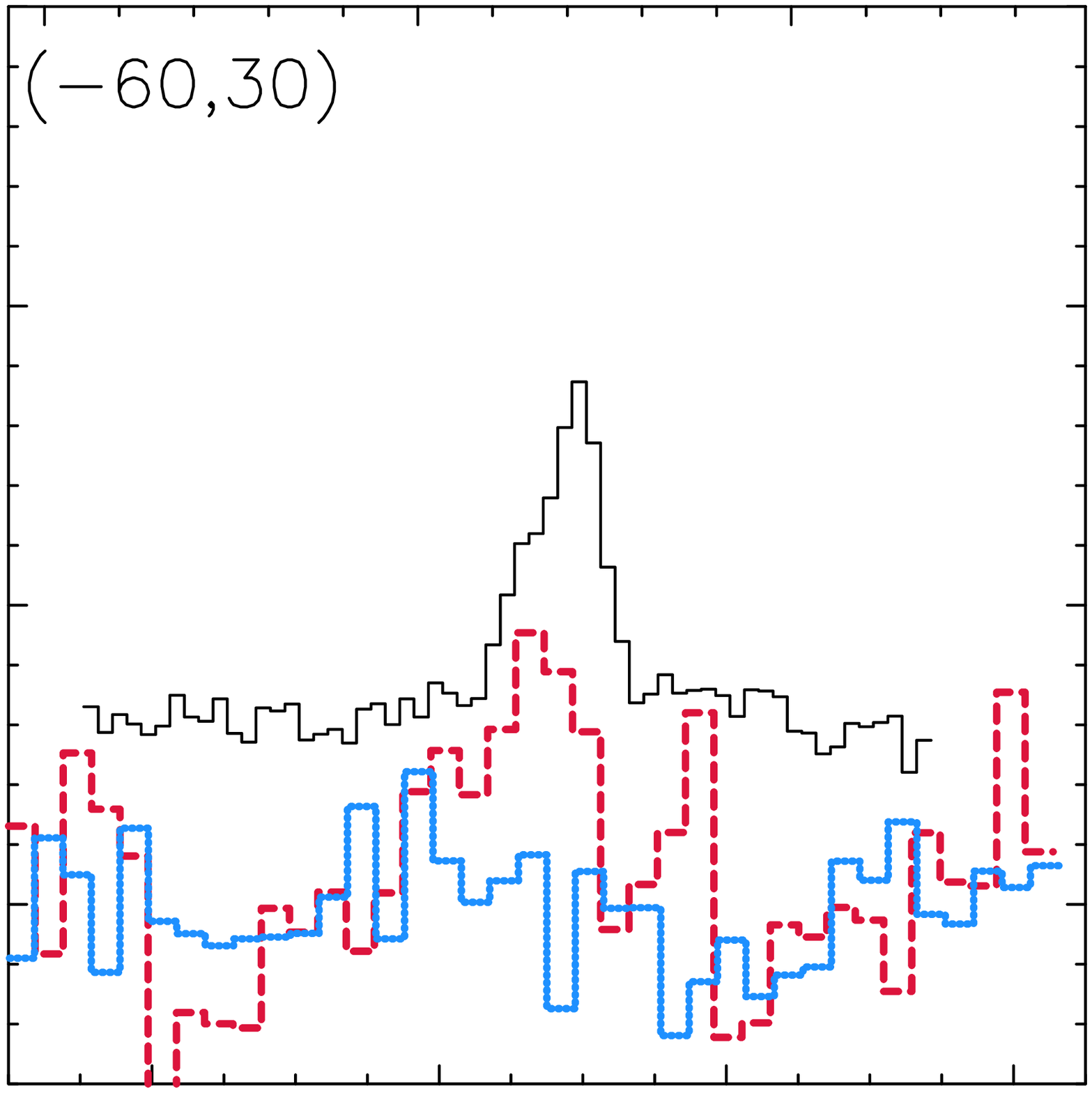}\\
\includegraphics[scale=0.168, angle=0]{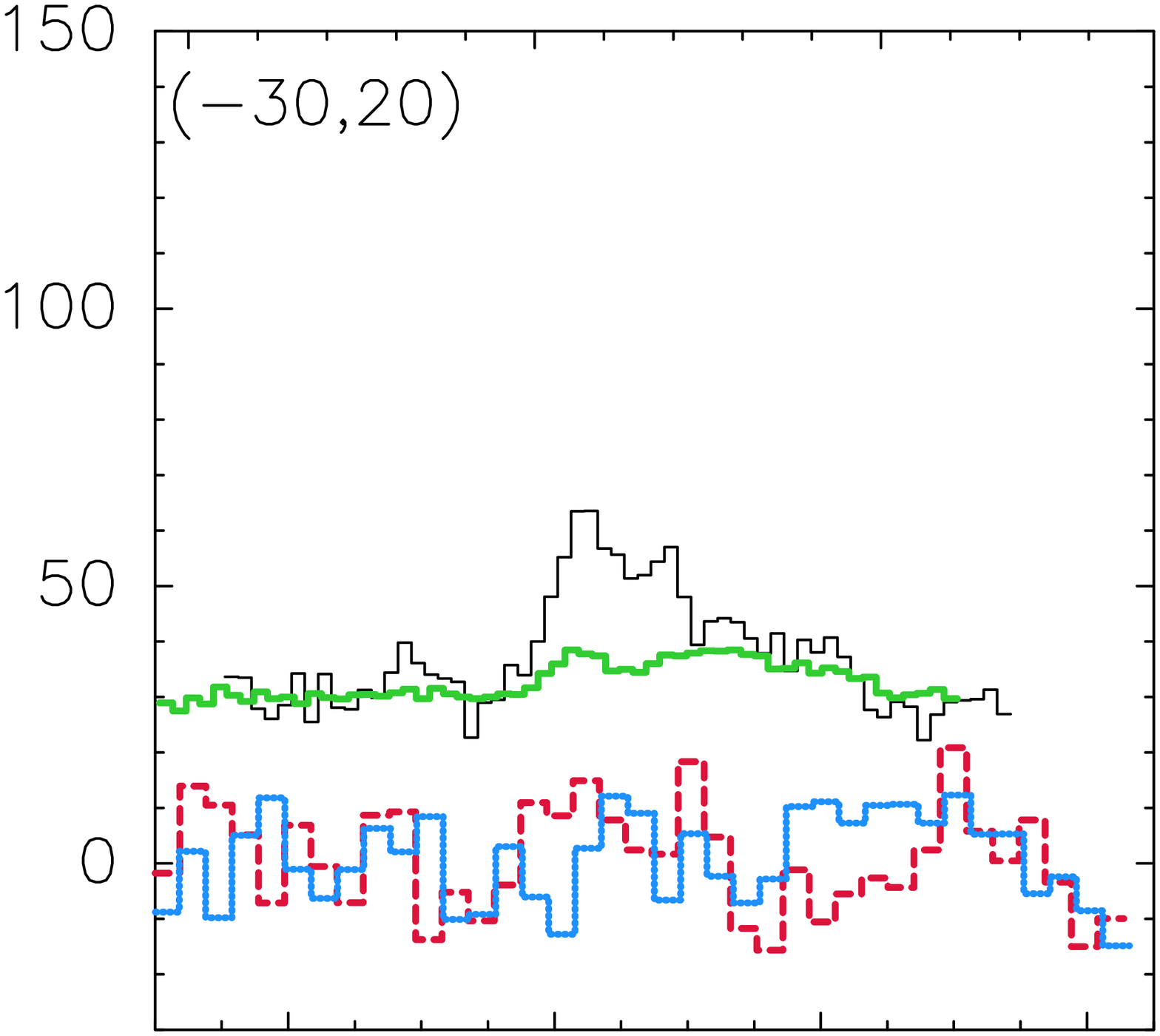}&
\includegraphics[scale=0.168, angle=0]{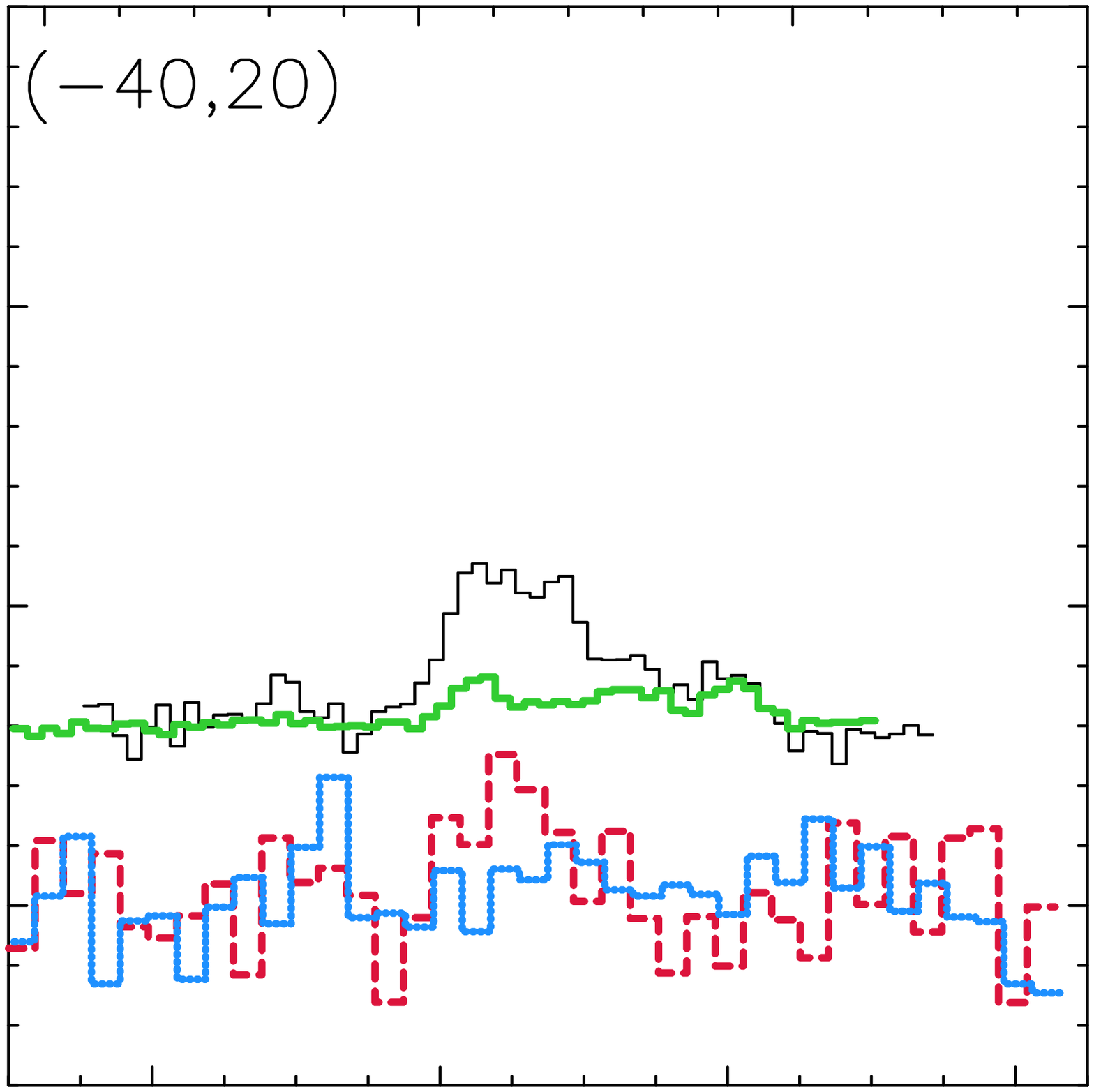}&
\includegraphics[scale=0.168, angle=0]{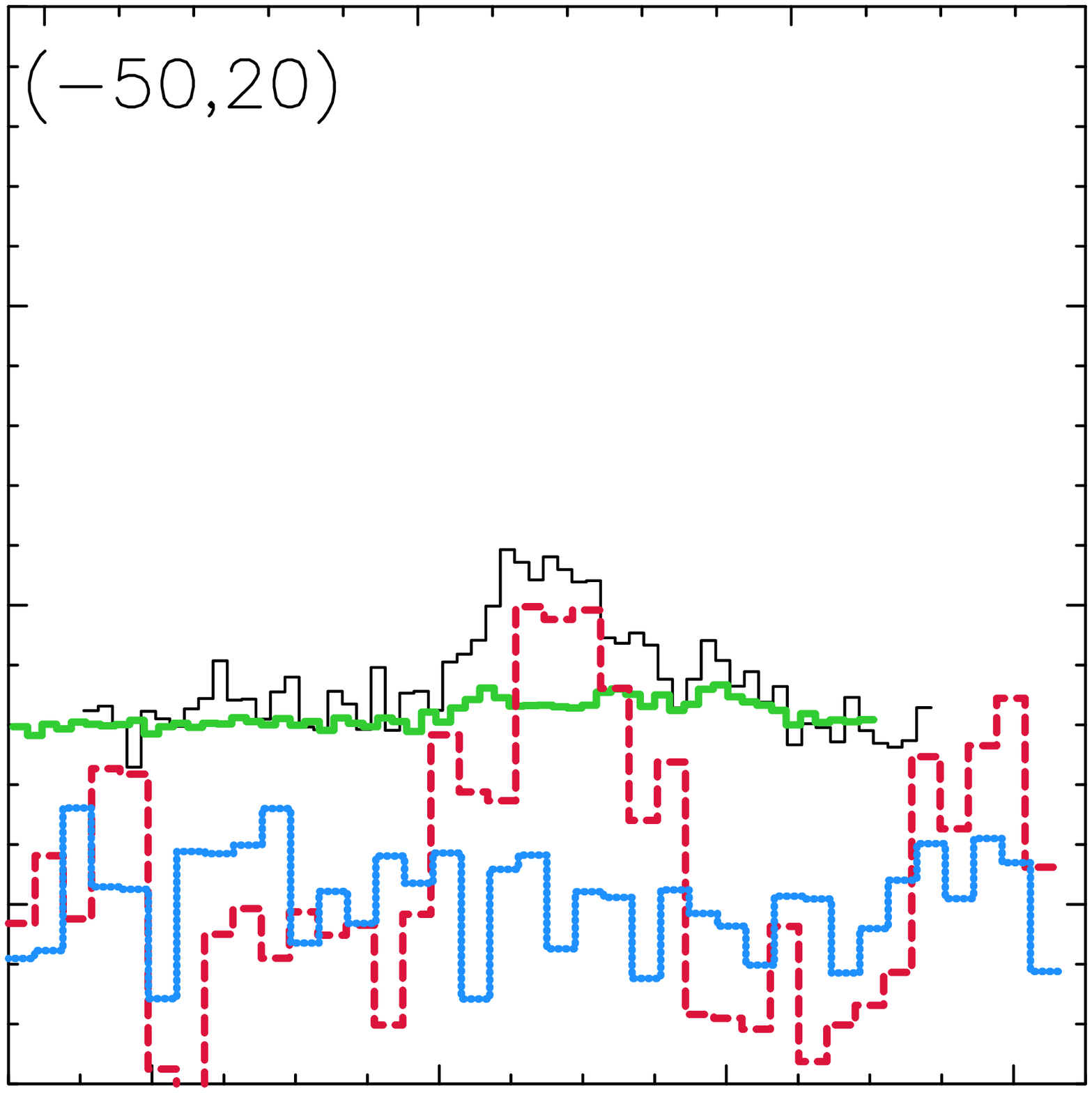}&
\includegraphics[scale=0.168, angle=0]{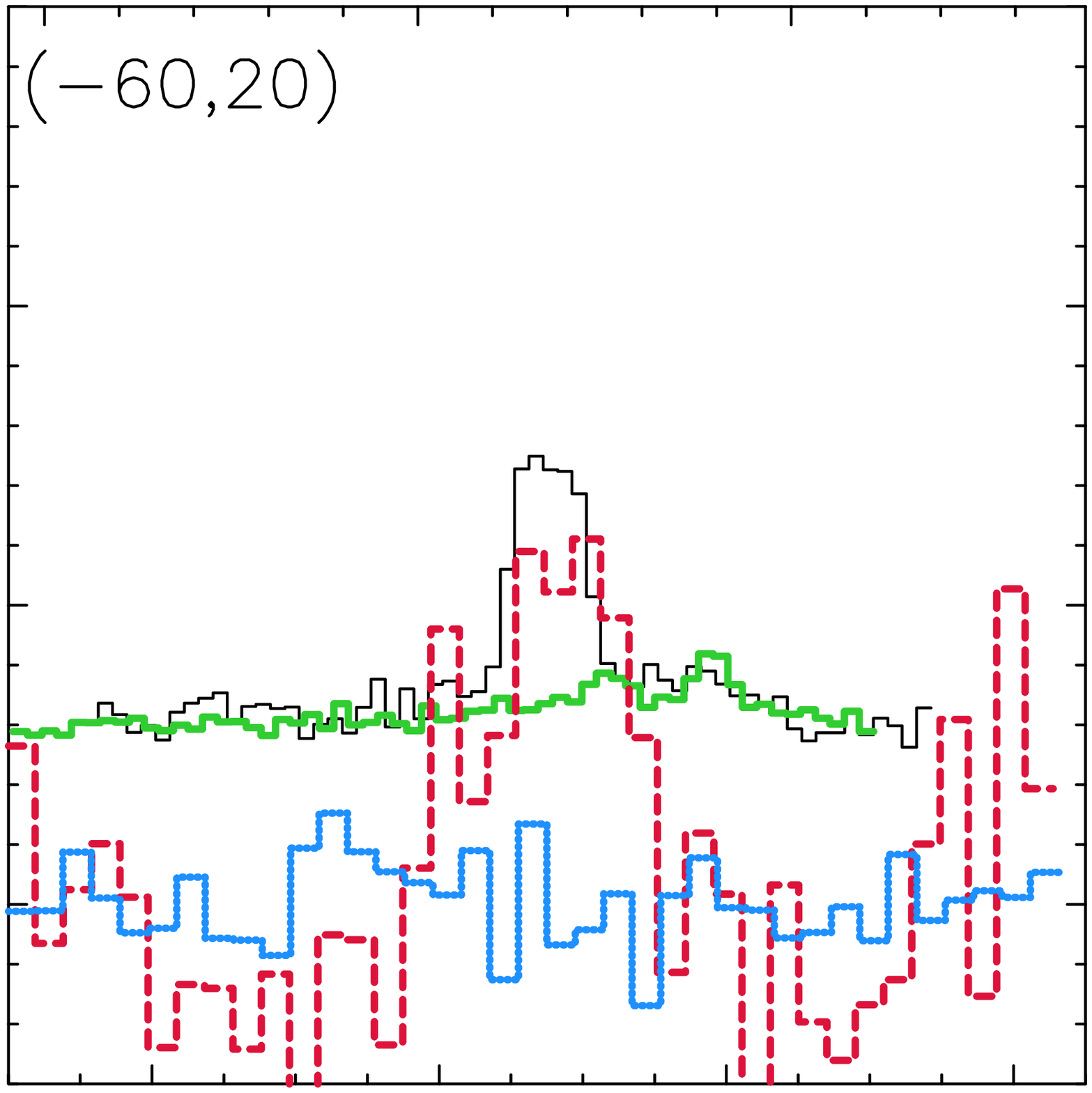}\\
\includegraphics[scale=0.168, angle=0]{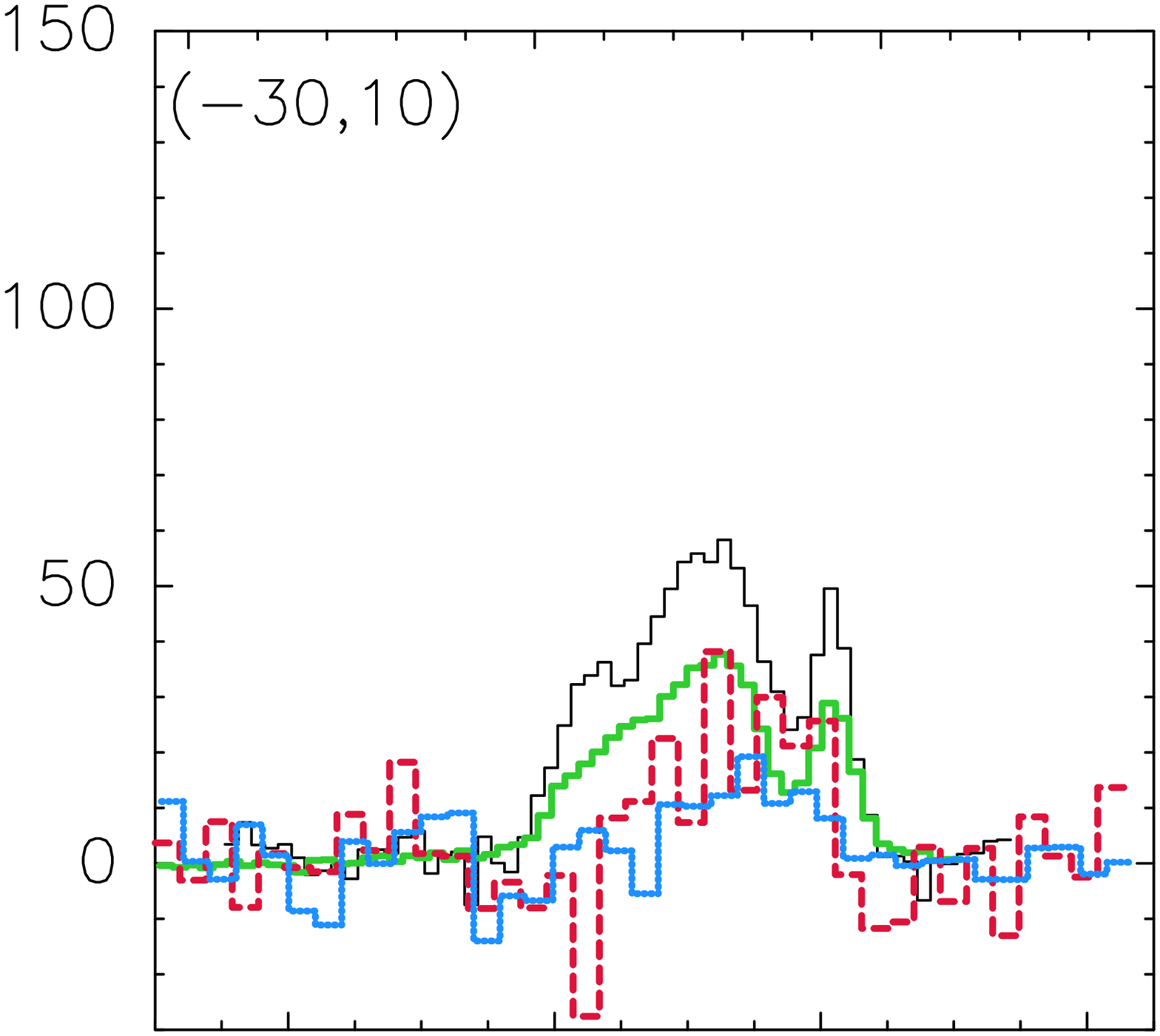}&
\includegraphics[scale=0.168, angle=0]{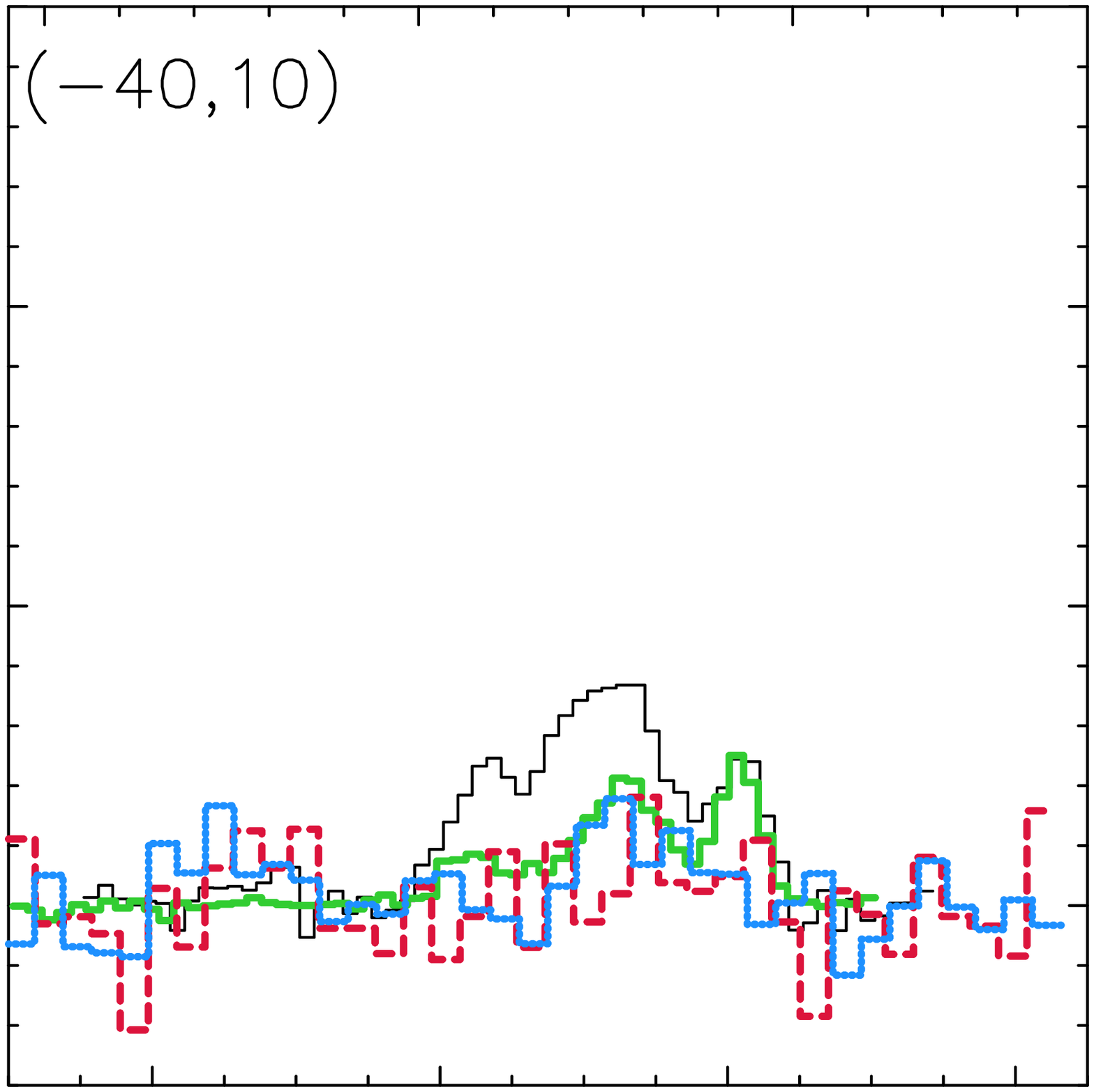}&
\includegraphics[scale=0.168, angle=0]{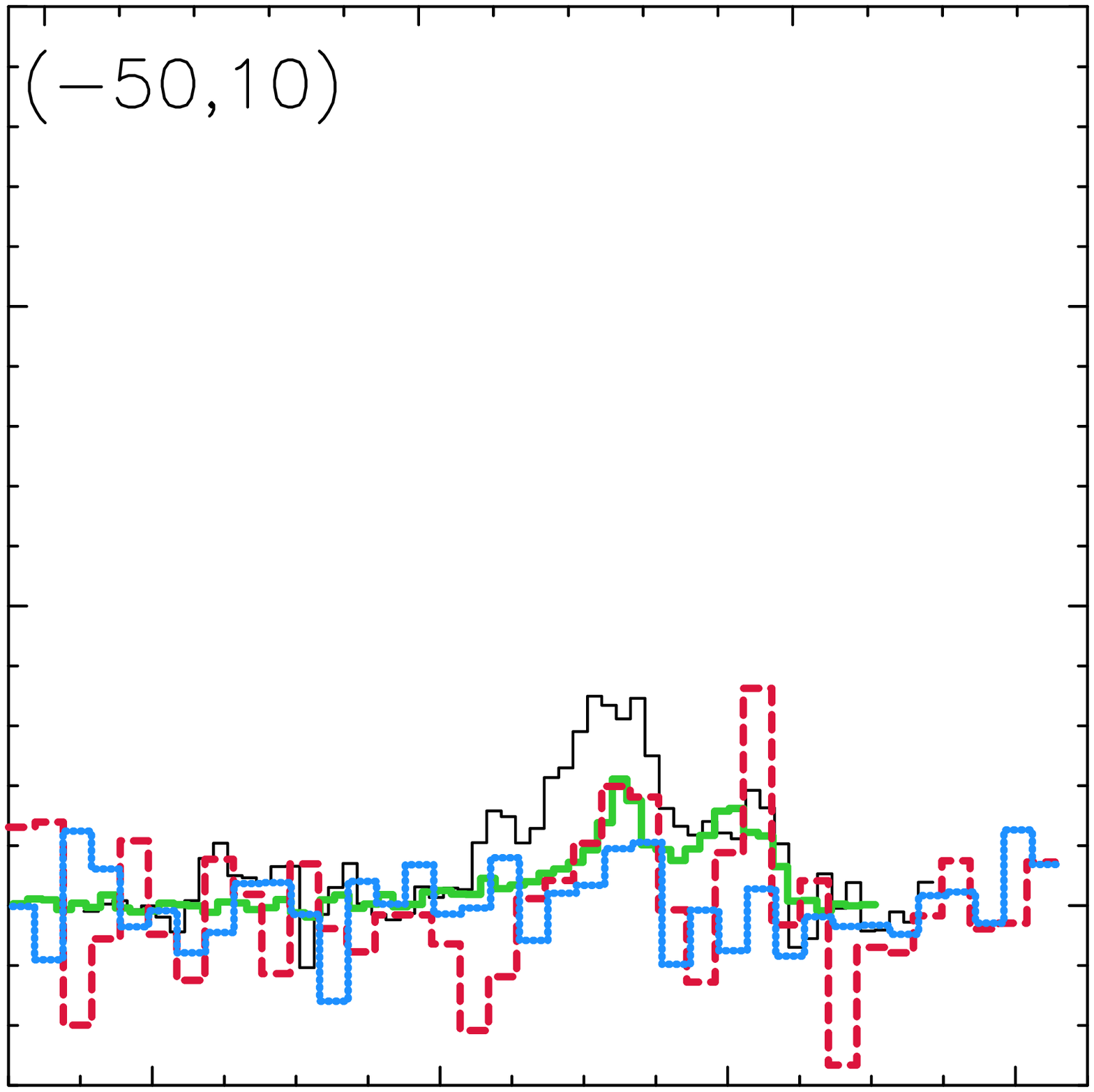}&
\includegraphics[scale=0.168, angle=0]{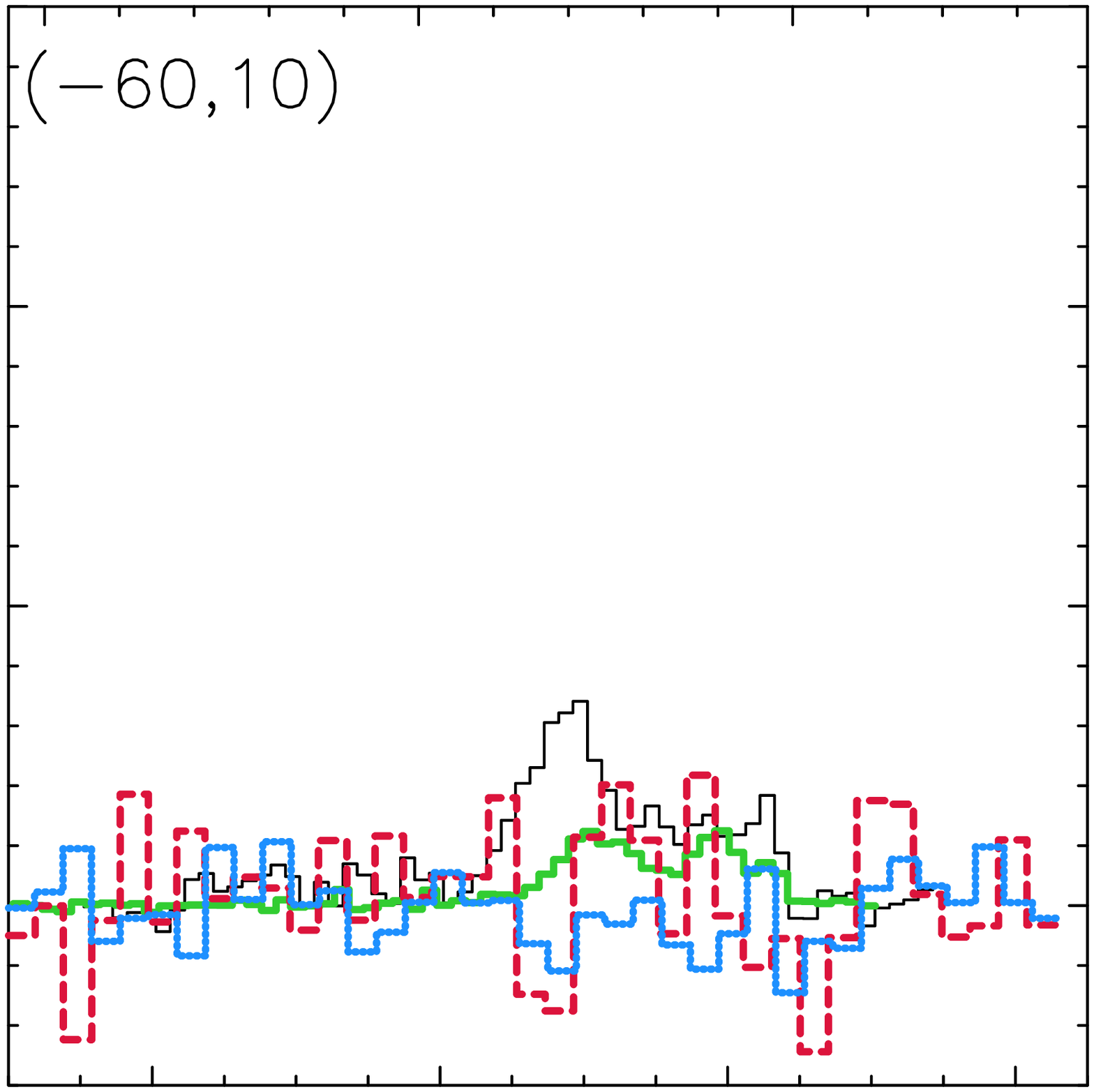}\\
\includegraphics[scale=0.168, angle=0]{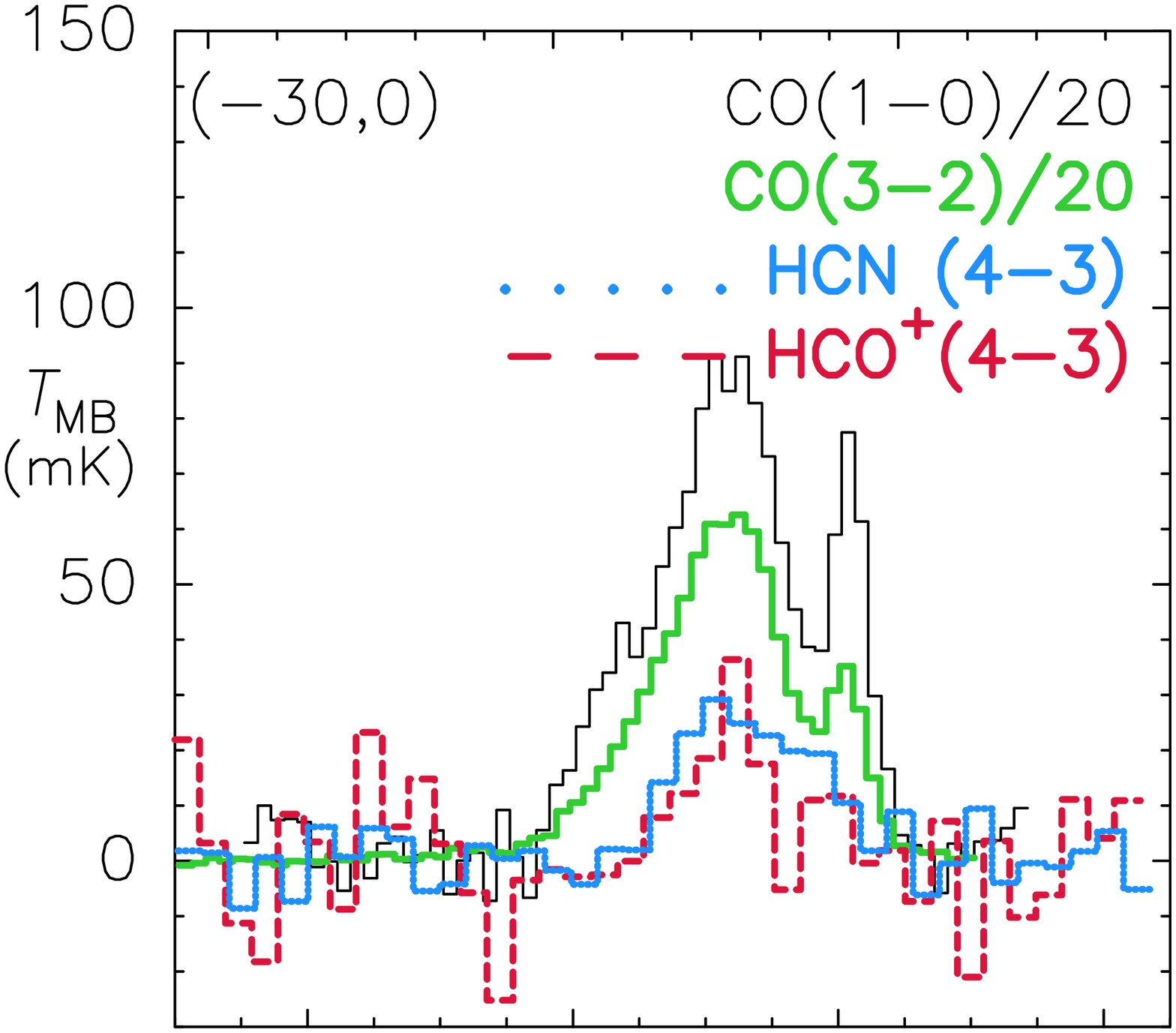}&
\includegraphics[scale=0.168, angle=0]{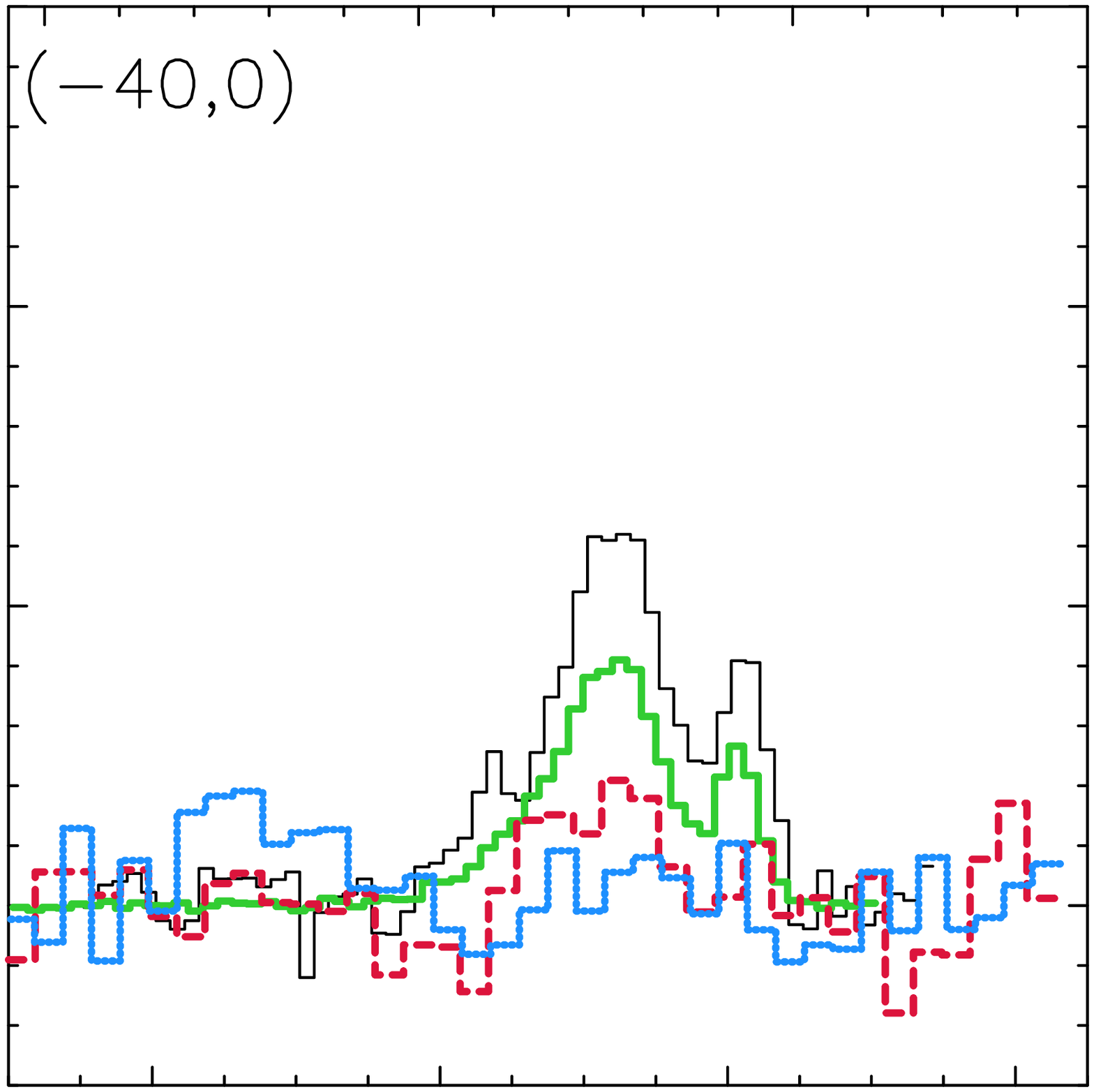}&
\includegraphics[scale=0.168, angle=0]{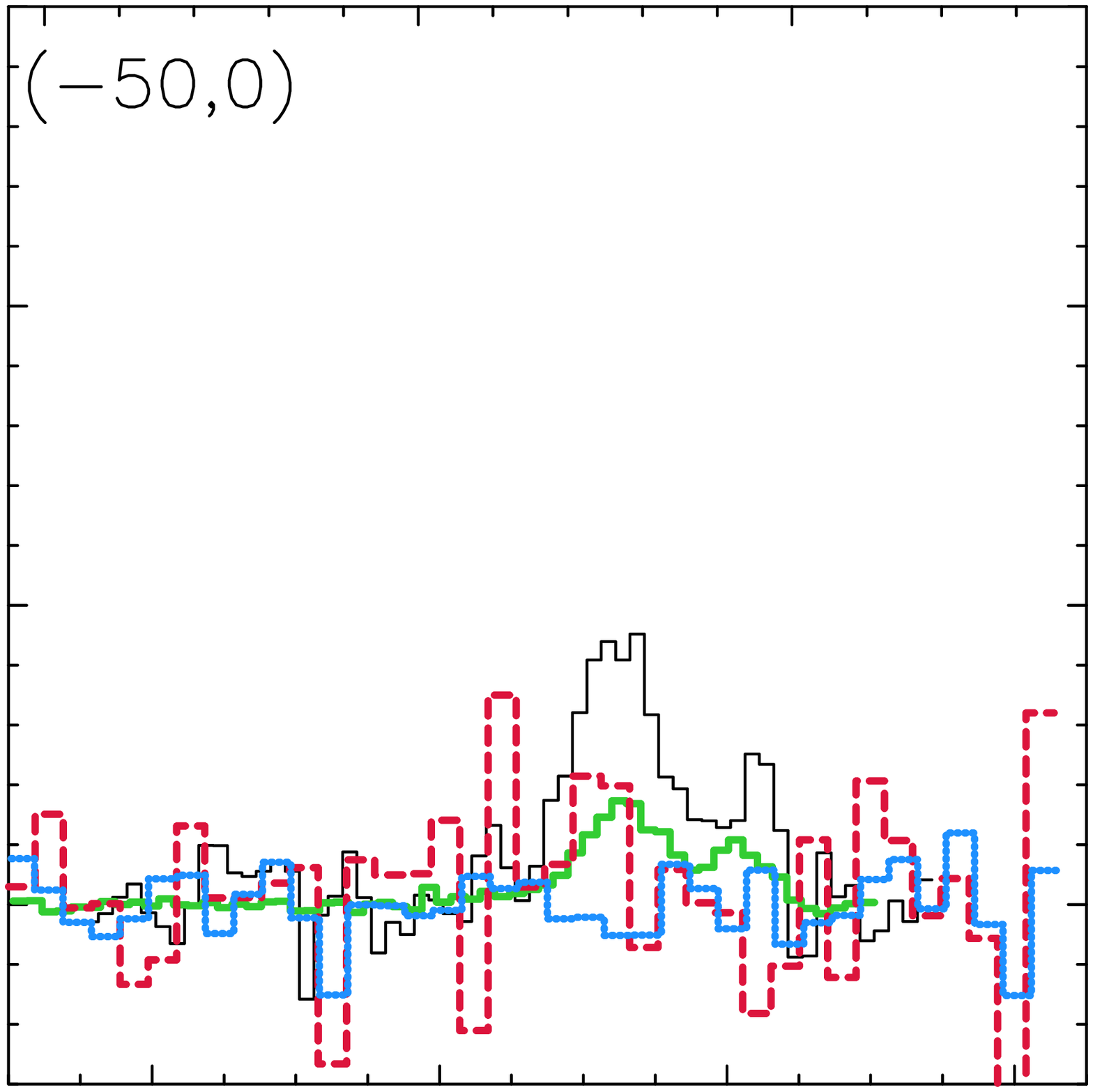}&
\includegraphics[scale=0.168, angle=0]{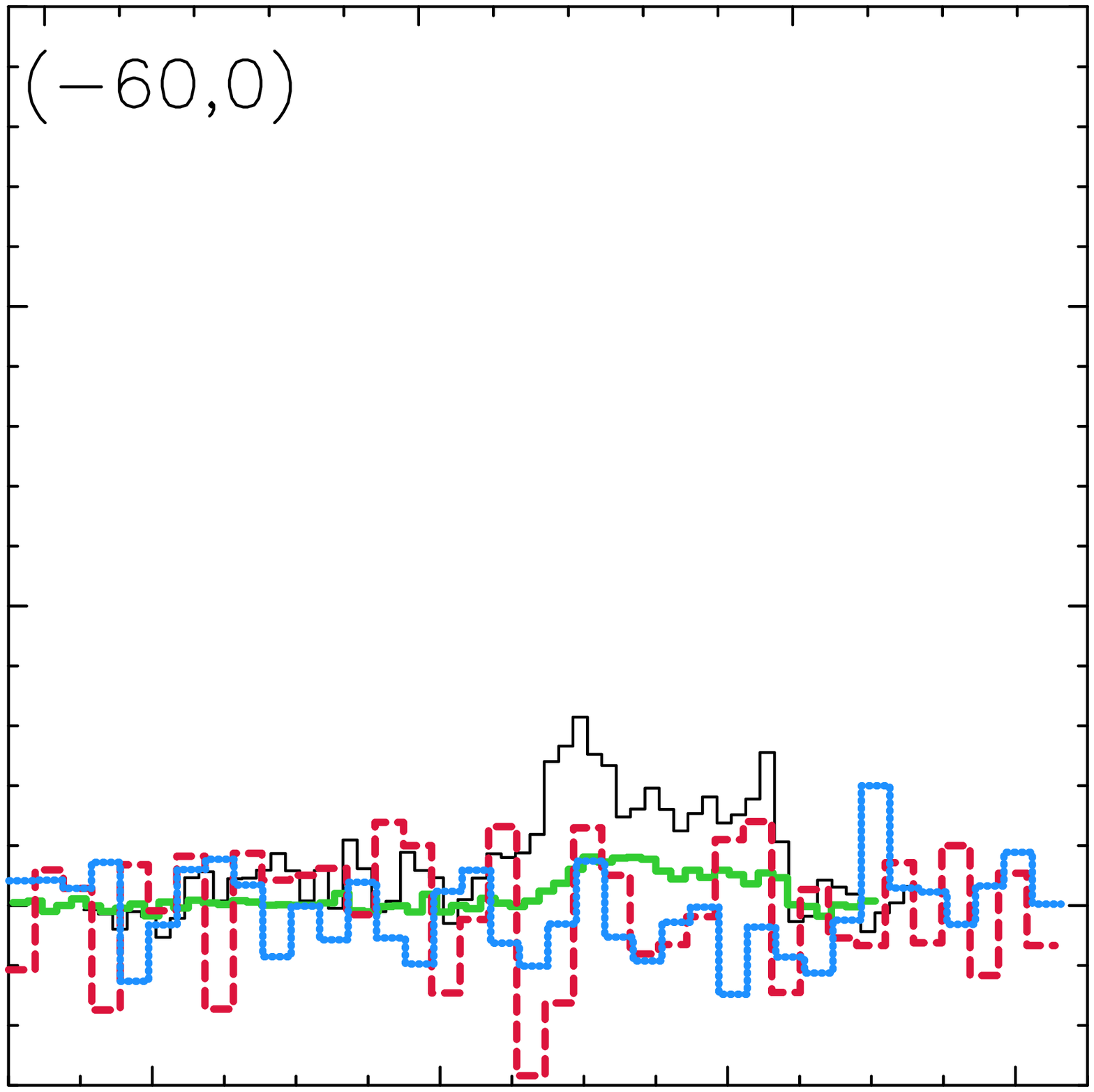}\\
\includegraphics[scale=0.168, angle=0]{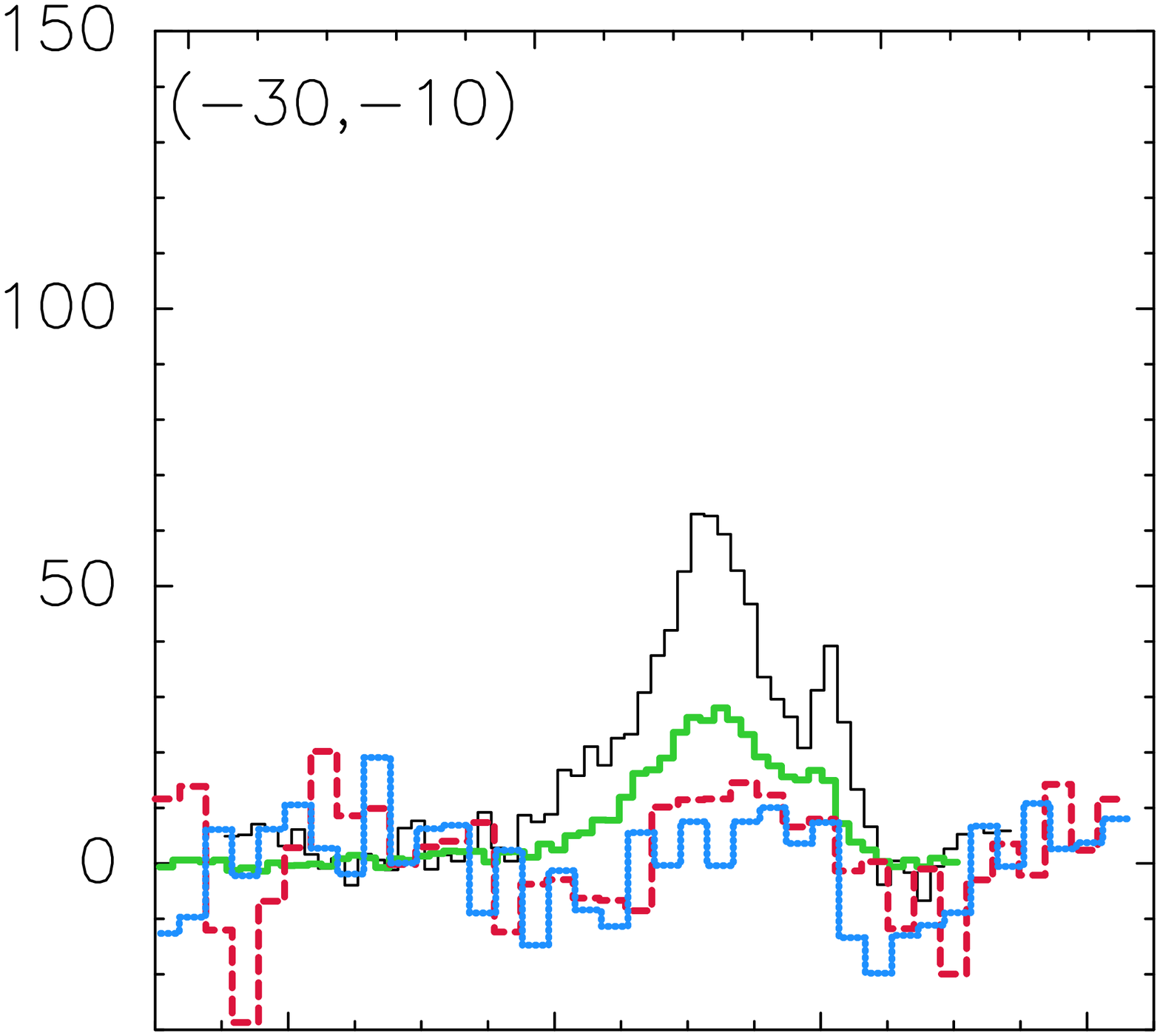}&
\includegraphics[scale=0.168, angle=0]{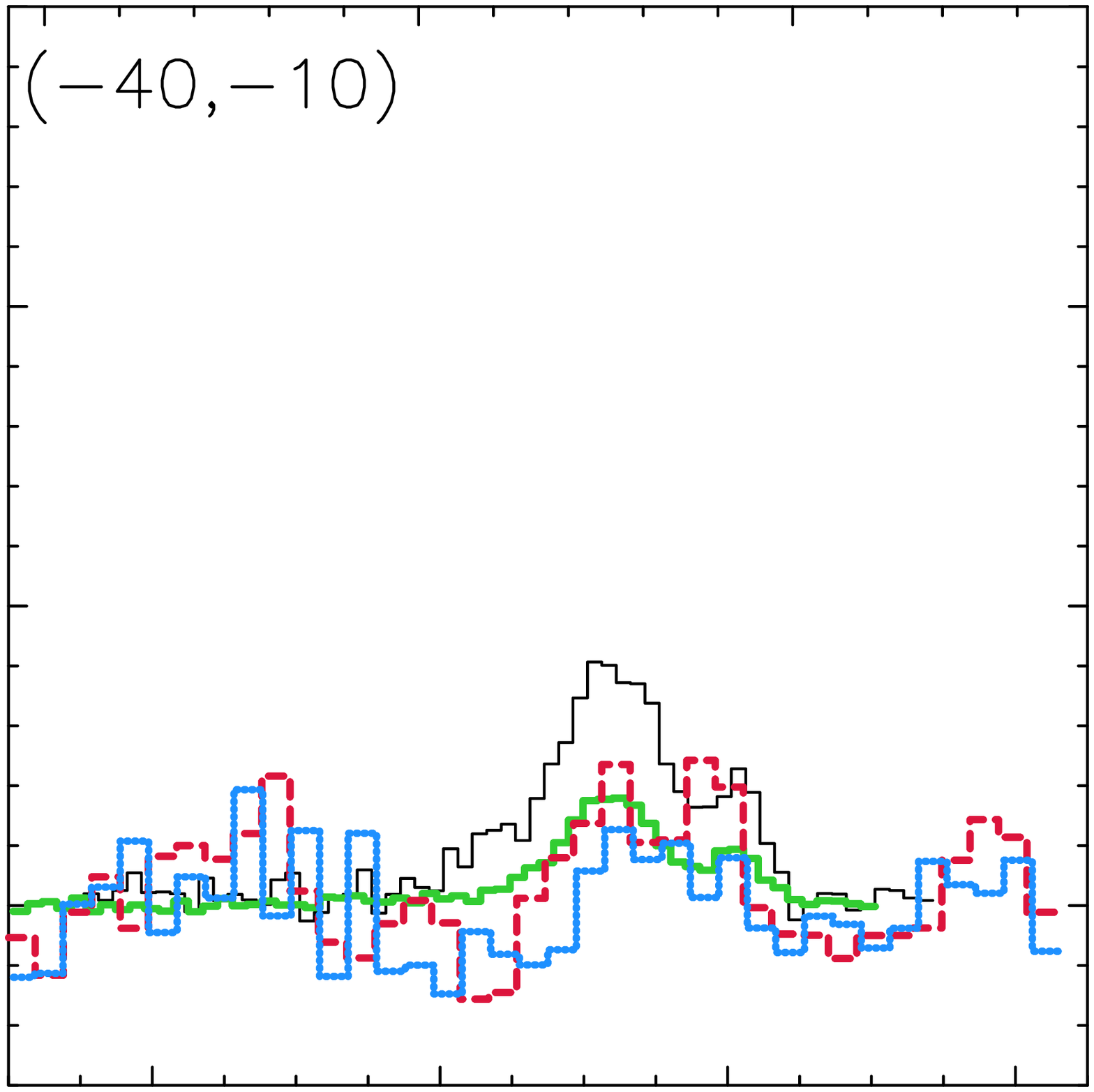}&
\includegraphics[scale=0.168, angle=0]{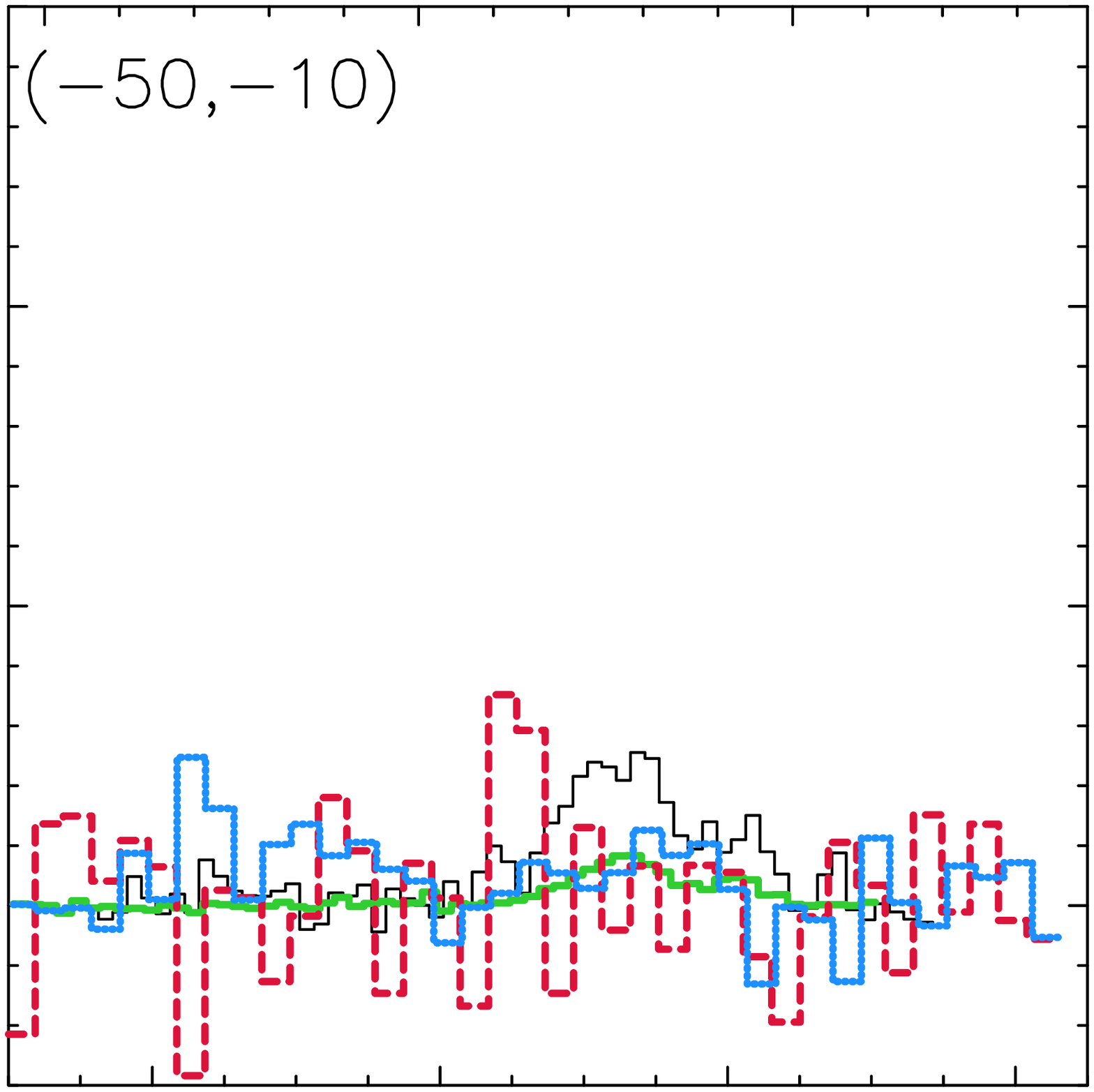}&
\includegraphics[scale=0.168, angle=0]{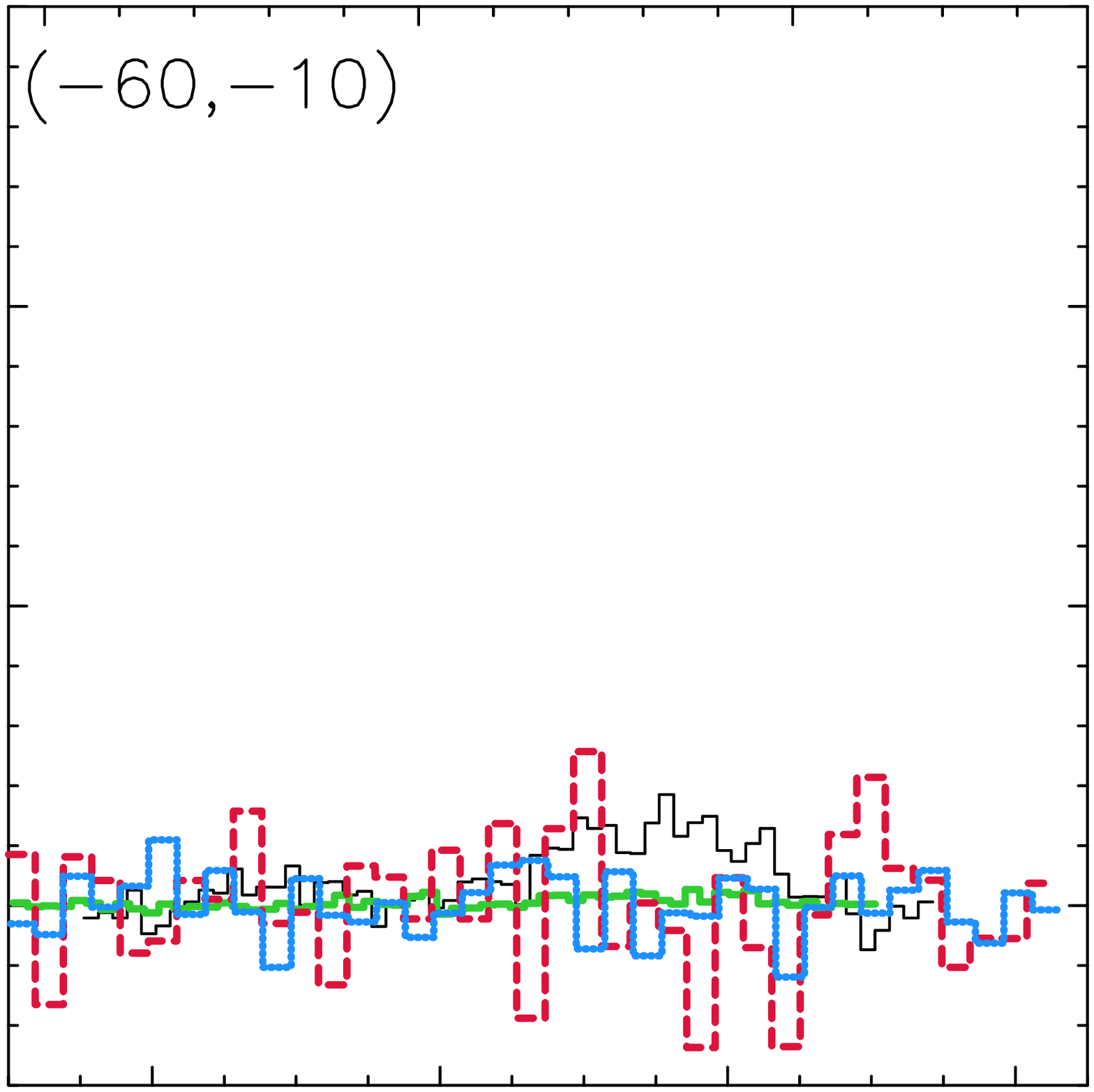}\\
\includegraphics[scale=0.168, angle=0]{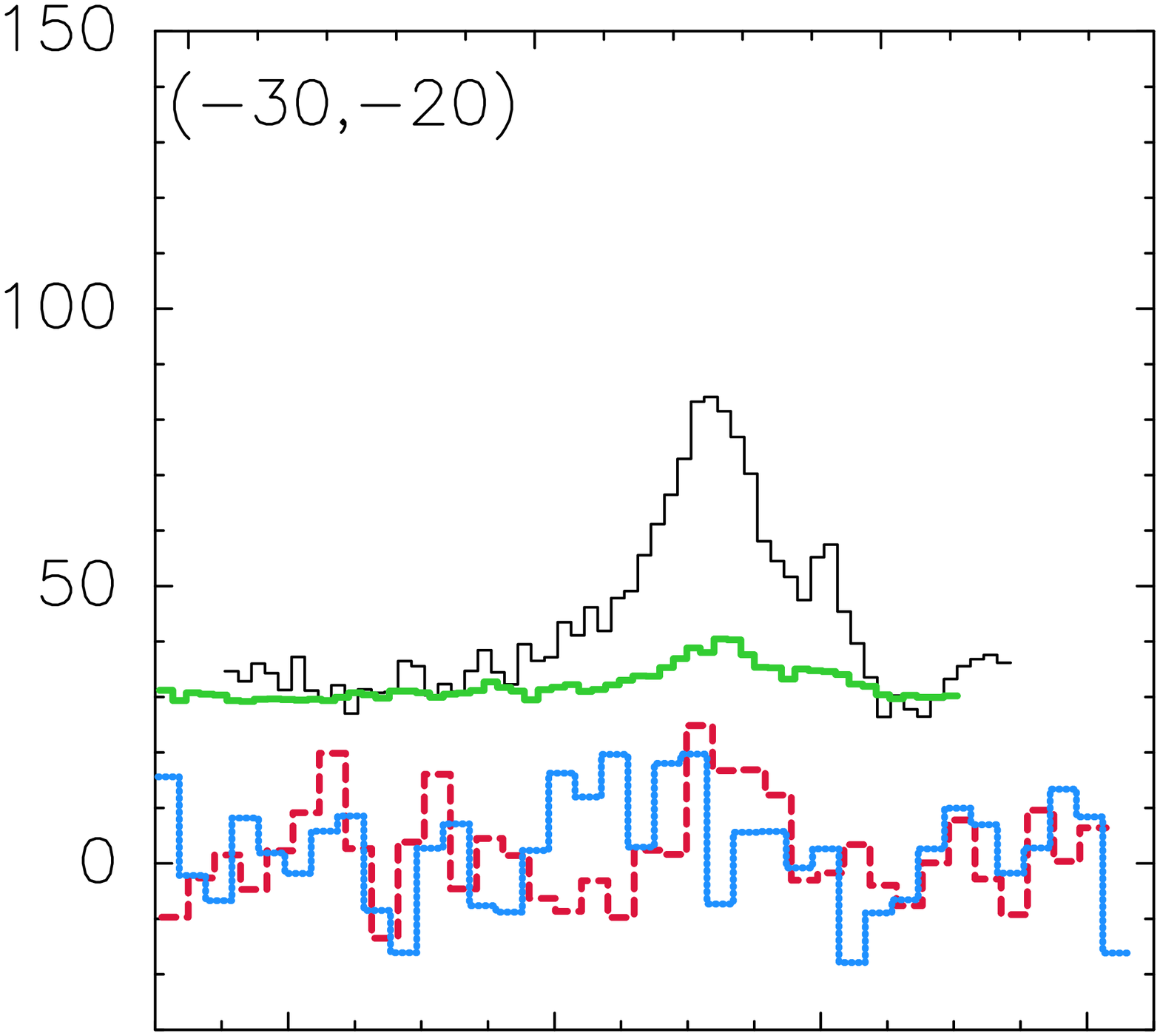}&
\includegraphics[scale=0.168, angle=0]{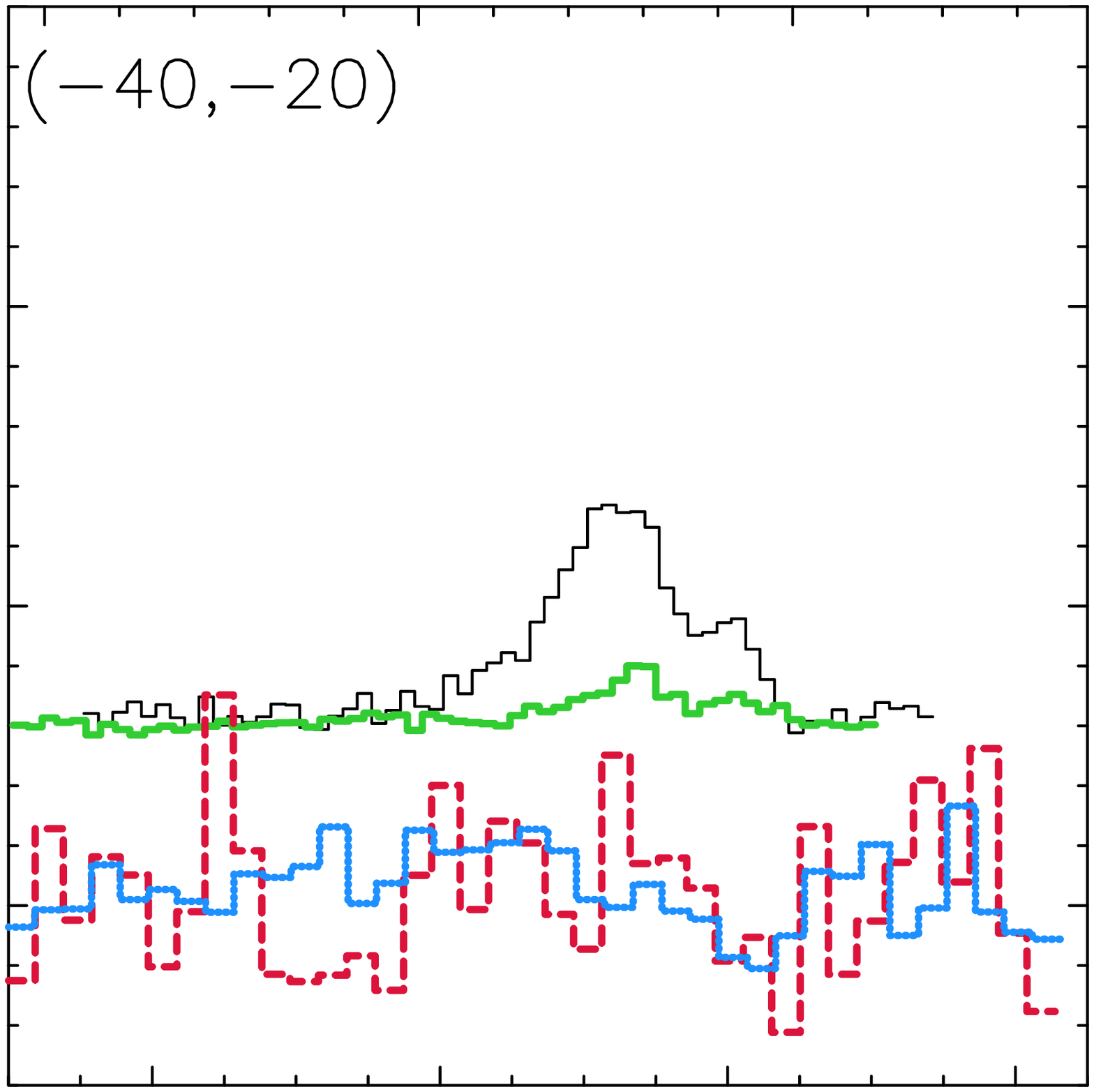}&
\includegraphics[scale=0.168, angle=0]{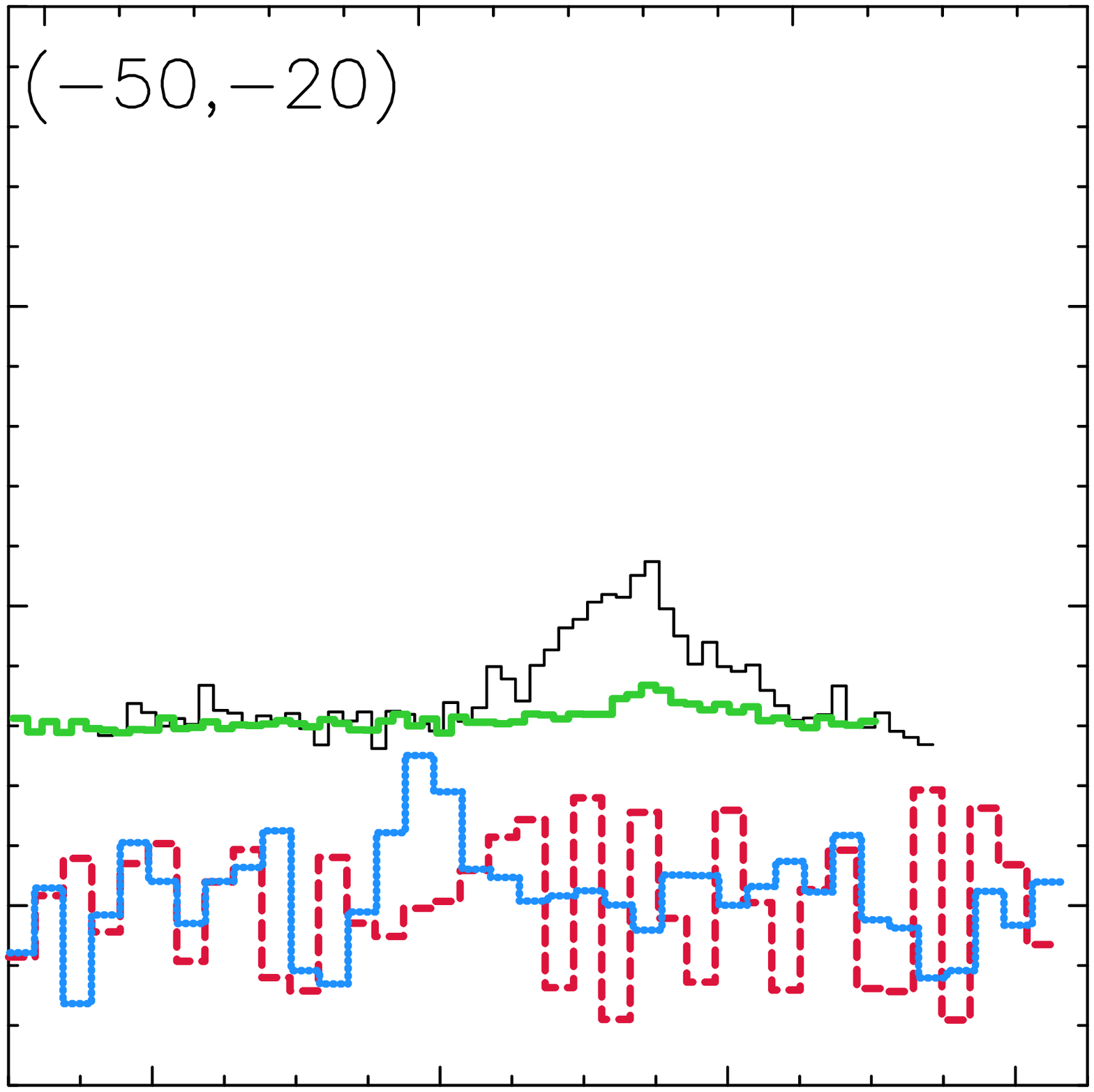}&
\includegraphics[scale=0.168, angle=0]{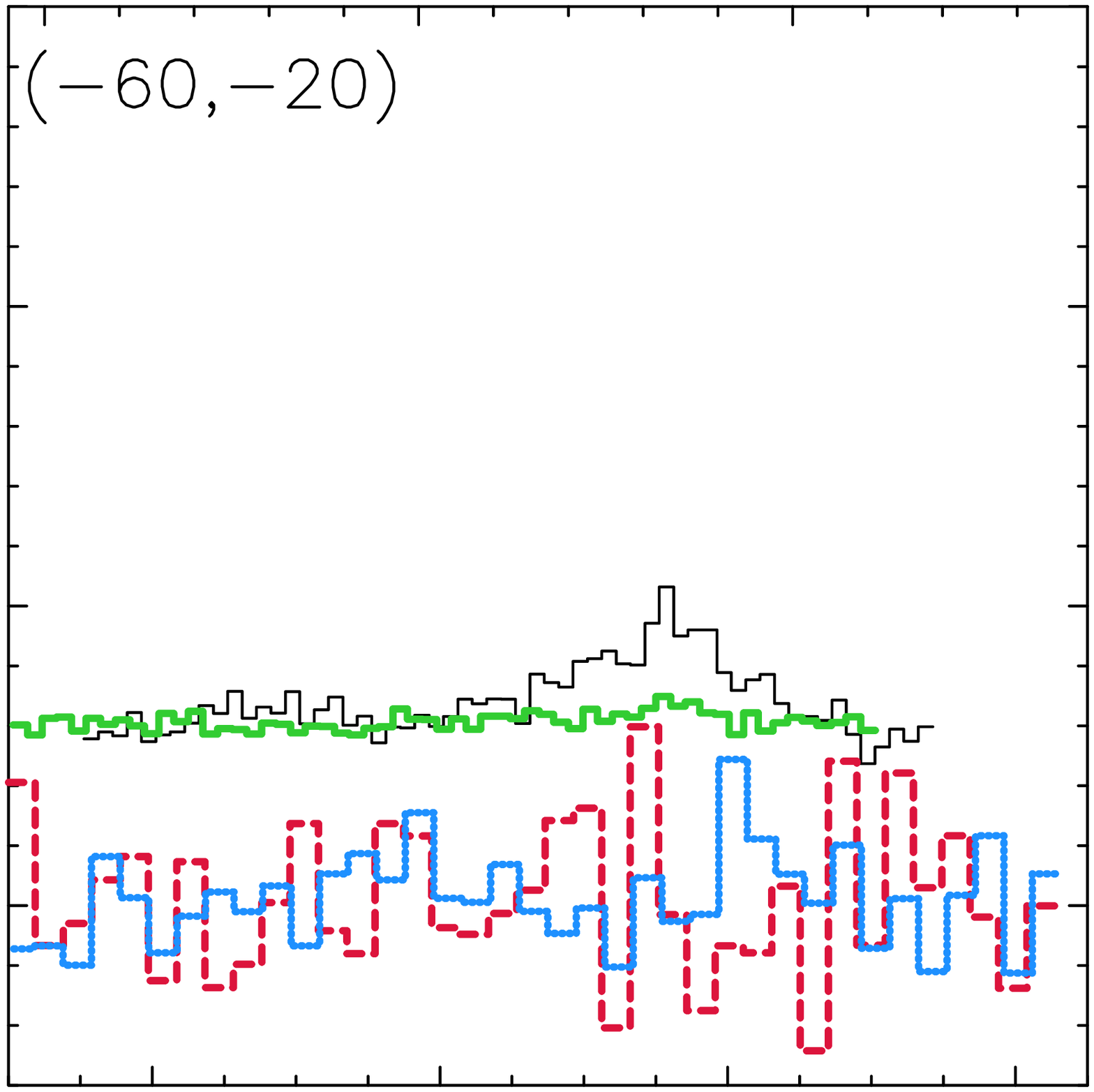}\\
\includegraphics[scale=0.168, angle=0]{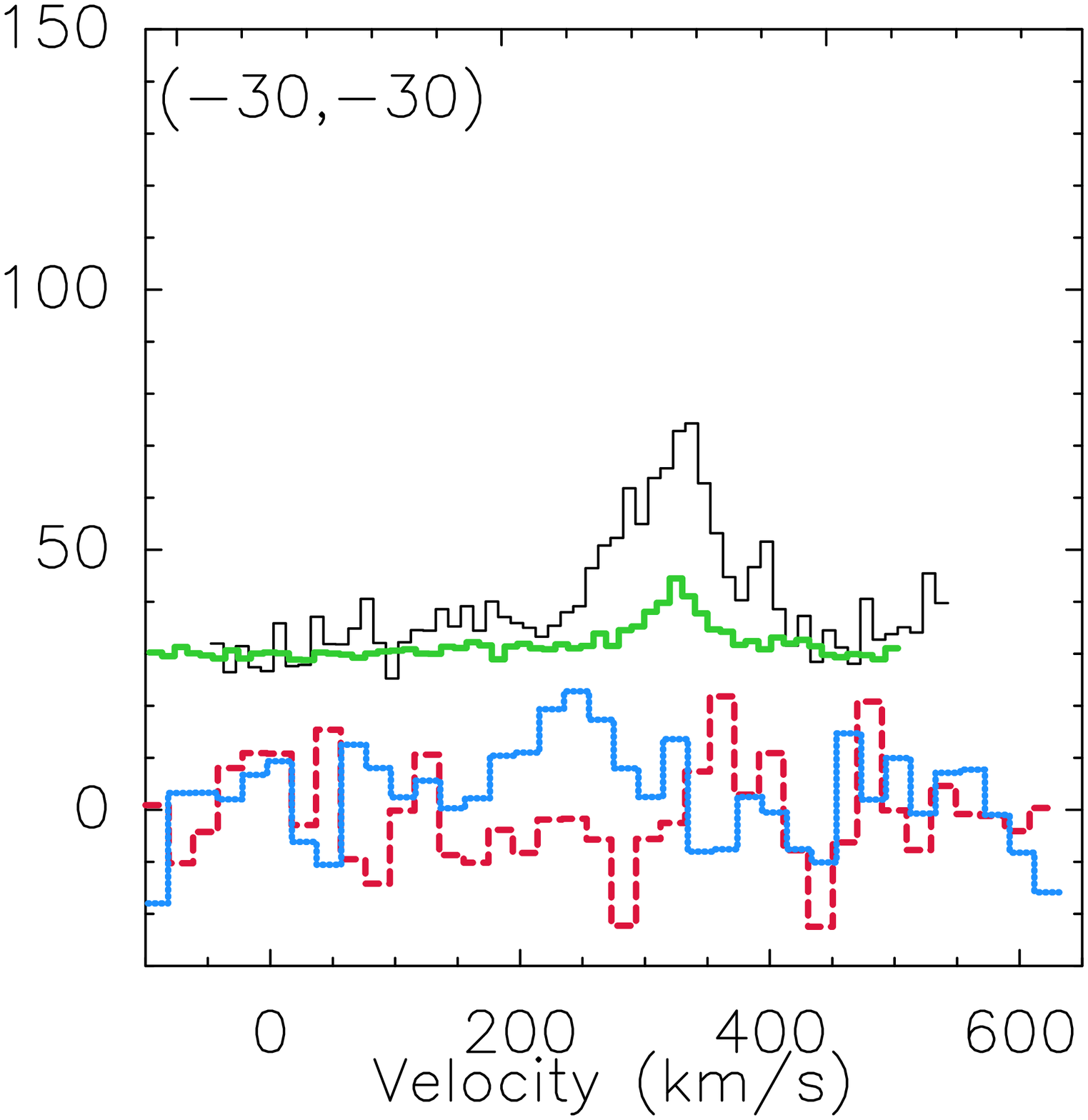}&
\includegraphics[scale=0.168, angle=0]{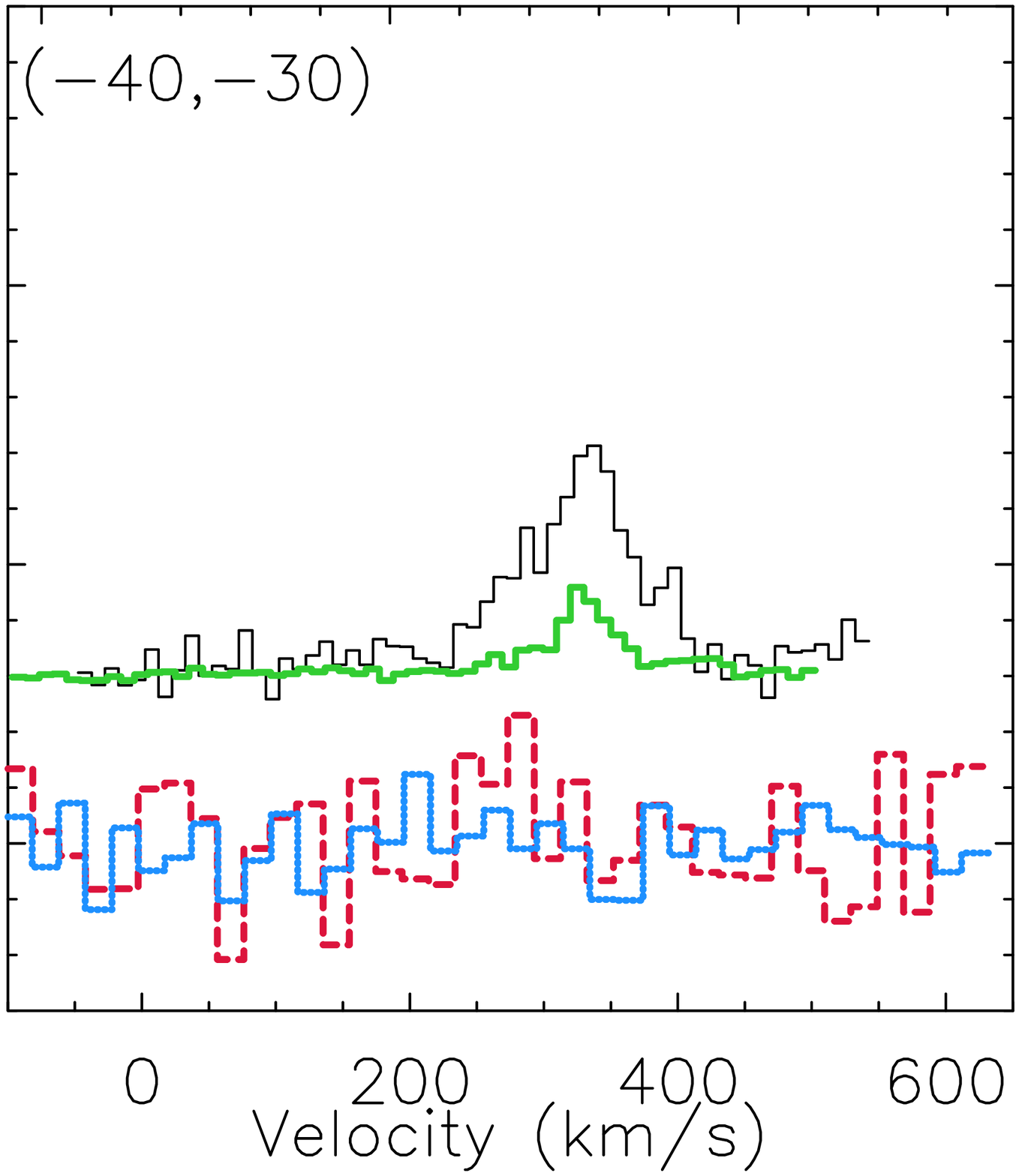}&
\includegraphics[scale=0.168, angle=0]{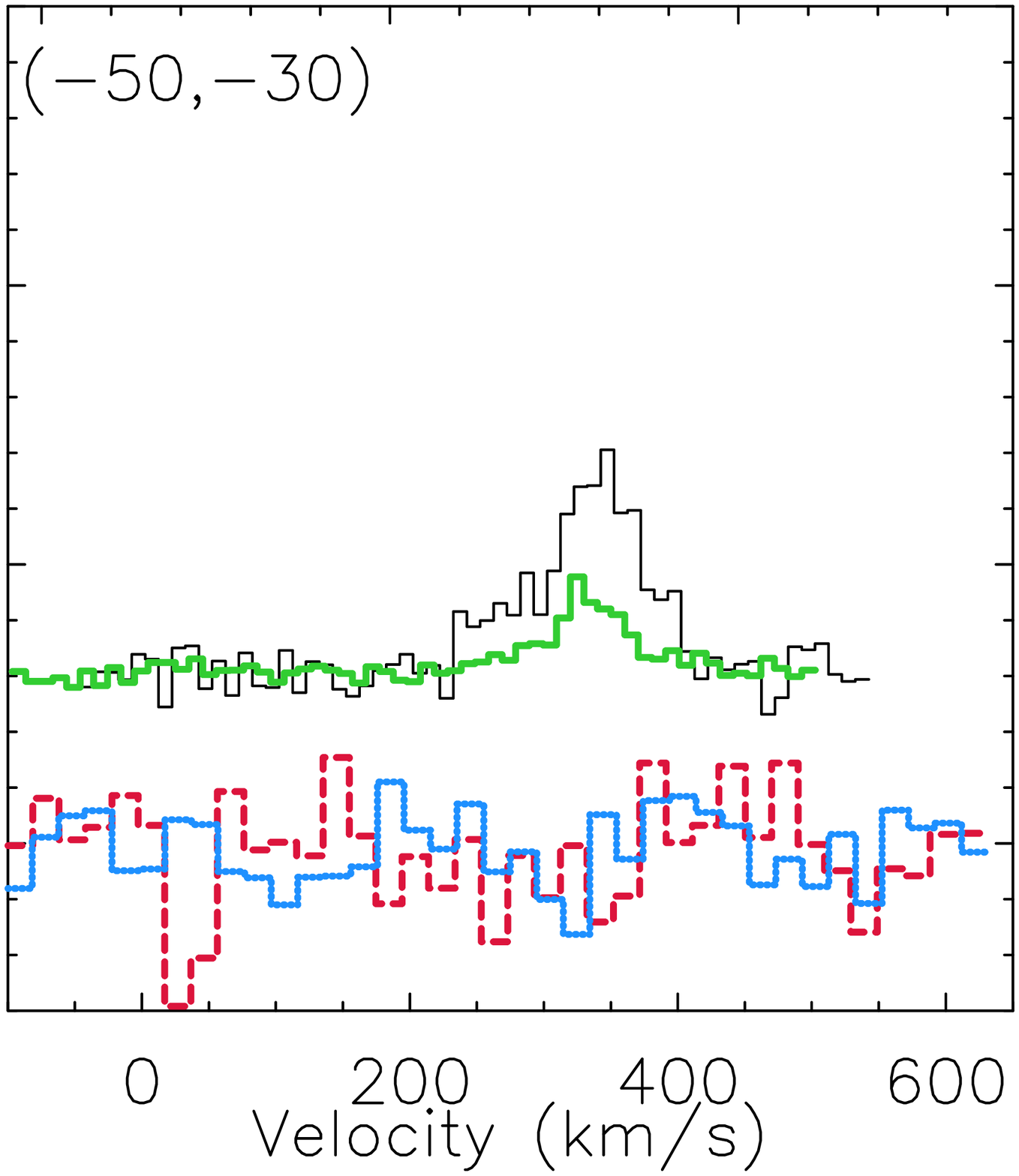}&
\includegraphics[scale=0.168, angle=0]{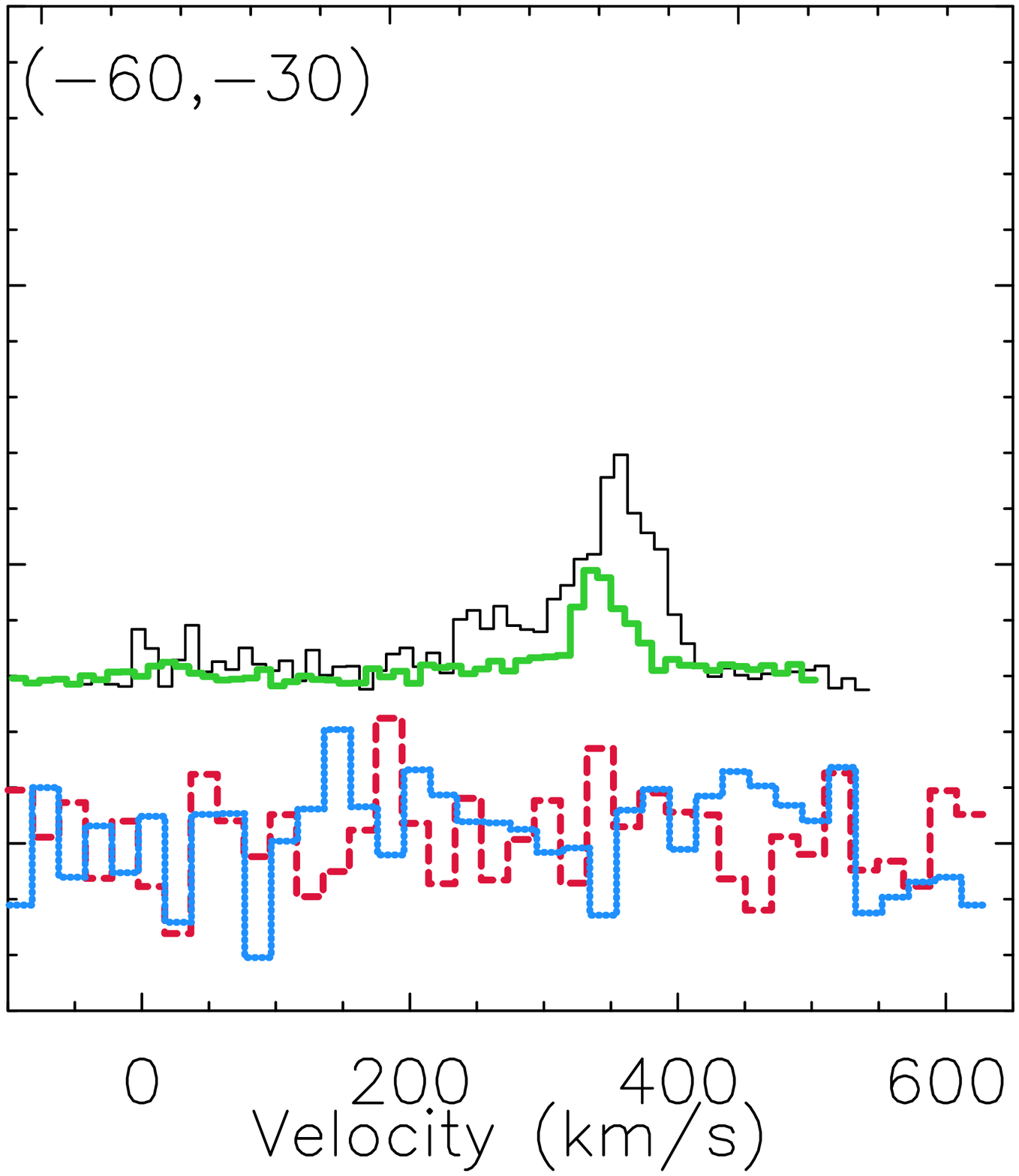}\\
\end{array}$
\end{center}
\begin{center}
\contcaption{Spectra of CO 1-0 (black), 
CO 3-2 (green), HCN 4-3
(blue) and HCO$^+$ 4-3 (red) emission in the central $\sim$ 1\,kpc region of NGC\,253.
}
\label{fig:spec2} 
\end{center} 
\end{figure*}

\begin{figure}
\begin{center}
\includegraphics[width=\linewidth, angle=0]{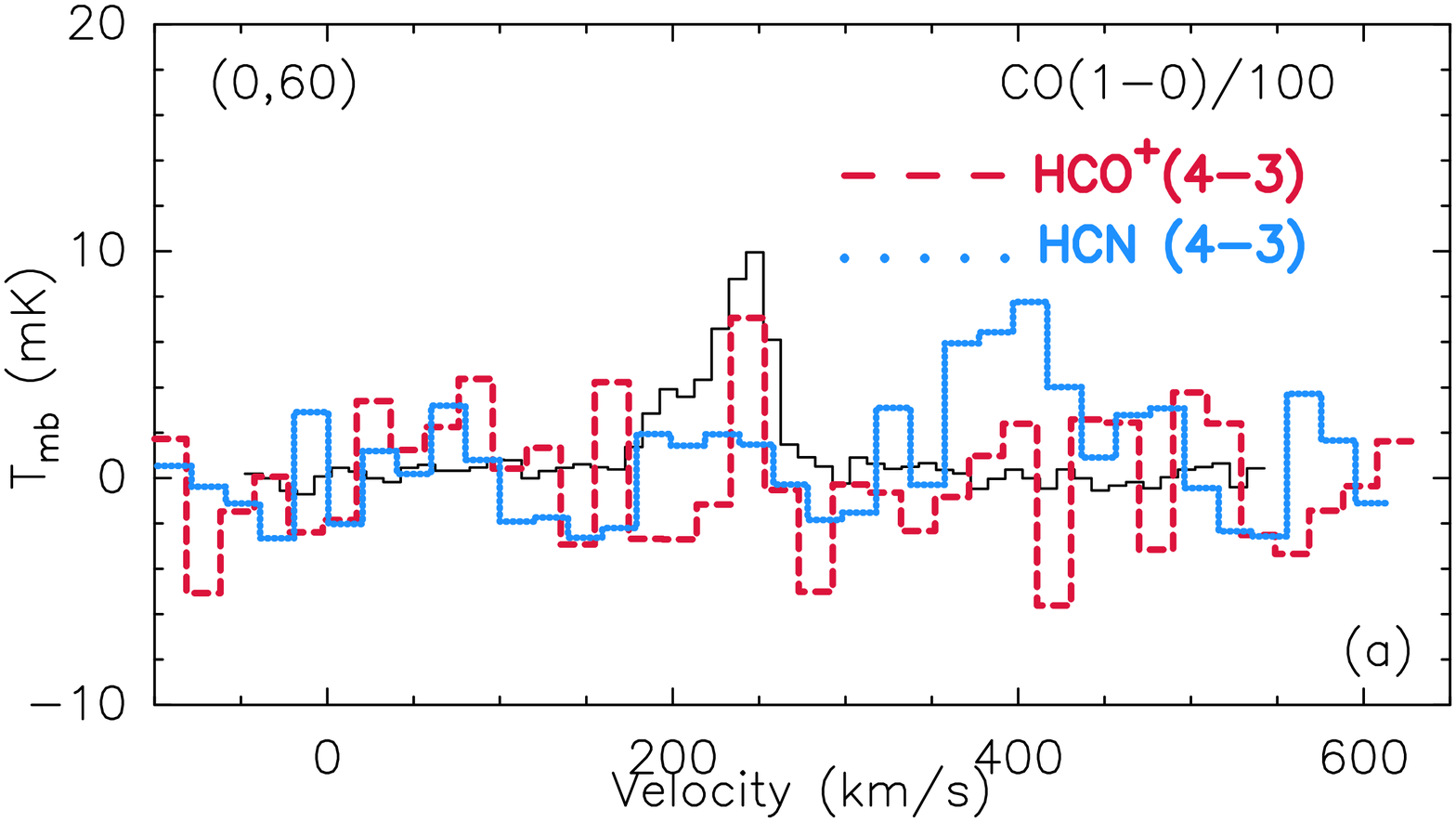}\\
\includegraphics[width=\linewidth, angle=0]{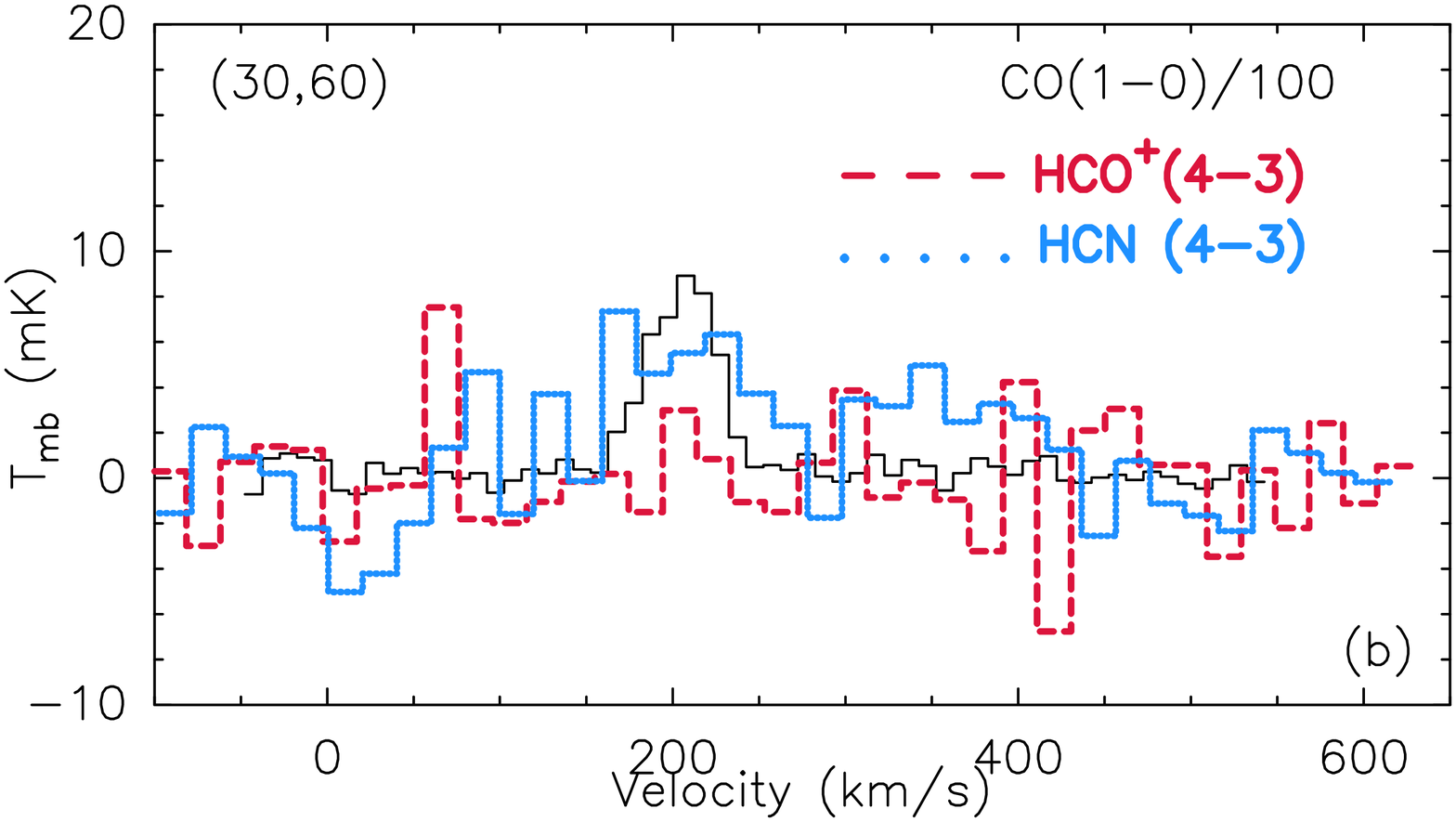}\\
\includegraphics[width=\linewidth, angle=0]{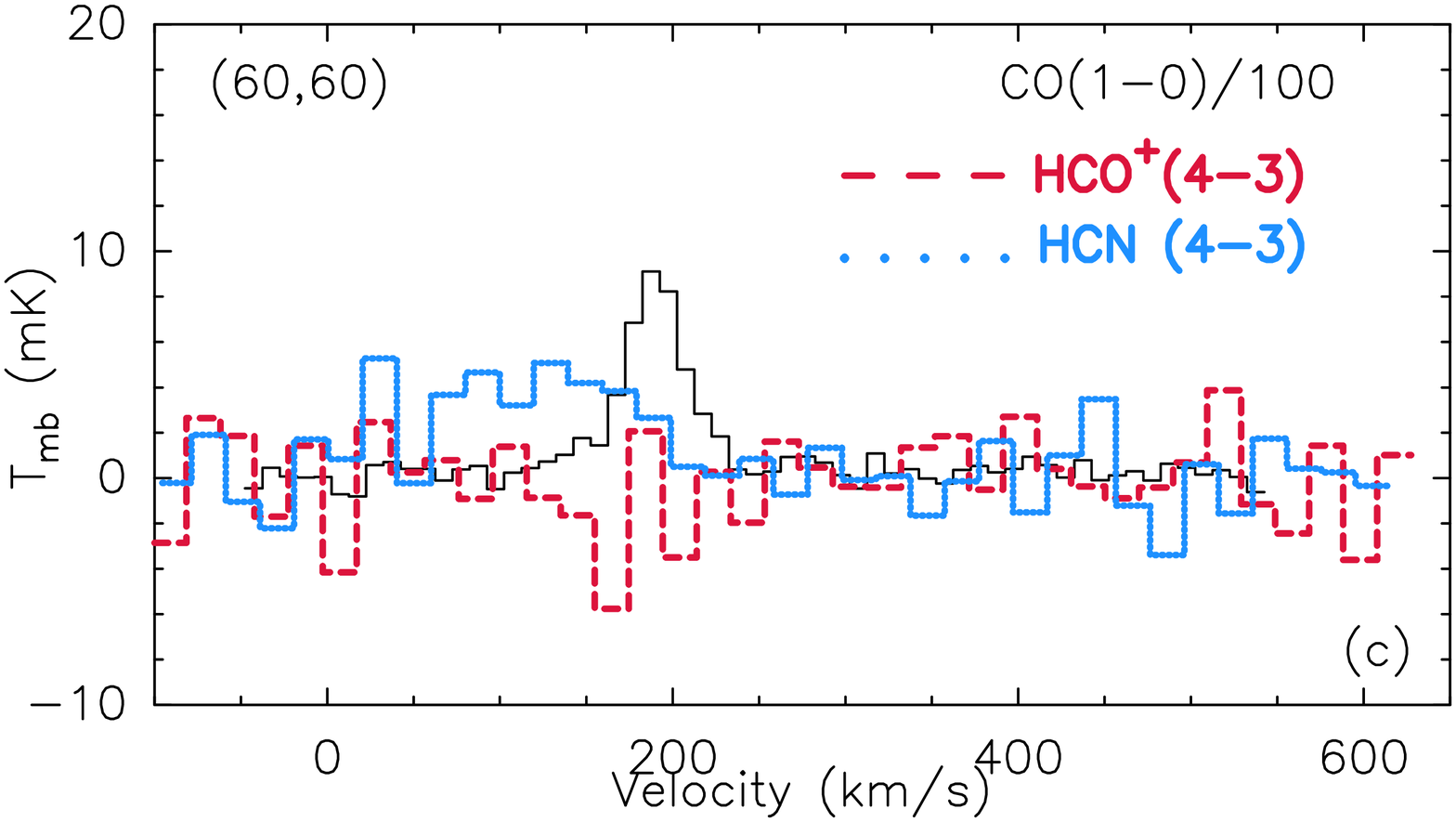}
\end{center}
\begin{center}
\caption{Examples of spectra obtained with stare-mode that are not shown in 
Fig.~\ref{fig:spec}.}
\label{fig:stare-spec} 
\end{center} 
\end{figure}

%---------------------- table 1 measurements -------------------------
% latex table generated in R 3.5.1 by xtable 1.8-3 package
% Tue Oct 29 23:02:16 2019
\begin{table*}
\caption{The full table is available online. This table lists the integrated intensities and line ratios. Rows are sorted according to the rows and then the columns of Figure 2. The last seven rows show data observed in JCMT-HARP's stare mode.
The errors inlude 10\% flux calibration uncertainty. The offset ($x,y$) is along the major and minor axes of NGC\,253, respectively. The last seven rows of the full table are positions observed only in
JCMT-HARP's stare mode. 
Part of their spectra are shown in Figure \ref{fig:spec}.
}
\label{tab:1}
\begin{tabular}{crrrr|cccc}
  \hline
offset & 
$I_{\rm HCN }$& 
$I_{\rm HCO^+}$ & 
$I_{\rm CO 1-0}$ & 
$I_{\rm CO 3-2}$ & 
\multirow{2}{*}{\large $\frac{\rm CO 3-2}{\rm CO 1-0}$} & 
\multirow{2}{*}{\large $\frac{\rm HCN 4-3}{\rm CO 1-0}$} & 
\multirow{2}{*}{\large $\frac{\rm HCO^+ 4-3}{\rm CO 1-0}$} & 
\multirow{2}{*}{\large $\frac{\rm HCN 4-3}{\rm HCO^+ 4-3}$} \\ 
\cline{2-5}
(arcsec) &
\multicolumn{4}{c}{(${\rm K\,km\,s^{-1}}$)}   \\
% \colnumbers
\hline 
  30,0 & 1.8$\pm$0.3 & 0.5$\pm$0.1 & 435$\pm$51 & 154.9$\pm$17.8 & 0.36$\pm$0.06 & 0.004$\pm$0.001 & 0.001$\pm$0.000 & 3.64$\pm$1.25 \\ 
  20,0 & 8.5$\pm$1.3 & 6.0$\pm$1.0 & 872$\pm$95 & 415.6$\pm$43.4 & 0.48$\pm$0.07 & 0.010$\pm$0.002 & 0.007$\pm$0.001 & 1.43$\pm$0.33 \\ 
  10,0 & 38.1$\pm$4.3 & 34.0$\pm$4.3 & 1100$\pm$118 & 883.6$\pm$90.9 & 0.80$\pm$0.12 & 0.035$\pm$0.005 & 0.031$\pm$0.005 & 1.12$\pm$0.19 \\ 
  0,0 & 60.1$\pm$6.4 & 74.6$\pm$8.3 & 1118$\pm$126 & 1001.2$\pm$103.2 & 0.90$\pm$0.14 & 0.054$\pm$0.008 & 0.067$\pm$0.011 & 0.80$\pm$0.12 \\ 
  -10,0 & 21.6$\pm$2.7 & 27.1$\pm$3.3 & 643$\pm$73 & 879.0$\pm$90.9 & 1.37$\pm$0.21 & 0.034$\pm$0.006 & 0.042$\pm$0.007 & 0.80$\pm$0.14 \\ 
  -20,0 & 9.0$\pm$1.4 & 10.3$\pm$1.5 & 502$\pm$57 & 371.3$\pm$38.8 & 0.74$\pm$0.11 & 0.018$\pm$0.003 & 0.020$\pm$0.004 & 0.88$\pm$0.19 \\ 
  -30,0 & 3.5$\pm$0.6 & 2.1$\pm$0.7 & 260$\pm$34 & 167.3$\pm$17.5 & 0.64$\pm$0.11 & 0.013$\pm$0.003 & 0.008$\pm$0.003 & 1.62$\pm$0.62 \\
\hline
%\multicolumn{9}{l}{ }\\
 \end{tabular}
\end{table*}

%------------------- table 2 physical parameters -----------------
% latex table generated in R 3.5.1 by xtable 1.8-3 package
% Tue Oct 29 23:18:48 2019
\begin{table*}
\caption{The full table is available online. This table lists the luminosities and dense-gas masses calculated based on Table \ref{tab:1}.
Rows are sorted according to the rows and then the columns of Figure \ref{fig:spec}. The last seven rows show data observed in JCMT-HARP's stare mode. The calibration uncertainty is included in the errors. For SFR the uncertainty on the conversion from \lir\ is not included. The last seven rows of the full table are positions observed only in JCMT-HARP's stare mode. Part of their spectra are shown in Figure \ref{fig:spec}. }
\label{tab:2}
\begin{tabular}{crrrrcccc}
\hline
offset & $L'_{\rm HCN}$ & $L'_{\rm HCO^+}$
& $L_{\rm IR}$ & SFR & $M_{\rm HCN}$
& $M_{\rm HCO^+}$ & $M_{\rm HCN}(G_0)$
& $M_{\rm HCO^+}(G_0)$ \\
(arcsec) &
\multicolumn{2}{c}{($10^4~{\rm K\,km\,s^{-1} pc^2}$)}  &
($10^7 L_{\sun}$) &
($10^{-3}\text{M}_{\sun} {\rm yr}^{-1}$) &
($10^6 \text{M}_{\sun} $) &
($10^6 \text{M}_{\sun} $) &
($10^6 \text{M}_{\sun} $) &
($10^6 \text{M}_{\sun} $) \\
\hline
  30,0 & 14.1$\pm$2.5 & 3.8$\pm$1.1 & 40.3$\pm$2.3 & 60$\pm$3.5 & 4.7$\pm$0.8 & 1.3$\pm$0.4 & 57.5$\pm$10.2 & 1.3$\pm$6.4 \\ 
  20,0 & 65.7$\pm$9.8 & 45.4$\pm$7.9 & 160.3$\pm$9.4 & 240$\pm$14.1 & 21.9$\pm$3.3 & 15.1$\pm$2.6 & 187.2$\pm$28.0 & 15.1$\pm$31.6 \\ 
  10,0 & 294.0$\pm$33.6 & 259.4$\pm$32.6 & 870.4$\pm$53.3 & 1306$\pm$79.9 & 98.0$\pm$11.2 & 86.5$\pm$10.9 & 523.4$\pm$60.4 & 86.5$\pm$81.4 \\ 
  0,0 & 463.8$\pm$49.7 & 569.1$\pm$63.5 & 1681.2$\pm$102.4 & 2522$\pm$153.5 & 154.6$\pm$16.6 & 189.7$\pm$21.2 & 706.9$\pm$76.6 & 189.7$\pm$135.7 \\ 
  -10,0 & 167.1$\pm$20.6 & 206.4$\pm$25.5 & 566.2$\pm$34.7 & 849$\pm$52.1 & 55.7$\pm$6.9 & 68.8$\pm$8.5 & 328.2$\pm$40.8 & 68.8$\pm$70.2 \\ 
  -20,0 & 69.7$\pm$10.5 & 78.4$\pm$11.7 & 64.0$\pm$3.7 & 96$\pm$5.6 & 23.2$\pm$3.5 & 26.1$\pm$3.9 & 258.4$\pm$39.3 & 26.1$\pm$61.1 \\ 
  -30,0 & 26.7$\pm$4.8 & 16.2$\pm$5.5 & 22.1$\pm$1.3 & 33$\pm$1.9 & 8.9$\pm$1.6 & 5.4$\pm$1.8 & 144.1$\pm$26.2 & 5.4$\pm$41.4 \\
\hline
%\multicolumn{9}{l}{}\\
\end{tabular}
\end{table*}

\section{Curve of growth and concentration index}
\begin{figure*}
%\begin{center}
\includegraphics[width=\textwidth]{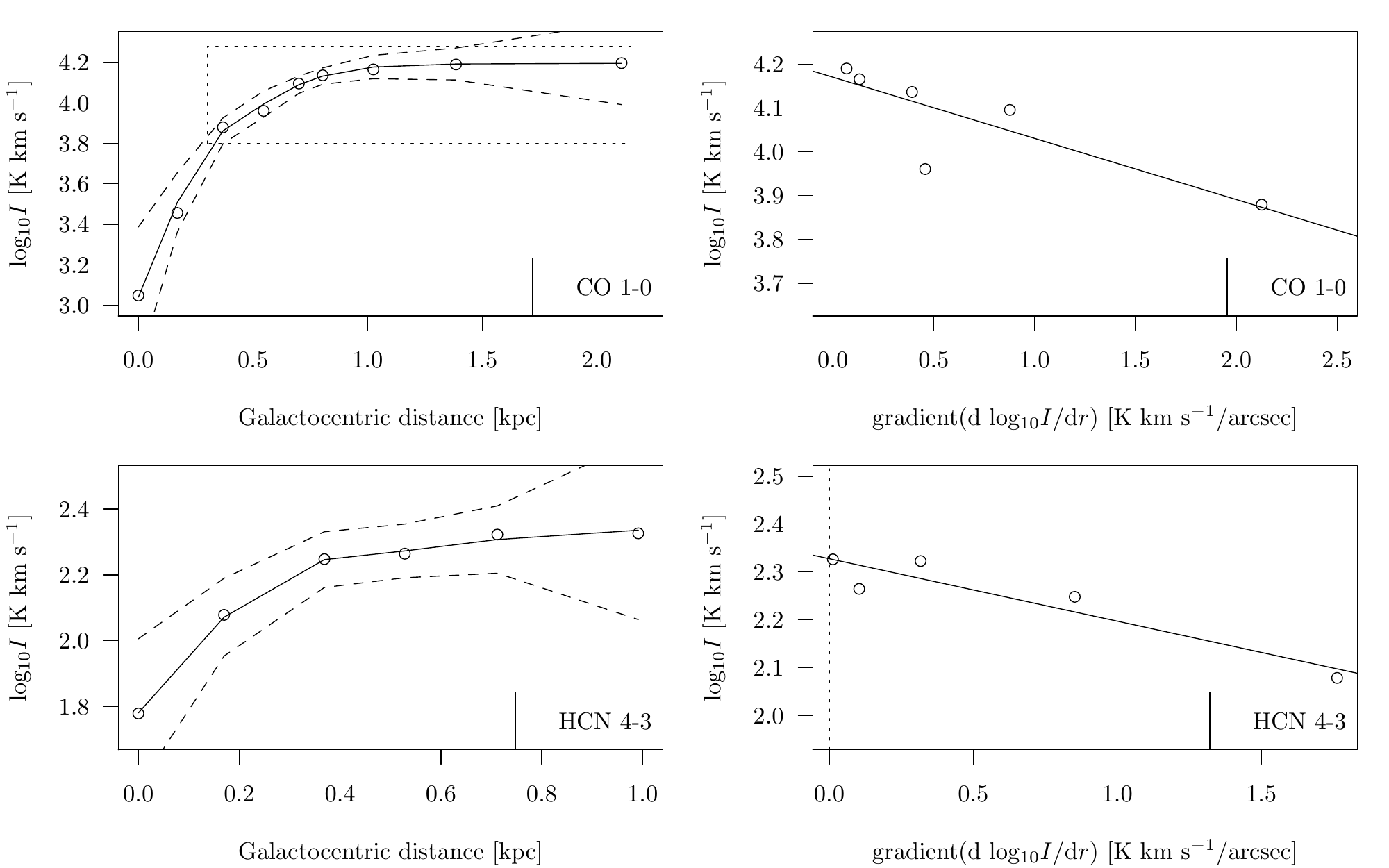}
\caption{Two examples of curve of growth (left column) and asymptotic intensities (right column). The solid and enclosing longer-dashed lines denote the fit and 95 per cent confidence interval. The dashed box in the top left panel indicates the seven data points 
that are used for the fits of the asymptotic intensity of CO 1-0. In the upper right panels the asymptotic intensity is derived from the inteception between the solid line (fit of the data) and the dashed line (zero gradient).
The asymptotic intensities are then used to derive the $r_{90}$ and $r_{50}$ of each tracer.
\label{fig:growth_curve}}
%\end{center} 
\end{figure*}

Fig.~\ref{fig:growth_curve} shows two examples (CO 1-0 and HCN 4-3) of how
we derive the curve of growth and the asymptotic intensities.
First we make weighted (by RMS) averaged intensities for those data points 
within every 0.17\,kpc bin along the inclination-corrected
radii, to obtain the smoothed radial profile. 
Second, the accumulated intensities inside each binned radius are calculated
for the curve of growth 
(in a logarithmic scale, left column in Fig.~\ref{fig:growth_curve}).
Third, we calculate the gradients of the accumulated intensities along the
curve of growth, d$m$/d$r$ (here $m = \text{log}_{10}$ $I$), and construct plots of
$m$ versus d$m$/d$r$ (right column in Fig.~\ref{fig:growth_curve}).
Finally on the plots we use linear fitting to get the intercept of $m$ at zero
gradient, i.e., the asymptotic intensity. Using the asymptotic intensity,
$r_{90}$ and $r_{50}$ can be fitted from the curve of growth.

\clearpage

\section{parameters used in \textsc{orac-dr}}

The \textsc{orac-dr} recipe used to reduce HCN 4-3 and \hcop\ 4-3 data of NGC\,253
is adjusted as following:
\begin{verbatim}
[REDUCE_SCIENCE_GRADIENT:NGC253]
BASELINE_REGIONS = -100:80,380:640
BASELINE_ORDER = 1
BASELINE_LINEARITY_LINEWIDTH = "80:380"
DESPIKE = 1
DESPIKE_BOX = 28
DESPIKE_CLIP = 3
DESPIKE_PER_DETECTOR = 1
HIGHFREQ_INTERFERENCE = 1
HIGHFREQ_INTERFERENCE_EDGE_CLIP = 3
HIGHFREQ_INTERFERENCE_THRESH_CLIP = 3
FREQUENCY_SMOOTH = 50	
\end{verbatim}
and the quality-assurance (QA) parameters applied in \textsc{orac-dr} are
as following:
\begin{verbatim}
[default]
BADPIX_MAP=0.1
TSYSBAD=1000
FLAGTSYSBAD=0.5
TSYSMAX=800
TSYSVAR=0.3
RMSVAR_RCP=0.5
RMSVAR_SPEC=0.2
RMSVAR_MAP=0.6
RMSTSYSTOL=0.15
RMSTSYSTOL_QUEST=0.15
RMSTSYSTOL_FAIL=0.2
RMSMEANTSYSTOL=1.0
CALPEAKTOL=0.2
CALINTTOL=0.2
RESTOL=1
RESTOL_SM=1
\end{verbatim}

%%%%%%%%%%%%%%%%%%%%%%%%%%%%%%%%%%%%%%%%%%%%%%%%%%

% Don't change these lines
\bsp	% typesetting comment
\label{lastpage}
\end{document}